\documentclass[twocolumn,
english,
aps,
pra,
10pt,
longbibliography,
superscriptaddress,
amsmath,
amssymb,
floatfix
]{revtex4-2}

\usepackage[utf8]{inputenc}
\usepackage[english]{babel}
\usepackage[T1]{fontenc}
\usepackage{amsmath}
\usepackage{amsfonts}
\usepackage{amssymb}
\usepackage{mathtools}
\usepackage{mathrsfs}
\usepackage{graphicx}
\usepackage{booktabs}
\usepackage[table]{xcolor}
\usepackage{multirow}
\usepackage{makecell}
\usepackage{siunitx}
\usepackage{rotating}
\usepackage{tabu}
\usepackage{braket}
\usepackage{placeins}
\usepackage{color}
\usepackage{listings}
\usepackage{pifont}
\usepackage{enumitem}
\usepackage{algorithmic}
\usepackage[colorlinks=true,urlcolor=blue,citecolor=blue,linkcolor=blue]{hyperref}

\lstdefinelanguage{json}{
    basicstyle=\ttfamily\small,
    numbers=left,
    numberstyle=\tiny,
    stepnumber=1,
    showstringspaces=false,
    breaklines=true,
    frame=single,
    literate=
     *{0}{{{\color{blue}0}}}{1}
      {1}{{{\color{blue}1}}}{1}
      {2}{{{\color{blue}2}}}{1}
      {3}{{{\color{blue}3}}}{1}
      {4}{{{\color{blue}4}}}{1}
      {5}{{{\color{blue}5}}}{1}
      {6}{{{\color{blue}6}}}{1}
      {7}{{{\color{blue}7}}}{1}
      {8}{{{\color{blue}8}}}{1}
      {9}{{{\color{blue}9}}}{1}
}

\begin{document}

\title{Setting angles in quantum approximate optimization at utility-scale}

\author{Maosheng Guo}
\altaffiliation{These authors contributed equally to this work.}
\affiliation{School of Mathematical and Statistical Sciences, Arizona State University, Tempe, Arizona 85287, USA}

\author{Joel Jurado Diaz}
\altaffiliation{These authors contributed equally to this work.}
\affiliation{School of Electrical, Computer, and Energy Engineering, Arizona State University, Tempe, Arizona 85287, USA}

\author{Anurag Ramesh}
\affiliation{Davidson School of Chemical Engineering, Purdue University, West Lafayette, Indiana 47906, USA}
\affiliation{USRA Research Institute for Advanced Computer Science (RIACS), CA, USA}

\author{Conrad J. Haupt}
\affiliation{IBM Quantum, IBM Research Europe - Zurich, Rüschlikon 8803, Switzerland}
\affiliation{Laboratory of Theoretical Physics of Nanosystems,
{\'E}cole Polytechnique F{\'e}d{\'e}rale de Lausanne, 1015 Lausanne,
Switzerland}

\author{Alberto Baiardi}
\affiliation{IBM Quantum, IBM Research Europe - Zurich, Rüschlikon 8803, Switzerland}

\author{Dimitrios Athanasakos}
\affiliation{Quantum Technologies Group, HSBC, Singapore, 117439, Singapore}

\author{M. Emre Sahin}
\affiliation{The Hartree Centre, STFC, Sci-Tech Daresbury, Warrington WA4 4AD, UK}

\author{Oscar Wallis}
\affiliation{The Hartree Centre, STFC, Sci-Tech Daresbury, Warrington WA4 4AD, UK}

\author{George Pennington}
\affiliation{The Hartree Centre, STFC, Sci-Tech Daresbury, Warrington WA4 4AD, UK}

\author{Christian Arenz}
\affiliation{School of Electrical, Computer, and Energy Engineering, Arizona State University, Tempe, Arizona 85287, USA}

\author{Sebastian Brandhofer}
\affiliation{IBM Quantum, IBM Research Europe - Ehningen, Ehningen 71139, Germany}

\author{Georgios Korpas}
\affiliation{Quantum Technologies Group, HSBC, Singapore, 117439, Singapore}
\affiliation{Czech Technical University in Prague, Prague 12135, The Czech Republic}

\author{Ieva \v{C}epait\.{e}}
\affiliation{Phasecraft Ltd, London, United Kingdom}

\author{J. A. Monta\~nez-Barrera}
\affiliation{J\"ulich Supercomputing Centre,
  Forschungszentrum J\"ulich, 52425 J\"ulich, Germany}

  \author{Jakub Marecek}
\affiliation{Czech Technical University in Prague, Prague 12135, The Czech Republic}
  
\author{Davide Venturelli}
\affiliation{USRA Research Institute for Advanced Computer Science (RIACS), CA, USA}

\author{Stephan Eidenbenz}
\affiliation{Los Alamos National Laboratory, Los Alamos, New Mexico 87544, USA}

\author{David E. Bernal Neira}
\affiliation{Davidson School of Chemical Engineering, Purdue University, West Lafayette, Indiana 47906, USA}

\author{Daniel J. Egger}
\altaffiliation{deg@zurich.ibm.com}
\affiliation{IBM Quantum, IBM Research Europe - Zurich, Rüschlikon 8803, Switzerland}

\begin{abstract}
The quantum approximate optimization algorithm (QAOA) is a powerful heuristic that seeks to solve combinatorial optimization problems using quantum hardware and classical optimization in tandem. 
Various methods exist to train the parameterized quantum circuits that serve as an ansatz in QAOA. 
However, which method works best to identify optimal angles for a given problem instance remains poorly understood, especially at utility-scale, i.e., $100$ qubits or more.
In this work, we address this challenge through utility-scale benchmarks from which we distill operational guidance for QAOA practitioners. 
First, we investigate approximation techniques, such as matrix product states and Pauli propagation, to find optimal angles.
Second, we train QAOA on small-scale representative problems and transfer the angles to larger ones. 
We then validate the results on quantum hardware for utility-scale problem instances that can be meaningfully executed. 
In this way, we identify insights for QAOA angle setting strategies that work best for problems at the utility scale, including as a function of resource cost for the search.
Crucially, the operational implications we draw from our benchmarks will help quantum optimization practitioners execute QAOA end-to-end pipelines efficiently on current and future hardware.
\end{abstract}

\maketitle

\section{Introduction}\label{sec:intro}

Quantum computers can tackle combinatorial optimization problems~\cite{Abbas2024}. As is the common for many application domains in classical optimization,   instances of combinatorial optimization are often  solved with heuristic algorithms on quantum computers, i.e., algorithms that do not have performance guarantees, but work well in practice.
The quantum approximate optimization algorithm (QAOA) is one such heuristic~\cite{Farhi2014}, with known performance guarantees in certain cases~\cite{Farhi2014, Wurtz2021a}.
To produce good solutions $x$ to a minimization problem $\min_x f(x)$, QAOA first maps $f(x)$ to a cost Hamiltonian $H_C$ with a ground state consisting of computational basis states that minimize $f$.
The ground state of $H_C$ is typically unknown or complex to produce. 
Therefore, QAOA creates a variational quantum circuit $\ket{\psi(\boldsymbol{\theta})}$ built from $p$ alternating layers of operators generated by $H_C$ and a mixer $H_M$ with an easy-to-prepare and known ground state.
QAOA angles $\boldsymbol{\theta}$ are optimized so that samples drawn from $\ket{\psi(\boldsymbol{\theta}^\star)}$ are candidate solutions $x$ to minimize $f(x)$.
Typically, optimal angles $\boldsymbol{\theta}^\star$ are found through a closed-loop optimization with quantum hardware to minimize $\langle H_C\rangle=\langle\psi(\boldsymbol{\theta})|H_C|\psi(\boldsymbol{\theta})\rangle$.

In practice, the closed-loop optimization of $\boldsymbol{\theta}$ is challenging.
First, the limited clock speed and availability of current quantum hardware make the optimization of $\boldsymbol{\theta}$ time-consuming and costly.
This impedes progress towards quantum advantage in optimization.
Second, at the quantum utility-scale, i.e., problems described by $\sim100$ or more qubits, the underlying quantum circuits cannot be simulated exactly on classical hardware.
Therefore, we cannot rely on an exact classical simulation of $\langle H_C\rangle$ to find $\boldsymbol{\theta}^\star$.
Third, the optimization landscape is non-trivial, making the optimization of $\boldsymbol{\theta}$ formally NP-hard~\cite{Bittel_2021}.
Even optimizing depth-one QAOA circuits, defined by only two angles, can be challenging due to local optima~\cite{Vijendran2025}.
As a result, many different angle setting strategies have been developed for QAOA to find~$\boldsymbol{\theta}^\star$.

Our work aims to create guidance for QAOA angle setting at utility-scale without a closed-loop with the quantum computer.
To do so, we first review the complexity aspects of finding $\boldsymbol{\theta}^\star$ in Sec.~\ref{sec:complexity}.
Second, we review existing QAOA angle setting methods and provide a taxonomy in Sec.~\ref{sec:back}.
This section also discusses strategies for evaluating $\langle H_C \rangle$. 
Next, we select and benchmark a set of QAOA angle setting methods focusing on QAOA depths $p\leq 10$.
We focus on small $p$, since noise limits the depth of the circuits that can be meaningfully executed today and in the near future.
This depth limitation is especially acute when the topology of the cost Hamiltonian is much more complex than the physical qubit couplings at hand.
The methodology of this benchmarking is presented in Sec.~\ref{sec:bench_method}.
The results, which include hardware runs on up to 144 qubits and depth-ten QAOA are presented in Sec.~\ref{sec:res}.
Based on this experience, we distill a set of operational implications for today's quantum approximate optimization practitioners in Sec.~\ref{sec:best}.
Practitioners in a hurry may jump directly to this section.
We conclude in Sec.~\ref{sec:conc}.

\begin{figure*}
    \centering
    \includegraphics[width=\textwidth, clip, trim=0 230 0 0]{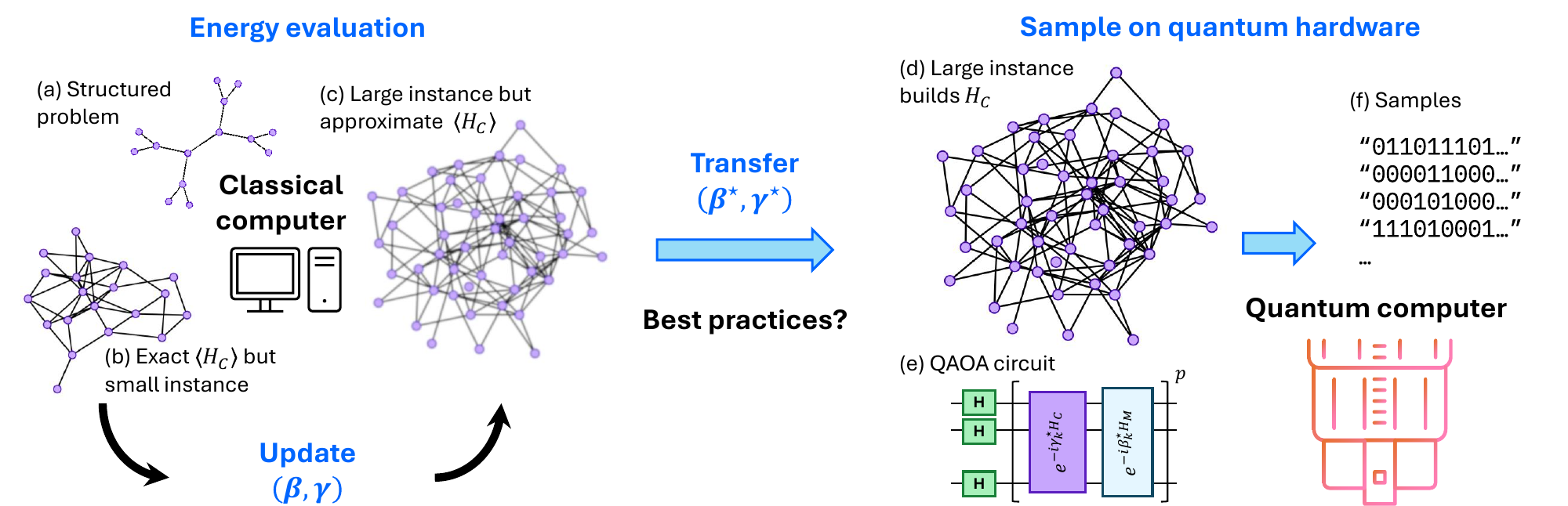}
    \caption{
    Utility-scale quantum approximate optimization.
    Good QAOA angles $\boldsymbol{\beta}^\star$ and $\boldsymbol{\gamma}^\star$ are found with classical computers.
    The angles may be optimized with a classical optimizer, and the energy evaluated on (a) representative structured problems~\cite{Wurtz2021a}, (b) small-scale representative problem instances with an exact energy evaluation, and (c) the utility-scale problem instance, but with an approximate energy evaluation.
    The utility-scale problem instance (d) informs the QAOA circuit (e) with the near-optimal angles $\boldsymbol{\beta}^\star$ and $\boldsymbol{\gamma}^\star$.
    (f) Only the sampling of good candidate solutions $x$, shown as bitstrings, is done on the quantum hardware.
    }
    \label{fig:strategies}
\end{figure*}

\section{Complexity Aspects of finding QAOA angles\label{sec:complexity}}
\subsection{From the complexity of MaxCut to the complexity of QAOA angles}

QAOA is most often studied with MaxCut as a benchmark, both because MaxCut admits a clean encoding as a two-local Ising Hamiltonian $H_C$, and because, from a theoretical computer science standpoint, MaxCut occupies a distinguished position: it is one of the few NP-hard problems for which polynomial-time approximability is essentially tight.
The Goemans--Williamson algorithm~\cite{goemans1995improved} rounds a semidefinite-programming (SDP) relaxation to produce a cut whose expected weight is at least
\begin{equation}
\alpha_{\rm GW}\;=\;\min_{\theta\in(0,\pi]}\frac{2}{\pi}\,\frac{\theta}{1-\cos\theta}\;\approx\;0.87856
\label{GWratio}
\end{equation}
times the optimum.
Conditional on the Unique Games Conjecture \cite{khot2010unique}, $\alpha_{\rm GW}$ is tight: no polynomial-time algorithm can approximate MaxCut to ratio $\alpha_{\rm GW}+\varepsilon$ for any $\varepsilon>0$.
Unconditionally, MaxCut admits no polynomial-time $(16/17+\varepsilon)$-approximation, $16/17\approx 0.9412$, unless $\textbf{P}=\textbf{NP}$.
If a quantum algorithm improved upon this threshold, it would solve an NP-Hard problem, which is considered unlikely within theoretical computer science.

These inapproximability results translate to the angle setting problem. 
The depth-$p$ QAOA objective
\begin{equation}
F_p(\boldsymbol{\theta})\;=\;\langle\psi(\boldsymbol{\theta})|H_C|\psi(\boldsymbol{\theta})\rangle,\qquad \boldsymbol{\theta}\in\mathbb{R}^{2p},
\label{eq:qaoa_objective}
\end{equation}
is a trigonometric polynomial in $\boldsymbol{\theta}$, cf.~\cite{sakos2026global},  whose global minimum $F_p^\star$ lower-bounds the true ground-state energy $E_0$. 
Suppose an angle-setting procedure, run in polynomial time, returned angles $\hat{\boldsymbol{\theta}}$ such that $F_p(\hat{\boldsymbol{\theta}})\leq F_p^\star+\Delta$.
Candidate solutions $x$ sampled from $\ket{\psi(\hat{\boldsymbol{\theta}})}$ would achieve an expected approximation ratio governed by $\Delta$ and by the intrinsic QAOA gap $E_0-F_p^\star$.
A polynomial-time algorithm that drove $\Delta$ to zero on all two-local Ising Hamiltonians $H_C$ would therefore approximate MaxCut, plausibly contradicting the inapproximability statements.
Thus, unless $\textbf{P}=\textbf{NP}$, one should not expect that a classical  polynomial-time algorithm can locate $\boldsymbol{\theta}^\star$ to arbitrarily small additive error on all instances of MaxCut, although there are regimes \cite{sakos2026global} when this is possible.

Bittel and Kliesch~\cite{Bittel_2021} make this observation rigorous and substantially strengthen it.
By reducing MaxCut to the classical optimization component of a generic variational quantum algorithm (VQA), they prove that the angle-setting problem is NP-hard, and that this hardness is robust: it persists when the ansatz uses only logarithmically many qubits, when it consists of a single layer, when it is built from free-fermionic gates, and when time evolution is taken to be continuous rather than digitized.
Specifically, assuming $\textbf{P}\neq\textbf{NP}$, no polynomial-time algorithm can guarantee an additive optimization error $\Delta < 1$ for every instance, and  $\Delta\geq(1-\alpha_{\max})/2$ for QAOA, where $\alpha_{\max}$ is the best polynomial-time approximation ratio attainable for the underlying combinatorial problem.
Importantly, the proof structure shows that the optimization landscape is riddled with persistent local minima whose values are bounded away from $F_p^\star$, where  gradient-based methods are guaranteed to get trapped, unless one considers some form of restarting~\cite{11389907} or noise injection~\cite{kungurtsev2024iteration}. 

The results of Ref.~\cite{Bittel_2021} have since been sharpened in two ways.
First, Bittel \emph{et al.}~\cite{bittelQCMA} prove that deciding the minimal depth $p$ required for a VQA -- and hence for QAOA -- to reach a prescribed target expectation value is $\textbf{QCMA}$-hard, ruling out efficient classical procedures for the depth-allocation problem under standard quantum complexity assumptions.
Second, Korpas \emph{et al.}~\cite{korpas2025undecidable} extends the NP-hardness of Bittel and Kliesch into outright undecidability in an idealized, noiseless setting.
The reduction maps the decision version of digitized VQA training to the solvability of a universal Diophantine equation in $58$ variables of degree $4$, thereby establishing equivalence to the halting problem.
The encoding embeds the Diophantine system into the structure of a variational circuit and aligns the cost function with a sum-of-squares polynomial.
Under mild assumptions, the construction implies that no finite-time algorithm can decide whether QAOA of depth $58$ achieves a target energy, and consequently that both the decision and the optimization versions of the digitized QAOA training problem are uncomputable in this regime.

\subsection{Structural obstructions: symmetry, locality, and pointwise dominance}

A complementary line of work exhibits obstructions that are intrinsic to the QAOA \emph{ansatz} at constant depth and even at depth $o(\log n)$---irrespective of how the angles are chosen.
In other words, even an oracle that hands the practitioner the global optimum $\boldsymbol{\theta}^\star$ does not rescue shallow QAOA on important problem families.
We organise these results by mechanism: symmetry protection, locality, pointwise dominance by classical algorithms, sub-logarithmic depth limitations from the overlap-gap property, and intrinsic degree dependence.
It will be useful to recall that MaxCut, Maximum Independent Set (MIS), and MAX-$k$-XOR are all instances of (valued) constraint satisfaction problems (CSPs), in which a finite arity-bounded predicate is imposed on each constraint and the objective is to maximize the total constraint weight; results stated below at the level of bounded-arity CSPs therefore apply uniformly to all three.

Bravyi \emph{et al.} ~\cite{Bravyi2020} observe that the QAOA state inherits any symmetry shared by the cost Hamiltonian $H_C$ and the mixer $H_M$.
For canonical QAOA with the transverse-field mixer $H_M=\sum_j X_j$ acting on $\ket{+}^{\otimes n}$, the state $\ket{\psi(\boldsymbol{\theta})}$ is invariant under the global $\mathbb{Z}_2$ spin-flip $\prod_j X_j$.
On any graph, this symmetry forces $\langle Z_i\rangle=0$ at every site, and constrains higher correlators.
For instances of MaxCut on bipartite, $D$-regular graphs (and more generally on triangle-free $D$-regular graphs), the authors leverage this symmetry, together with the locality of constant-depth QAOA, to prove that for any fixed $p$ the approximation ratio achieved by QAOA is bounded away from the optimum by a margin that does \emph{not} shrink as $D$ or $n$ grow.
Specifically, on these instances, the Goemans--Williamson algorithm provably outperforms QAOA at any constant level $p$.
The classical $\alpha_{\rm GW}\approx 0.878$ guarantee is unmatched by any constant-depth QAOA schedule, no matter how the angles are tuned.

Bravyi \emph{et al.}~\cite{BravyiKlieschKoenigTang2022} refine and extends this picture to the more general MAX-$k$-CUT (approximate graph colouring) problem.
The mechanism is a light-cone argument: the depth-$p$ operator $\prod_{k=1}^{p}U_M(\beta_k)U_C(\gamma_k)$ propagates information across at most $p$ edges of the interaction graph, and so the reduced density matrix on any vertex is determined by its $p$-neighbourhood alone.
On graphs whose local structure is essentially tree-like -- as in random regular and high-girth families -- this neighbourhood is statistically homogeneous across sites, and the output of constant-depth QAOA behaves as a stochastic local algorithm in the sense of Hatami \emph{et al.}~\cite{Hatami2014}.
Such local algorithms are provably unable to break the symmetry of certain instances, and the paper shows that the standard, non-recursive QAOA fails to solve approximate graph colouring on most regular bipartite graphs at any constant $p$. 
Thus, even if angle-setting were tractable, constant-depth QAOA would remain structurally inadequate on important classes of MaxCut and MAX-$k$-CUT instances.
Barak and Marwaha~\cite{BarakMarwaha2022} sharpen the picture further by reframing the comparison in instance-by-instance terms rather than in terms of worst-case ratios.

Chou \emph{et al.}~\cite{ChouLoveSandhuShi2021} extend locality-based obstructions from \emph{constant} depth to $\Omega(\log n)$ depth by reasoning about random graphs with the overlap-gap property \cite{Mezard2005,Achlioptas2011,Gamarnik2014,Gamarnik2019,Gamarnik2021}, where the intersection of any nearly optimal independent sets is either large or small.
Consider a sparse random $D$-regular graph. If $2p$ is small relative to $\log n/\log D$, distant measurement outcomes are nearly independent, and the graphs have \cite{Gamarnik2014,Gamarnik2019,Gamarnik2021} the overlap-gap property (OGP). 
Refs.~\cite{Gamarnik2014,Gamarnik2019,Gamarnik2021} consider a class of \emph{generic local algorithms} for bounded-arity valued-CSP and show that these fail to achieve an arbitrarily good approximation, with probability tending to one, on those random valued CSPs whose solution geometry exhibits the OGP.
Since MaxCut, MIS, and MAX-$k$-XOR all fit into the bounded-arity valued-CSP framework, the OGP-based obstruction is fundamentally a statement about CSPs, not only about MaxCut.
Chou \emph{et al.}~\cite{ChouLoveSandhuShi2021} then show that 
QAOA at depth $p<\varepsilon\log n$ falls into this class for an absolute constant $\varepsilon>0$, since its light cone at site $i$ touches at most $D^p=n^{O(\varepsilon)}$ variables on bounded $D$-degree instances. 
More recently, Chen \emph{et al.}~\cite{ChenHuangMarwaha2023} generalized the results to \emph{all} quantum circuits of depth up to $\varepsilon\log n$ for bounded-degree instances of valued CSPs.
This holds for the \emph{best possible} choice of angles, and so cannot be repaired by improving the angle-setting procedure.

\subsection{Sidestepping structural obstructions}

The obstructions above are not generic to all variants of QAOA.
They apply to QAOA in its original form and to procedures that train angles on a per-instance basis.
At least two modifications sidestep the symmetry and locality  barriers in distinct ways, and may inherit the approximation guarantees of best-in-class classical algorithms.
Specifically, the pointwise-dominance conjecture \cite{BarakMarwaha2022} would not apply.

Sakos \emph{et al.}~\cite{sakos2026global} provide a partial mitigation strategy. 
They identify a regime, where the angle-setting problem becomes tractable, in the sense of a fully polynomial randomized approximation scheme (FPRAS). 
For every $\varepsilon>0$, a polynomial-time algorithm returns an $\varepsilon$-approximate global optimum of $F_p$ with high probability, in runtime and query complexity polynomial in $1/\varepsilon$ and in the number of qubits. 
However, this regime is limited to poly-depth parameterized quantum circuits with a constant number of angles. This regime contains fixed-depth QAOA, breaks the symmetry barrier, but does not break  the locality barrier. 

Recursive QAOA \cite{Bravyi2020,BravyiKlieschKoenigTang2022} remedies the symmetry and locality obstructions.
At each round, one runs a low-depth QAOA on the current instance, measures all two-point correlators $M_{ij}=\langle Z_iZ_j\rangle_{\boldsymbol{\theta}^\star}$ for edges $(i,j)\in E$, selects the pair $(i^\star,j^\star)$ with the largest $|M_{ij}|$, and \emph{eliminates} a variable.
The reduced instance has one fewer variable; the procedure recurses until the remaining problem is small enough to be solved exactly.
Because each elimination step is informed by global measurements on the full system, RQAOA is no longer a constant-depth local algorithm \cite{Hatami2014}, and the locality lower bounds of Refs.~\cite{BravyiKlieschKoenigTang2022, BarakMarwaha2022} do not apply.
Numerically, RQAOA matches or exceeds the performance of Goemans--Williamson on some ensembles, where standard QAOA provably fails.

Warm-starting~\cite{Egger2021warmstartingquantum} replaces the symmetric initial state $\ket{+}^{\otimes n}$ by one biased toward the solution of a classical relaxation of the problem.
For MaxCut, one first solves the Goemans--Williamson SDP, obtains relaxed values $x_i^{\rm rel}\in[0,1]$, and prepares the product state
\begin{equation}
\ket{\psi_0}\;=\;\bigotimes_{i=1}^{n}\bigl(\sqrt{1-x_i^{\rm rel}}\,\ket{0}+\sqrt{x_i^{\rm rel}}\,\ket{1}\bigr).
\label{eq:psi0}
\end{equation}
Thanks to an alternative mixer, warm-started QAOA then reproduces the relaxed solution $x_i^{\rm rel}$ at zero variational angles. 
Subsequent steps inherit the classical $\alpha_{\rm GW}$ guarantees.
Any $\hat{\boldsymbol{\theta}}$, including the trivial choice $\hat{\boldsymbol{\theta}}=0$, returns a candidate solution whose expected ratio is at least $\alpha_{\rm GW}$.
Alternative warm-starting strategies have been studied by \cite{tate2023warm,truger2024warm,tate2025warm}.
Crucially, warm-starting breaks both the global $\mathbb{Z}_2$ symmetry~\cite{Bravyi2020} 
and the locality bottleneck of~\cite{BravyiKlieschKoenigTang2022, BarakMarwaha2022}, because $\ket{\psi_0}$ already encodes the global problem structure derived from a non-local SDP.

The complexity-theoretic landscape sketched above motivates our study. 
The objective function value, be it $\langle H_C\rangle$ or $f(x)$, as a function of the angles features many local minima -- even at depth one and for problem instances on logarithmically many qubits.
This makes finding $\boldsymbol{\theta}^\star$ hard.
Since the exact angle setting is computationally hard, any utility-scale workflow utilizing QAOA must rely on heuristics.
There, empirical guidance, distilled from systematic benchmarks rather than asymptotic worst-case bounds, is indispensable.
The remainder of this paper focuses on the empirical guidance for standard QAOA~\cite{Farhi2014}.
The methodology and guidance we distill also helps inform angle setting methods for QAOA variant that overcome symmetry and locality obstructions, such as warm-start and R-QAOA.


\section{Methods to find QAOA angles\label{sec:back}}

We consider algorithms that prepare a state and sample from it to minimize a classical objective function $f(x)\in \mathbb{R}$ with $x\in\{0,1\}^n$. 
The canonical QAOA~\cite{Farhi2014} creates a state 
\begin{align}
    \ket{\psi(\boldsymbol{\theta})}=\prod_{k=1}^pU_M(\beta_k)U_C(\gamma_k)\ket{+}^{\otimes n}
\end{align}
by alternating $p$ layers of the cost-function unitary $U_C(\gamma) = e^{-i\gamma H_C}$ with the transverse-field mixing unitary $U_M(\beta) = \prod_j e^{-i\beta X_j}$.
Setting the QAOA angles $\boldsymbol{\theta}=(\beta_1, ..., \beta_p, \gamma_1, ...,\gamma_p)$ at the utility scale can be done in multiple ways.
First, certain methods infer the angles from the problem structure, see Fig.~\ref{fig:strategies}(a), or draw on physics-inspired QAOA schedules.
Second, one may train the angles on exact simulations of small-scale problems representative of the utility-scale problem, see Fig.~\ref{fig:strategies}(b).
Finally, the energy evaluation in the training may only approximate the underlying quantum circuits of the utility-scale problem, for instance, with matrix product states (MPS)~\cite{Schollwoeck2011_DMRG, Stoudenmire2020_Limits-QC} or Pauli propagation (PP)~\cite{rudolph2025pauli, rall2019simulation}, see Fig.~\ref{fig:strategies}(c).
The resulting angles are then transferred to the utility-scale problem.
Crucially, when such methods are employed, the candidate solutions $x$ to $\min_x f(x)$ cannot be generated during the optimization of $\boldsymbol{\theta}^\star$.
The candidate solutions are thus only produced by the final sampling executed on the quantum computer.

\subsection{Survey and Taxonomy\label{sec:overview}}

We now survey QAOA angle setting methods known to us and classify them into a taxonomy with four main classes, namely, \emph{physics inspired}, \emph{parameter transfer}, \emph{iterative}, and \emph{machine learning}.
A given method may feature in multiple classes within this taxonomy, see Tab.~\ref{tab:qaoa_taxonomy}.

\begin{table*}
\caption{Taxonomy of QAOA angle setting methods. A method may belong to multiple classes of the taxonomy. 
A thick green tick-mark {\color{green!50!black}\ding{52}} highlights the main idea behind the method while a thin black tickmark \ding{51} indicates the other classes in the taxonomy that the method relies upon.
\emph{Dimension reduction} methods attempt to optimize the QAOA angles in a smaller search space than the standard $2p$ space.
The \emph{Max size} column indicates the problem class considered and the largest number of qubits in parentheses.
An asterisk * indicates that multiple problem classes are considered.
\emph{Database required} methods need good QAOA angles pre-trained on other instances.
Methods that evaluate $\langle H_C\rangle$ are flagged in the last column.
\label{tab:qaoa_taxonomy}
}
\centering
\begin{tabular}{l l r r r r r r r r}\hline\hline
Name & Refs. & \shortstack{Physics \\ Inspired} & \shortstack{Parameter \\ Transfer} & \shortstack{Iterative\\ $p\to p+1$} & \shortstack{Machine \\ Learning} & \shortstack{Dimension\\reduction} & \shortstack{Max \\ size} & \shortstack{Database \\ required} & \shortstack{$\langle H_C\rangle$ \\ needed} \\ \hline
TQA & \cite{ Willsch2022,sack2021quantum} & {\color{green!50!black}\ding{52}} & & & & & MC(12) & & (\ding{51})\\
Linear ramps & \cite{Montanezbarrera2024} & {\color{green!50!black}\ding{52}} & & & & & *(109) & & (\ding{51})\\
SGIR-QAOA & \cite{Mcdowall2026} & {\color{green!50!black}\ding{52}} & & & & & MIS(30) & & (\ding{51})\\
Smooth angles & \cite{boulebnane2025}  & {\color{green!50!black}\ding{52}} & & & & & SK(20) & \ding{51} \\ 
CD-QAOA & \cite{WurtzLove2022CDQAOA} & {\color{green!50!black}\ding{52}} & \ding{51} & \ding{51} & & & MC(14) \\
FALQON & \cite{Magann2022FALQON} & {\color{green!50!black}\ding{52}} & & & & & MC(20) & & \\ \hline
Parameter concentration & \cite{Akshay2021, brandao2018, Shaydulin2019, Galda2021, Montanezbarrera2024Transfer, pelofske2026evaluatinglimitsqaoaparameter, Pelofske_2024_QAOA_scaling, Barron2024, Kotil_2025} & & {\color{green!50!black}\ding{52}} & & & & Ising(156) & \ding{51}\\
Fixed angles & \cite{Wurtz2021a, Wurtz2021} & & {\color{green!50!black}\ding{52}} & & & & MC(16) & \ding{51} \\
Weighted transfer & \cite{Shaydulin2023, Sureshbabu2024, cepaite2025quantum} & & {\color{green!50!black}\ding{52}} & & & & MC(20) & & \ding{51}\\
Fuzzy clustering & \cite{Acampora2023} & & {\color{green!50!black}\ding{52}} & & (\ding{51}) & & MC(14) & \ding{51} \\
QAOA-PCA & \cite{Parry2025} & & {\color{green!50!black}\ding{52}} & & (\ding{51}) & \ding{51} & MC(8) & \ding{51} \\
Layer selective & \cite{venturelli2024transfer} & & {\color{green!50!black}\ding{52}} & & & \ding{51} & MC(18) & \ding{51} & \ding{51} \\ 
Proxy-QAOA & \cite{Sud2024} & & {\color{green!50!black}\ding{52}} & & & & MC(20) & & \\
DARBO & \cite{Cheng2024} & & {\color{green!50!black}\ding{52}} & & & & MC(16) & & \ding{51} \\
Radial basis & \cite{Oleary2025} & & {\color{green!50!black}\ding{52}} & & & & Ising(127) & & \ding{51} \\ \hline
Layer-QAOA & \cite{Campos2021} & & & {\color{green!50!black}\ding{52}} & & \ding{51} & -(10) & & \ding{51}\\
Interp./Fourier & \cite{Zhou2020} & & & {\color{green!50!black}\ding{52}} & & & MC(22) & & \ding{51}\\
Iterative Interpolation & \cite{apte2025}  & & & {\color{green!50!black}\ding{52}} & & \ding{51} & SK(28) & & \ding{51}\\
Transition States & \cite{Sack2023, Medina2024} & & & {\color{green!50!black}\ding{52}} & & & MC(22) & & \ding{51}\\ 
JuliQAOA & \cite{Golden_2023, Golden2023b} & & & {\color{green!50!black}\ding{52}} & \ding{51} & & *(18) & & \ding{51}\\ 
Iter. SSR-MCTS & \cite{Agirre2025} & & & {\color{green!50!black}\ding{52}} & \ding{51} & & 3-SAT(11) & & \ding{51}\\ \hline
QAOA-GPT & \cite{Tyagin2025}  & & & & {\color{green!50!black}\ding{52}} & & MC(14)  & \ding{51} & \\
Clustering & \cite{Acampora2023, Moussa2022}& & \ding{51} & & {\color{green!50!black}\ding{52}} & & MC(18) & \ding{51} & \\
Iterative ML & \cite{Xie2023} & & & \ding{51} & {\color{green!50!black}\ding{52}} & & MC(8) & \ding{51} & \\
$p>1$ from $p=1$ & \cite{Alam2020} & & & (\ding{51}) & {\color{green!50!black}\ding{52}} & & MC(8) & \ding{51} & \\ 
RL \& KDE based ML & \cite{Khairy2020} & & \ding{51} & & {\color{green!50!black}\ding{52}} & & MC(8) & \ding{51} & \ding{51} \\
RL-RQAOA & \cite{Patel2024} & & & & {\color{green!50!black}\ding{52}} & & Ising(30) & & \ding{51} \\
Diffusion & \cite{meng2024parameter} & & & & {\color{green!50!black}\ding{52}} & & MC(8) & \ding{51} & \\
QLSTM & \cite{Chen2025} & & \ding{51} & & {\color{green!50!black}\ding{52}} & & MC(16) & & \ding{51} \\ 
NN & \cite{Amosy2024-xm} & & & & {\color{green!50!black}\ding{52}} & & MC(16) & \ding{51} & \\
GATs & \cite{xu2025qaoa} & & \ding{51} & & {\color{green!50!black}\ding{52}} & & MIS(25)  & \ding{51} & \\  \hline\hline
\end{tabular}
\end{table*}

\paragraph{Physics-inspired methods} typically base their QAOA angles on annealing schedules, such as linear ramps~\cite{sack2021quantum, Willsch2022, Montanezbarrera2024, Sakai2025}.
For example, the Trotterized Quantum Annealing approach~\cite{sack2021quantum} initializes $\boldsymbol{\beta}$ and $\boldsymbol{\gamma}$ with increasing and decreasing linear ramps, respectively, see App.~\ref{app:tqa_lr}.
By contrast, in linear-ramp QAOA, $\boldsymbol{\beta}$ and $\boldsymbol{\gamma}$ are set to a linear ramp with different slopes without any further optimization~\cite{Montanezbarrera2024}.
Ref.~\cite{Mcdowall2026} extends the linear-ramps by incorporating information about the spectral gap of the instantaneous Hamiltonian into the QAOA angles.
The authors propose to make the resulting Spectral Gap Informed Ramps (SGIR) QAOA scalable with an extrapolation.
Going beyond linear ramps may be possible by drawing connections to optimized annealing schedules~\cite{Das2005, Mukherjee2015}.
Crucially, there is a strong connection between quantum annealing and the QAOA with gradually varying angles~\cite{boulebnane2025}.
This allows the design of schedules for a deep QAOA by extrapolating optimized schedules at low depth. 
Several QAOA angle schedules arise from optimal-control and counterdiabatic ideas. 
Counterdiabatic (CD)-QAOA uses approximate adiabatic gauge potentials to match Trotter terms and suppress diabatic excitations~\cite{WurtzLove2022CDQAOA}.
Furthermore, optimal control allows us to contrast QA and QAOA to reason about the ideal schedule to solve an optimization problem~\cite{bapat2018bang,Brady2021BangAnnealBang,Venuti2021PontryaginOpen}.
The feedback-based Algorithm for Quantum Optimization (FALQON) applies a measurement–feedback law that updates angles layer-by-layer without a classical optimizer and guarantees monotonic improvement with depth~\cite{Magann2022FALQON, Li2024FALQONSim, Rahman2024FALQONQCBO}.

\paragraph{Parameter transfer methods} rely on the clustering of optimal QAOA angles for similar problem instances~\cite{Akshay2021, brandao2018, Shaydulin2019, Galda2021, Montanezbarrera2024Transfer, pelofske2026evaluatinglimitsqaoaparameter, Pelofske_2024_QAOA_scaling, Barron2024, Kotil_2025} and classes~\cite{Montanezbarrera2024Transfer}.
This clustering is understood by the structure of the sub-graphs seen by QAOA~\cite{Akshay2021, brandao2018, Shaydulin2019, Galda2021}.
For example, in MaxCut on $k$ random-regular graphs, angles transfer well between graphs with even or odd degree $k$ but transfer poorly from even to odd $k$ and vice versa~\cite{Galda2021}.
This can be overcome by considering the symmetries of MaxCut~\cite{Lyngfelt2025}.
The fixed angle conjecture is a form of parameter transfer in which angles are found for MaxCut worst case graphs and transfered to random regular graphs~\cite{Wurtz2021a, Wurtz2021}, see App.~\ref{app:fixed_angles}.
While the result does not hold for general graphs, the fixed angles, provided by Wurtz and Lykov~\cite{WurtzGithub}, may serve as initial points that an optimizer can refine.
Furthermore, the fixed angles method also works on other problem classes such as MIS~\cite{Wybo2025}.
QAOA angles also transfer from small to large problem instances of the same class~\cite{pelofske2026evaluatinglimitsqaoaparameter}.
Crucially, the optimized angles may scale with system size $n$, as observed, e.g., for low autocorrelation binary sequences where $\gamma^\star$ scales as $1/n$~\cite{Shaydulin2024}.
One can also transfer angles from an unweighted problem instance to a weighted one by rescaling $\boldsymbol{\gamma}$ based on $H_C$~\cite{Shaydulin2023, Sureshbabu2024, cepaite2025quantum}.
Recent research also considers re-optimizing angles after a transfer at a reduced cost, e.g., QAOA-PCA~\cite{Parry2025} discussed in App.~\ref{app:qaoa_pca}, and re-optimizing a subset of the layers~\cite{venturelli2024transfer}. 
These methods can result in good QAOA angles in fewer iterations of the classical optimizer compared to directly re-optimizing all QAOA angles.
Methods that find $\boldsymbol{\theta}^\star$ with approximate models of the QAOA cost function, efficient to compute classically, i.e., surrogate models, can be seen as a form of parameter transfer~\cite{Sud2024, Cheng2024, Oleary2025}.
For example, Ref.~\cite{Sud2024} introduces a classical proxy for QAOA to find $\boldsymbol{\theta}^\star$.
DARBO~\cite{Cheng2024}, a trust-region Bayesian optimization for QAOA, uses fewer circuit calls and shows robustness under finite-shot noise in superconducting-hardware.
The shot efficiency of surrogate models also helps train QAOA angles when hardware access is limited~\cite{Oleary2025}.

\paragraph{Iterative methods} typically initialize an optimization of depth $p+1$ angles from (near) optimal depth $p$ angles.
This approach often results in smooth schedules, which usually perform well.
The first iteration may be done at depth-one, where the energy can be efficient to compute classically~\cite{Egger2021warmstartingquantum} and 2D grid scans of $(\beta, \gamma)$ give a clear picture of the optimization landscape~\cite{Farhi2014}. 
Such grid scans are made efficient by a reduction to a one-dimensional search over $\gamma$ in which the optimal $\beta(\gamma)$ is computed analytically~\cite{Vijendran2025}.
Crucially, optimizing only the last QAOA layer at depth $p+1$ saturates the performance of QAOA~\cite{Campos2021}.
Notable iterative methods include Interp. and Fourier~\cite{Zhou2020}, see App.~\ref{app:interp_fourier}.
The Iterative Interpolation method~\cite{apte2025} generalizes the Fourier approach by expressing the QAOA schedules in a different basis, such as Chebyshev or Legendre polynomials. 
Furthermore, Ref.~\cite{apte2025} introduces a dimensionality reduction in which low-order modes are optimized first.
Alternatively, a greedy iterative angle optimization scheme can be built on transition states~\cite{Sack2023}.
Here, the initial point $\smash{\boldsymbol{\theta}^{(0)}_{p+1}}$ of the optimizer at depth-$p+1$ is a local optimum $\boldsymbol{\theta}^\star_p$ of depth-$p$ QAOA such that $\smash{\boldsymbol{\theta}^{(0)}_{p+1}}$ has a saddle point in the optimization landscape.
Therefore, the optimizer can reach a state at depth $p+1$ with a lower energy than the state at depth $p$.
Lower bounds on the resulting energy improvement are available from quartic expansions of the QAOA cost function~\cite{Medina2024}.
JuliQAOA~\cite{Golden_2023} is an angle optimization package based on a statevector energy computation. 
It leverages a basin hopping optimization algorithm in which initial angles at depth-$p+1$ are extrapolated from depth-$p$ angles according to $\boldsymbol{\theta}_p^{(0)}=(\beta_1,...,\beta_p,\beta_p, \gamma_1,...,\gamma_p,\gamma_p)$~\cite{Golden2023b}.
Agirre \emph{et al.}~\cite{Agirre2025} map the QAOA angle optimization to a discrete optimization task, which they address with a Monte-Carlo Tree Search algorithm on a restricted search space.
The method is iterative since the optimal angles at depth $p$ inform the restricted search space at depth $p+1$.
ADAPT-QAOA~\cite{zhu2022adaptive} iteratively builds the ansatz by appending problem-specific mixer generators chosen from an operator pool using gradient information.

\paragraph{Machine learning methods} use models, trained on existing optimized angles and problem instances, to predict good QAOA angles.
Here, the quality of the QAOA angles in the training dataset is crucial to obtain a model capable of inferring high-quality angles~\cite{Liang2024, Tyagin2025}.
Machine learning can be used in multiple ways.
First, unsupervised learning can help identify QAOA angles obtained from a parameter transfer from similar problem instances~\cite{Acampora2023, Moussa2022}.
Here, similarity is determined by a clustering of the problem instances already present in the dataset.
Second, the angle prediction can be iterative.
Xie \emph{et al.}~\cite{Xie2023} create a convolutional neural network that predicts depth $p+1$ angles based on depth-$p$.
Alam \emph{et al.}~\cite{Alam2020} explore various regression models to predict initial angles at a depth $p>1$ given optimal angles at depth $p=1$.
These initial angles typically reduce the number of iterations needed by the classical optimizer to find $\boldsymbol{\theta}^\star$.
Third, reinforcement learning (RL) has also been explored.
For instance, Khairy \emph{et al.}~\cite{Khairy2020} use RL and kernel density estimation to infer QAOA angles.
In RL-QAOA~\cite{Patel2024}, a RL agent finds both the $(\beta, \gamma)$ and the recursive-QAOA variable substitution policy~\cite{Bravyi2020}.
Meng \emph{et al.}~\cite{meng2024parameter} frame angle setting as a generative task and train denoising diffusion probabilistic models to sample high‑quality starting angles to further optimize.
Chen \emph{et al.}~\cite{Chen2025} introduce a quantum Long Short Term Memory (LSTM) to optimize the QAOA angles.
Amosy \emph{et al.}~\cite{Amosy2024-xm} show that a fully connected neural network, accepting the adjacency matrix as input, can predict QAOA angles for MaxCut without subsequent iterations of an optimizer.
Xu \emph{et al.}~\cite{xu2025qaoa} train a graph attention network to identify problems similar to an unseen one to then transfer QAOA angles.

\subsection{Energy Evaluation\label{sec:evaluation}}

Not all QAOA angle setting methods require evaluating and optimizing the energy $\langle H_C\rangle$.
\emph{Physics-inspired methods} do not need $\langle H_C\rangle$ unless they optimize the ramps, for instance.
Some \emph{parameter transfer methods} can do without $\langle H_C\rangle$. 
\emph{Machine learning methods} rely on a database of pre-trained angles which may be designed with methods that evaluate $\langle H_C\rangle$. 
If $\langle H_C\rangle$ is needed, it can be found with exact statevector simulations for small qubit numbers, sampling quantum hardware, or by sampling or direct computation via approximate methods such as MPS and PP, that we now describe.

\subsubsection{Hardware energy evaluation\label{sec:hw_training}}

The optimization of QAOA angles can be executed on quantum hardware, but at a high cost and only if the resulting circuits fall within the device's noise limits.
Estimating $\langle H_C\rangle$ and sampling candidate solutions~$x$ both require expressing the circuits that prepare $\ket{\psi(\boldsymbol{\theta})}$ into hardware native instructions.
SWAP gates are needed when the topology of $H_C$ does not match the physical qubit connectivity. 
Here, it is essential to exploit the commutative nature of the gates in $e^{-i\gamma H_C}$ to obtain compact circuits executable within qubit coherence times.
This is achievable with predetermined networks of SWAP gates~\cite{Weidenfeller2022}.
The number of required layers of SWAP gates for non-dense problems can be minimized by carefully choosing which decision variable is mapped to which qubit~\cite{Matsuo2023}.
The expectation value $\langle H_C\rangle$ can be error mitigated with methods such as probabilistic error cancellation~\cite{Temme2017} or amplification~\cite{Kim2023} or even machine learning~\cite{Sack2024}.
Crucially, error mitigation only impacts the quality of the expecation values, but not the final sampling.
Indeed, it is harder to error mitigate samples than expectation values.
Instead, it is often better to simply draw $\sqrt{\gamma}$ more samples where $\gamma$ measures the noise strength in the quantum circuit~\cite{Barron2024} and post-select on the best samples.

\begin{figure}
    \centering
    \includegraphics[width=0.95\columnwidth]{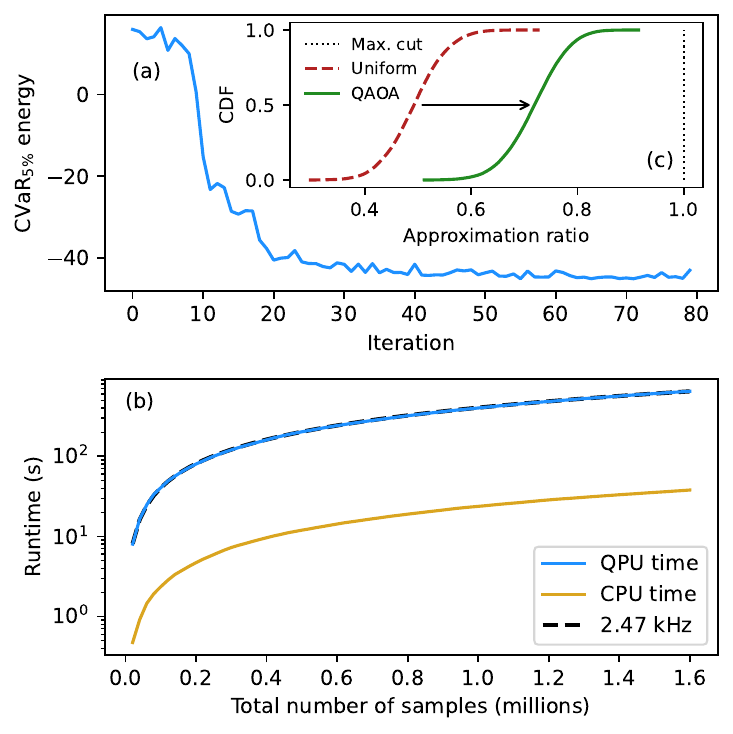}
    \caption{Convergence of a 100 qubit depth-four QAOA on \emph{ibm\_boston}.
    (a) Energy of the 5\% best samples as a function of the number of COBYLA iterations.
    (b) Duration required to obtain the samples from the QPU as measured by the user (QPU time) and the time it took to compute the objective to minimize from the samples (CPU time).
    The inset (c) shows the improvement in the cumulative distribution of the quantum samples (green solid curve) over uniform sampling (red dashed curve).
    The best measured quantum sampled achieves a $91.5\%$ approximation ratio.
    }
    \label{fig:hardware_conv}
\end{figure}

As an example, we train on \emph{ibm\_boston} a utility-scale depth-four QAOA defined on a 100-node weighted MaxCut.
The underlying QAOA circuit has $N_\text{CZ}=1980$ two-qubit gates applied in 40 layers.
To extract a sizable signal from the quantum samples we optimize $\langle H_C\rangle$ computed on the 5\% best samples $x$ as measured by the MaxCut objective $f(x)$~\cite{Barron2024}, see App.~\ref{app:hardware_train}.
The optimizer thus sees a clear signal resulting in high-quality samples, see Fig.~\ref{fig:hardware_conv}(a) and~(c).
From the user's perspective, the $20{\rm k}$ samples are generated at a rate of $\sim 2.47~\rm{kHz}$, see App.~\ref{app:hardware_train}.
Including the classical processing of the samples the whole training took 11.4~minutes, see Fig.~\ref{fig:hardware_conv}(b).
This speed is enabled by a fast execution stack, low single- and two-qubit gate times, and an efficient classical processing of the samples.

\subsubsection{Tensor networks\label{sec:tns}}

Tensor networks (TNs) are numerical tools to classically approximate quantum circuits through a fixed \textit{ansatz}.
Different TN \textit{ans\"atze} are tailored to accurately represent circuits with different entanglement structures.
For instance, MPS have a one-dimensional topology that efficiently represent short-range correlated circuits.
Conversely, a circuit displaying strong quantum correlations between any pair of qubits is hard to efficiently encode as an MPS, but can be compactly represented through multidimensional TNs. 
The key advantage of MPSs over other TNs is that a circuit can be simulated with an MPS using relatively simple numerical tools.
Throughout this work we leverage MPS energy evaluators based on Quimb~\cite{Gray2018} and Qiskit-Aer~\cite{Qiskit}.
Other approaches are possible, see App.~\ref{app:tn}.

A product state can be represented as a very compact MPS with a bond dimension, i.e., the parameter that tunes the MPS representation power, of 1.
Single-qubit gates can be simulated exactly without increasing the bond dimension.
On the other hand, every application of a two-qubit gate increases the bond dimension.
For deep circuits, composed of hundreds of two-qubit gates, the associated growth of the bond dimension makes storing and manipulating the exact MPS representation of the target quantum circuit computationally prohibitive.
Fortunately, in practice one works with approximate representations of the circuit, obtained by compressing the MPS through a series of singular value decompositions.
The accuracy of the resulting approximation can be bounded from below~\cite{Stoudenmire2020_Limits-QC,Ayral2023_DMRG-FiniteFidelity}, see App.~\ref{app:tn}.
Importantly, sampling from a given TN state and calculating the expectation value of some operator do not have the same complexity~\cite{Watanabe2026}.
For MPS and loop-free TNs efficient methods for both tasks exist.
However, this is not true for TNs with loops, for which sampling is harder than calculating expectation values, see App.~\ref{app:tn_sample}.
This motivates schemes where the QAOA angles are computed classically, and the quantum computer samples candidate solutions once the angles are bound in the circuit.

\begin{figure}[ht]
    \centering
    \includegraphics[width=\columnwidth, trim=0 3cm 0 0, clip]{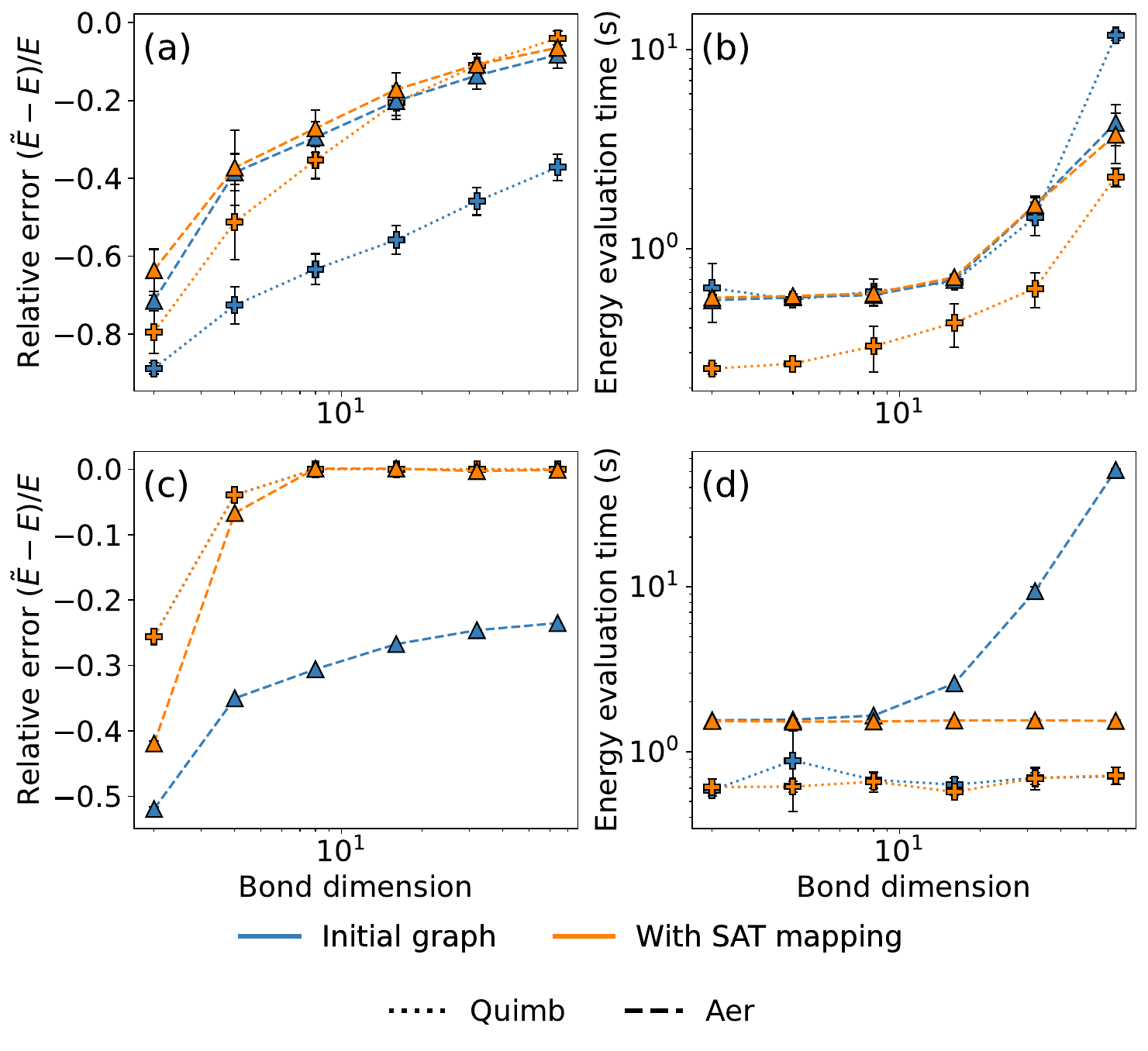}
    \caption{
    Impact of the SAT mapping on five random instances of three-regular graphs with 40 nodes (a) and (b), and 100-node graphs (c) and (d). 
    Panels (a) and (c) show the relative error between the MPS energies and the exact energies. Panels (b) and (d) show the duration of the energy evaluation for different bond dimensions.
    Each marker is the average of five instances and the error bars show the standard deviation.
    }
    \label{fig:mps_sat}
\end{figure}

When simulating quantum circuits with TNs, each qubit maps to a tensor.
This mapping can impact the simulation accuracy.
For MPSs, the accuracy is maximal when qubits that, as a result of the quantum circuit, are strongly entangled are close to each other in the one-dimensional MPS chain.
For QAOA, the wave function entanglement structure reflects the topology of the unitary $e^{-i\gamma H_C}$.
We can therefore classically optimize the mapping of the qubits onto the TN based on $H_C$.
This situation is similar to mapping decision variables to physical qubits when executing QAOA on quantum hardware. 
Ref.~\cite{Matsuo2023} formulates this mapping as a boolean satisfiability problem (SAT) and we refer to this as the \emph{SAT mapping}.
We now illustrate how the SAT mapping increases the accuracy of the MPS computation and reduces its runtime.
We evaluate the energy $\tilde{E}$ of depth-one QAOA with an MPS for five random three-regular graphs with 40 nodes and graphs with 100 nodes built from a line and two SWAP layers, see App.~\ref{app:graphs}.
The QAOA angles $\beta$ and $\gamma$ are near optimal, found based on a grid scan of the energy landscape.
The accuracy of the MPS energies is obtained by comparison to the exact ones~\cite{Egger2021warmstartingquantum}.
With a SAT mapping, the relative accuracy and runtime of the MPS evaluator at fixed bond dimension is higher than without, see Fig.~\ref{fig:mps_sat}.
Therefore, improved MPS energy evaluations are possible by carefully mapping the problem to the topology of the TN.

\subsubsection{Pauli propagation}\label{sec:pauli_prop}

An alternative to TNs is PP which computes expectation values by backpropagating the target observable through the circuit in the Heisenberg picture~\cite{rudolph2025pauli, rall2019simulation}.
It tracks how Pauli operators transform when conjugated by the gate operators.
This process builds a tree of Pauli paths, whose sum yields the total backpropagated observable.
Pauli propagation can be faster and more precise at simulating expectation values than TNs when the Pauli decomposition of the backpropagated observable is sparse~\cite{rudolph2025pauli, Dowling2024Magic}.
Clifford circuits are trivial to simulate with PP since, by definition, the backpropagation does not increase the complexity of the observable.
Conversely, Clifford circuits may generate highly-entangled states and thus prove hard to compute using TNs.
Therefore, since magic, i.e., non-stabilizerness measures the non-Cliffordness of a circuit, it helps predict whether a circuit can be simulated efficiently with PP.
Moreover, PP is agnostic to the topology of the underlying Hamiltonian, making it more practical for 2D and 3D systems for which only few efficient TN-based methods exist~\cite{Begusic2025SparsePauli}.

\begin{figure}
    \centering
    \includegraphics[width=\columnwidth, clip, trim=10 0 10 0]{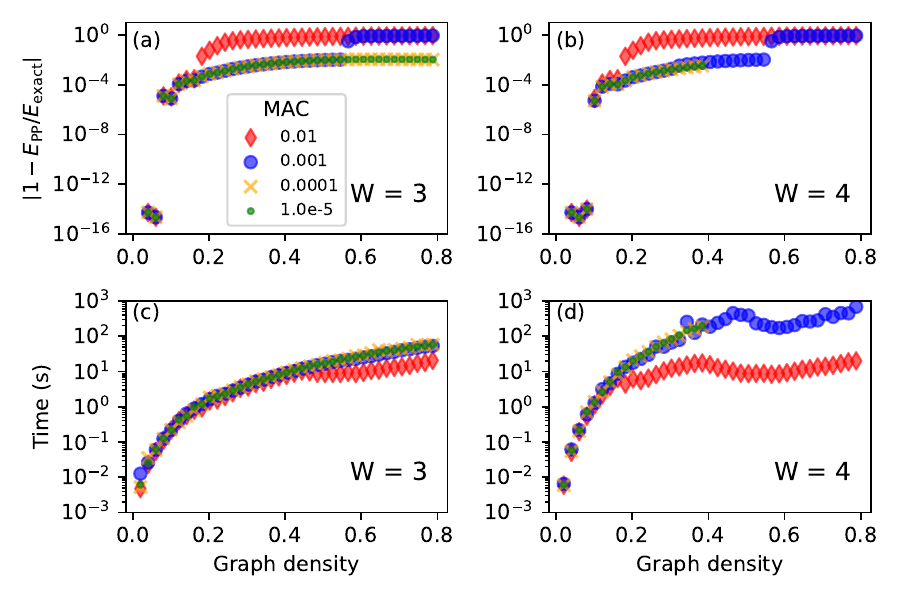}
    \caption{Pauli propagation as an evaluator of QAOA state energy for random $k$-regular graphs with $n = 100$ nodes and $p = 1$ layer. 
    The node degree $k$ is changed to vary the graph density $k/(n-1)$. 
    Panels (a) and (b) show the absolute error in energy estimated by Pauli Propagation compared to the exact energy. 
    Panels (c) and (d) show duration of each computation for a given graph density and set of parameters $W$ and MAC. The energy for graphs with density above $40\%$ is not computed for $W = 4$ and low $\textrm{MAC}\leq 10^{-4}$ are not presented in the plots as these were too time- and memory-intensive.}
    \label{fig:pp_numerics}
\end{figure}

The primary challenge of PP is the exponential branching of the propagation tree due to non-Clifford gates.
Two complementary heuristic strategies help truncate this tree. 
First, \emph{Pauli weight truncation}, removes Pauli strings with more than $W$ non-identity Paulis.
These high-weight terms often have negligible overlap with typical QAOA initial states, such as product states. 
A rigorous error bound exists because the expectation value of any Pauli string with respect to a product state is bounded in magnitude by one; thus, the error introduced is at most the sum of the discarded coefficients.
Second, \emph{coefficient pruning} discards Pauli strings whose coefficients fall below a chosen threshold during propagation. 
The resulting approximation error is again bounded by the total magnitude of the discarded coefficients, which can be efficiently monitored during the computation. 
This ensures a computationally tractable approximation with a tunable trade‑off between runtime and accuracy.

We illustrate PP and the impact of these approximations with the Julia package PauliPropagation.jl~\cite{rudolph2025pauli}.
We compute the energies of states generated by depth-one QAOA on random-regular graphs $G(V,E)$ with $|V| = 100$ nodes and graph densities up to 0.8 for MaxCut.
Depth-one QAOA is implemented as it allows for efficient computation of the state energies~\cite{Ozaeta_2022}.
This enables a classical optimization of the QAOA angles and provides a benchmark energy for Pauli propagation.

We explore both weight truncation and coefficient pruning, labeled `mean absolute coefficient' (MAC). 
For $p = 1$ no correlators with a weight beyond four contribute to the expectation values. 
Thus, we report simulations keeping only Pauli strings of weight $W = 3$ and lower in Fig.~\ref{fig:pp_numerics}(a) and (c), or $W = 4$ and lower in Fig.~\ref{fig:pp_numerics}(b) and (d). 
As expected, increasing the truncation parameter $W$ does not impact the resulting expectation values significantly.
A term in the tree is discarded if its coefficient is below the $\rm MAC$.
Here, we use $\rm MAC \in [10^{-5}, 10^{-2}]$ which affects both the efficiency and accuracy of PP more than the weight truncation.
Computations with $W = 4$ and $\rm MAC \leq 10^{-4}$ require a runtime exceeding $10^3$ seconds and are thus not shown.

PP performs best on sparse graphs both in terms of computation time, see Fig.~\ref{fig:pp_numerics}(c) and (d), and the absolute error in energy $\left|1 - E_{\rm PP}/E_{\rm exact}\right|$, see Fig.~\ref{fig:pp_numerics}(a) and (b).
Here, $E_{\rm PP}$ is the state energy estimated by PP and $E_{\rm exact}$ is the true expected state energy.
Both are obtained with the optimal QAOA angles.
This is expected, since sparse graphs have few terms in the cost Hamiltonian leading to few trees with few branches.
For graphs with edge densities below $10\%$, PP estimates the energy of the output state of each circuit with less than $0.1\%$ error in under 1 second, regardless of the truncation parameters. 
The computation time grows exponentially with density. 
Above a threshold density near $0.6$ the error of the estimated energy is above $80\%$ error for $W = 3$ and $\rm MAC \geq 10^{-4}$. 
When $W = 4$ the computations require more than $10^3$ seconds for graphs with a density above $0.4$. 
Nevertheless, these simulations demonstrate that PP can estimate QAOA expectation values and thus optimize QAOA angles at utility-scale without a QPU.


\section{Benchmarking methodology\label{sec:bench_method}}

Since performance benchmarks guide best-practices, we study multiple QAOA angle setting methods with utility-scale problems as target.
Many of the methods discussed in Sec.~\ref{sec:overview} are typically tested on MaxCut problems, often on random regular graphs, and at sizes that are simulatable classically with exact methods, see Tab.~\ref{tab:qaoa_taxonomy}.
Ref.~\cite{Katial2025} compares different angle setting strategies via instance space analysis on twelve-node MaxCut problems spanning a broad range of graphs.
Here, we benchmark (i)~at a small-scale with exact energy evaluation, and, crucially, (ii)~at utility-scale, by transferring angles from either small to large instances or from a training based on approximate energy evaluation.
As metrics, we employ quality, as measured by the energy, and the duration required to find $\boldsymbol{\beta}$ and $\boldsymbol{\gamma}$.

We benchmark the Interp., Fourier, TQA, recursive transition states (RTS), linear ramps, and fixed angles methods with various energy evaluation tools.
The details of each method are in App.~\ref{app:param_methods}.
The fixed angles, derived for random $k$-regular graphs, are those of Wurtz and Lykov~\cite{Wurtz2021}.
We employ them on non-regular problem instances by choosing the set of fixed angles with a regularity closest to the average node degree of the non-regular problem.
The TQA, linear ramps, and fixed angles methods are carried out with and without optimization.
Throughout this work, a $\dagger$ superscript indicates that the slopes or fixed angles are not optimized to the problem instance.
A $\star$ superscript indicates that each QAOA angle is optimized to the problem instance.
For TQA and linear ramps, the absence of a superscript indicates that only the slopes are optimized.
We define the methods in JSON files for transparency and run them with the QAOA Training Pipeline~\cite{training_pipeline}, see App.~\ref{app:training_pipeline}.
Since we are interested in the performance of the angle setting methods and not the impact of the classical optimizer we do not benchmark the hyperparameters of the optimizer, if present.
Instead, when an optimization is required we employ COBYLA with a small initial step of $0.2$.

We benchmark with the MaxCut, the Maximum Independent Set (MIS) and the Low Autocorrelation Binary Sequence (LABS) problem classes, see App.~\ref{app:problem_models}.
The MaxCut and MIS problem instances are initialized from graphs stored in a database we create. 
We consider unweighted random-regular and Erd\H{o}s-R\'{e}nyi graphs, weighted graphs built by applying swaps to a line of nodes and weighted heavy-hex graph, see App.~\ref{app:graphs}. 
To extend beyond quadratic optimization we also consider the LABS problem. 
Unlike the graph-based MaxCut and MIS, the LABS objective is uniquely defined for each system size $n$. 
LABS is currently prohibitive for noisy hardware due to its many four-local interactions and resulting deep circuits~\cite{Dragoi2025}. 
Consequently, we report only statevector simulations. 

\begin{table*}[!htbp]
    \centering
    \begin{tabular}{l|cccc|cccc|}
\toprule
& \multicolumn{4}{c|}{Approximation Ratio} & \multicolumn{4}{c|}{Num. Instances} \\
            & ER & HH & LB & RR & ER & HH & LB & RR \\
\midrule
\multicolumn{9}{c}{\rule{0pt}{8pt}MPS (Aer)} \\
\midrule
Fixed Angles$^\star$ & {\cellcolor[HTML]{000000}} \color[HTML]{F1F1F1} {\cellcolor{white}} \color{black} - & {\cellcolor[HTML]{8A59A8}} \color[HTML]{F1F1F1} $89.7\pm 3.9$ & {\cellcolor[HTML]{6F0A6B}} \color[HTML]{F1F1F1} $96.4\pm 1.0$ & {\cellcolor[HTML]{4F004D}} \color[HTML]{F1F1F1} $98.5\pm 1.4$ & {\cellcolor[HTML]{000000}} \color[HTML]{F1F1F1} {\cellcolor{white}} \color{black} - & {\cellcolor[HTML]{FDE4E1}} \color[HTML]{000000} $20$ [$39$--$144$] & {\cellcolor[HTML]{FEEAE7}} \color[HTML]{000000} $14$ [$100$] & {\cellcolor[HTML]{FEEEEA}} \color[HTML]{000000} $11$ [$40$--$100$] \\
Fixed Angles$^\dagger$ & {\cellcolor[HTML]{000000}} \color[HTML]{F1F1F1} {\cellcolor{white}} \color{black} - & {\cellcolor[HTML]{8C6AB1}} \color[HTML]{F1F1F1} $88.3\pm 3.6$ & {\cellcolor[HTML]{821982}} \color[HTML]{F1F1F1} $94.5\pm 2.3$ & {\cellcolor[HTML]{5A0457}} \color[HTML]{F1F1F1} $97.8\pm 3.3$ & {\cellcolor[HTML]{000000}} \color[HTML]{F1F1F1} {\cellcolor{white}} \color{black} - & {\cellcolor[HTML]{FDE4E1}} \color[HTML]{000000} $20$ [$39$--$144$] & {\cellcolor[HTML]{FEEAE7}} \color[HTML]{000000} $14$ [$100$] & {\cellcolor[HTML]{FEEEEA}} \color[HTML]{000000} $11$ [$40$--$100$] \\
Fourier$^\star$ & {\cellcolor[HTML]{9FBCDA}} \color[HTML]{000000} $81.2\pm 6.1$ & {\cellcolor[HTML]{8C91C4}} \color[HTML]{F1F1F1} $85.1\pm 4.8$ & {\cellcolor[HTML]{A3BFDC}} \color[HTML]{000000} $80.8\pm 8.9$ & {\cellcolor[HTML]{8A54A6}} \color[HTML]{F1F1F1} $90.1\pm 5.9$ & {\cellcolor[HTML]{FDE4E1}} \color[HTML]{000000} $20$ [$40$--$50$] & {\cellcolor[HTML]{FDE4E1}} \color[HTML]{000000} $20$ [$39$--$144$] & {\cellcolor[HTML]{FCC0BF}} \color[HTML]{000000} $50$ [$40$--$100$] & {\cellcolor[HTML]{FEEEEA}} \color[HTML]{000000} $11$ [$40$--$100$] \\
Interp.$^\star$ & {\cellcolor[HTML]{831D85}} \color[HTML]{F1F1F1} $94.2\pm 4.0$ & {\cellcolor[HTML]{88459F}} \color[HTML]{F1F1F1} $91.3\pm 4.6$ & {\cellcolor[HTML]{7C0E78}} \color[HTML]{F1F1F1} $95.5\pm 2.4$ & {\cellcolor[HTML]{8948A0}} \color[HTML]{F1F1F1} $91.1\pm 3.3$ & {\cellcolor[HTML]{FDE4E1}} \color[HTML]{000000} $20$ [$40$--$50$] & {\cellcolor[HTML]{FDE4E1}} \color[HTML]{000000} $20$ [$39$--$144$] & {\cellcolor[HTML]{FCC0BF}} \color[HTML]{000000} $50$ [$40$--$100$] & {\cellcolor[HTML]{FEEEEA}} \color[HTML]{000000} $11$ [$40$--$100$] \\
Linear Ramp$^\star$ & {\cellcolor[HTML]{6E096A}} \color[HTML]{F1F1F1} $96.5\pm 0.8$ & {\cellcolor[HTML]{8B60AC}} \color[HTML]{F1F1F1} $89.2\pm 4.9$ & {\cellcolor[HTML]{7C0E78}} \color[HTML]{F1F1F1} $95.5\pm 1.7$ & {\cellcolor[HTML]{5F055C}} \color[HTML]{F1F1F1} $97.4\pm 2.2$ & {\cellcolor[HTML]{FDE4E1}} \color[HTML]{000000} $20$ [$40$--$50$] & {\cellcolor[HTML]{FDE4E1}} \color[HTML]{000000} $20$ [$39$--$144$] & {\cellcolor[HTML]{FCC0BF}} \color[HTML]{000000} $50$ [$40$--$100$] & {\cellcolor[HTML]{FEEEEA}} \color[HTML]{000000} $11$ [$40$--$100$] \\
Linear Ramp & {\cellcolor[HTML]{831A83}} \color[HTML]{F1F1F1} $94.4\pm 1.8$ & {\cellcolor[HTML]{873B99}} \color[HTML]{F1F1F1} $92.1\pm 1.8$ & {\cellcolor[HTML]{831C84}} \color[HTML]{F1F1F1} $94.3\pm 2.0$ & {\cellcolor[HTML]{740B70}} \color[HTML]{F1F1F1} $96.0\pm 2.0$ & {\cellcolor[HTML]{FDE4E1}} \color[HTML]{000000} $20$ [$40$--$50$] & {\cellcolor[HTML]{FDE4E1}} \color[HTML]{000000} $20$ [$39$--$144$] & {\cellcolor[HTML]{FCC0BF}} \color[HTML]{000000} $50$ [$40$--$100$] & {\cellcolor[HTML]{FEEEEA}} \color[HTML]{000000} $11$ [$40$--$100$] \\
Linear Ramp$^\dagger$ & {\cellcolor[HTML]{8A54A6}} \color[HTML]{F1F1F1} $90.1\pm 2.1$ & {\cellcolor[HTML]{94A6CE}} \color[HTML]{F1F1F1} $83.3\pm 1.5$ & {\cellcolor[HTML]{88419D}} \color[HTML]{F1F1F1} $91.7\pm 3.8$ & {\cellcolor[HTML]{9FBCDA}} \color[HTML]{000000} $81.2\pm 7.5$ & {\cellcolor[HTML]{FDE4E1}} \color[HTML]{000000} $20$ [$40$--$50$] & {\cellcolor[HTML]{FDE4E1}} \color[HTML]{000000} $20$ [$39$--$144$] & {\cellcolor[HTML]{FCC0BF}} \color[HTML]{000000} $50$ [$40$--$100$] & {\cellcolor[HTML]{FEEEEA}} \color[HTML]{000000} $11$ [$40$--$100$] \\
Recursive TS$^\star$ & {\cellcolor[HTML]{770C73}} \color[HTML]{F1F1F1} $95.8\pm 1.8$ & {\cellcolor[HTML]{8C93C4}} \color[HTML]{F1F1F1} $85.0\pm 5.1$ & {\cellcolor[HTML]{831F86}} \color[HTML]{F1F1F1} $94.1\pm 2.5$ & {\cellcolor[HTML]{842289}} \color[HTML]{F1F1F1} $93.9\pm 4.6$ & {\cellcolor[HTML]{FDE4E1}} \color[HTML]{000000} $20$ [$40$--$50$] & {\cellcolor[HTML]{FDE4E1}} \color[HTML]{000000} $20$ [$39$--$144$] & {\cellcolor[HTML]{FCC0BF}} \color[HTML]{000000} $50$ [$40$--$100$] & {\cellcolor[HTML]{FEEEEA}} \color[HTML]{000000} $11$ [$40$--$100$] \\
TQA$^\star$ & {\cellcolor[HTML]{8B5DAA}} \color[HTML]{F1F1F1} $89.4\pm 2.0$ & {\cellcolor[HTML]{8C70B3}} \color[HTML]{F1F1F1} $87.9\pm 2.9$ & {\cellcolor[HTML]{8A58A8}} \color[HTML]{F1F1F1} $89.8\pm 3.0$ & {\cellcolor[HTML]{7C0E78}} \color[HTML]{F1F1F1} $95.5\pm 3.5$ & {\cellcolor[HTML]{FDE4E1}} \color[HTML]{000000} $20$ [$40$--$50$] & {\cellcolor[HTML]{FDE4E1}} \color[HTML]{000000} $20$ [$39$--$144$] & {\cellcolor[HTML]{FCC0BF}} \color[HTML]{000000} $50$ [$40$--$100$] & {\cellcolor[HTML]{FEEEEA}} \color[HTML]{000000} $11$ [$40$--$100$] \\
TQA & {\cellcolor[HTML]{8C80BB}} \color[HTML]{F1F1F1} $86.5\pm 2.1$ & {\cellcolor[HTML]{8C79B8}} \color[HTML]{F1F1F1} $87.1\pm 2.5$ & {\cellcolor[HTML]{8C8FC2}} \color[HTML]{F1F1F1} $85.4\pm 4.4$ & {\cellcolor[HTML]{863394}} \color[HTML]{F1F1F1} $92.7\pm 4.6$ & {\cellcolor[HTML]{FDE4E1}} \color[HTML]{000000} $20$ [$40$--$50$] & {\cellcolor[HTML]{FDE4E1}} \color[HTML]{000000} $20$ [$39$--$144$] & {\cellcolor[HTML]{FCC0BF}} \color[HTML]{000000} $50$ [$40$--$100$] & {\cellcolor[HTML]{FEEEEA}} \color[HTML]{000000} $11$ [$40$--$100$] \\
TQA$^\dagger$ & {\cellcolor[HTML]{9AB4D6}} \color[HTML]{000000} $82.0\pm 4.4$ & {\cellcolor[HTML]{A6C2DD}} \color[HTML]{000000} $80.4\pm 1.6$ & {\cellcolor[HTML]{C8DAEA}} \color[HTML]{000000} $76.9\pm 6.1$ & {\cellcolor[HTML]{909ECA}} \color[HTML]{F1F1F1} $84.0\pm 5.8$ & {\cellcolor[HTML]{FDE4E1}} \color[HTML]{000000} $20$ [$40$--$50$] & {\cellcolor[HTML]{FDE4E1}} \color[HTML]{000000} $20$ [$39$--$144$] & {\cellcolor[HTML]{FCC0BF}} \color[HTML]{000000} $50$ [$40$--$100$] & {\cellcolor[HTML]{FEEEEA}} \color[HTML]{000000} $11$ [$40$--$100$] \\
\cline{1-9}
\multicolumn{9}{c}{\rule{0pt}{8pt}MPS (Quimb)} \\
\midrule
Fixed Angles$^\star$ & {\cellcolor[HTML]{000000}} \color[HTML]{F1F1F1} {\cellcolor{white}} \color{black} - & {\cellcolor[HTML]{000000}} \color[HTML]{F1F1F1} {\cellcolor{white}} \color{black} - & {\cellcolor[HTML]{6E2706}} \color[HTML]{F1F1F1} $98.1$ & {\cellcolor[HTML]{692606}} \color[HTML]{F1F1F1} $98.4\pm 1.2$ & {\cellcolor[HTML]{000000}} \color[HTML]{F1F1F1} {\cellcolor{white}} \color{black} - & {\cellcolor[HTML]{000000}} \color[HTML]{F1F1F1} {\cellcolor{white}} \color{black} - & {\cellcolor[HTML]{FFF7F3}} \color[HTML]{000000} $1$ [$100$] & {\cellcolor[HTML]{FEECE9}} \color[HTML]{000000} $12$ [$40$--$100$] \\
Fixed Angles$^\dagger$ & {\cellcolor[HTML]{000000}} \color[HTML]{F1F1F1} {\cellcolor{white}} \color{black} - & {\cellcolor[HTML]{000000}} \color[HTML]{F1F1F1} {\cellcolor{white}} \color{black} - & {\cellcolor[HTML]{742905}} \color[HTML]{F1F1F1} $97.7$ & {\cellcolor[HTML]{742905}} \color[HTML]{F1F1F1} $97.7\pm 2.2$ & {\cellcolor[HTML]{000000}} \color[HTML]{F1F1F1} {\cellcolor{white}} \color{black} - & {\cellcolor[HTML]{000000}} \color[HTML]{F1F1F1} {\cellcolor{white}} \color{black} - & {\cellcolor[HTML]{FFF7F3}} \color[HTML]{000000} $1$ [$100$] & {\cellcolor[HTML]{FEECE9}} \color[HTML]{000000} $12$ [$40$--$100$] \\
Fourier$^\star$ & {\cellcolor[HTML]{FED36F}} \color[HTML]{000000} $79.6\pm 6.1$ & {\cellcolor[HTML]{FEE596}} \color[HTML]{000000} $77.4\pm 10.9$ & {\cellcolor[HTML]{EB6F14}} \color[HTML]{F1F1F1} $88.3\pm 5.6$ & {\cellcolor[HTML]{CB4B02}} \color[HTML]{F1F1F1} $91.8\pm 9.3$ & {\cellcolor[HTML]{FDE4E1}} \color[HTML]{000000} $20$ [$40$--$50$] & {\cellcolor[HTML]{FDD8D5}} \color[HTML]{000000} $31$ [$39$--$144$] & {\cellcolor[HTML]{FEEEEB}} \color[HTML]{000000} $10$ [$100$] & {\cellcolor[HTML]{FEECE9}} \color[HTML]{000000} $12$ [$40$--$100$] \\
Interp.$^\star$ & {\cellcolor[HTML]{782A05}} \color[HTML]{F1F1F1} $97.4\pm 1.5$ & {\cellcolor[HTML]{F5841E}} \color[HTML]{F1F1F1} $86.6\pm 10.1$ & {\cellcolor[HTML]{7C2C05}} \color[HTML]{F1F1F1} $97.1\pm 0.6$ & {\cellcolor[HTML]{BB4403}} \color[HTML]{F1F1F1} $92.9\pm 5.3$ & {\cellcolor[HTML]{FDE4E1}} \color[HTML]{000000} $20$ [$40$--$50$] & {\cellcolor[HTML]{FDD8D5}} \color[HTML]{000000} $31$ [$39$--$144$] & {\cellcolor[HTML]{FEEEEB}} \color[HTML]{000000} $10$ [$100$] & {\cellcolor[HTML]{FEECE9}} \color[HTML]{000000} $12$ [$40$--$100$] \\
Linear Ramp$^\star$ & {\cellcolor[HTML]{782A05}} \color[HTML]{F1F1F1} $97.5\pm 0.7$ & {\cellcolor[HTML]{FEB23F}} \color[HTML]{000000} $82.7\pm 10.3$ & {\cellcolor[HTML]{8E3104}} \color[HTML]{F1F1F1} $95.9\pm 1.0$ & {\cellcolor[HTML]{812D05}} \color[HTML]{F1F1F1} $96.8\pm 2.1$ & {\cellcolor[HTML]{FDE4E1}} \color[HTML]{000000} $20$ [$40$--$50$] & {\cellcolor[HTML]{FDD8D5}} \color[HTML]{000000} $31$ [$39$--$144$] & {\cellcolor[HTML]{FEEEEB}} \color[HTML]{000000} $10$ [$100$] & {\cellcolor[HTML]{FEECE9}} \color[HTML]{000000} $12$ [$40$--$100$] \\
Linear Ramp & {\cellcolor[HTML]{882F05}} \color[HTML]{F1F1F1} $96.4\pm 0.9$ & {\cellcolor[HTML]{FEB340}} \color[HTML]{000000} $82.6\pm 9.6$ & {\cellcolor[HTML]{A13804}} \color[HTML]{F1F1F1} $94.6\pm 1.1$ & {\cellcolor[HTML]{A03704}} \color[HTML]{F1F1F1} $94.7\pm 3.1$ & {\cellcolor[HTML]{FDE4E1}} \color[HTML]{000000} $20$ [$40$--$50$] & {\cellcolor[HTML]{FDD8D5}} \color[HTML]{000000} $31$ [$39$--$144$] & {\cellcolor[HTML]{FEEEEB}} \color[HTML]{000000} $10$ [$100$] & {\cellcolor[HTML]{FEECE9}} \color[HTML]{000000} $12$ [$40$--$100$] \\
Linear Ramp$^\dagger$ & {\cellcolor[HTML]{9B3504}} \color[HTML]{F1F1F1} $95.0\pm 1.3$ & {\cellcolor[HTML]{FFEFAC}} \color[HTML]{000000} $75.6\pm 7.6$ & {\cellcolor[HTML]{C44802}} \color[HTML]{F1F1F1} $92.2\pm 2.5$ & {\cellcolor[HTML]{FEBA46}} \color[HTML]{000000} $82.1\pm 10.8$ & {\cellcolor[HTML]{FDE4E1}} \color[HTML]{000000} $20$ [$40$--$50$] & {\cellcolor[HTML]{FDD8D5}} \color[HTML]{000000} $31$ [$39$--$144$] & {\cellcolor[HTML]{FEEEEB}} \color[HTML]{000000} $10$ [$100$] & {\cellcolor[HTML]{FEECE9}} \color[HTML]{000000} $12$ [$40$--$100$] \\
Recursive TS$^\star$ & {\cellcolor[HTML]{782A05}} \color[HTML]{F1F1F1} $97.4\pm 0.9$ & {\cellcolor[HTML]{FFFFE5}} \color[HTML]{000000} $70.8\pm 13.9$ & {\cellcolor[HTML]{AC3D03}} \color[HTML]{F1F1F1} $93.8\pm 1.4$ & {\cellcolor[HTML]{AB3C03}} \color[HTML]{F1F1F1} $94.0\pm 2.6$ & {\cellcolor[HTML]{FDE4E1}} \color[HTML]{000000} $20$ [$40$--$50$] & {\cellcolor[HTML]{FDD8D5}} \color[HTML]{000000} $31$ [$39$--$144$] & {\cellcolor[HTML]{FEEEEB}} \color[HTML]{000000} $10$ [$100$] & {\cellcolor[HTML]{FEECE9}} \color[HTML]{000000} $12$ [$40$--$100$] \\
TQA$^\star$ & {\cellcolor[HTML]{CB4B02}} \color[HTML]{F1F1F1} $91.8\pm 2.9$ & {\cellcolor[HTML]{FEC859}} \color[HTML]{000000} $80.8\pm 9.3$ & {\cellcolor[HTML]{BB4403}} \color[HTML]{F1F1F1} $92.9\pm 1.8$ & {\cellcolor[HTML]{8B3005}} \color[HTML]{F1F1F1} $96.1\pm 2.2$ & {\cellcolor[HTML]{FDE4E1}} \color[HTML]{000000} $20$ [$40$--$50$] & {\cellcolor[HTML]{FDD8D5}} \color[HTML]{000000} $31$ [$39$--$144$] & {\cellcolor[HTML]{FEEEEB}} \color[HTML]{000000} $10$ [$100$] & {\cellcolor[HTML]{FEECE9}} \color[HTML]{000000} $12$ [$40$--$100$] \\
TQA & {\cellcolor[HTML]{EE7617}} \color[HTML]{F1F1F1} $87.7\pm 1.8$ & {\cellcolor[HTML]{FED36F}} \color[HTML]{000000} $79.6\pm 9.2$ & {\cellcolor[HTML]{D75908}} \color[HTML]{F1F1F1} $90.5\pm 2.6$ & {\cellcolor[HTML]{BC4503}} \color[HTML]{F1F1F1} $92.8\pm 5.2$ & {\cellcolor[HTML]{FDE4E1}} \color[HTML]{000000} $20$ [$40$--$50$] & {\cellcolor[HTML]{FDD8D5}} \color[HTML]{000000} $31$ [$39$--$144$] & {\cellcolor[HTML]{FEEEEB}} \color[HTML]{000000} $10$ [$100$] & {\cellcolor[HTML]{FEECE9}} \color[HTML]{000000} $12$ [$40$--$100$] \\
TQA$^\dagger$ & {\cellcolor[HTML]{FEA332}} \color[HTML]{000000} $83.9\pm 2.8$ & {\cellcolor[HTML]{FFF4B6}} \color[HTML]{000000} $74.8\pm 6.3$ & {\cellcolor[HTML]{FC9427}} \color[HTML]{000000} $85.1\pm 4.3$ & {\cellcolor[HTML]{FEB643}} \color[HTML]{000000} $82.4\pm 8.3$ & {\cellcolor[HTML]{FDE4E1}} \color[HTML]{000000} $20$ [$40$--$50$] & {\cellcolor[HTML]{FDD8D5}} \color[HTML]{000000} $31$ [$39$--$144$] & {\cellcolor[HTML]{FEEEEB}} \color[HTML]{000000} $10$ [$100$] & {\cellcolor[HTML]{FEECE9}} \color[HTML]{000000} $12$ [$40$--$100$] \\
\cline{1-9}
\multicolumn{9}{c}{\rule{0pt}{8pt}PP} \\
\midrule
Fixed Angles$^\star$ & {\cellcolor[HTML]{000000}} \color[HTML]{F1F1F1} {\cellcolor{white}} \color{black} - & {\cellcolor[HTML]{0A703A}} \color[HTML]{F1F1F1} $94.2\pm 0.6$ & {\cellcolor[HTML]{2C8F4B}} \color[HTML]{F1F1F1} $90.7\pm 1.9$ & {\cellcolor[HTML]{11753D}} \color[HTML]{F1F1F1} $93.5\pm 1.6$ & {\cellcolor[HTML]{000000}} \color[HTML]{F1F1F1} {\cellcolor{white}} \color{black} - & {\cellcolor[HTML]{FDD9D6}} \color[HTML]{000000} $30$ [$39$--$144$] & {\cellcolor[HTML]{FEEEEB}} \color[HTML]{000000} $10$ [$100$] & {\cellcolor[HTML]{FEECE9}} \color[HTML]{000000} $12$ [$40$--$100$] \\
Fixed Angles$^\dagger$ & {\cellcolor[HTML]{000000}} \color[HTML]{F1F1F1} {\cellcolor{white}} \color{black} - & {\cellcolor[HTML]{258745}} \color[HTML]{F1F1F1} $91.5\pm 1.3$ & {\cellcolor[HTML]{5AB76A}} \color[HTML]{F1F1F1} $86.7\pm 1.7$ & {\cellcolor[HTML]{177B3F}} \color[HTML]{F1F1F1} $92.9\pm 1.6$ & {\cellcolor[HTML]{000000}} \color[HTML]{F1F1F1} {\cellcolor{white}} \color{black} - & {\cellcolor[HTML]{FDD9D6}} \color[HTML]{000000} $30$ [$39$--$144$] & {\cellcolor[HTML]{FEEEEB}} \color[HTML]{000000} $10$ [$100$] & {\cellcolor[HTML]{FEECE9}} \color[HTML]{000000} $12$ [$40$--$100$] \\
Fourier$^\star$ & {\cellcolor[HTML]{258745}} \color[HTML]{F1F1F1} $91.5\pm 1.9$ & {\cellcolor[HTML]{268846}} \color[HTML]{F1F1F1} $91.3\pm 3.1$ & {\cellcolor[HTML]{64BC6F}} \color[HTML]{F1F1F1} $86.0\pm 2.2$ & {\cellcolor[HTML]{2C8F4B}} \color[HTML]{F1F1F1} $90.7\pm 2.4$ & {\cellcolor[HTML]{FDE4E1}} \color[HTML]{000000} $20$ [$40$--$50$] & {\cellcolor[HTML]{FDD9D6}} \color[HTML]{000000} $30$ [$39$--$144$] & {\cellcolor[HTML]{E23E99}} \color[HTML]{F1F1F1} $111$ [$40$--$100$] & {\cellcolor[HTML]{FEE6E3}} \color[HTML]{000000} $18$ [$40$--$100$] \\
Interp.$^\star$ & {\cellcolor[HTML]{086E3A}} \color[HTML]{F1F1F1} $94.4\pm 1.9$ & {\cellcolor[HTML]{016937}} \color[HTML]{F1F1F1} $95.0\pm 0.7$ & {\cellcolor[HTML]{359C53}} \color[HTML]{F1F1F1} $89.6\pm 1.7$ & {\cellcolor[HTML]{2F934D}} \color[HTML]{F1F1F1} $90.4\pm 2.5$ & {\cellcolor[HTML]{FDE4E1}} \color[HTML]{000000} $20$ [$40$--$50$] & {\cellcolor[HTML]{FDD9D6}} \color[HTML]{000000} $30$ [$39$--$144$] & {\cellcolor[HTML]{E23E99}} \color[HTML]{F1F1F1} $111$ [$40$--$100$] & {\cellcolor[HTML]{FEE6E3}} \color[HTML]{000000} $18$ [$40$--$100$] \\
Linear Ramp$^\star$ & {\cellcolor[HTML]{006435}} \color[HTML]{F1F1F1} $95.6\pm 1.2$ & {\cellcolor[HTML]{11753D}} \color[HTML]{F1F1F1} $93.5\pm 0.6$ & {\cellcolor[HTML]{2D914B}} \color[HTML]{F1F1F1} $90.6\pm 2.5$ & {\cellcolor[HTML]{197C40}} \color[HTML]{F1F1F1} $92.7\pm 1.5$ & {\cellcolor[HTML]{FDE4E1}} \color[HTML]{000000} $20$ [$40$--$50$] & {\cellcolor[HTML]{FDD9D6}} \color[HTML]{000000} $30$ [$39$--$144$] & {\cellcolor[HTML]{E23E99}} \color[HTML]{F1F1F1} $111$ [$40$--$100$] & {\cellcolor[HTML]{FEE6E3}} \color[HTML]{000000} $18$ [$40$--$100$] \\
Linear Ramp & {\cellcolor[HTML]{056C39}} \color[HTML]{F1F1F1} $94.7\pm 1.2$ & {\cellcolor[HTML]{167A3F}} \color[HTML]{F1F1F1} $93.0\pm 0.7$ & {\cellcolor[HTML]{379E54}} \color[HTML]{F1F1F1} $89.4\pm 2.4$ & {\cellcolor[HTML]{218242}} \color[HTML]{F1F1F1} $91.9\pm 1.5$ & {\cellcolor[HTML]{FDE4E1}} \color[HTML]{000000} $20$ [$40$--$50$] & {\cellcolor[HTML]{FDD9D6}} \color[HTML]{000000} $30$ [$39$--$144$] & {\cellcolor[HTML]{E23E99}} \color[HTML]{F1F1F1} $111$ [$40$--$100$] & {\cellcolor[HTML]{FEE6E3}} \color[HTML]{000000} $18$ [$40$--$100$] \\
Linear Ramp$^\dagger$ & {\cellcolor[HTML]{0D733C}} \color[HTML]{F1F1F1} $93.9\pm 1.1$ & {\cellcolor[HTML]{70C275}} \color[HTML]{000000} $85.2\pm 1.2$ & {\cellcolor[HTML]{49AF61}} \color[HTML]{F1F1F1} $87.7\pm 2.1$ & {\cellcolor[HTML]{43AC5E}} \color[HTML]{F1F1F1} $88.1\pm 1.9$ & {\cellcolor[HTML]{FDE4E1}} \color[HTML]{000000} $20$ [$40$--$50$] & {\cellcolor[HTML]{FDD9D6}} \color[HTML]{000000} $30$ [$39$--$144$] & {\cellcolor[HTML]{E23E99}} \color[HTML]{F1F1F1} $111$ [$40$--$100$] & {\cellcolor[HTML]{FEE6E3}} \color[HTML]{000000} $18$ [$40$--$100$] \\
Recursive TS$^\star$ & {\cellcolor[HTML]{2F934D}} \color[HTML]{F1F1F1} $90.4\pm 1.2$ & {\cellcolor[HTML]{359C53}} \color[HTML]{F1F1F1} $89.6\pm 1.5$ & {\cellcolor[HTML]{51B365}} \color[HTML]{F1F1F1} $87.2\pm 2.2$ & {\cellcolor[HTML]{3EA75A}} \color[HTML]{F1F1F1} $88.6\pm 1.9$ & {\cellcolor[HTML]{FDE4E1}} \color[HTML]{000000} $20$ [$40$--$50$] & {\cellcolor[HTML]{FDD9D6}} \color[HTML]{000000} $30$ [$39$--$144$] & {\cellcolor[HTML]{E23E99}} \color[HTML]{F1F1F1} $111$ [$40$--$100$] & {\cellcolor[HTML]{FEE6E3}} \color[HTML]{000000} $18$ [$40$--$100$] \\
TQA$^\star$ & {\cellcolor[HTML]{268846}} \color[HTML]{F1F1F1} $91.3\pm 2.5$ & {\cellcolor[HTML]{228343}} \color[HTML]{F1F1F1} $91.8\pm 0.7$ & {\cellcolor[HTML]{6BC072}} \color[HTML]{000000} $85.6\pm 3.0$ & {\cellcolor[HTML]{248644}} \color[HTML]{F1F1F1} $91.6\pm 1.6$ & {\cellcolor[HTML]{FDE4E1}} \color[HTML]{000000} $20$ [$40$--$50$] & {\cellcolor[HTML]{FDD9D6}} \color[HTML]{000000} $30$ [$39$--$144$] & {\cellcolor[HTML]{E23E99}} \color[HTML]{F1F1F1} $111$ [$40$--$100$] & {\cellcolor[HTML]{FEE6E3}} \color[HTML]{000000} $18$ [$40$--$100$] \\
TQA & {\cellcolor[HTML]{7EC97B}} \color[HTML]{000000} $84.4\pm 2.4$ & {\cellcolor[HTML]{31974F}} \color[HTML]{F1F1F1} $90.0\pm 0.9$ & {\cellcolor[HTML]{DCF1A5}} \color[HTML]{000000} $77.5\pm 7.8$ & {\cellcolor[HTML]{379E54}} \color[HTML]{F1F1F1} $89.4\pm 3.0$ & {\cellcolor[HTML]{FDE4E1}} \color[HTML]{000000} $20$ [$40$--$50$] & {\cellcolor[HTML]{FDD9D6}} \color[HTML]{000000} $30$ [$39$--$144$] & {\cellcolor[HTML]{E23E99}} \color[HTML]{F1F1F1} $111$ [$40$--$100$] & {\cellcolor[HTML]{FEE6E3}} \color[HTML]{000000} $18$ [$40$--$100$] \\
TQA$^\dagger$ & {\cellcolor[HTML]{8BCE81}} \color[HTML]{000000} $83.5\pm 2.7$ & {\cellcolor[HTML]{A9DB8C}} \color[HTML]{000000} $81.5\pm 1.7$ & {\cellcolor[HTML]{F1FAB5}} \color[HTML]{000000} $75.0\pm 7.2$ & {\cellcolor[HTML]{66BD70}} \color[HTML]{F1F1F1} $85.9\pm 1.8$ & {\cellcolor[HTML]{FDE4E1}} \color[HTML]{000000} $20$ [$40$--$50$] & {\cellcolor[HTML]{FDD9D6}} \color[HTML]{000000} $30$ [$39$--$144$] & {\cellcolor[HTML]{E23E99}} \color[HTML]{F1F1F1} $111$ [$40$--$100$] & {\cellcolor[HTML]{FEE6E3}} \color[HTML]{000000} $18$ [$40$--$100$] \\
\cline{1-9}
\multicolumn{9}{c}{\rule{0pt}{8pt}SV} \\
\midrule
Fixed Angles$^\star$ & {\cellcolor[HTML]{03446A}} \color[HTML]{F1F1F1} $97.4\pm 1.4$ & {\cellcolor[HTML]{000000}} \color[HTML]{F1F1F1} {\cellcolor{white}} \color{black} - & {\cellcolor[HTML]{03446A}} \color[HTML]{F1F1F1} $97.5\pm 1.3$ & {\cellcolor[HTML]{023858}} \color[HTML]{F1F1F1} $98.7\pm 1.0$ & {\cellcolor[HTML]{FEE7E4}} \color[HTML]{000000} $17$ [$11$--$20$] & {\cellcolor[HTML]{000000}} \color[HTML]{F1F1F1} {\cellcolor{white}} \color{black} - & {\cellcolor[HTML]{FEE9E5}} \color[HTML]{000000} $16$ [$10$--$20$] & {\cellcolor[HTML]{FBAFBA}} \color[HTML]{000000} $60$ [$10$--$20$] \\
Fixed Angles$^\dagger$ & {\cellcolor[HTML]{056EAD}} \color[HTML]{F1F1F1} $92.0\pm 3.6$ & {\cellcolor[HTML]{000000}} \color[HTML]{F1F1F1} {\cellcolor{white}} \color{black} - & {\cellcolor[HTML]{034F7D}} \color[HTML]{F1F1F1} $96.3\pm 1.6$ & {\cellcolor[HTML]{023E62}} \color[HTML]{F1F1F1} $98.0\pm 1.0$ & {\cellcolor[HTML]{FEE7E4}} \color[HTML]{000000} $17$ [$11$--$20$] & {\cellcolor[HTML]{000000}} \color[HTML]{F1F1F1} {\cellcolor{white}} \color{black} - & {\cellcolor[HTML]{FEE9E5}} \color[HTML]{000000} $16$ [$10$--$20$] & {\cellcolor[HTML]{FBAFBA}} \color[HTML]{000000} $60$ [$10$--$20$] \\
Fourier$^\star$ & {\cellcolor[HTML]{CDD0E5}} \color[HTML]{000000} $78.1\pm 13.1$ & {\cellcolor[HTML]{EDE7F2}} \color[HTML]{000000} $74.2\pm 12.3$ & {\cellcolor[HTML]{D8D7E9}} \color[HTML]{000000} $76.8\pm 11.3$ & {\cellcolor[HTML]{ADC1DD}} \color[HTML]{000000} $80.7\pm 14.0$ & {\cellcolor[HTML]{F769A1}} \color[HTML]{F1F1F1} $92$ [$10$--$20$] & {\cellcolor[HTML]{FDE4E1}} \color[HTML]{000000} $20$ [$12$--$21$] & {\cellcolor[HTML]{51006C}} \color[HTML]{F1F1F1} $180$ [$10$--$20$] & {\cellcolor[HTML]{49006A}} \color[HTML]{F1F1F1} $184$ [$10$--$20$] \\
Interp.$^\star$ & {\cellcolor[HTML]{034D79}} \color[HTML]{F1F1F1} $96.5\pm 1.1$ & {\cellcolor[HTML]{045D92}} \color[HTML]{F1F1F1} $94.7\pm 1.0$ & {\cellcolor[HTML]{04588A}} \color[HTML]{F1F1F1} $95.4\pm 1.9$ & {\cellcolor[HTML]{056DAC}} \color[HTML]{F1F1F1} $92.1\pm 6.8$ & {\cellcolor[HTML]{F769A1}} \color[HTML]{F1F1F1} $92$ [$10$--$20$] & {\cellcolor[HTML]{FDE4E1}} \color[HTML]{000000} $20$ [$12$--$21$] & {\cellcolor[HTML]{51006C}} \color[HTML]{F1F1F1} $180$ [$10$--$20$] & {\cellcolor[HTML]{49006A}} \color[HTML]{F1F1F1} $184$ [$10$--$20$] \\
Linear Ramp$^\star$ & {\cellcolor[HTML]{034B76}} \color[HTML]{F1F1F1} $96.7\pm 1.1$ & {\cellcolor[HTML]{04598C}} \color[HTML]{F1F1F1} $95.3\pm 0.8$ & {\cellcolor[HTML]{045788}} \color[HTML]{F1F1F1} $95.5\pm 1.6$ & {\cellcolor[HTML]{034165}} \color[HTML]{F1F1F1} $97.8\pm 1.2$ & {\cellcolor[HTML]{F769A1}} \color[HTML]{F1F1F1} $92$ [$10$--$20$] & {\cellcolor[HTML]{FDE4E1}} \color[HTML]{000000} $20$ [$12$--$21$] & {\cellcolor[HTML]{51006C}} \color[HTML]{F1F1F1} $180$ [$10$--$20$] & {\cellcolor[HTML]{49006A}} \color[HTML]{F1F1F1} $184$ [$10$--$20$] \\
Linear Ramp & {\cellcolor[HTML]{045B8E}} \color[HTML]{F1F1F1} $95.0\pm 1.4$ & {\cellcolor[HTML]{045E93}} \color[HTML]{F1F1F1} $94.6\pm 0.9$ & {\cellcolor[HTML]{04649E}} \color[HTML]{F1F1F1} $93.5\pm 1.9$ & {\cellcolor[HTML]{034F7D}} \color[HTML]{F1F1F1} $96.2\pm 1.7$ & {\cellcolor[HTML]{F769A1}} \color[HTML]{F1F1F1} $92$ [$10$--$20$] & {\cellcolor[HTML]{FDE4E1}} \color[HTML]{000000} $20$ [$12$--$21$] & {\cellcolor[HTML]{51006C}} \color[HTML]{F1F1F1} $180$ [$10$--$20$] & {\cellcolor[HTML]{49006A}} \color[HTML]{F1F1F1} $184$ [$10$--$20$] \\
Linear Ramp$^\dagger$ & {\cellcolor[HTML]{0A73B2}} \color[HTML]{F1F1F1} $91.3\pm 2.1$ & {\cellcolor[HTML]{73A9CF}} \color[HTML]{F1F1F1} $84.8\pm 4.3$ & {\cellcolor[HTML]{167BB6}} \color[HTML]{F1F1F1} $90.5\pm 3.2$ & {\cellcolor[HTML]{0569A5}} \color[HTML]{F1F1F1} $92.8\pm 3.3$ & {\cellcolor[HTML]{F769A1}} \color[HTML]{F1F1F1} $92$ [$10$--$20$] & {\cellcolor[HTML]{FDE4E1}} \color[HTML]{000000} $20$ [$12$--$21$] & {\cellcolor[HTML]{51006C}} \color[HTML]{F1F1F1} $180$ [$10$--$20$] & {\cellcolor[HTML]{49006A}} \color[HTML]{F1F1F1} $184$ [$10$--$20$] \\
Recursive TS$^\star$ & {\cellcolor[HTML]{045B8E}} \color[HTML]{F1F1F1} $95.0\pm 2.0$ & {\cellcolor[HTML]{0569A5}} \color[HTML]{F1F1F1} $92.8\pm 2.4$ & {\cellcolor[HTML]{056BA7}} \color[HTML]{F1F1F1} $92.5\pm 2.2$ & {\cellcolor[HTML]{04629A}} \color[HTML]{F1F1F1} $93.8\pm 3.1$ & {\cellcolor[HTML]{F769A1}} \color[HTML]{F1F1F1} $92$ [$10$--$20$] & {\cellcolor[HTML]{FDE4E1}} \color[HTML]{000000} $20$ [$12$--$21$] & {\cellcolor[HTML]{51006C}} \color[HTML]{F1F1F1} $180$ [$10$--$20$] & {\cellcolor[HTML]{49006A}} \color[HTML]{F1F1F1} $184$ [$10$--$20$] \\
TQA$^\star$ & {\cellcolor[HTML]{04598C}} \color[HTML]{F1F1F1} $95.3\pm 1.3$ & {\cellcolor[HTML]{0569A5}} \color[HTML]{F1F1F1} $92.8\pm 1.9$ & {\cellcolor[HTML]{0567A1}} \color[HTML]{F1F1F1} $93.2\pm 2.5$ & {\cellcolor[HTML]{034B76}} \color[HTML]{F1F1F1} $96.7\pm 1.5$ & {\cellcolor[HTML]{F769A1}} \color[HTML]{F1F1F1} $92$ [$10$--$20$] & {\cellcolor[HTML]{FDE4E1}} \color[HTML]{000000} $20$ [$12$--$21$] & {\cellcolor[HTML]{51006C}} \color[HTML]{F1F1F1} $180$ [$10$--$20$] & {\cellcolor[HTML]{49006A}} \color[HTML]{F1F1F1} $184$ [$10$--$20$] \\
TQA & {\cellcolor[HTML]{0872B1}} \color[HTML]{F1F1F1} $91.4\pm 3.7$ & {\cellcolor[HTML]{2081B9}} \color[HTML]{F1F1F1} $89.8\pm 3.3$ & {\cellcolor[HTML]{7DACD1}} \color[HTML]{F1F1F1} $84.1\pm 8.2$ & {\cellcolor[HTML]{056FAE}} \color[HTML]{F1F1F1} $91.9\pm 4.6$ & {\cellcolor[HTML]{F769A1}} \color[HTML]{F1F1F1} $92$ [$10$--$20$] & {\cellcolor[HTML]{FDE4E1}} \color[HTML]{000000} $20$ [$12$--$21$] & {\cellcolor[HTML]{51006C}} \color[HTML]{F1F1F1} $180$ [$10$--$20$] & {\cellcolor[HTML]{49006A}} \color[HTML]{F1F1F1} $184$ [$10$--$20$] \\
TQA$^\dagger$ & {\cellcolor[HTML]{3790C0}} \color[HTML]{F1F1F1} $88.2\pm 2.8$ & {\cellcolor[HTML]{B5C4DF}} \color[HTML]{000000} $80.0\pm 5.1$ & {\cellcolor[HTML]{ACC0DD}} \color[HTML]{000000} $80.8\pm 8.1$ & {\cellcolor[HTML]{2C89BD}} \color[HTML]{F1F1F1} $89.0\pm 3.9$ & {\cellcolor[HTML]{F769A1}} \color[HTML]{F1F1F1} $92$ [$10$--$20$] & {\cellcolor[HTML]{FDE4E1}} \color[HTML]{000000} $20$ [$12$--$21$] & {\cellcolor[HTML]{51006C}} \color[HTML]{F1F1F1} $180$ [$10$--$20$] & {\cellcolor[HTML]{49006A}} \color[HTML]{F1F1F1} $184$ [$10$--$20$] \\
\cline{1-9}
\bottomrule
\end{tabular}

    \caption{Average approximation ratio and number of graph instances for different methods and graph types, with depth-ten QAOA.
    Graph types shown are Erd\H{o}s-R\'{e}nyi (ER), heavy-hex (HH), line-based (LB), and random-regular (RR).
    The average approximation ratios are shown, with their standard deviation, over all graph instances of a given type.
    The number of evaluated graph instances are shown with the range of number of nodes in square brackets.
    Combinations of methods and graph types that were not evaluated are denoted with a hyphen.
    Methods that employed QAOA angle optimization are marked with a star, e.g., Fixed Angle$^\star$.
    Cells are colored such that higher values are darker.
    Purple, orange, green, and blue denote Matrix Product State (MPS) with Qiskit Aer, MPS with Quimb, Pauli-Propagation (PP), and Statevector (SV) energy-evaluation methods for approximation ratios, respectively.
    Pink/purple is used for all cells denoting the number of instances, regardless of the evaluation method.
    }
    \label{tab:summary_p10}
\end{table*}


\section{Benchmarking results\label{sec:res}}

In total, we benchmarked 34 methods to find QAOA angles for the different problem instances resulting in running 79\,908 trainer chains on classical hardware, see App.~\ref{app:training_pipeline}.
We focus on quality in Sec.~\ref{sec:results_quality}, and then validate the angles on quantum hardware in Sec.~\ref{sec:results_validation}.
Sec.~\ref{sec:results_ressources} presents a resource cost analysis.

\subsection{Quality\label{sec:results_quality}}

We compare the quality of angle training methods from the perspective of the (possibly approximated) energy.
The methods are comparable as long as the same energy evaluator is used.
We report approximation ratios for MaxCut problems by converting the energy $\langle H_C\rangle$ to a cut value $C=\langle H_C\rangle+\frac{1}{2}\sum_{(i,j)\in E}w_{ij}$ where $w_{ij}$ is the weight of edge $(i,j)\in E$.
Next, $C$ is converted to an approximation ratio using the minimum and maximum cut values found with CPLEX.

\begin{figure}[htbp]
    \centering
    \includegraphics[width=\columnwidth]{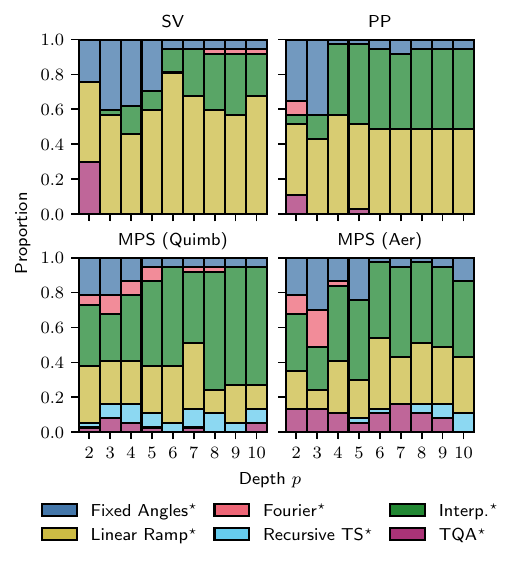}
    \caption{
    Proportion of the best methods for MaxCut over a set of 37 graphs as function of QAOA depth.
    The only methods not present in the legend are unoptimized variants, e.g., Fixed Angles vs. Fixed Angles$^\star$.
    A method is the \emph{best} for a given graph and depth if it achieves the highest energy over all methods.
    The lack of fixed angles for some instances is recorded as a failure of the method.
    The SV panel aggregates data over ER, HH, LB, and RR graphs with sizes 20, 21, 20, and 20, respectively.
    The PP and MPS panels aggregate data over ER, HH, LB, and RR graphs with sizes 50, 144, 100, and 100, respectively.
    }
    \label{fig:best_methods}
\end{figure}

First, we test the different methods on two sets of $37$ graphs with different energy evaluation methods and QAOA depths up to $p=10$.
The first and second sets correspond to \textit{small} and \textit{large} graphs, for SV and PP, MPS (Quimb), and MPS (Aer), respectively.
Each set of $37$ instances has $10$ Erd\H{o}s-R\'enyi (ER) graphs with $20\%$ edge probability, $10$ heavy-hex (HH) graphs, $10$ line-based (LB) graphs with $1$ through $10$ swap layers, and $7$ random-regular (RR) graphs with degrees $d=3,4,\ldots,9$.
The number of nodes per graph-type for the small set (large set) are $20$ ($50$), $21$ ($144$), $20$ ($100$), and $20$ ($100$) nodes, respectively.
For each graph we run the six selected angle setting methods and report the fraction of graphs for which a method gave the best QAOA angles as measured by the energy of the (possibly approximate) evaluator, see Fig.~\ref{fig:best_methods}.

Linear Ramps and Interp. provide the best angles for most instances, across all energy evaluation methods.
Fourier, TQA, and RTS only do well on a few instances.
The fixed angles, optimized to each instance, perform well when they are available, but $75.4\%$ of the $666$ graph-depth combinations did not have fixed angles in our database, primarily for $p\geq{}4$.
Comparing the two best methods for the SV, MPS (Quimb), and MPS (Aer) energy evaluation methods, we observe that the best method's lead is significantly larger for random-regular graphs than for other graph types.
For example, Interp.$^\star$ energies are up to $10\%$ smaller than Linear Ramp$^\star$ energies for the statevector energy evaluator, for all graph types other than random-regular where they are up to about $50\%$ smaller.
In some cases, this lead for random-regular graphs is larger for lower degrees.
Additional analysis is in App.~\ref{app:best_method_distribution}.

Next, we broaden the pool of graph instances and present $p=10$ results in Tab.~\ref{tab:summary_p10}.
Here, a parameter transfer from fixed angles based on the average node degree performs very well, when available, since the resulting approximation ratios are often above 90\%, see MPS and PP in Tab.~\ref{tab:summary_p10}.
Furthermore, Interp. and Linear Ramps also perform well.
TQA and Fourier underperform the other methods.
The approximation ratios of TQA and Linear Ramps increase as we first optimize the slopes of $(\boldsymbol{\beta}, \boldsymbol{\gamma})$ and then each individual QAOA angle, as also observed in Ref.~\cite{eichenseher2025pattern}.
However, this performance gain requires a longer runtime.
Finally, RTS did not produce QAOA angles for utility-scale problems at the maximum considered depth $p=10$. 
Here, the energy approximation takes too long to optimize angles starting from the $2p+1$ transition states.
For problem instances with 20 nodes or fewer a statevector energy evaluation is possible. 
Here, starting the QAOA angle optimization from fixed angles, when available, and further optimizing them also performs the best, see SV in Tab.~\ref{tab:summary_p10}.
By design, these angles perform well on random-regular graphs and optimizing them to each graph instance slightly increases the approximation ratio.
Interestingly, such fixed angles also serve as good initial points for an optimization on Erd\H{o}s-R\'enyi and line-based graphs, and lead to better angles than, for example, Interp.
Additional analysis of the SV results is in App.~\ref{app:sv_results}.

\begin{figure}[h]
    \centering
    \includegraphics[width=0.99\linewidth]{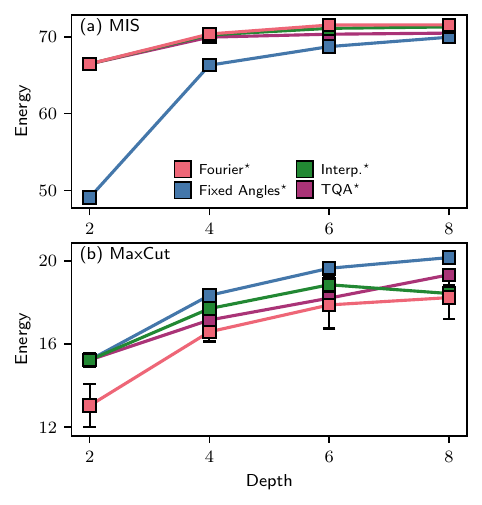}
    \caption{Comparison of the achieved energy values for the Fourier, Fixed Angles, Interp., and TQA methods, and the PP evaluation method with different results for MIS (a) and MaxCut (b). 
    We use ten 40-node random three-regular graphs. 
    The markers and error bars indicate the mean and standard deviation, respectively, of the energy over the ten instances.}
    \label{fig:mis_vs_maxcut}
\end{figure}

The results in Tab.~\ref{tab:summary_p10} are for MaxCut problems.
We now change the problem instances to MIS.
The angle setting methods behave differently on MIS than MaxCut, see Fig.~\ref{fig:mis_vs_maxcut}.
First, the fixed angles do not perform well on MIS even after an optimization (blue lines).
This is not surprising since the fixed angles in Ref.~\cite{WurtzGithub} are designed for MaxCut.
These angles could be rescaled to enhance their performance following Ref.~\cite{Sureshbabu2024}.
On MaxCut the Fourier method performs the worst while on MIS it is among the best performing methods along with TQA as measured by the energy, see Fig.~\ref{fig:mis_vs_maxcut}.
The data thus show that different methods can have a different performance depending on the problem class.

To extend our benchmarks beyond quadratic optimization, we evaluate the angle-setting methods on the LABS problem~\cite{bernasconi1987,Mertens_1996,Packebusch_2016} by evaluating them in terms of the average energy, average normalized Merit Factor $MF_\text{norm} = \langle MF \rangle / MF_\text{opt}$, and Time-to-Solution (TTS), see App.~\ref{app:labs}. 
The Fourier and Interp. methods consistently perform the best across all system sizes and depths, yielding virtually identical results, see Tab.~\ref{tab:labs_mftts} for $p=50$ QAOA.  
A direct comparison of the optimized angle schedules across all LABS sizes $n$ confirms that Fourier and Interp. converge to the same QAOA angles. 
The Linear Ramp method is also competitive: it produces a lower $MF_\text{norm}$ but requires only $\sim 5\%$ of the compute time of the iterative methods.
This is interesting since it parameterizes the full schedule with just two slopes $\Delta_\beta$ and $\Delta_\gamma$. 
We initialize the slopes with only four values $\{0.01, 0.05, 0.1, 0.2\}$, applied identically to $\Delta_\beta$ and $\Delta_\gamma$.
Further optimizing each angle in the ramps, Linear Ramps$^\star$, brings a small improvement in $MF_\text{norm}$. 
A finer or asymmetric grid over $(\Delta_\beta, \Delta_\gamma)$ might close the remaining gap with the iterative methods.
Crucially, the iterative methods are very sensitive to the $p = 1$ initialization. 
The depth-one landscape is $\pi$-periodic with extremely narrow basins of attraction, see App.~\ref{app:labs}, and a poor initial point can lead to severe phase over-rotation at higher depths.
Linear Ramp sidesteps this issue by parametrizing the schedule globally rather than building it layer-by-layer, making it an attractive default for LABS and, more broadly, for higher-order problems with similarly rugged landscapes.

\begin{table*}[t]
\centering
\caption{
Normalized Merit factor ($MF_{\text{norm}} = \langle MF \rangle / MF_{\text{opt}}$) and Time-to-Solution (TTS) of the different angle setting methods as a function of the number of variables $n$ in the LABS Hamiltonian. 
Data is provided for a QAOA depth of $p=50$. 
RTS results are for $p=10$. 
Here, LR stands for Linear Ramps and LR/LR$^\star$ and TQA/TQA$^\star$ represent the best results across all scanned initial slopes.
Results are reported in the format $(MF_{\text{norm}}, TTS)$.
Grey cells highlight the highest $MF_{\text{norm}}$.
}
\label{tab:labs_mftts}
\footnotesize
\setlength{\tabcolsep}{4.5pt} 
\begin{tabular}{l *{8}{c}}
\hline\hline
& \multicolumn{8}{c}{Node count ($n$)} \\
Method & 14 & 15 & 16 & 17 & 18 & 19 & 20 & 21 \\ 
\midrule
Fourier     & (0.97, 1.1) & (0.67, 7.7) & (0.81, 4.7) & (0.86, 4.3) & (0.69, 15.9)  & (0.71, 61.9) & (0.56, 156) & (0.50, 554) \\
Interp.      & \cellcolor[HTML]{C0C0C0}(0.97, 1.1) & \cellcolor[HTML]{C0C0C0}(0.68, 5.5) & \cellcolor[HTML]{C0C0C0}(0.83, 4.5) & \cellcolor[HTML]{C0C0C0}(0.86, 4.2) & \cellcolor[HTML]{C0C0C0}(0.69, 15.3) & \cellcolor[HTML]{C0C0C0}(0.71, 56.5) & \cellcolor[HTML]{C0C0C0}(0.56, 151) & \cellcolor[HTML]{C0C0C0}(0.50, 513) \\
Trivial     & (0.81, 2.5) & (0.55, 26.2) & (0.71, 13.5) & (0.78, 11.4) & (0.59, 38.9)         & (0.59, 114)  & (0.48, 406) & (0.42, $1 \cdot 10^3$) \\
RTS         & (0.66, 7.0) & (0.43, 75.4) & (0.58, 40.4) & (0.66, 45.8) & (0.46,   160)       & (0.48, 573)  & (0.39, $1 \cdot 10^3$) & (0.35, $2 \cdot 10^3$) \\
TQA         & (0.27, 289) & (0.19, $1 \cdot 10^3$) & (0.26, $1 \cdot 10^3$) & (0.29, $2 \cdot 10^3$) & (0.20, $1 \cdot 10^4$) & (0.21, $3 \cdot 10^5$) & (0.17, $1 \cdot 10^5$) & (0.15, $2 \cdot 10^5$) \\
TQA$^\star$ & (0.29, 171) & (0.19, $2 \cdot 10^3$) & (0.26, 655) & (0.29, $2 \cdot 10^3$) & (0.20, $1 \cdot 10^4$) & (0.21, $3 \cdot 10^5$) & (0.17, $5 \cdot 10^4$) & (0.15, $1 \cdot 10^5$) \\
LR          & (0.79, 2.7) & (0.53, 23.6) & (0.71, 14.5) & (0.80, 12.8) & (0.59, 34.9) & (0.59, 103) & (0.48, 321) & (0.44, 699) \\
LR$^\star$  & (0.81, 2.3) & (0.56, 21.0) & (0.73, 12.8) & (0.80, 10.4) & (0.60, 30.8) & (0.61, 103) & (0.49, 338) & (0.45, 925) \\ 
\hline \hline
\end{tabular}
\end{table*}

\subsection{Validation\label{sec:results_validation}}

Sec.~\ref{sec:results_quality} compares one angle setting method to another without telling us how good the QAOA angles optimized with MPS or PP are on an ideal quantum computer.
We now validate the QAOA angles by looking at problems that either lie on the edge of classical simulability or that have low enough density to be meaningfully executed at utility-scale on quantum hardware.

\subsubsection{Exact simulability\label{sec:exact_sv_40q}}

\begin{figure}
    \centering
    \includegraphics[width=0.96\columnwidth]{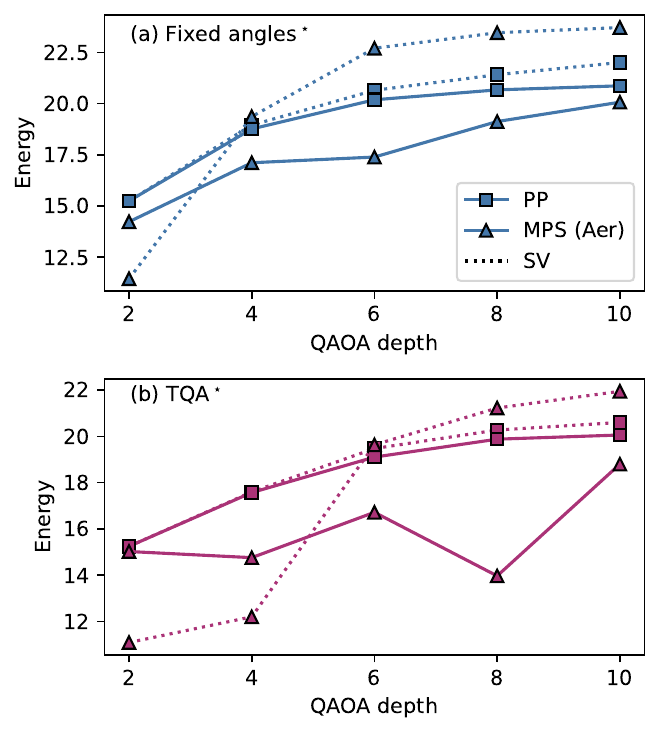}
    \caption{Energy versus QAOA layers comparison of PP and MPS approximation methods (solid lines) and exact solution (dotted lines) of QAOA simulation for a 40-qubit three-regular MaxCut problem. 
    The parameters in (a) and (b) are optimized fixed angles and TQA, respectively.}
    \label{fig:pp_mps_sv}
\end{figure}

We compare the energy of QAOA angles optimized
with PP and MPS to exact results.
Here, we consider a 40-qubit MaxCut problem on a three-regular graph.
The angles are found by optimizing fixed angles, see Fig.~\ref{fig:pp_mps_sv}(a), and TQA schedules, see Fig.~\ref{fig:pp_mps_sv}(b), to the specific problem instance with MPS (solid lines with triangles) and PP (solid lines with squares).
Here, the MPS bond dimension is 32 and the PP maximum weight and truncation threshold are 6 and $10^{-4}$, respectively.
The optimized energy of the approximate methods is compared to the exact state vector simulation of the QAOA circuit (dotted lines).
The full state vector simulations are run using JUQCS-G software \cite{Willsch2022} on JUWELS Booster, a cluster of 3744 NVIDIA A100 Tensor Core GPUs, integrated into the modular supercomputer JUWELS~\cite{Krause2019, JuwelsClusterBooster}.
The simulation of a 40-qubit problems requires approximately 256 compute nodes, utilizing 1,024 NVIDIA A100 GPUs, and takes about five minutes to complete on average.

At first glance, the MPS-based optimization appears to yield lower energy values than the PP approach, compare the triangular and square markers with solid lines in Fig.~\ref{fig:pp_mps_sv}(a). 
However, evaluating the same QAOA angles with an exact state-vector simulation (dashed lines) shows that the MPS energies are generally underestimated. 
Importantly, this underestimation does not prevent the optimizer from identifying high-quality angles. 
On the contrary, the angles obtained with MPS lead to exact energies that become competitive with, and eventually improve upon, those found with PP as the depth increases.

The optimization of the fixed angles method with the Qiskit Aer MPS energy evaluation was $1.5\pm0.8$ times faster than the optimization with PP.
For the TQA method, this factor was $2.9\pm1.6$.
Therefore, on average the MPS optimization was faster than the PP based optimization.
While PP follows the true energy more closely than the MPS data, this improved accuracy does not necessarily translate into better QAOA angles at larger depth. 
These observations indicate that even in the absence of exact energies, the MPS method can guide the search towards the region of optimal angles in less time for the considered instance. 
Similar behavior is observed for angles found with TQA, see Fig.~\ref{fig:pp_mps_sv}(b).

\subsubsection{Hardware\label{sec:hardware}}

\begin{figure*}
    \centering
    \includegraphics[width=0.95\textwidth, clip, trim=0 5 0 10]{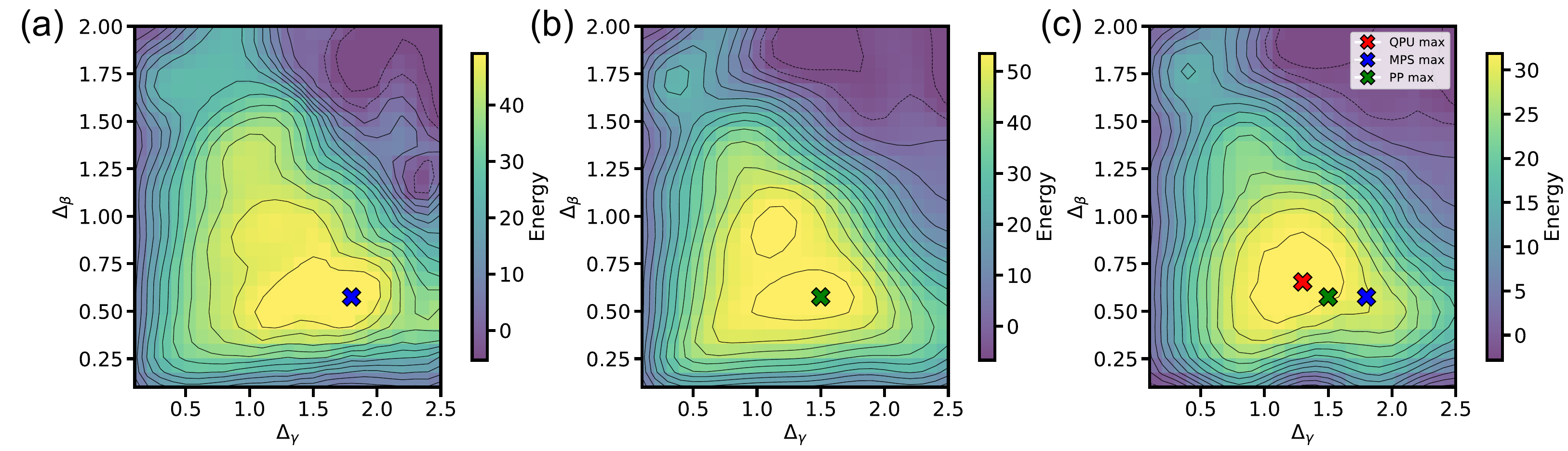}
    \caption{
    Energy landscapes for a 144-qubit heavy-hex MaxCut problem with Linear Ramp at depth $p=5$. 
    (a) MPS with bond dimension 32 and SAT mapping. 
    (b) PP with W=4 and MAC=0.0001. 
    (c) QPU \emph{ibm\_fez} with 1000 shots per $(\Delta_\beta, \Delta_\gamma)$ configuration.
    The crosses indicate the location of the maximum energy identified by each method.  
    }
    \label{fig:evaluators_landscape}
\end{figure*}

In some cases, we can assess the quality of the QAOA angles obtained with MPS and PP with quantum hardware at utility-scale.
We exemplify this first with Linear Ramps on a depth-five QAOA for a 144-qubit heavy-hex weighted MaxCut problem.
We compute the QAOA energy landscape with an MPS and PP for $\Delta_\beta\in[0, 2]$ and $\Delta_\gamma\in[0,2.5]$, see Fig.~\ref{fig:evaluators_landscape}(a) and (b), respectively.
The energy is computed over 625 individual data points. 
The MPS and PP simulations run on a compute node with two AMD EPYC 7402 processors (96 CPUs), requiring 42.2 minutes to complete both simulations. 
The MPS bond dimension is 32 and the PP maximum Pauli weight and coefficient are $4$ and $0.0001$, respectively.
We then compare these energy landscapes to one measured on the \emph{ibm\_fez} QPU by sampling 1000 shots per datapoint which took approximately 3 minutes in total, see Fig.~\ref{fig:evaluators_landscape}(c).
Here, the PP evaluator is a better approximation to the energy than the MPS. 
The PP landscape is closer to the QPU's landscape than the MPS. 
App.~\ref{app:example_hw} gives a second quantum hardware validation example for a MaxCut problem on a 100 node three-regular graph at depth-one where an exact energy evaluation is possible classically.

Despite noise, we may still see a sufficient signal to distinguish whether a given set of angles is better than another one.
We first validate the performance of the QAOA angles up to $p=10$ obtained from the different methods on ten weighted MaxCut problems with hardware native heavy-hex graphs with 144 nodes and $\mathcal{N}(0, 1)$ edge weights.
These problem instances are the simplest to evaluate on hardware since their topology matches the qubit connectivity.
For each circuit we gather 4096 samples $x$ from \emph{ibm\_boston} and compute the cut value $f(x)$.
We average these cut values and report them as an approximation ratio obtained by solving the combinatorial optimization problem exactly with CPLEX.
We then report this \emph{hardware approximation ratio} against the \emph{estimated approximation ratio} obtained from the approximate energy seen in the training.

\begin{figure*}
    \centering
    \includegraphics[width=0.95\textwidth, clip, trim=0 0 0 30]{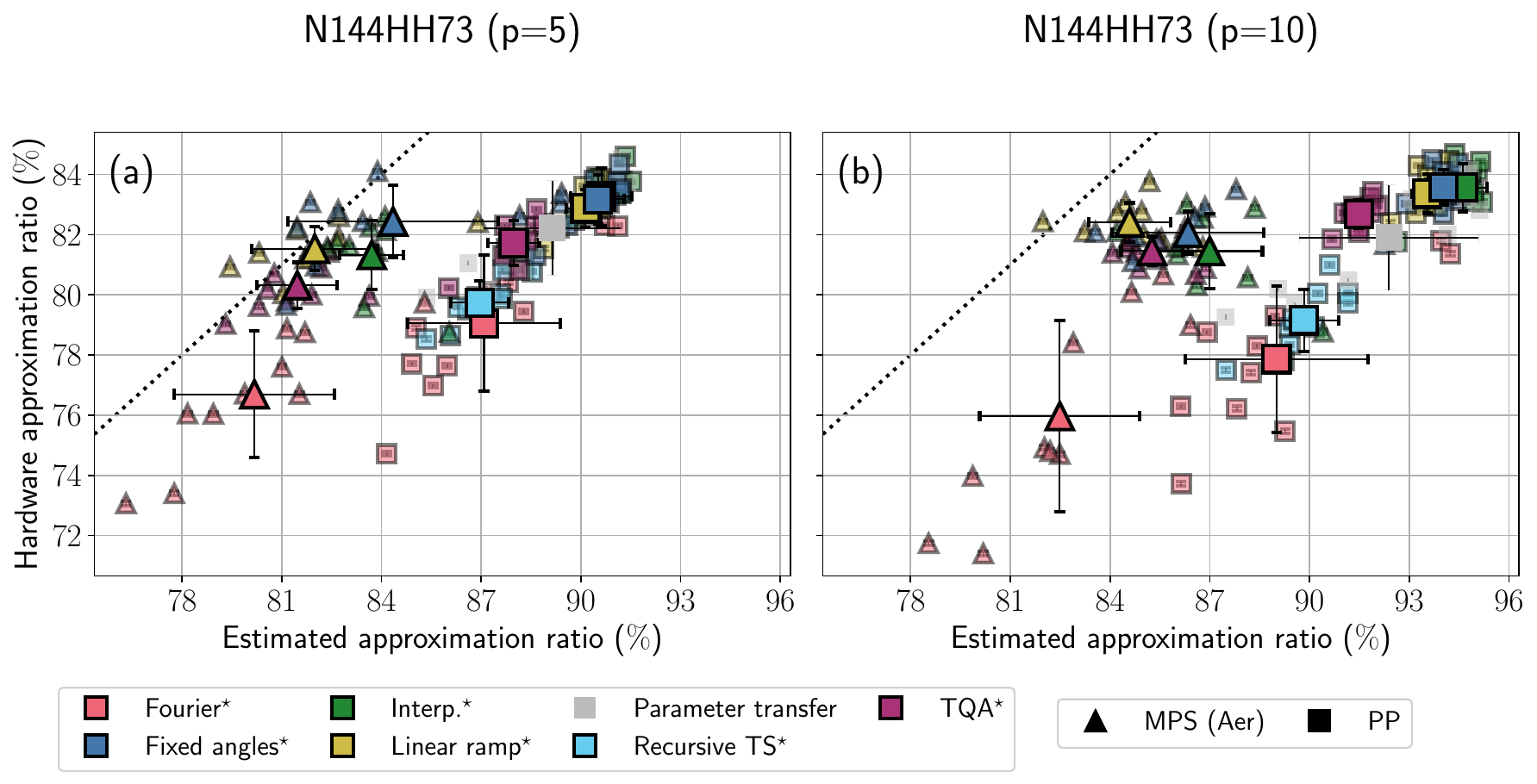}
    \caption{
    Hardware validation on \emph{ibm\_boston} for ten 144-node heavy-hex graphs (small markers) using QAOA at depths $p=5$ (a) and $p=10$ (b), evaluated with the Qiskit Aer MPS with a max bond dimension of 40 and Pauli propagation with a max weight of 6 (fixed angles interpret the graphs as having a degree of three). 
    The large markers with error bars show the mean and standard deviation of the ten instances.
    The dashed line indicates where the hardware approximation ratio is equal to the estimated approximation ratio.
    For the MPS, the correlation coefficients between the hardware and the estimated approximation ratio are 0.66 and 0.61 for $p=5$ and $10$, respectively.
    For PP, these numbers are 0.93 and 0.89.
    }
    \label{fig:144hh510_hw}
\end{figure*}

We observe a positive correlation between the estimated ratios and the hardware ratios, although the hardware ratios are lower than their estimated counterparts due to noise, see data for $p=5$ and $10$ in Fig.~\ref{fig:144hh510_hw}.
Interp. with PP (green squares) and reoptimized fixed angles with PP (dark blue squares) perform the best in terms of both the estimated and the hardware approximation ratios.
A transfer of angles from small-scale instances to these utility-scale ones (grey squares) performs well, but does not beat Interp. with PP.
The Fourier methods with the MPS and PP evaluators (red triangles and squares, respectively) exhibit a large variance and result in low hardware approximation ratios compared to the other methods.
Crucially, the Fourier method with PP energy evaluation reports a higher estimated approximation ratio than any MPS-based method, see Fig.~\ref{fig:144hh510_hw}.
However, the QPU data show that this outperformance is fake since the hardware approximation ratios of the Fourier method are lower than all other methods, irrespective of the energy evaluator.

\begin{table*}[ht]
\centering
\resizebox{\textwidth}{!}{%
\begin{tabular}{l|ccccccccccccc|cccccccccccc|}
\cline{2-26}
& \multicolumn{13}{c|}{Estimated} & \multicolumn{12}{c|}{Hardware} \\
\hline

Methods(p=5) & 1. & 2. & 3. & 4. & 5. & 6. & 7. & 8. & 9. & 10. & 11. & 12. & Median & 2. & 3. & 4. & 5. & 6. & 7. & 8. & 9. & 10. & 11. & 12. & Median \\
\hline

1. (I, PP)$^\star$ & -- & 1 & 1 & 3.4e-1 & {\cellcolor[HTML]{00441B}} \color[HTML]{F1F1F1} 7.4e-4 & {\cellcolor[HTML]{00441B}} \color[HTML]{F1F1F1} 1.7e-6 & {\cellcolor[HTML]{00441B}} \color[HTML]{F1F1F1} 4.1e-5 & {\cellcolor[HTML]{00441B}} \color[HTML]{F1F1F1} 7.4e-13 & {\cellcolor[HTML]{00441B}} \color[HTML]{F1F1F1} 1.9e-11 & {\cellcolor[HTML]{00441B}} \color[HTML]{F1F1F1} 1.2e-17 & {\cellcolor[HTML]{00441B}} \color[HTML]{F1F1F1} 7.5e-19 & {\cellcolor[HTML]{00441B}} \color[HTML]{F1F1F1} 7.6e-21 & 91 & 1 & 1 & 2.6e-1 & {\cellcolor[HTML]{309950}} \color[HTML]{F1F1F1} 8.9e-3 & {\cellcolor[HTML]{00441B}} \color[HTML]{F1F1F1} 1.8e-10 & {\cellcolor[HTML]{00441B}} \color[HTML]{F1F1F1} 7.6e-10 & {\cellcolor[HTML]{00441B}} \color[HTML]{F1F1F1} 4.4e-4 & 8.5e-1 & {\cellcolor[HTML]{309950}} \color[HTML]{F1F1F1} 1.3e-3 & {\cellcolor[HTML]{00441B}} \color[HTML]{F1F1F1} 3.9e-8 & {\cellcolor[HTML]{00441B}} \color[HTML]{F1F1F1} 9.5e-15 & 83.4 \\
2. (FA, PP)$^\star$ & -- & -- & 1 & 6.2e-1 & {\cellcolor[HTML]{309950}} \color[HTML]{F1F1F1} 2.6e-3 & {\cellcolor[HTML]{00441B}} \color[HTML]{F1F1F1} 7.4e-6 & {\cellcolor[HTML]{00441B}} \color[HTML]{F1F1F1} 1.6e-4 & {\cellcolor[HTML]{00441B}} \color[HTML]{F1F1F1} 4.3e-12 & {\cellcolor[HTML]{00441B}} \color[HTML]{F1F1F1} 1.0e-10 & {\cellcolor[HTML]{00441B}} \color[HTML]{F1F1F1} 7.0e-17 & {\cellcolor[HTML]{00441B}} \color[HTML]{F1F1F1} 4.4e-18 & {\cellcolor[HTML]{00441B}} \color[HTML]{F1F1F1} 4.4e-20 & 90.8 & -- & 1 & 4.6e-1 & {\cellcolor[HTML]{B5E1AE}} \color[HTML]{000000} 2.0e-2 & {\cellcolor[HTML]{00441B}} \color[HTML]{F1F1F1} 6.2e-10 & {\cellcolor[HTML]{00441B}} \color[HTML]{F1F1F1} 2.6e-9 & {\cellcolor[HTML]{309950}} \color[HTML]{F1F1F1} 1.1e-3 & 1 & {\cellcolor[HTML]{309950}} \color[HTML]{F1F1F1} 3.1e-3 & {\cellcolor[HTML]{00441B}} \color[HTML]{F1F1F1} 1.3e-7 & {\cellcolor[HTML]{00441B}} \color[HTML]{F1F1F1} 3.4e-14 & 83.4 \\
3. (LR, PP)$^\star$ & -- & -- & -- & 1 & 6.2e-2 & {\cellcolor[HTML]{00441B}} \color[HTML]{F1F1F1} 5.1e-4 & {\cellcolor[HTML]{309950}} \color[HTML]{F1F1F1} 6.9e-3 & {\cellcolor[HTML]{00441B}} \color[HTML]{F1F1F1} 7.6e-10 & {\cellcolor[HTML]{00441B}} \color[HTML]{F1F1F1} 1.6e-8 & {\cellcolor[HTML]{00441B}} \color[HTML]{F1F1F1} 1.5e-14 & {\cellcolor[HTML]{00441B}} \color[HTML]{F1F1F1} 9.2e-16 & {\cellcolor[HTML]{00441B}} \color[HTML]{F1F1F1} 9.0e-18 & 90.3 & -- & -- & 1 & 1.3e-1 & {\cellcolor[HTML]{00441B}} \color[HTML]{F1F1F1} 1.7e-8 & {\cellcolor[HTML]{00441B}} \color[HTML]{F1F1F1} 6.8e-8 & {\cellcolor[HTML]{B5E1AE}} \color[HTML]{000000} 1.2e-2 & 1 & {\cellcolor[HTML]{B5E1AE}} \color[HTML]{000000} 2.8e-2 & {\cellcolor[HTML]{00441B}} \color[HTML]{F1F1F1} 2.9e-6 & {\cellcolor[HTML]{00441B}} \color[HTML]{F1F1F1} 1.2e-12 & 82.9 \\
4. (PT, PP) & -- & -- & -- & -- & 4.9e-1 & {\cellcolor[HTML]{B5E1AE}} \color[HTML]{000000} 1.2e-2 & 1.0e-1 & {\cellcolor[HTML]{00441B}} \color[HTML]{F1F1F1} 6.8e-8 & {\cellcolor[HTML]{00441B}} \color[HTML]{F1F1F1} 1.3e-6 & {\cellcolor[HTML]{00441B}} \color[HTML]{F1F1F1} 1.8e-12 & {\cellcolor[HTML]{00441B}} \color[HTML]{F1F1F1} 1.2e-13 & {\cellcolor[HTML]{00441B}} \color[HTML]{F1F1F1} 1.1e-15 & 89.8 & -- & -- & -- & 1 & {\cellcolor[HTML]{00441B}} \color[HTML]{F1F1F1} 3.4e-5 & {\cellcolor[HTML]{00441B}} \color[HTML]{F1F1F1} 1.1e-4 & 8.5e-1 & 1 & 1 & {\cellcolor[HTML]{309950}} \color[HTML]{F1F1F1} 2.6e-3 & {\cellcolor[HTML]{00441B}} \color[HTML]{F1F1F1} 5.4e-9 & 82.4 \\
5. (TQA, PP)$^\star$ & -- & -- & -- & -- & -- & 1 & 1 & {\cellcolor[HTML]{00441B}} \color[HTML]{F1F1F1} 5.5e-4 & {\cellcolor[HTML]{309950}} \color[HTML]{F1F1F1} 5.3e-3 & {\cellcolor[HTML]{00441B}} \color[HTML]{F1F1F1} 6.1e-8 & {\cellcolor[HTML]{00441B}} \color[HTML]{F1F1F1} 4.7e-9 & {\cellcolor[HTML]{00441B}} \color[HTML]{F1F1F1} 5.6e-11 & 88 & -- & -- & -- & -- & {\cellcolor[HTML]{309950}} \color[HTML]{F1F1F1} 3.6e-3 & {\cellcolor[HTML]{309950}} \color[HTML]{F1F1F1} 9.7e-3 & 1 & 1 & 1 & 1.1e-1 & {\cellcolor[HTML]{00441B}} \color[HTML]{F1F1F1} 1.6e-6 & 81.8 \\
6. (RTS, PP)$^\star$ & -- & -- & -- & -- & -- & -- & 1 & 6.4e-2 & 3.1e-1 & {\cellcolor[HTML]{00441B}} \color[HTML]{F1F1F1} 4.2e-5 & {\cellcolor[HTML]{00441B}} \color[HTML]{F1F1F1} 4.3e-6 & {\cellcolor[HTML]{00441B}} \color[HTML]{F1F1F1} 7.6e-8 & 86.9 & -- & -- & -- & -- & -- & 1 & 5.2e-2 & {\cellcolor[HTML]{00441B}} \color[HTML]{F1F1F1} 2.6e-6 & {\cellcolor[HTML]{B5E1AE}} \color[HTML]{000000} 2.2e-2 & 1 & 9.9e-1 & 79.8 \\
7. (F, PP)$^\star$ & -- & -- & -- & -- & -- & -- & -- & {\cellcolor[HTML]{309950}} \color[HTML]{F1F1F1} 7.3e-3 & 5.1e-2 & {\cellcolor[HTML]{00441B}} \color[HTML]{F1F1F1} 1.7e-6 & {\cellcolor[HTML]{00441B}} \color[HTML]{F1F1F1} 1.5e-7 & {\cellcolor[HTML]{00441B}} \color[HTML]{F1F1F1} 2.2e-9 & 86.6 & -- & -- & -- & -- & -- & -- & 1.1e-1 & {\cellcolor[HTML]{00441B}} \color[HTML]{F1F1F1} 9.5e-6 & 5.2e-2 & 1 & 6.2e-1 & 79.2 \\
8. (I, MPSAer)$^\star$ & -- & -- & -- & -- & -- & -- & -- & -- & 1 & 4.9e-1 & 1.5e-1 & {\cellcolor[HTML]{B5E1AE}} \color[HTML]{000000} 1.3e-2 & 83.7 & -- & -- & -- & -- & -- & -- & -- & 2.6e-1 & 1 & 7.1e-1 & {\cellcolor[HTML]{00441B}} \color[HTML]{F1F1F1} 5.6e-5 & 81.6 \\
9. (FA, MPSAer)$^\star$ & -- & -- & -- & -- & -- & -- & -- & -- & -- & 1.2e-1 & {\cellcolor[HTML]{B5E1AE}} \color[HTML]{000000} 3.0e-2 & {\cellcolor[HTML]{309950}} \color[HTML]{F1F1F1} 1.7e-3 & 83.1 & -- & -- & -- & -- & -- & -- & -- & -- & 4.6e-1 & {\cellcolor[HTML]{00441B}} \color[HTML]{F1F1F1} 2.7e-4 & {\cellcolor[HTML]{00441B}} \color[HTML]{F1F1F1} 2.9e-10 & 82.7 \\
10. (LR, MPSAer)$^\star$ & -- & -- & -- & -- & -- & -- & -- & -- & -- & -- & 1 & 1 & 81.7 & -- & -- & -- & -- & -- & -- & -- & -- & -- & 4.2e-1 & {\cellcolor[HTML]{00441B}} \color[HTML]{F1F1F1} 1.7e-5 & 81.4 \\
11. (TQA, MPSAer)$^\star$ & -- & -- & -- & -- & -- & -- & -- & -- & -- & -- & -- & 1 & 81.5 & -- & -- & -- & -- & -- & -- & -- & -- & -- & -- & 9.6e-2 & 80.1 \\
12. (F, MPSAer)$^\star$ & -- & -- & -- & -- & -- & -- & -- & -- & -- & -- & -- & -- & 80.4 & -- & -- & -- & -- & -- & -- & -- & -- & -- & -- & -- & 76.7 \\
\hline

Methods(p=10) & 1. & 2. & 3. & 4. & 5. & 6. & 7. & 8. & 9. & 10. & 11. & 12. & Median & 2. & 3. & 4. & 5. & 6. & 7. & 8. & 9. & 10. & 11. & 12. & Median \\
\hline

1. (I, PP)$^\star$ & -- & 7.6e-1 & 7.9e-2 & {\cellcolor[HTML]{B5E1AE}} \color[HTML]{000000} 1.8e-2 & {\cellcolor[HTML]{00441B}} \color[HTML]{F1F1F1} 1.9e-5 & {\cellcolor[HTML]{00441B}} \color[HTML]{F1F1F1} 3.7e-9 & {\cellcolor[HTML]{00441B}} \color[HTML]{F1F1F1} 1.7e-10 & {\cellcolor[HTML]{00441B}} \color[HTML]{F1F1F1} 2.4e-16 & {\cellcolor[HTML]{00441B}} \color[HTML]{F1F1F1} 1.6e-18 & {\cellcolor[HTML]{00441B}} \color[HTML]{F1F1F1} 5.5e-22 & {\cellcolor[HTML]{00441B}} \color[HTML]{F1F1F1} 7.4e-24 & {\cellcolor[HTML]{00441B}} \color[HTML]{F1F1F1} 1.5e-26 & 94.9 & 1 & 1 & {\cellcolor[HTML]{309950}} \color[HTML]{F1F1F1} 4.3e-3 & 2.0e-1 & {\cellcolor[HTML]{00441B}} \color[HTML]{F1F1F1} 1.3e-13 & {\cellcolor[HTML]{00441B}} \color[HTML]{F1F1F1} 2.6e-14 & {\cellcolor[HTML]{00441B}} \color[HTML]{F1F1F1} 4.2e-6 & {\cellcolor[HTML]{00441B}} \color[HTML]{F1F1F1} 5.6e-4 & {\cellcolor[HTML]{00441B}} \color[HTML]{F1F1F1} 4.6e-7 & {\cellcolor[HTML]{B5E1AE}} \color[HTML]{000000} 1.7e-2 & {\cellcolor[HTML]{00441B}} \color[HTML]{F1F1F1} 3.9e-16 & 83.5 \\
2. (FA, PP)$^\star$ & -- & -- & 1 & 7.5e-1 & {\cellcolor[HTML]{309950}} \color[HTML]{F1F1F1} 9.9e-3 & {\cellcolor[HTML]{00441B}} \color[HTML]{F1F1F1} 8.7e-6 & {\cellcolor[HTML]{00441B}} \color[HTML]{F1F1F1} 5.5e-7 & {\cellcolor[HTML]{00441B}} \color[HTML]{F1F1F1} 1.5e-12 & {\cellcolor[HTML]{00441B}} \color[HTML]{F1F1F1} 9.8e-15 & {\cellcolor[HTML]{00441B}} \color[HTML]{F1F1F1} 3.1e-18 & {\cellcolor[HTML]{00441B}} \color[HTML]{F1F1F1} 3.7e-20 & {\cellcolor[HTML]{00441B}} \color[HTML]{F1F1F1} 6.3e-23 & 94.1 & -- & 1 & {\cellcolor[HTML]{309950}} \color[HTML]{F1F1F1} 3.7e-3 & 1.8e-1 & {\cellcolor[HTML]{00441B}} \color[HTML]{F1F1F1} 1.0e-13 & {\cellcolor[HTML]{00441B}} \color[HTML]{F1F1F1} 2.0e-14 & {\cellcolor[HTML]{00441B}} \color[HTML]{F1F1F1} 3.4e-6 & {\cellcolor[HTML]{00441B}} \color[HTML]{F1F1F1} 4.8e-4 & {\cellcolor[HTML]{00441B}} \color[HTML]{F1F1F1} 3.7e-7 & {\cellcolor[HTML]{B5E1AE}} \color[HTML]{000000} 1.5e-2 & {\cellcolor[HTML]{00441B}} \color[HTML]{F1F1F1} 3.1e-16 & 83.3 \\
3. (LR, PP)$^\star$ & -- & -- & -- & 1 & 2.4e-1 & {\cellcolor[HTML]{309950}} \color[HTML]{F1F1F1} 1.1e-3 & {\cellcolor[HTML]{00441B}} \color[HTML]{F1F1F1} 1.0e-4 & {\cellcolor[HTML]{00441B}} \color[HTML]{F1F1F1} 6.6e-10 & {\cellcolor[HTML]{00441B}} \color[HTML]{F1F1F1} 5.4e-12 & {\cellcolor[HTML]{00441B}} \color[HTML]{F1F1F1} 1.7e-15 & {\cellcolor[HTML]{00441B}} \color[HTML]{F1F1F1} 2.1e-17 & {\cellcolor[HTML]{00441B}} \color[HTML]{F1F1F1} 3.2e-20 & 93.7 & -- & -- & {\cellcolor[HTML]{B5E1AE}} \color[HTML]{000000} 3.3e-2 & 8.9e-1 & {\cellcolor[HTML]{00441B}} \color[HTML]{F1F1F1} 4.0e-12 & {\cellcolor[HTML]{00441B}} \color[HTML]{F1F1F1} 7.7e-13 & {\cellcolor[HTML]{00441B}} \color[HTML]{F1F1F1} 6.8e-5 & {\cellcolor[HTML]{309950}} \color[HTML]{F1F1F1} 6.2e-3 & {\cellcolor[HTML]{00441B}} \color[HTML]{F1F1F1} 8.2e-6 & 1.1e-1 & {\cellcolor[HTML]{00441B}} \color[HTML]{F1F1F1} 1.3e-14 & 83.2 \\
4. (PT, PP) & -- & -- & -- & -- & 6.8e-1 & {\cellcolor[HTML]{309950}} \color[HTML]{F1F1F1} 7.0e-3 & {\cellcolor[HTML]{00441B}} \color[HTML]{F1F1F1} 7.6e-4 & {\cellcolor[HTML]{00441B}} \color[HTML]{F1F1F1} 8.2e-9 & {\cellcolor[HTML]{00441B}} \color[HTML]{F1F1F1} 7.6e-11 & {\cellcolor[HTML]{00441B}} \color[HTML]{F1F1F1} 2.5e-14 & {\cellcolor[HTML]{00441B}} \color[HTML]{F1F1F1} 3.0e-16 & {\cellcolor[HTML]{00441B}} \color[HTML]{F1F1F1} 4.7e-19 & 93.6 & -- & -- & -- & 1 & {\cellcolor[HTML]{00441B}} \color[HTML]{F1F1F1} 2.8e-5 & {\cellcolor[HTML]{00441B}} \color[HTML]{F1F1F1} 7.1e-6 & 1 & 1 & 4.9e-1 & 1 & {\cellcolor[HTML]{00441B}} \color[HTML]{F1F1F1} 2.0e-7 & 82.4 \\
5. (TQA, PP)$^\star$ & -- & -- & -- & -- & -- & 6.8e-1 & 2.1e-1 & {\cellcolor[HTML]{00441B}} \color[HTML]{F1F1F1} 3.9e-5 & {\cellcolor[HTML]{00441B}} \color[HTML]{F1F1F1} 6.2e-7 & {\cellcolor[HTML]{00441B}} \color[HTML]{F1F1F1} 3.7e-10 & {\cellcolor[HTML]{00441B}} \color[HTML]{F1F1F1} 5.1e-12 & {\cellcolor[HTML]{00441B}} \color[HTML]{F1F1F1} 7.9e-15 & 91.5 & -- & -- & -- & -- & {\cellcolor[HTML]{00441B}} \color[HTML]{F1F1F1} 7.1e-8 & {\cellcolor[HTML]{00441B}} \color[HTML]{F1F1F1} 1.6e-8 & 5.9e-2 & 9.5e-1 & {\cellcolor[HTML]{B5E1AE}} \color[HTML]{000000} 1.4e-2 & 1 & {\cellcolor[HTML]{00441B}} \color[HTML]{F1F1F1} 2.9e-10 & 82.7 \\
6. (RTS, PP)$^\star$ & -- & -- & -- & -- & -- & -- & 1 & {\cellcolor[HTML]{B5E1AE}} \color[HTML]{000000} 2.8e-2 & {\cellcolor[HTML]{309950}} \color[HTML]{F1F1F1} 1.3e-3 & {\cellcolor[HTML]{00441B}} \color[HTML]{F1F1F1} 2.4e-6 & {\cellcolor[HTML]{00441B}} \color[HTML]{F1F1F1} 4.8e-8 & {\cellcolor[HTML]{00441B}} \color[HTML]{F1F1F1} 1.1e-10 & 89.8 & -- & -- & -- & -- & -- & 1 & {\cellcolor[HTML]{B5E1AE}} \color[HTML]{000000} 1.7e-2 & {\cellcolor[HTML]{00441B}} \color[HTML]{F1F1F1} 2.7e-4 & 7.4e-2 & {\cellcolor[HTML]{00441B}} \color[HTML]{F1F1F1} 4.5e-6 & 1 & 79.1 \\
7. (F, PP)$^\star$ & -- & -- & -- & -- & -- & -- & -- & 1.6e-1 & {\cellcolor[HTML]{B5E1AE}} \color[HTML]{000000} 1.1e-2 & {\cellcolor[HTML]{00441B}} \color[HTML]{F1F1F1} 3.5e-5 & {\cellcolor[HTML]{00441B}} \color[HTML]{F1F1F1} 8.6e-7 & {\cellcolor[HTML]{00441B}} \color[HTML]{F1F1F1} 2.4e-9 & 88.3 & -- & -- & -- & -- & -- & -- & {\cellcolor[HTML]{309950}} \color[HTML]{F1F1F1} 6.5e-3 & {\cellcolor[HTML]{00441B}} \color[HTML]{F1F1F1} 7.3e-5 & {\cellcolor[HTML]{B5E1AE}} \color[HTML]{000000} 3.0e-2 & {\cellcolor[HTML]{00441B}} \color[HTML]{F1F1F1} 1.1e-6 & 1 & 77.8 \\
8. (I, MPSAer)$^\star$ & -- & -- & -- & -- & -- & -- & -- & -- & 1 & 2.1e-1 & {\cellcolor[HTML]{B5E1AE}} \color[HTML]{000000} 2.0e-2 & {\cellcolor[HTML]{00441B}} \color[HTML]{F1F1F1} 2.8e-4 & 86.7 & -- & -- & -- & -- & -- & -- & -- & 1 & 1 & 5.5e-1 & {\cellcolor[HTML]{00441B}} \color[HTML]{F1F1F1} 3.7e-4 & 81.5 \\
9. (FA, MPSAer)$^\star$ & -- & -- & -- & -- & -- & -- & -- & -- & -- & 8.5e-1 & 2.4e-1 & {\cellcolor[HTML]{309950}} \color[HTML]{F1F1F1} 8.8e-3 & 85.8 & -- & -- & -- & -- & -- & -- & -- & -- & 1 & 1 & {\cellcolor[HTML]{00441B}} \color[HTML]{F1F1F1} 2.5e-6 & 81.9 \\
10. (TQA, MPSAer)$^\star$ & -- & -- & -- & -- & -- & -- & -- & -- & -- & -- & 1 & 4.4e-1 & 84.8 & -- & -- & -- & -- & -- & -- & -- & -- & -- & 1.8e-1 & {\cellcolor[HTML]{309950}} \color[HTML]{F1F1F1} 2.4e-3 & 81.5 \\
11. (LR, MPSAer)$^\star$ & -- & -- & -- & -- & -- & -- & -- & -- & -- & -- & -- & 1 & 84.7 & -- & -- & -- & -- & -- & -- & -- & -- & -- & -- & {\cellcolor[HTML]{00441B}} \color[HTML]{F1F1F1} 2.7e-8 & 82.3 \\
12. (F, MPSAer)$^\star$ & -- & -- & -- & -- & -- & -- & -- & -- & -- & -- & -- & -- & 82.3 & -- & -- & -- & -- & -- & -- & -- & -- & -- & -- & -- & 74.9 \\
\hline

\end{tabular}
}
\caption{Conover–Iman post-hoc test with Holm correction on estimated (left) and hardware (right) ratios generated from 10 144-node heavy-hex graphs at $p=5$ (top) and $p=10$ (bottom). \raisebox{0.7ex}{\colorbox[HTML]{B5E1AE}{\hspace{1em}}} indicates p-values less than 0.05, \raisebox{0.7ex}{\colorbox[HTML]{309950}{\hspace{1em}}} indicates p-values less than 0.01, and \raisebox{0.7ex}{\colorbox[HTML]{00441B}{\hspace{1em}}} indicates p-values less than 0.001.}
\label{tab:N144HH73_5_10_100pc_hw}
\end{table*}

At $p=10$ there is still enough signal from the hardware to differentiate some of the QAOA angle training methods, see Fig.~\ref{fig:144hh510_hw}(b).
There is an overall increase of both the estimated ratios and the hardware ratios compared to $p=5$. 
The methods still have a similar qualitative performance relative to each other as for $p=5$.
However, most of them are closer to each other in terms of hardware approximation ratios due to the increased noise from the deeper circuits.
We see that the PP evaluator consistently overestimates the hardware approximation ratio.
If we instead compute the hardware approximation ratio over the 5\% best samples, as measured by $f(x)$, we observe a better agreement between the PP evaluation and the QPU samples, data in App.~\ref{app:stat_test}.
Since an average over the tail of the sample distribution provides better estimates of noiseless expectation values~\cite{Barron2024} these results suggest that PP is a good energy evaluator.
A comparison of the hardware approximation ratio to the estimated approximation ratio for ten 100-node line-based graphs, i.e., built from a line and two SWAP layers, corroborate this suggestion.
Here, the MPS estimated approximation ratio correlate negatively to the QPU estimated approximation ratio, i.e., -0.15 and -0.43 for $p=2$ and $6$ respectively. 
By contrast, the PP correlations are positive with 0.87 and 0.40 for $p=2$ and $6$ respectively, see App.~\ref{app:hw_line_based}.
Since MPS evaluators struggle on instances with small cycles~\cite{Rudolph2025_BP-QuantumCircuit}, the 193 small length-four cycles in these line-based graphs may explain this effect.
By comparison, the smallest cycles in heavy-hex graphs have size twelve.

To test whether the different angle training methods have a statistically significant impact on the estimated and the hardware approximation ratios separately, we perform a Kruskal–Wallis statistical test followed by Conover-Iman posthoc test with Holm correction, see App.~\ref{app:stat_test}.
Two methods are statistically indistinguishable from each other when their p-value is close to one.
At $p=5$ the test distinguishes most of the estimated approximation ratios, see the top-left subtable in Tab.~\ref{tab:N144HH73_5_10_100pc_hw}.
The hardware approximation ratios are harder to tell apart.
For example, Interp.$^\star$, Fixed Angles$^\star$, TQA$^\star$ (all with PP) and Fixed Angles$^\star$ with MPS are statistically indistinguishable since they have high p-values, see the top-right subtable Tab.~\ref{tab:N144HH73_5_10_100pc_hw}.
Pairs like Interp.$^\star$ with PP and Fourier$^\star$ with MPS have a meaningful difference.
In the $p=10$ data the methods are distinguishable from one another in terms of their estimated approximation ratios, see bottom panel in Tab.~\ref{tab:N144HH73_5_10_100pc_hw}.
However, on hardware many methods are indistinguishable.
Therefore, when the quantum circuits have a depth and width close to the noise limits of the QPU the best methods become indistinguishable from each other and it matters little if the QAOA angles were trained with the MPS or PP.

\subsection{Resource cost analysis\label{sec:results_ressources}}

\begin{figure*}[t]
    \centering
    \includegraphics[width=0.95\textwidth]{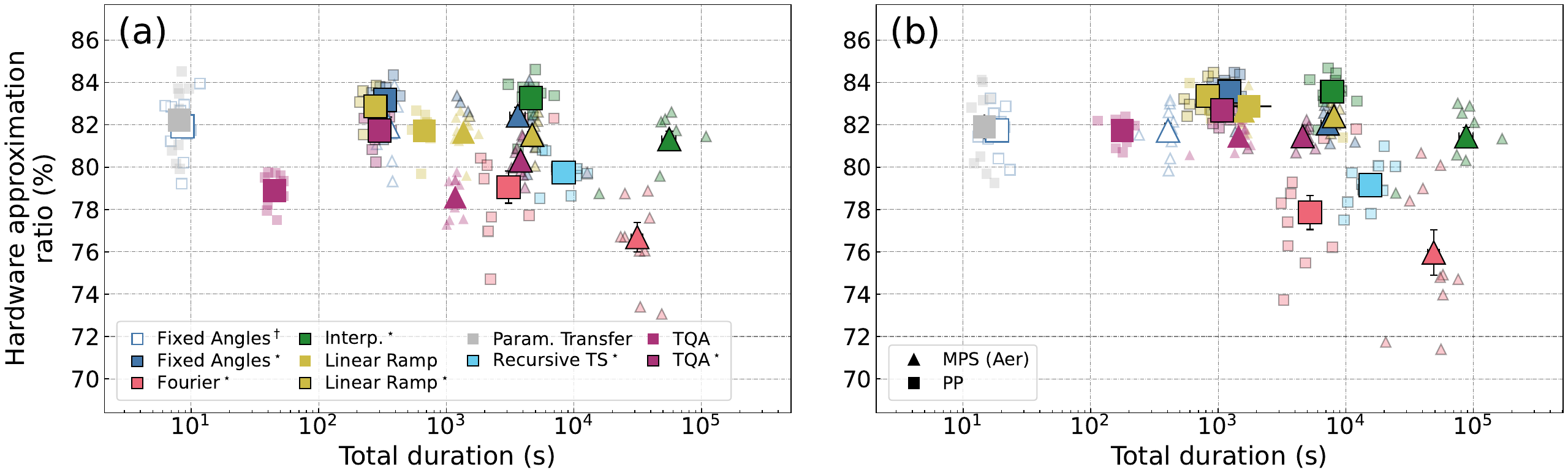}
    \caption{Resource-Cost analysis for ten 144-node heavy-hex graphs (light small markers) on \emph{ibm\_boston} at depths (a) $p=5$ and (b) $p=10$.
    The larger markers show the average of the ten instances.
    Each point represents an angle-setting strategy.
    The $x$-axis shows the Total Duration, while the  $y$-axis shows the achieved approximation ratio.
    Methods that optimize the QAOA angles are marked with a star, e.g., Fixed Angle$^\star$. Methods that have some method-parameter optimization are indicated without any superscript. All other methods are indicated with a $\dagger$.}
    \label{fig:performance}
\end{figure*}

We compare the computational burden of running angle setting methods, with the Stochastic Benchmark framework~\cite{Neira2024}. It evaluates QAOA as a parameterized stochastic optimization solver whose performance depends on algorithmic parameters and stochastic sampling from quantum and classical hardware.
This framework applied to our case analyzes QAOA across a distribution of problem instances to quantify how solver performance varies with parameter choices and available computational resources.
By performing repeated experiments across representative problem instances and parameter settings, the framework statistically characterizes solver performance and identifies parameter-setting \emph{strategies}, i.e. prescriptions on how to use the angle setting methods independently from the knowledge of the individual problem instance, that balance exploration, i.e., testing new parameters, and exploitation, i.e., sampling the optimized solver to determine the best found solution $x$.

The framework computes a visualization referred to as a \emph{Window Sticker}, summarizing the expected performance of each angle-setting strategy as a function of the computational budget. 
We define the \textit{computational budget} of each angle setting method as the \textit{total duration} $T_{total} = T_{AS}+T_{QPU}$ to be consumed at runtime, i.e., once the problem instance to be solved is defined.
Here, $T_{AS}$ is the time spent by the angle setting strategy. 
It corresponds mostly to the time spent by the classical optimizer (COBYLA) running $N$ iterations to determine the optimal QAOA angles for a given method.
Here, this is often bottlenecked by the time required to classically evaluate the expected energy $t_{\langle \rangle}$.
Conversely, the QPU execution time $T_{QPU}$ is the time spent gathering $Q$ samples from the quantum hardware $T_{QPU} = Q t_{shot}$, where each shot has a duration $t_{shot}$. 

To apply the Stochastic Benchmark framework to utility-scale instances, we analyze the training performance on heavy-hex graphs with $144$ qubits.
The final sampling is done on \emph{ibm\_boston} with $Q=4096$ shots.
The reported angle setting time of iterative methods, such as Interp. and Fourier, is the cumulative training time required to reach the target QAOA depth, i.e. $T_{AS}=N\sum_{p^\prime=1\dots p} t_{\langle \rangle}^{(p^\prime)}$ instead of $N t_{\langle \rangle}^{(p)}$ for the other non-iterative methods.
Consequently, iterative methods have larger running costs since they optimize the QAOA angles at multiple depths, see App.~\ref{app:cost_analysis}. 

\begin{figure*}[t]
\includegraphics[width=0.95\textwidth]{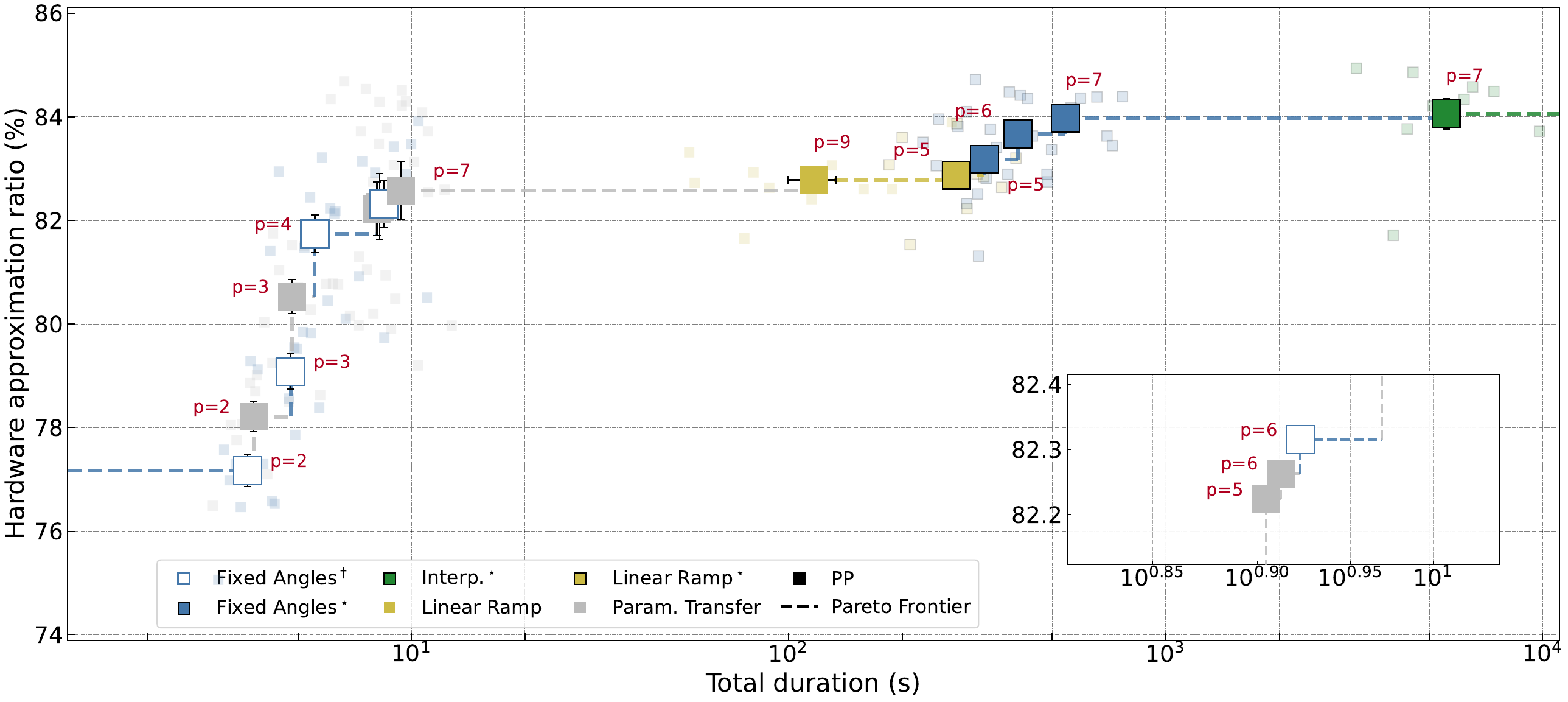}
\caption{Recommendation plot showing the Pareto frontier across angle-setting strategies. 
The $x$-axis represents the total duration, and the  $y$-axis shows the achieved approximation ratio. 
The frontier identifies the strategies that provide the best trade-offs between computational budget and solution quality.
Methods that employed QAOA angle optimization are marked with a star, e.g., Fixed Angle$^\star$. Methods that have some method-parameter optimization are indicated without any superscript. All other methods are indicated with a $\dagger$.
\label{fig:recommendation}}
\end{figure*}

The quality metric is the \emph{hardware approximation ratio} achieved on \emph{ibm\_boston}.
Figure~\ref{fig:performance} visualizes the trade-off between the computational budget ($x$-axis) and the quality metric ($y$-axis) for QAOA depths five and ten for different combinations of angle setting methods and energy evaluators.
Physics-inspired methods such as linear ramp and structure-informed methods such as parameter transfer and Fixed Angles in combination with PP remain among the fastest angle-setting methods, completing within small computational budgets of under $10^3$ seconds. 
These methods achieve approximation ratios ranging from $82\%$ to $85\%$, providing competitive baseline solutions at low cost.
In contrast, iterative angle training methods incur substantially higher computational costs with total durations often exceeding $10^4$ seconds due to the repeated circuit evaluations, see also App.~\ref{app:cost_analysis}. 
Furthermore, when evaluated on a QPU, the improvement in the hardware approximation ratio is modest at best.
Importantly, while the quality advantage of PP over MPS becomes less pronounced on hardware due to noise, PP has a lower computational budget than MPS.

The Pareto frontier shows the best perfoming strategy among the ones analyzed for each budget of time allowed, see Fig.~\ref{fig:recommendation}. 
Low-cost strategies such as parameter transfer and Fixed Angles remain on the frontier, while higher-cost iterative methods such as Interp. provide only incremental gains.
This indicates diminishing returns from extensive angle optimization when performance is evaluated directly on quantum hardware.
At smaller computational budgets, parameter transfer and Fixed Angles without a reoptimization of the QAOA angles provide the fastest route to obtaining reasonably high-quality solutions.
As the computational budget increases, methods such as Linear Ramp and Fixed Angles with full QAOA angle reoptimization, combined with PP, achieve the highest approximation ratios and therefore become preferable when solution quality is the primary objective.

The Goemans-Williamson algorithm with its $\alpha_{\rm GW}$ guarantee, see Eq.~\eqref{GWratio}, puts the time scales in Fig.~\ref{fig:recommendation} in perspective.
It finds the global minimizers in all but a handful of instances within 10 rounds of the hyperplane rounding. 
The computation of the SDP relaxation and the 10 rounds of the hyperplane rounding can be executed within 10 seconds per instance even on a standard laptop (Apple MacBook M3), see App. \ref{app:example_gw}.

We also estimate what the stochastic-benchmark analysis of performance would look like, if the QAOA angles were trained on quantum hardware, by emulating the sampling with the MPS (Aer) simulator.
For each candidate parameter setting strategy we define a shot resolution for the expectation values, through a resource budget breakdown:
\begin{equation}
    T_{\text{proxy}} = N (t_\text{qc} + M \, t_{\text{shot}}) + Q \, t_{\text{shot}}.
    \label{eq:tproxy}
\end{equation}
Here, $N$ is the number of COBYLA iterations or its cumulative sum across sub-depths for recursive strategies such as Interp., $M$ is the number of shots used per COBYLA iteration, and $Q$ is the final number of shots once the angles are set.
Each iteration incurs a fixed penalty $t_\text{qc}$ to send the circuits to the backend and prepare them for execution.
Since Fig.~\ref{fig:hardware_conv} reports $(t_\text{qc}+Mt_\text{shot})^{-1}=2.47~\textrm{kHz}$ for $M=20\,000$ on \emph{ibm\_boston}, we set $t_{\text{shot}}=1/(10\,\text{kHz})= 100\,\mu\text{s}$ and $t_{\text{qc}}= 6~\mathrm{s}$. 
We neglect pre-processing times, such as fixed-angles look-up and parameter transfer times, as they are small compared to the other time scales.

\begin{figure*}[htb!]
    \centering
    \includegraphics[width=0.95\textwidth]{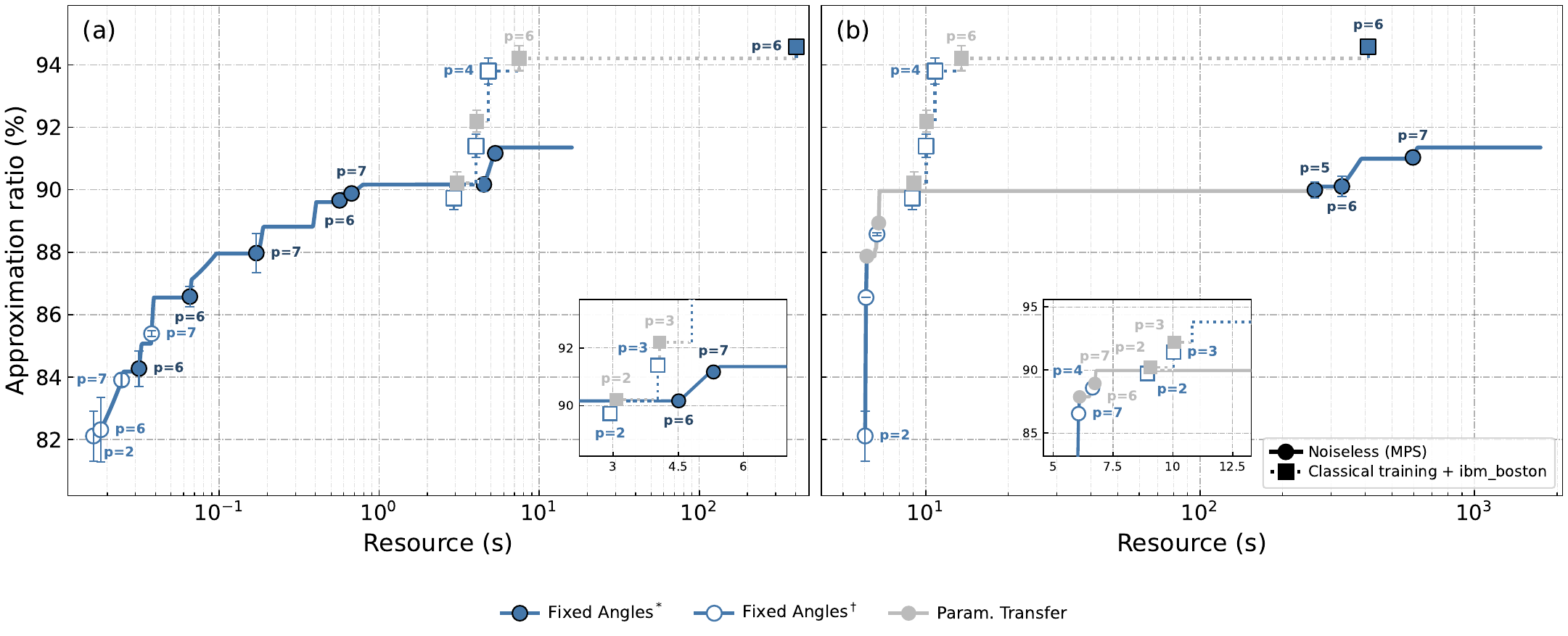}
    \caption{Actionable Pareto frontier for 144-node heavy-hex MaxCut on ten test instances under two resource models.
    (a)~Shot time only ($t_\text{qc}=0~\mathrm{s}$).
    (b)~An overhead of~$t_\text{qc}=6~\mathrm{s}$ applied to every submitted circuit, Eq.~\eqref{eq:tproxy}.
    Both panels are rebuilt from the same simulated points under their own cost model, with every family capped at $Q \le 10^4$ so that all of them share a common sampling grid.
    Line color shows which method configuration is the best deployable strategy at that resource level.
    Solid circle-marked lines are the MPS estimation, and dotted square-marked lines are the measured real-hardware frontier on \emph{ibm\_boston}.
    The hardware resource is classical training wall-clock time plus $Q\,t_\text{shot}$.
    Markers sit where a strategy supersedes the previous one on the frontier, and error bars show the $\pm 1$ standard error over the ten instances.
    The densest cluster of transitions between strategies is enlarged in the inset.
    Methods that optimize the QAOA angles are marked with a star, e.g. Fixed Angles$^\star$. Methods that have some method-parameter optimization are indicated without any superscript. All other methods are indicated with a $\dagger$.}
    \label{fig:pareto-overlay}
\end{figure*}

To determine the recommended strategy, we run optimizations for each of the settings in a grid corresponding to the cartesian product of parameter combinations $N \times M$ and the final fixed number of shots $Q$.
For Fixed Angles$^\star$, Linear Ramp, and Interp$^\star$, we perform a sweep over $N\in\{10, 20, 40, 50, 60, 75, 80, 100, 150\}$, $M\in\{10, 25, 50, 100, 200, 500, 1000\}$, and $Q\in\{100, 200, 250, 500, 1000, 2500, 5000, 10000\}$. 
Conversely, because Fixed Angles$^\dagger$ and Parameter Transfer are training-free, they utilize only the grid $Q$ for evaluation. 
This analysis covers all strategies shown on the Pareto frontier of Fig.~\ref{fig:recommendation}.
We ran the strategies using the methodology of Sec.~\ref{sec:bench_method} for these families on 30 heavy-hex-144 MaxCut instances split into 20 training and 10 test instances.
We fit an \emph{actionable prescription curve}~\cite{Neira2024} per method on the training instances and evaluate it on the same ten instances from Fig.~\ref{fig:recommendation}. 
The Pareto frontier in Fig.~\ref{fig:pareto-overlay} shows the best approximation ratio achievable with at most the given resource cost.
The underlying data is provided in App.~\ref{app:cost_analysis}.

In Fig.~\ref{fig:pareto-overlay} we compare the frontier with and without latency to show how $t_\text{qc}$ reorders the strategies.
Fig.~\ref{fig:pareto-overlay}(a) sets $t_\text{qc}=0~\mathrm{s}$, an ideal world in which circuits are prepared for execution in negligible time, and Fig.~\ref{fig:pareto-overlay}(b) sets $t_\text{qc}=6~\mathrm{s}$ in line with current observations.
This latency shifts the cross-over point between executing angle optimization on classical or quantum hardware.

In Fig.~\ref{fig:pareto-overlay}(a), Fixed Angles$^\dagger$ dominates the frontier at the smaller budgets, consistent with it being the fastest angle-setting method in Fig.~\ref{fig:performance}.
Tailoring $\boldsymbol{\beta}$ and $\boldsymbol{\gamma}$ to each problem instance initialized from fixed angles, i.e., Fixed Angles$^\star$, dominates the Pareto front for the rest of the range.
Parameter Transfer and Linear Ramp never dominate a segment of this frontier.
In Fig.~\ref{fig:pareto-overlay}(b), Parameter Transfer is best for intermediate budgets since it submits a single circuit, and Fixed Angles$^\star$ reclaims the frontier only at the largest budgets.
The best approximation ratio is the same in both panels.
However, the budget needed to reach it grows by roughly two orders of magnitude when latency is considered.
The hardware ratios in Fig.~\ref{fig:pareto-overlay} exceed those in Fig.~\ref{fig:recommendation} because Fig.~\ref{fig:pareto-overlay} scores each run by its best observed bitstring, matching the simulated curves, whereas Fig.~\ref{fig:recommendation} scores by expectation value of the observed bitstrings.

The methodology behind Fig.~\ref{fig:pareto-overlay} consumed $5.2\times10^{9}$ simulated shots across $17$ strategies and $1.6\times10^{7}$ circuit submissions, of which the $9$ strategies on the frontier account for roughly a fifth.
On \emph{ibm\_boston}, with $t_\text{shot}=100,\mu\mathrm{s}$ and $t_\text{qc}=6\mathrm{s}$, the full campaign would require $145$ hours of sampling but nearly three years of submission overhead, making it impractical under these conditions. 
The total runtime is therefore dominated by the submission latency per circuit evaluation. 
Reducing this latency would make campaigns of this scale feasible on quantum hardware.

\section{Discussion\label{sec:best}}

We now distil the operational implications of our benchmarking to help researchers and industry practitioners interested in utility-scale QAOA.
Training QAOA angles at this scale is possible with approximate energy evaluation methods such as MPS and PP.
MPS based methods achieve a higher quality and shorter runtime when the topology of the problem is mapped to the tensor line of the MPS, e.g., as done in Ref.~\cite{Matsuo2023} for hardware, see Sec.~\ref{sec:tns}. 
Furthermore, for low-connectivity problem graphs without short cycles MPS-based QAOA simulations remain efficient under small bond dimension~\cite{Watanabe2026}.
Crucially, a method that appears good from the perspective of a training with an approximate energy evaluation may underperform on quantum hardware.
Therefore, practitioners must carefully validate the performance of their angles, ideally, on the QPU as done in Sec.~\ref{sec:exact_sv_40q} and Sec.~\ref{sec:hardware}.
Whether large and dense problem instances can be trained offline at utility-scale is still an open question. First, current hardware is too noisy and approximate methods may take too long to run to obtain the needed accuracy.
Such conditions favour approaches like a parameter transfer.

As the depth and width of the QAOA circuits come close to the noise limits of the hardware, small differences between similar training methods become unmeasurable, see Sec.~\ref{sec:hardware}.
Therefore, when investigating QAOA close to hardware limits, it may not be necessary to use the angle setting method with the highest theoretical quality.
This may allow practitioners to select good methods while favoring runtime over quality.

Parameter transfer methods, and in particular the fixed angle approach~\cite{Wurtz2021}, are very powerful.
When applied to the problems for which they are designed, they perform well without any optimization making them fast, see App.~\ref{app:sv_results}.
Tailoring fixed angles to the problem instance increases the approximation ratio.
Therefore, practitioners may find value in developing fixed angles tailored to their problem class, if possible.
In addition, fixed angles can serve as good initial points for optimizers on problem instances for which they were not designed. 
This is exemplified by transferring fixed angles for regular graphs to Erd\H{o}s-R\'{e}nyi graphs to initialize a COBYLA optimization.
Rescaling fixed angles may help transfer them to other problem classes~\cite{Sureshbabu2024}. 
When transferring angles from smaller instances, performance improves with increasing circuit depth, but eventually saturates once the smaller training instances themselves reach optimality, although this limitation can be mitigated by training on more representative problem sizes~\cite{Watanabe2026}.

Iterative methods like Interp. often result in high quality angles at the cost of a large runtime.
Crucially, for problems with sharp local optima, such as LABS, the iterative methods are very sensitive to the quality of the $p=1$ angles.
Fortunately, low-order problems are efficient to evaluate at depth-one~\cite{Egger2021warmstartingquantum}.

Crucially, our comparison of MaxCut against MIS reveals that methods that underperform on one problem class may perform well on another one.
This is the case for the Fourier method.
These observations are not impacted by the type of energy evaluation employed.
Therefore, when practitioners deal with a new problem class testing several methods at small-scale is reasonable before spending significant resources at utility-scale.

Finally, in a practical setting, similar optimization problems are often solved on a regular basis.
This favours methods based on a parameter transfer or a database of known good angles.
This database can be built-up with time where more ressources are spent optimizing the angles of the first few instances encountered.
In settings where similar optimization problems are solved repeatedly, benchmarking can rank angle-setting methods and help decide how to allocate resources. 

Ultimately the performance estimation for real-world production systems should follow a statistical analysis like the one we presented in section \ref{sec:results_ressources}. 
However, a ``window sticker'' such as the one presented in Fig.\ref{fig:pareto-overlay} is cheap to display but expensive to earn.  
Access to abundant quantum wall-clock time and futher engineering advances are required for practical operational benchmarks for utility scale quantum optimization.


\section{Conclusions \label{sec:conc}}

Our work studies methods to find good QAOA angles for utility-scale problems where the energy cannot be computed exactly with brute-force classical methods.
Here, the closed-loop with the QPU can be too resource-intensive and costly.
Therefore, we focus on methods that either transfer QAOA angles or initialize an optimizer close to good known extrema in the $(\boldsymbol{\beta}, \boldsymbol{\gamma})$ space and classically refine them.
First, we survey the literature of QAOA angle setting methods. 
They are often tested at a scale where the energy $\langle H_C\rangle$ can be computed exactly.
We present a taxonomy of these methods to help practitioners navigate the tools at their disposal.
Second, we test these methods at utility-scale.
We evaluate their quality from the perspective of the approximate energy evaluation and then validate it on the QPU.
The largest problem we study is a depth-ten QAOA on 144 qubits.
We provide a methodology, based on a statistical test, to differentiate the angle setting methods based on their quality.
Finally, we show a Pareto front of time optimal QAOA angle setting methods.
Here, methods that rely on parameter transfer or fixed angles without further optimization perform the best and can be typically obtained at the cost of an $\mathcal{O}(1)$ look-up.
For MaxCut, the approximation ratio of these angles can be increased by 1-2\% at a large cost in training duration.

Crucially, some angle-setting methods are very time-consuming owing to the classical estimation of $\langle H_C\rangle$.
To address these limitations, we estimate the actionable Pareto front of the QAOA angle setting methods executed on the quantum hardware by treating an MPS sampler as a quantum backend.
We set realistic sampling and pre-processing times based on the quantum hardware.
For the sparse heavy-hex instances, the advantage of training angles on QPUs depends on classical latencies during circuit execution.
These latencies will change with hardware improvements and design choices.
Therefore, benchmarking or estimating runtimes is recommended before committing to training angles on classical or quantum hardware.
For instance, our results indicate that QAOA angles for dense problems may train faster on quantum than classical hardware.

Standard MaxCut problems are easy to solve classically in a short time.
In comparison, the methods that optimize the QAOA angles have a long runtime.
Going beyond MaxCut, Ref.~\cite{koch2025quantum} proposes a set of ten hard benchmarks which include MIS and LABS to track the progress of quantum optimization.
A further problem class to explore is multi-objective optimization~\cite{Kotil_2025}. It can quickly become hard classically as the number of objectives increases.
Here, a quantum approach also relies on QAOA circuits which need good angles.

In summary, we focus on standard QAOA with depths $p\leq 10$ and problems with a topology close to that of the quantum hardware.
This allows us to validate performance on the QPU.
As the quantum hardware continues to improve, QAOA angle setting methods for deeper QAOA and for denser optimization problems will surely also be studied.
Crucially, the insights we derive should transfer to QAOA variants such as warm-start and R-QAOA that overcome the obstacles that standard QAOA faces even at depth $o(\log n)$.
A study of angle setting methods in this context may also provide value and could exploit, for example, the recursive nature of algorithms like R-QAOA.

\section{Data availability}

The data supporting these findings are publicly available on Github at \href{https://github.com/Quantum-Working-Groups/QAOA-Parameter-Setting}{QAOA Parameter Setting}.

\section{Acknowledgments}

We acknowledge the use of IBM Quantum Credits for this work.
J.J.D acknowledges support by the Fulbright U.S. Student Program, sponsored by the U.S. Department of State and the Comisión Fulbright, Spain.
This work was supported as a part of NCCR SPIN, a National Centre of Competence in Research, funded by the Swiss National Science Foundation (grant number 225153).
S.E. acknowledges support by NNSA’s Advanced Simulation and
Computing Beyond Moore’s Law Project at Los Alamos
National Laboratory. 
J.M. has been supported by
the Czech Science Foundation (23-07947S). 
C.A. acknowledges support from the National Science Foundation NSF (Grant
No. 2231328). D.V. and A.R. acknowledge funding from NSF CCF 1918549 and used resources available through the National Research Platform (NRP) at the
University of California, San Diego. NRP has been developed, and is supported in part, by funding
from National Science Foundation, from awards 1730158, 1540112, 1541349, 1826967, 2112167, 2100237, and 2120019, as well as additional funding from community partners.
J.A.M.B. acknowledges the Gauss Centre for Supercomputing e.V. (www.gauss-centre.eu) for providing computing time on the GCS Supercomputer JUWELS Booster at Jülich Supercomputing Centre (JSC).

\appendix

\section{QAOA angles training methods\label{app:param_methods}}

We now explain in detail the QAOA angle setting methods that we test in Sec.~\ref{sec:res}.
We initialize recursive methods, such as Interp., Fourier, and RTS, with $p=1$ QAOA angles found by a two-dimensional grid scan followed by a SciPy optimization with COBYLA around the best point in the grid.
Here, we always use an efficient and exact energy evaluation designed for $p=1$~\cite{Egger2021warmstartingquantum}.

\subsection{Trivial Layer-Extension \label{app:trivial}}

The iterative layer-extension method with a trivial initialization, starts the optimization at depth $p+1$ using the converged angles $\boldsymbol{\beta}_p^\star$ and $\boldsymbol{\gamma}_p^\star$ from depth $p$. 
To maintain the energy level achieved at the previous depth, the new variational angles are initialized at zero
\begin{align}
\boldsymbol{\beta}^{0}_{p+1} =& \left([\boldsymbol{\beta}_p^\star]_1, \dots, [\boldsymbol{\beta}_p^\star]_p, 0\right), \\
\boldsymbol{\gamma}_{p+1}^{0} =& \left([\boldsymbol{\gamma}_p^\star]_1, \dots, [\boldsymbol{\gamma}_{p}^\star]_p, 0\right).
\label{eq:trivial}
\end{align}
Here, $[\boldsymbol
v]_i$ is the $i^\text{th}$ element of the vector $\boldsymbol v$.
Under this method the new $p+1$ layer initially acts as an identity operator, ensuring that the local optimizer begins from a stable point with an energy expectation value equal to the $p^\text{th}$ optimum. 

\subsection{Interp. and Fourier\label{app:interp_fourier}}

Based on observations that QAOA angles slowly and continuously vary between circuit depths $p$ and $p+1$, Zhou et al.~\cite{Zhou2020} propose the Interp. and Fourier methods.
Interp. linearly interpolates the angles obtained at depth $p$ to find an initial point for the optimization at depth $p+1$. 
This procedure starts from depth-one.
Given a depth-$p$ locally optimal schedule $\boldsymbol{\gamma}^\star_p$, the initial point  $\boldsymbol{\gamma}^0_{p+1}$ for the optimization at depth $p+1$ is
\begin{equation}
[\boldsymbol\gamma^0_{p+1}]_i = \frac{i-1}{p}[\boldsymbol\gamma^\star_{p}]_{i-1} + \frac{p-i+1}{p}[\boldsymbol\gamma^\star_{p}]_{i}
\label{eq:INTERP_G}
\end{equation}
for $i=1, 2,...,p+1$. 
Here, $[\boldsymbol
v]_i$ is the $i^\text{th}$ element of the vector $\boldsymbol v$, and $[\boldsymbol\gamma^\star_{p}]_0\equiv [\boldsymbol\gamma^\star_{p}]_{p+1}\equiv 0$. 
The expression for $\boldsymbol\beta$ is analogous to Eq.~(\ref{eq:INTERP_G}). 
The iteration terminates once a specified target QAOA depth is reached.

Alternatively, the Fourier method expresses the QAOA angle vectors in the sine-cosine basis as
\begin{align}
\left[\boldsymbol\gamma_p\right]_i =& \sum_{k=1}^q u_k\sin\left(\left(k-\frac{1}{2}\right)\left(i-\frac{1}{2}\right)\frac{\pi}{p}\right),~\text{and} \\
\left[\boldsymbol\beta_p\right]_i =& \sum_{k=1}^q v_k\cos\left(\left(k-\frac{1}{2}\right)\left(i-\frac{1}{2}\right)\frac{\pi}{p}\right).
\end{align}
Here, $q$ is the number of frequency components.
The standard version of the method corresponds to the $p=q$ case.
The QAOA angle optimization then takes place over $(\boldsymbol{u}, \boldsymbol{v})$.
The Fourier method generates an initial point for the optimization at $p+1$ by adding a high-frequency component, initialized at zero, to the optimal arrays such that $\boldsymbol{u}^0_{p+1} = (\boldsymbol u^\star_p,0)$ and $\boldsymbol{v}^0_{p+1} = (\boldsymbol v^\star_p,0)$. 
In this work, we chose $p=q$, therefore, the number of frequency components $q$ grows with $p$.
Then, the optimization starts from this initial point until it converges to a new local optimum $(\boldsymbol u^\star_{p+1},\boldsymbol v^\star_{p+1})$.

Zhou \emph{et al.}~\cite{Zhou2020} introduce multiple versions of the method, labeled Fourier[$q$, $R$], motivated by the observation that the Fourier method can converge to suboptimal angles.
Here, $q\in\mathbb{N}$ is the number of frequency components, and $R\in\mathbb{N}$ controls the number of random perturbations added to escape local optima.
The optimizations for depth $p+1$ thus starts from $R+1$ additional initial points generated by adding random perturbations to the best local optimum found at depth $p$. 
Specifically, at the beginning of the depth $p+1$ optimization, and for random perturbation~$r$
\begin{equation}
\boldsymbol{u}^{0, r}_{p+1} = 
\begin{cases}
    (\boldsymbol{u}^{\star}_{p},0), r=0\\
    (\boldsymbol{u}^{\star}_{p} + \alpha\boldsymbol{u}^{P,r}_{p},0), 1\leq r\leq R
    \end{cases}
\end{equation}

\begin{equation}
\boldsymbol{v}^{0, r}_{p+1} = 
\begin{cases}
    (\boldsymbol{v}^{\star}_{p},0), r=0\\
    (\boldsymbol{v}^{\star}_{p} + \alpha\boldsymbol{v}^{P,r}_{p},0), 1\leq r\leq R.
    \end{cases} 
\end{equation}
The arrays $\boldsymbol{u}_p^{P,r}$ and $\boldsymbol{v}_p^{P,r}$ are drawn from Gaussian distributions
\begin{equation}
    [\boldsymbol{u}^{P,r}_p]_k \sim \mathcal{N}(0,[\boldsymbol{u}^{\star}_p]^2_k)
\end{equation}
\begin{equation}
    [\boldsymbol{v}^{P,r}_p]_k \sim \mathcal{N}(0,[\boldsymbol{v}^{\star}_p]^2_k).
\end{equation}
Here, $\alpha$ is a parameter that regulates how much the random perturbation affects the value. 

The authors report a comparison between the Interp. method and different configurations for Fourier, specifically, Fourier[$p$, 0], Fourier[$p$, 10], and Fourier[5, 10].
The instance is a 14-node weighted three-regular graph. 
They observed that below depth $p \lesssim 20$, all methods and configurations have almost identical characteristics. 
Therefore, we use Fourier[$p$, 0].
Furthermore, since we explore depths $p \lesssim 20$, the other configurations would have a similar performance.

\begin{figure}
    \centering
    \includegraphics[width=\linewidth]{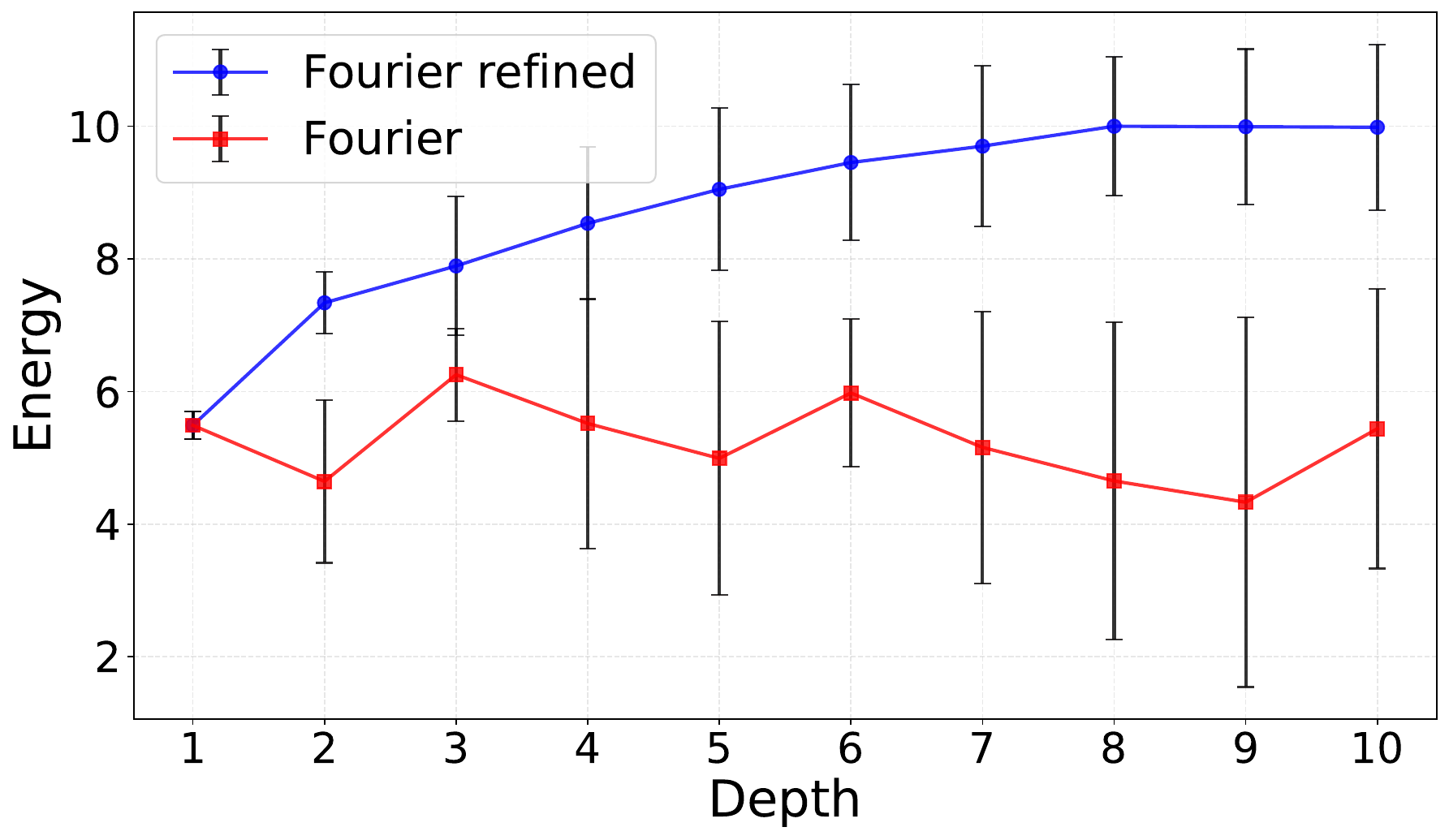}
    \caption{
    Improvements to Fourier by increasing the resolution of the optimizer.
    Fourier refined (blue dots) uses a finer grid scan, $\textrm{maxiter}=500$, and $\textrm{rhobeg}=0.01$. Fourier (red squares) uses a coarse grid scan, $\textrm{maxiter}=100$, and $\textrm{rhobeg}=0.2$.  
    The markers and error bars indicate the mean and standard deviation, respectively, of the energy for MaxCut over ten 20-node random three-regular instances, the same as in Fig.~\ref{fig:mis_vs_maxcut}.}
    \label{fig:comparison_fourier_refined}
\end{figure}

On MaxCut, for example, the Fourier method has a poor performance compared to the other methods, see Fig.~\ref{fig:144hh510_hw} and \ref{fig:mis_vs_maxcut} of the main text. 
To fix this, we (i) increase the resolution of the initial grid search from 15 points in $[0, \pi]$ to $100$ points in $[0,\frac{2}{\sqrt2}\pi]$, (ii) reduce COBYLA's initial step size rhobeg from $0.2$ to $0.01$, and (iii) increase the maximum allowed number of iterations from $100$ to $500$.
With these changes, the Fourier results greatly improve in quality, see Fig.~\ref{fig:comparison_fourier_refined}, and in accordance with the results reported by Zhou et al.~\cite{Zhou2020}. 
However, the run time of the method at depth 10 increases from approximately an average of 40.16 average number of energy evaluations accounting for grid search and Fourier training to 1259.15 evaluations for the grid search and Fourier training with the refined parameters.
By comparison, Interp. on a lower resolution required on average only 46.31 energy evaluations for the same depth.

\subsection{Trotterised quantum annealing and linear ramps\label{app:tqa_lr}}

A first-order Trotter discretization of the quantum annealing algorithm, with the Hamiltonian $H(t) = (1-t/T)H_M + (t/T)H_C$ is equivalent to a QAOA with the linear angle schedule
\begin{equation}
\label{eq:tqa_angles}
    \gamma_i = \frac{i}{p}\Delta t \qquad \beta_i = \left(1-\frac{i}{p}\right)\Delta t,
\end{equation}
see Ref.~\cite{sack2021quantum}.
Here, the discretized time steps  are $t_i=i\Delta t$ with $i=1,...,p$.
In our work we use the schedule presented in Eq.~(\ref{eq:tqa_angles}) directly as QAOA angles and refer to the angle setting method as the Trotterized Quantum Annealing (TQA) method.
Ref.~\cite{sack2021quantum} suggests that $\Delta t=0.75$ is a good value to use.

The Linear Ramp protocol~\cite{Montanezbarrera2024} is similar to the TQA method. 
However, Linear Ramp decouples the slopes of the $\beta$ and $\gamma$ angles.
Therefore, the QAOA angles are
\begin{equation}
\beta_i = \left(1-\frac{i}{p}\right)\Delta_\beta,
\qquad
\gamma_i = \frac{i+1}{p}\Delta_\gamma,
\end{equation}
for $i = 0, \ldots, p-1$. 
The trainable parameters of LR-QAOA are thus $\Delta_\beta$ and $\Delta_\gamma$.

In TQA and Linear Ramp, the schedules can be taken as is with a fixed $\Delta t$ and $(\Delta_\beta, \Delta_\gamma)$, respectively, and independently of the considered problem instance.
The results associated to this approach are indicated as TQA$^\dagger$ and Linear Ramp$^\dagger$ in Tab.~\ref{tab:summary_p10} of the main text.
The parameters $\Delta t$ and $(\Delta_\beta, \Delta_\gamma)$ can be optimized for each considered problem instance in an attempt to improve the performance of the linear schedules.
This is indicated by the absence of superscript, i.e., TQA and Linear Ramp, in Tab.~\ref{tab:summary_p10}.
Finally, after $\Delta t$ or $(\Delta_\beta, \Delta_\gamma)$ are optimized, we can further refine each angle of the QAOA schedule.
Results derived with this type of optimization are shown as TQA$^\star$ and Linear Ramp$^\star$ in Tab.~\ref{tab:summary_p10}.

\subsection{Recursive Transition States}

In Ref.~\cite{Sack2023}, Sack \emph{et al.} propose a recursive initialization method of the QAOA angles for depth $p+1$ by using the angles of depth $p$ to create Transition States (TS) in the $p+1$-depth energy landscape.
A TS is a saddle point with a unique negative curvature direction.
The TS at depth $p+1$ are
\begin{equation}
\label{eq:ts}
\begin{split}
\Gamma_{TS}^{p+1}(i,j)
&=
(\beta_1^\star,\ldots,\beta_{j-1}^\star,0,\beta_j^\star,\ldots,\beta_p^\star, \\
&\qquad
\gamma_1^\star,\ldots,\gamma_{i-1}^\star,0,\gamma_i^\star,\ldots,\gamma_p^\star)
\end{split}
\end{equation}
with $(j=i \ \text{or}\ j=i+1 \ \forall i\in[1,p]\bigr)
\ \text{or}\ i=j=p+1$.
The TS at $p+1$ thus have the same energy as the optimal found at depth $p$, as the angles $(\boldsymbol{\beta}^\star, \boldsymbol{\gamma}^\star)$ for depth $p$ are used, and the additional positions are initialized to 0.
As the $2p+1$ transition states are unstable saddle points and have the same energy as QAOA for depth $p$, initializing an optimization algorithm from one of these TS, or a neighborhood, will result in a state with lower energy. 
Therefore, Sack \emph{et al.} propose a greedy iterative procedure that runs the Broyden-Fletcher-Goldfarb-Shanno (BFGS) algorithm from each of the $2p+1$ transition states and selects the best resulting set of angles.
Based on these angles, the method moves to the next QAOA depth.

\subsection{Fixed angles\label{app:fixed_angles}}

We now discuss the fixed angles developed by Wurtz and Lykov for unweighted MaxCut on $d$-regular graphs~\cite{Wurtz2021}. 
First, the expectation of the QAOA cost function is a sum of $|E|$ individual edge contributions
\begin{align}
    F_p^{\mathcal{G}}(\gamma,\beta)=&\sum_{(i,j)\in E}f^{(ij)}_p(\gamma,\beta)\\
    =&\sum_{(i,j)\in E}\langle\psi(\beta,\gamma) |Z_iZ_j|\psi(\beta,\gamma)\rangle.
\end{align}
Here, each $f^{(ij)}_p(\gamma,\beta)$ is computed on a subgraph which only includes vertices at a distance $p$ from the vertices $i$ or $j$. 
This locality implies that QAOA cannot distinguish between graphs with cycles of size $>2p+1$.
This suggests that worst case graphs have no such small cycles~\cite{Wurtz2021a}.
Since the only subgraphs of a graph with no small cycles are tree graphs, which have no cycles, the optimized cost function of a worst-case graph
$\mathcal{G}_*=(V^*,E^*)$ is
\begin{equation}
    F_p^{\mathcal{G}_*}(\gamma,\beta)=\sum_{(i,j)\in E^*}f^{(ij)}_p(\gamma,\beta)=M^*f^\text{tree}_p(\gamma,\beta),
\end{equation}
where $M^*=|E^*|$. 
Therefore, the tree graph's cost function $f^\text{tree}_p(\gamma,\beta)$ provides the approximation ratio of the worst-case graph and the performance guarantee. 
This leads to two conjectures by Wurtz and Love~\cite{Wurtz2021a}.
\begin{enumerate}
    \item Large Loop conjecture: Worst-case graphs for fixed $p$ are bipartite and have no cycles less than $2p + 2$.
    \item Fixed Angle conjecture: Any graph evaluated at fixed angles optimal to the tree subgraph will have an approximation ratio larger than the guarantee.
\end{enumerate}
The fixed angles $\{\gamma,\beta\}_\text{tree}$, optimal to the tree graph, thus act as \emph{universally good} angles with good, but not maximum, performance on any MaxCut on a graph with the corresponding regularity.

To provide evidence for the conjectures, Wurtz and Lykov computed the fixed angles for unweighted regular graphs of varying degree. For $p \geq 11$, the performance guarantee for QAOA on
three-regular graphs is larger than the guarantee of the Goemans-Williamson algorithm, the best general-purpose
MaxCut solver with a performance guarantee. 
For all three regular graphs with $\leq 16$ vertices, they observe that no instances violate the performance guarantee. 
Additionally, the fixed angles are usually very close to global optima, with the typical difference between the fixed angle approximation ratio and global approximation ratio being $< 0.003$ for $p = 2$.
This suggests that the fixed angles are in the same basin of attraction as the global optima.
Therefore, it can be preferable to directly use the fixed angles rather than reoptimize them for minimal gain.

For non-regular graphs we employ the fixed angles of the $d$-regular graph with degree $d$ closest to the average node degree.
Crucially, fixed angles
may not be available for arbitrary $d$.
For example, the average node degree of Erd\H{o}s-R\'{e}nyi graphs increases with graph order $|V|$ and edge probability and the fixed angles for such high-degree graphs are not in the database~\cite{WurtzGithub}.

\subsection{Parameter transfer}

In Sec.~\ref{sec:results_validation} of the main text, we transfer angles from small-scale heavy-hex and line-based problems to larger ones.
The parameter transfer is done as follows.
(i)~We construct a database of optimized QAOA angles containing 12 and 39 nodes heavy-hex graphs and line-based graphs with 10 to 22 nodes.
The QAOA angles for the  $\leq 22$ and 39 node graphs are found with an exact and approximate energy evaluation, respectively.
We run all the methods described in Sec.~\ref{sec:res} and chose the best angles, as measured by the energy.  
A data point $(G_i, \boldsymbol{\theta}_i^\star)$, consisting of a graph $G_i$ and its optimized QAOA angles $\boldsymbol{\theta}^\star_i$ is added to the database by computing a set of $m$ features $f(G_i)\in\mathbb{R}^m$.
We use the QAOA depth and the average degree of the nodes in $G_i$ as the $m=2$ features. 
We chose the second feature due to the empirical observation that graphs with similar properties result in similar QAOA angles.
(ii)
Since QAOA angles for similar graphs tend to cluster, we group the entries of the database.
Here, graphs with similar features $f$ are assigned to the same cluster. 
This clustering is done with the DBSCAN algorithm~\cite{ester1996density, pedregosa2011scikit}.
It detects clusters based on density and has two parameters \texttt{Eps} and \texttt{MinPts}. 
\texttt{Eps} defines the size of the neighborhood $N_\text{eps}(f) = \{f_i: ||f_i-f||\leq \texttt{Eps}\}$ of a point $f$.
\texttt{MinPts} defines the number of points inside a cluster by the condition $|N_\text{eps}(f)|\geq \texttt{MinPts}$. 
If this condition is satisfied by a database entry $f$, then a cluster is formed by the set of features $\{f_i: \exists f_1 = f, f_2,...,f_k,...,f_i \text{ such that } f_{k+1}\in N_\text{eps}(f_k) \text{ and }  N_\text{eps}(f_k) \geq \texttt{MinPts}\}$.
(iii) To transfer angles to a new unseen graph $G'$ we find the cluster $f_c$ in the database whose features are closest to $f(G')$ as measured by $f_c={\rm argmin}_{f_i}||f_i-f(G')||$, for a fixed circuit depth.
(iv) Finally, we average the QAOA angles of the graphs in the database stored under $f_c$, and use the resulting averaged schedule as the angles for $G'$. 

\subsection{Principal Component QAOA\label{app:qaoa_pca}}

Inspired by clustering and transferability observations~\cite{Akshay2021, galda2023similarity}, QAOA-PCA~\cite{Parry2025} uses principle component analysis to reduce the dimensionality of the QAOA angle space.
For a given depth $p$, QAOA-PCA requires a training set of precomputed optimal angles for $N$ different problem instances, given as an $N \times 2p$ matrix $X$. 
The $2p \times 2p$ covariance matrix $\Sigma=\frac{1}{N-1}X^TX$ is built from $X$ and diagonalised. 
In the eigenbasis of $\Sigma$, the optimal angles in the training set have a large variance along the dominant eigenvectors, the principal components.
Conversely, eigenvectors with small eigenvalues exhibit little variation. 
They may be redundant and fixed to their mean values.
By expressing the QAOA angles in this basis, i.e., $(\beta_1,~...,~\beta_p,~\gamma_1,~...,~\gamma_p)=\sum_{i=1}^{2p}\alpha_i \boldsymbol{v}_i$, QAOA-PCA reduces the dimensionality of the search space by only optimising the coefficients $\alpha_i$ of $m<2p$ principal components.
Ref.~\cite{Parry2025} studies QAOA-PCA for the MaxCut problem on graphs with up to eight vertices while reducing the angle count by at least $50\%$.
QAOA-PCA converges in fewer iterations than standard QAOA, at the cost of a slightly lower approximation ratio. 

\begin{figure}[!htbp]
    \centering
    \includegraphics[width=\columnwidth]{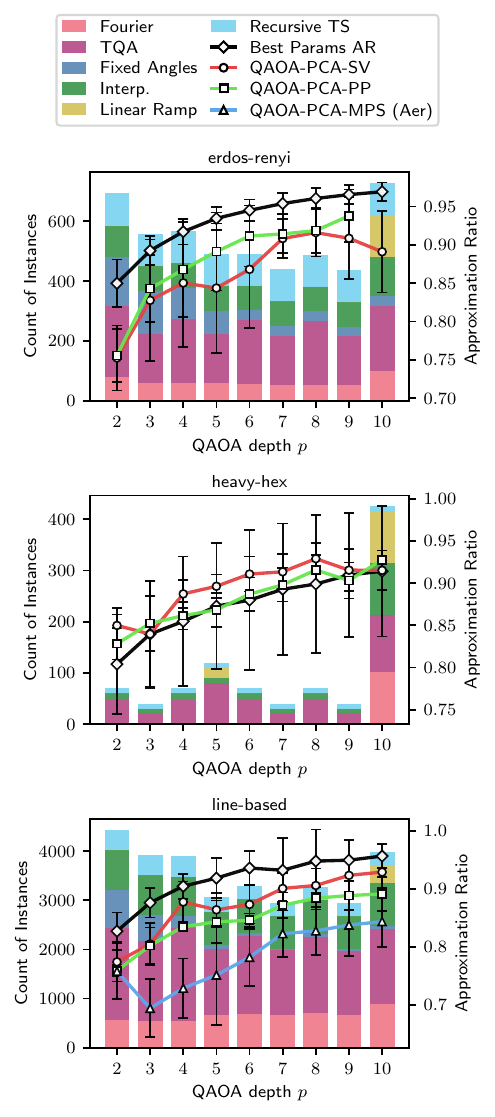}
    \caption{
    Performance of QAOA-PCA across the Erd\H{o}s--R\'enyi, heavy-hex, and line-based graph families for MaxCut. 
    The line plots show the approximation ratio as a function of QAOA depth for the best-angles baseline (red) and the PCA-based variants using state-vector (blue), Pauli-propagation (green), and MPS-Aer simulators (purple). 
    The stacked histograms show the proportion of the angle setting methods that contribute to the PCA basis.}
    \label{fig:qaoa_pca_results}
\end{figure}

We evaluate QAOA-PCA for depths $p=2,\dots,10$ on a per-graph-family basis.
I.e., we construct separate PCA databases for Erd\H{o}s--R\'enyi, heavy-hex, and line-based graphs. 
The reduced search space is thus learnt only from optimized QAOA angles associated with the same class of instances. 
Therefore, for each family and each depth $p$, we define the PCA basis from the best optimized angle vectors of the Fourier, TQA, fixed angle, Interp., Linear Ramp, and RTS methods, red lines in Fig.~\ref{fig:qaoa_pca_results}. 
The stacked histograms in Fig.~\ref{fig:qaoa_pca_results} show their relative contributions. 
We retain $p$ principal components at depth $p$, thus reducing the number of trainable variables by 50\%. 
We optimized the principal components with COBYLA with \texttt{maxiter}=100, and \texttt{rhobeg}=0.2.
We use the state-vector, Qiskit Aer MPS, and PP evaluators. 
The state-vector calculations are restricted to instances with up to $20$ nodes.
The MPS- and PP-based experiments happen on the same instances used for the corresponding baseline methods to compare with standard QAOA training.

QAOA-PCA improves the approximation ratio with increasing depth, while optimizing half as many angles as standard QAOA, see Fig.~\ref{fig:qaoa_pca_results}. 
The best-angle baseline, red line in Fig.~\ref{fig:qaoa_pca_results}, is better than QAOA-PCA except for heavy-hex.
Nevertheless, QAOA-PCA still preserves a substantial fraction of the achievable performance.
Therefore, within a fixed graph family, much of the relevant structure of the QAOA angle landscape is captured by a comparatively small number of principal directions.
Interestingly, QAOA-PCA managed to find better angles than the other methods for the heavy-hex graphs.
This suggestes that QAOA-PCA may be used to further refine QAOA angles in some cases.

\section{Tensor Networks\label{app:tn}}

As highlighted in the main text, MPS-based TN approaches are mainly limited to one-dimensional systems.
Two-dimensional TNs provide in principle more expressive ans\"atze, which can encode circuits with multi-dimensional correlations.
However, the lack of efficient algorithms for simulating quantum circuits using such TNs has so far limited their practical applications.
The key issue underlying two-dimensional TNs is that contracting TNs with a topology that includes loops, such as square-lattices, is classically hard.
For this reason, the SVD-based compression schemes in MPS-based simulations can hardly be expanded to two-dimensional systems.
Belief propagation (BP) has recently become popular~\cite{Tindall2023_BP-TN} to overcome this problem.
It thus simulates the circuit based on the equation obtained for loop-less TNs and applies it to a TNs with loops.
BP has been applied to simulate utility-scale quantum circuits~\cite{Rudolph2025_BP-QuantumCircuit}, including QAOA ones~\cite{Luchnikov2024_BP-Annealing}.
Notably, BP becomes increasingly inaccurate as the circuit topology deviates from one‑dimensional structures: the method yields very accurate representations of quantum circuits with a heavy-hex topology, but becomes inaccurate when applied to circuits with a square-topology~\cite{Rudolph2025_BP-QuantumCircuit}.
Moreover, although heuristic accuracy estimates exist~\cite{Rudolph2025_BP-QuantumCircuit}, they are not as rigorous as the same bounds available for MPSs.

\subsection{Time complexity in MPS computations}

We analyze here the time complexity of MPS-based simulations by calculating $\langle H_C\rangle$ for depth-10 QAOA, using a bond dimension of 24. 
$H_C$ is the MaxCut Ising Hamiltonian.
The underlying graphs are the four categories presented in App.~\ref{app:graphs}.
For reference, we also report the PP runtime with a maximum Pauli weight of five and minimum coefficient of $10^{-4}$.
In total, we evaluate 5084 instances on the Sol supercomputer~\cite{HPC:ASU23}.
Before evaluating $\langle H_C\rangle$ with an MPS we relabel the nodes in the graphs following the SAT mapping of Ref.~\cite{Matsuo2023}.

\begin{figure}[ht]
    \centering
    \includegraphics[width=\columnwidth]{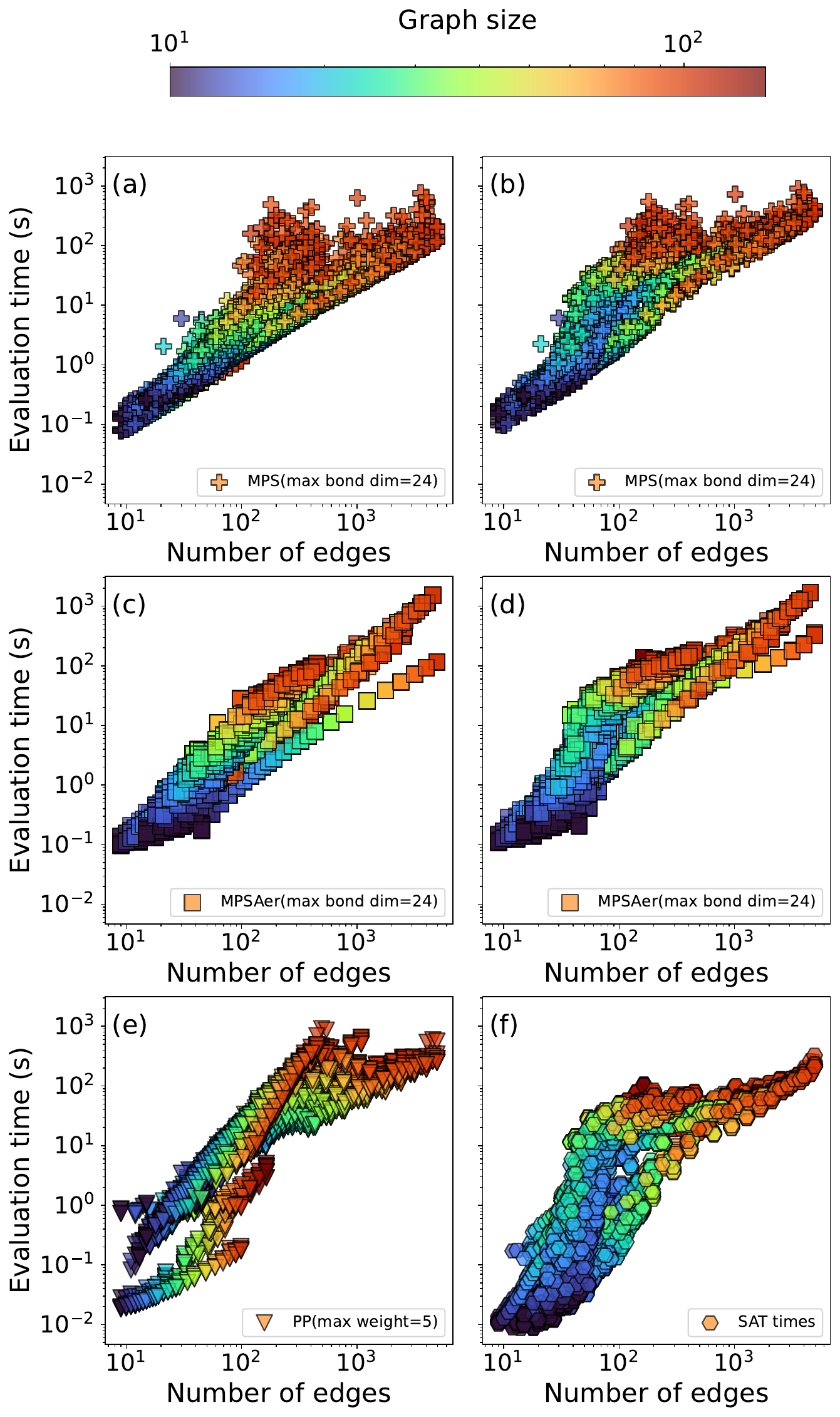}
    \caption{Evaluation time of $p=10$ QAOA as a function of the number of edges in the graphs.
    Each data point is a different graph with the color showing the number of nodes.
    Panels (a) and (b) show the Quimb-based MPS without and with the duration of the SAT remapping, respectively. 
    Panels (c) and (d) show the Qiskit Aer-based MPS without and with the duration of the SAT remapping, respectively.
    (e) evaluation time of PP. 
    (f) duration of the SAT remapping.}
    \label{fig:eval_time_num_edges}
\end{figure}

The evaluation time has a polynomial scaling with the number of edges in the graph. 
For MPS-based and PP-based evaluators the computational cost increases noticeably with graph size and density, see Fig.~\ref{fig:eval_time_num_edges}(a), (c), and (e).
For instance, computing $\langle H_C\rangle$ for a fully connected graph with 100 nodes typically takes 100 to 1000 seconds.
Dense utility-scale instances are thus challenging to train with approximate methods simply based on an evaluation time argument. 
Crucially, this does not even consider that the quality of the estimated $\langle H_C\rangle$ decreases, for a given bond dimension (for MPS simulations) or for a given coefficient truncation (for PP simulations), as the graph density increases.

We observed that the Quimb-based backend is faster than the Aer-based one for 3891 graphs out of the 5084.
Pauli propagation performs favorably for smaller graphs, often yielding lower evaluation times than both MPS approaches. However, its cost grows faster with the number of edges, eventually exceeding that of the MPS evaluators for larger graphs, see Fig.~\ref{fig:eval_time_num_edges}(e).
The runtime of the SAT remapping increases rapidly at smaller graph sizes before reaching a plateau, see Fig.~\ref{fig:eval_time_num_edges}(f). 
Although the time to run this remapping is comparable to the MPS evaluation time, this overhead is incurred only once per graph instance when optimizing $\boldsymbol{\beta}$ and $\boldsymbol{\gamma}$, making it a reasonable preprocessing method.

\subsection{Fidelity bounds in MPS computations}
\label{app:mps_bounds}

\begin{figure}
    \centering
    \includegraphics[width=0.95\columnwidth, clip, trim=7 0 6 0]{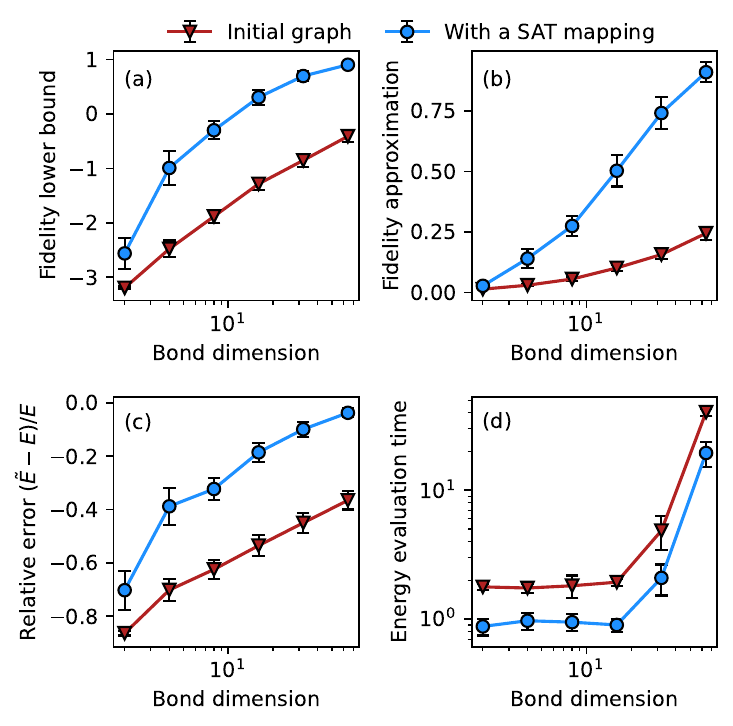}
    \caption{
    Impact on the energy evaluation from mapping the QAOA variables onto the MPS one-dimensional lattice for $p=1$ QAOA. 
    Data are obtained on a set of five random-three-regular graphs with 40 nodes and presented as function of the bond dimensoin.
    Panels (a) and (b) show $F_\text{bound}$ and $F_\text{approx}$, respectively.
    (c) Relative error in energy between the MPS $\tilde{E}$ and the exact energy $E$.
    (d) Runtime of the Quimb-based MPS simulations.
    \label{fig:mps_sat_bounds}}
\end{figure}

We now show two metrics from Ref.~\cite{Stoudenmire2020_Limits-QC} to gauge the accuracy of an MPS-based simulation.
The first metric 
\begin{equation}
  F_\text{bound} = 1 - 2 \sum_{i=1}^{N_g} \sqrt{1 - \sum_{j=1}^{m} \lambda_{j,i}^2 } \, ,
  \label{eq:FidelityBound}
\end{equation}
lower-bounds the fidelity of the simulation~\cite{Stoudenmire2020_Limits-QC}.
Here, $N_g$ is the number of two-qubit gates of the circuit, $m$ is the bond dimension, and $\left\{ \lambda_{j,i} \right\}_{j=1}^{m}$ are the $m$  singular values retained when applying the SVD compression after applying the $i^\text{th}$ gate.
This lower bound can be very loose and can even become negative for low bond-dimension simulation, see Fig.~\ref{fig:mps_sat_bounds}(a).
The second metric
\begin{equation}
  F_\text{approx} = \prod_{i=1}^{N_g} \left( \sum_{j=1}^{m} \lambda_{j,i}^2 \right)
  \label{eq:FidelityApprox}
\end{equation}
estimates the fidelity of an MPS-based simulation.
Our numerical simulations, reported in the upper panel of Fig.~\ref{fig:mps_sat_bounds}, confirm that $F_\text{approx}$ yields a very accurate estimate of the fidelity of an MPS-based computation~\cite{Stoudenmire2020_Limits-QC}.
Notably, these metrics also show the impact of the SAT mapping: applying this preprocessing step increases both $F_\text{bound}$ and $F_\text{approx}$, see Fig.~\ref{fig:mps_sat_bounds}. 
This indicates that appyling the SAT mapping increases the accuracy of the underlying MPS simulation for a given bond dimension.

\subsection{Sampling and energy evaluation}
\label{app:tn_sample}

For a given TN state $\vert \Psi_\text{TN} \rangle$, calculating the average of an operator $\langle \Psi_\text{TN} \vert O \vert \Psi_\text{TN} \rangle$ and sampling from the probability distribution $p_{\sigma} = \left| \langle \sigma \vert \Psi_\text{TN} \rangle \right|^2$ do not have the same complexity.
For loop-free TNs (such as MPS) efficient methods for executing both steps exists~\cite{Vidal2012_PerfectSampling}, with a computational cost scaling polynomially with the bond dimension.
For TNs with loops, calculating both generic expectation values and sampling from the underlying probability distribution is, in principle, NP-hard.
Heuristic schemes for expectation value calculations exist such as, for instance, the BP approach introduced above or methods based on light-cone simplifications.
This is, however, not true for sampling.
Sampling from the probability distribution induced by a TN requires evaluating the expectation value associated with observables that are non-local and, therefore, cannot leverage light-cone simplifications.
Moreover, even if BP-based sampling schemes have been devised, they are less efficient than their expectation value evaluation counterparts~\cite{Luchnikov2024_BP-Annealing,Rudolph2025_BP-QuantumCircuit}.

\section{QAOA Training pipeline \label{app:training_pipeline}}

The QAOA angles in this work are trained  with the QAOA Training Pipeline~\cite{training_pipeline}.
This open-source software package is a collection of \texttt{Trainers} and \texttt{Evaluators} that implement some of the QAOA training methods in the literature.
The pipeline defines a QAOA training method as a chain of multiple trainers that run sequentially, improving the QAOA angles at the different steps using the methodologies described in the main text. 
These trainers may require evaluating the energy with an \texttt{Evaluator}.
The training pipeline implements multiple evaluator based on Statevector (SV) simulation, Pauli Propagation (PP), or Matrix Product States (MPS).
These evaluators are good for different tasks. 
For example, the Statevector evaluator gives the exact energy.
However, its computational cost scales exponentially with the size of the problem instance. 
PP and MPS are approximate energy evaluation methods, and each has its own trade-offs, as discussed in Sec.~\ref{sec:evaluation}.

To make the training reproducible and transparent, the QAOA Training pipeline defines QAOA training methods through JSON configuration files. 
These files specify the trainer chain, their parameters, and the evaluators along with their parameters. 

\section{Problem models\label{app:problem_models}}

We now briefly show how we model the maximum cut (MaxCut), Maximum Independent Set (MIS) and Low Autocorrelation Binary Sequence (LABS) problems.
In MaxCut we seek a partitioning $V_0\cap V_1=\emptyset
$ of the nodes $V=V_0\cup V_1$ of a graph $G=(V, E)$ such that the sum of the edge weights connecting $V_0$ and $V_1$ is maximum.
For each node $i\in V$ we associate a binary decision variable $x_i$ such that $x_i=j$ if $i\in V_j$ for $j=0,1$.
Then the objective of MaxCut is to maximize
\begin{align}\label{eqn:max_cut}
    \max_{x\in\{0, 1\}^n}\sum_{(i,j)\in E} w_{ij} x_i(1-x_j)
\end{align}
where $w_{ij}$ is the weight of edge $(i,j)\in E$.

MaxCut is a quadratic unconstrained binary optimization (QUBO) problem, as opposed to MIS which has constraints.
In MIS we seek the largest sub-set of nodes $S\subset V$ for a graph $G=(V, E)$ such that no two nodes in $S$ share an edge in $E$.
Mathematically, we associate a decision variable $x_i$ to node $i\in V$ such that $x_i=1$ if $i\in S$ and $0$ otherwise and optimize
\begin{align}\label{eqn:mis1}
    & \max_{x\in\{0, 1\}^n}\sum_{i\in V}x_i \\
    \text{such that}\quad & x_i+x_j\leq 1~\forall~(i,j)\in E. \label{eqn:mis2}
\end{align}
To tackle this problem with QAOA we convert it to a QUBO by squaring the constraints and adding them to the objective with a Lagrange multiplier set to two.
This ensures that the ground states of the resulting QUBO are maximum independent sets.
Finally, to convert the MaxCut and MIS QUBOs to Hamiltonian cost operators $H_C$ we replace each binary variable $x_i$ by a spin variable $z_i\in\{-1, 1\}$ such that $2x_i=1-z_i$ and then promote $z_i$ to Pauli spin operators $Z_i$.
By contrast to MaxCut, the MIS cost-operator has both one- and two-local terms representing the objective and constraints, respectively.
This can make the MIS QAOA energy landscape more complex than the MaxCut one.

The LABS problem class consists of finding a sequence $S = (s_1, s_2, \dots, s_n)$ with $s_i \in \{-1, +1\}$ that minimizes the sum of squared off-peak autocorrelations. 
For a given lag $k \in \{1, \dots, n-1\}$, the autocorrelation is $C_k(S) = \sum_{i=1}^{n-k} s_i s_{i+k}$. 
The objective is to minimize the energy function $E(S) = \sum_{k=1}^{n-1} C_k^2(S)$, which is inversely proportional to the merit factor $MF = n^2 / 2E(S)$.
Unlike the quadratic forms of MaxCut and MIS, the LABS objective is quartic since
\begin{equation}
    H_C = \sum_{k=1}^{n-1} \left( \sum_{i=1}^{n-k} Z_i Z_{i+k} \right)^2 = \sum_{k=1}^{n-1} \sum_{i,j=1}^{n-k} Z_i Z_{i+k} Z_j Z_{j+k}.
    \label{eq:labs_hamiltonian}
\end{equation}
This Hamiltonian has many quartic interactions which are challenging to implement on noisy hardware~\cite{Dragoi2025}.

\section{Graph instances\label{app:graphs}}

Here, we provide details on the graphs that initialize the MaxCut and MIS problem instances.
We built a database with four types of graphs.
(i)~Unweighted random $k$-regular graphs with $k\in\{3, 4,...,9\}$.
We include these graphs since certain training methods, such as the fixed angles, are specifically designed for them.
(ii)~Unweighted Erd\H{o}s-R\'{e}nyi graphs with edge probabilities of 20\%, 30\%, 40\% or 50\% are included because their random structure allows us to benchmark how well the QAOA angle setting methods perform when a given regularity is not present.
(iii)~Weighted line-based graphs with edge weights drawn uniformly from the set $\{-1, 1\}$.
Here, a graph $G_{\ell}=(V, E_\ell)$ with $|V|=n$ nodes is built from a line graph $G_{0}=(V, E_0)$ with $E_0=\{(i,i+1)~|~i,i+1\in V\}$ to which we apply $\ell$ layers of the line SWAP network, defined in Ref.~\cite{Weidenfeller2022}, to generate additional edges in $E_\ell$.
At $k=n-2$ the resulting graph is a fully-connected graph with edge weights of either $-1$ or $+1$, thus resembling a low-resolution Sherrington-Kirkpatrick model.
These graphs are included because their energy evaluation becomes difficult at scale as $\ell$ increases. 
Furthermore, such graphs produce QAOA circuits that are easy to implement on superconducting qubit hardware when built from a few SWAP layers by leveraging Ref.~\cite{Weidenfeller2022}.
(iv) Finally, we include weighted hardware native heavy-hex graphs with uniform weights drawn from $\mathcal{N}(0, 1)$.
These graphs allow us to study the QAOA angle setting methods at utility-scale on quantum hardware since they do not require routing with SWAP gates~\cite{Montanezbarrera2024}.
For MaxCut, the edge weights enter the QUBO through Eq.~(\ref{eqn:max_cut}).
For MIS, we ignore the edge weights, see Eq.~(\ref{eqn:mis2}).

\begin{figure*}[t]
    \centering
    \includegraphics[width=0.95\textwidth]{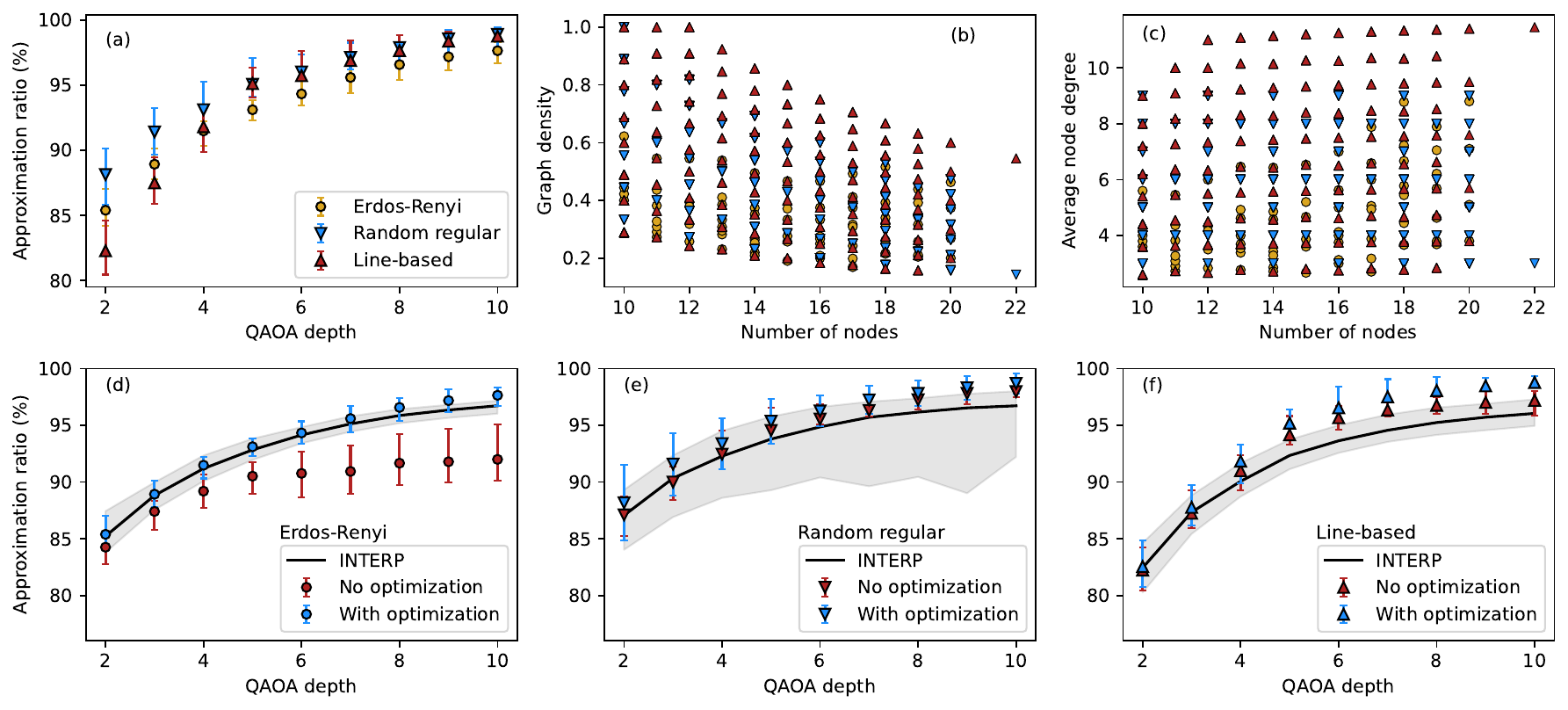}
    \caption{
    Performance of the fixed angles methodology with and without re-optimization.
    Energies are evaluated with state vector evaluators.
    In all panels the solid lines and markers show median values and shaded areas and error bars show the 25\% and 75\% quantiles.
    (a) approximation ratio for the three graph families to which the fixed angles are applied.
    (b) and (c) show scatter plots of graph edge density and the average degree of nodes, respectively, as function of the number of nodes.
    (d), (e), and (f) show the approximation ratio for Erd\H{o}s-R\'{e}nyi, Random $k$-regular, and line-based graphs as a function of QAOA depth.}
    \label{fig:sv_analysis}
\end{figure*}

\section{Quality analysis}

\subsection{Goemans-Williamson as a baseline\label{app:example_gw}}

A genuinely challenging baseline for QAOA to beat on MaxCut is the Goemans-Williamson algorithm \cite{goemans1995improved}, which is fully classical and runs in polynomial-time. For the heavy-hex, Erd\H{o}s-R\'{e}nyi, random-regular, and line-based instances considered throughout the paper, we generate results for 10, 100, 1000, and 10000 rounds of the Goemans-Williamson randomized hyperplane rounding.
Often, we see that the cost of the cuts obtained in 10, 100, 1000, and 10000 randomized hyperplane rounding rounds coincide with the global minimizer computed using IBM ILOG CPLEX, an exponential-time classical algorithm.  For 10-node graphs, the runtime is under 0.01 seconds. For 100-node graphs, the runtime grows to 0.1--0.5s. On the 144-node heavy-hex graphs, the runtime can be as long as 25.45 seconds. All times are measured on Apple MacBook Pro with an M3 processor and include both solving the SDP relaxation and the rounding. 

\subsection{Statevector results\label{app:sv_results}}

In this section we show a deeper analysis of the results in Sec.~\ref{sec:results_quality} obtained with state vectors for small-scale instances, i.e. $n\leq 22$ nodes.
The approximation ratio of MaxCut for Erd\H{o}s-R\'{e}nyi, random-regular, and line-based graphs steadily increases with depth $p$, see Fig.~\ref{fig:sv_analysis}(a).
The density and the average node degree of the graphs included in this data are shown in Fig.~\ref{fig:sv_analysis}(b) and (c).
Crucially, we observe that the fixed angles need to be optimized to each problem instance of Erd\H{o}s-R\'{e}nyi graphs to beat the performance of, for exmaple, Interp. which we find performs well in general.
For the random-regular and line-based graph instances the fixed angles beat Interp. even without optimization.

\begin{figure}[tb]
    \centering
    \includegraphics[width=\linewidth]{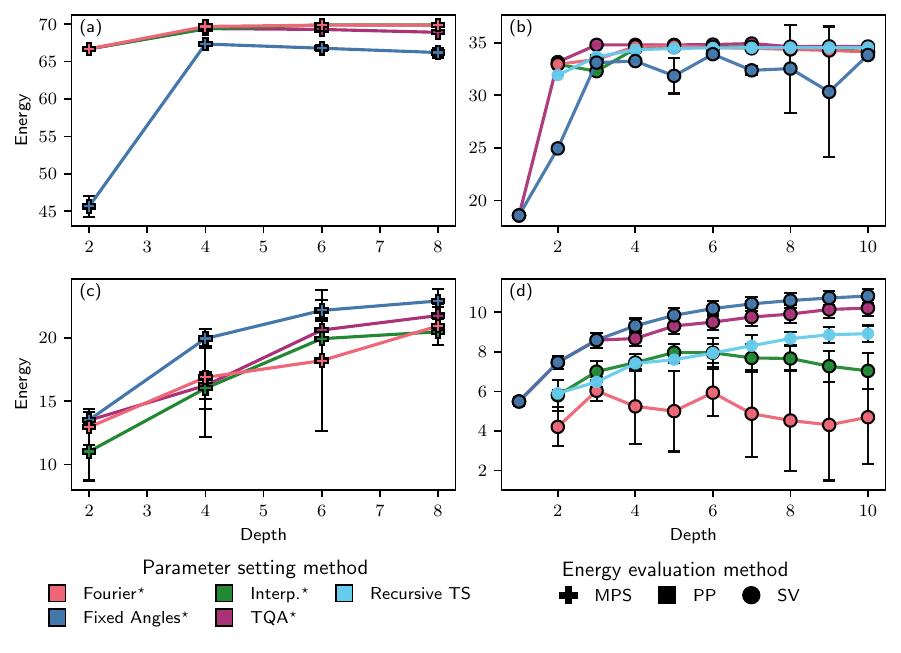}
    \caption{Energy achieved by different combinations of angle setting methods and energy evaluators, for MIS (a) and (b), and MaxCut (c) and (d). 
    Panels (a) and (c) show results obtained with an MPS energy evaluators on ten 40-node random three-regular graph instances. 
    Panels (b) and (d) show results obtained with state vector simulators on ten 20-node random three-regular graphs. 
    The markers and error bars indicate the mean and standard deviation, respectively, of the energy over the ten instances.}
    \label{fig:mis_vs_maxcut_full}
\end{figure}

\subsection{Additional MIS and MaxCut data}

Here, we first present results with MPS and Statevector energy evaluation for MIS and MaxCut problems based on three-regular graphs, complementing the PP data shown in Sec.~\ref{sec:results_quality}. 
MPS have a similar behavior to the PP results shown in Sec.~\ref{sec:results_quality}.
Fourier$^\star$ is best in terms of energy for MIS, and the Fixed Angles$^\star$ is best for MaxCut, see Fig.~\ref{fig:mis_vs_maxcut_full}. 
For the MaxCut Statevector results, we see that the best method is Fixed Angles$^\star$. 
This is expected, as these angles are designed for random regular graphs. 
However, TQA$^\star$ and Fourier$^\star$ are among the best performing methods for MIS. 
We also observe poor performance for Fourier$^\star$ and Interp.$^\star$ on MaxCut on the evaluated instances. 
A better performance could be obtained by refining the optimizer, see App.~\ref{app:interp_fourier}. 

\begin{figure}[ht]
    \centering
    \includegraphics[width=\columnwidth, trim={0 0 0 2cm},
    clip]{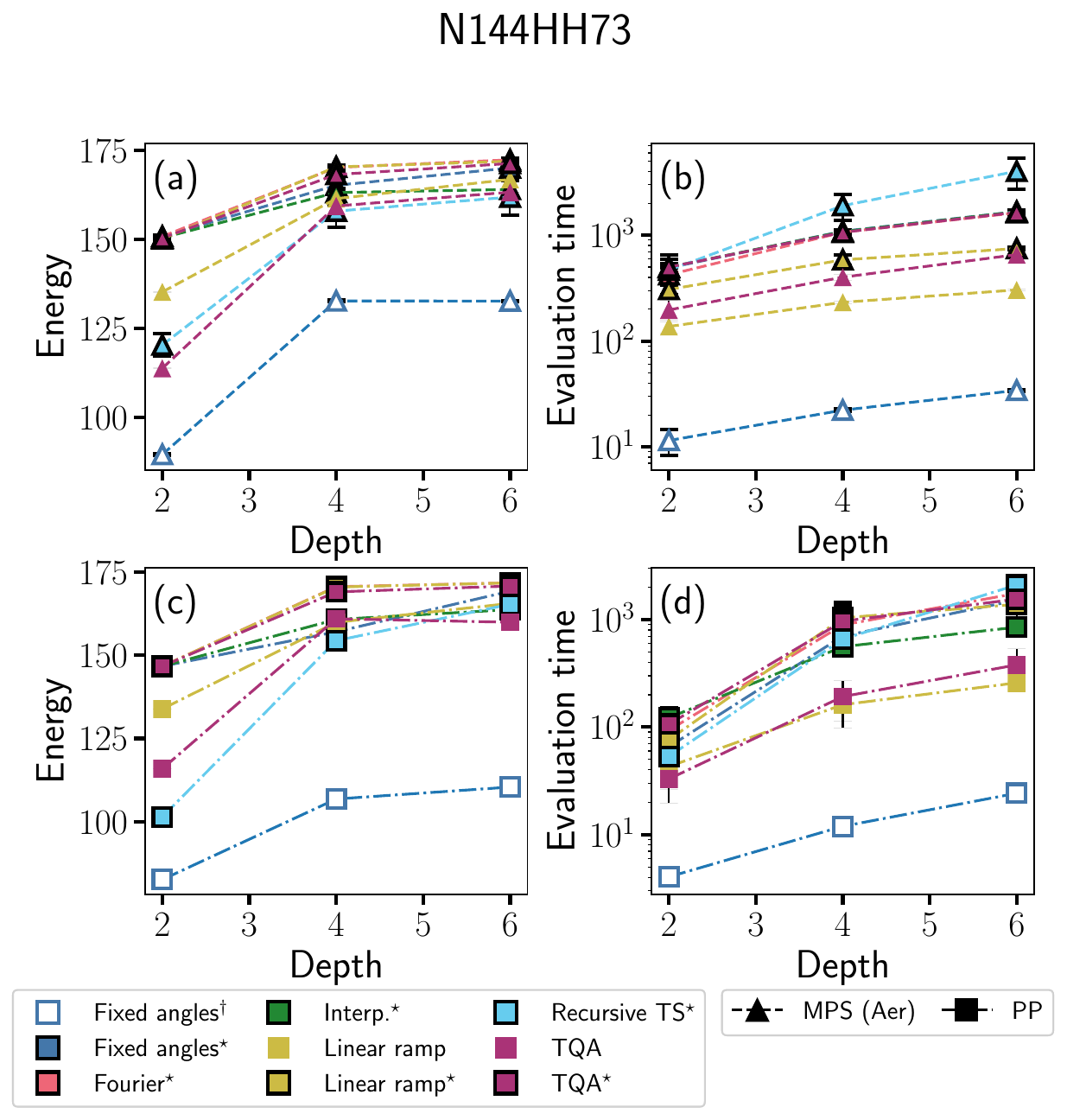}
    \caption{
    (a) and (c) MIS energy obtained from MPS (Aer), with a bond dimension of 24, and PP simulations with a max weight of 4, respectively.
    Markers are the average of ten 144-node heavy-hex graphs.
    (b) and (d) show the runtime of the QAOA angle setting methods with MPS and PP, respectively.
    }
    \label{fig:misN144HH73}
\end{figure}

Second, we test the training methods for MIS on utility scale problems on heavy-hex graphs with 144 nodes.
Similar to Sec.~\ref{sec:results_quality}, the non-optimized Fixed Angles performs the worst, see hollow blue triangles/squares in Fig.~\ref{fig:misN144HH73}.
By contrast, Fourier$^\star$ and Linear Ramp$^\star$ perform the best, see red and yellow markers, respectively, in Fig.~\ref{fig:misN144HH73}. 
Overall, the training process for the MIS problem is longer than for MaxCut on the same graphs.
A summary of a selection of MIS results is given in Tab.~\ref{tab:summary_p06_mis}, with a similar structure to Tab.~\ref{tab:summary_p10}.
Given the longer training process, the table shows fewer graphs overall.
Many of these graphs do not have entries in the Fixed Angles database, and could not be run.

\begin{table*}[p]
    \centering
    \begin{tabular}{l|cccc|}
\toprule
            & ER & HH & LB & RR \\
\midrule
\multicolumn{5}{c}{\rule{0pt}{8pt}MPS (Aer)} \\
\midrule
Fourier$^\star$ & {\cellcolor[HTML]{84248A}} \color[HTML]{F1F1F1} $65882.6\pm 2605.2$ & {\cellcolor[HTML]{BED2E6}} \color[HTML]{000000} $17238.8\pm 41.9$ & {\cellcolor[HTML]{A8C3DE}} \color[HTML]{000000} $24117.0\pm 532.0$ & {\cellcolor[HTML]{BDD1E5}} \color[HTML]{000000} $17396.2\pm 68.4$ \\
Interp.$^\star$ & {\cellcolor[HTML]{8C89C0}} \color[HTML]{F1F1F1} $41477.0\pm 10764.6$ & {\cellcolor[HTML]{C0D4E6}} \color[HTML]{000000} $16407.2\pm 251.6$ & {\cellcolor[HTML]{ACC6DF}} \color[HTML]{000000} $22849.9\pm 266.4$ & {\cellcolor[HTML]{BDD1E5}} \color[HTML]{000000} $17439.4\pm 59.8$ \\
Linear Ramp$^\star$ & {\cellcolor[HTML]{852D90}} \color[HTML]{F1F1F1} $63828.8\pm 1920.9$ & {\cellcolor[HTML]{BED2E6}} \color[HTML]{000000} $17199.0\pm 19.1$ & {\cellcolor[HTML]{A7C2DD}} \color[HTML]{000000} $24453.7\pm 62.5$ & {\cellcolor[HTML]{BDD1E5}} \color[HTML]{000000} $17474.2\pm 19.2$ \\
Linear Ramp & {\cellcolor[HTML]{8947A0}} \color[HTML]{F1F1F1} $58099.4\pm 5174.1$ & {\cellcolor[HTML]{BFD3E6}} \color[HTML]{000000} $16689.6\pm 9.6$ & {\cellcolor[HTML]{A8C3DE}} \color[HTML]{000000} $24088.6\pm 117.4$ & {\cellcolor[HTML]{BED2E6}} \color[HTML]{000000} $16943.9\pm 14.9$ \\
Linear Ramp$^\dagger$ & {\cellcolor[HTML]{C9DBEA}} \color[HTML]{000000} $13260.6\pm 3107.3$ & {\cellcolor[HTML]{C4D7E8}} \color[HTML]{000000} $15020.7\pm 16.9$ & {\cellcolor[HTML]{B5CCE3}} \color[HTML]{000000} $19940.6\pm 24.3$ & {\cellcolor[HTML]{C4D7E8}} \color[HTML]{000000} $15035.5\pm 18.1$ \\
Recursive TS$^\star$ & {\cellcolor[HTML]{84278C}} \color[HTML]{F1F1F1} $65385.0\pm 2303.4$ & {\cellcolor[HTML]{C1D4E7}} \color[HTML]{000000} $16192.4\pm 543.3$ & {\cellcolor[HTML]{A8C3DE}} \color[HTML]{000000} $24319.1\pm 40.1$ & {\cellcolor[HTML]{BED2E6}} \color[HTML]{000000} $17263.4\pm 56.7$ \\
TQA$^\star$ & {\cellcolor[HTML]{96ACD2}} \color[HTML]{F1F1F1} $32061.9\pm 11764.8$ & {\cellcolor[HTML]{BED2E6}} \color[HTML]{000000} $17136.6\pm 34.3$ & {\cellcolor[HTML]{A8C3DE}} \color[HTML]{000000} $24234.3\pm 142.4$ & {\cellcolor[HTML]{BDD1E5}} \color[HTML]{000000} $17302.5\pm 39.1$ \\
TQA & {\cellcolor[HTML]{CBDCEB}} \color[HTML]{000000} $12617.7\pm 2917.7$ & {\cellcolor[HTML]{C0D4E6}} \color[HTML]{000000} $16331.0\pm 4.1$ & {\cellcolor[HTML]{B1C9E1}} \color[HTML]{000000} $21160.4\pm 21.2$ & {\cellcolor[HTML]{BFD3E6}} \color[HTML]{000000} $16776.6\pm 9.0$ \\
TQA$^\dagger$ & {\cellcolor[HTML]{EEF5F9}} \color[HTML]{000000} $-326.3\pm 8564.5$ & {\cellcolor[HTML]{C1D4E7}} \color[HTML]{000000} $15953.9\pm 8.8$ & {\cellcolor[HTML]{B5CCE3}} \color[HTML]{000000} $19842.5\pm 23.5$ & {\cellcolor[HTML]{BFD3E6}} \color[HTML]{000000} $16726.1\pm 7.6$ \\
\cline{1-5}
\multicolumn{5}{c}{\rule{0pt}{8pt}MPS (Quimb)} \\
\midrule
Fourier$^\star$ & {\cellcolor[HTML]{B34003}} \color[HTML]{F1F1F1} $65029.4\pm 4379.9$ & {\cellcolor[HTML]{FEE28E}} \color[HTML]{000000} $17218.3\pm 28.1$ & {\cellcolor[HTML]{FECC61}} \color[HTML]{000000} $24645.5\pm 21.4$ & {\cellcolor[HTML]{FEE28E}} \color[HTML]{000000} $17161.4\pm 323.8$ \\
Interp.$^\star$ & {\cellcolor[HTML]{FEECA4}} \color[HTML]{000000} $11970.4\pm 10406.1$ & {\cellcolor[HTML]{FEE799}} \color[HTML]{000000} $14789.6\pm 641.0$ & {\cellcolor[HTML]{FED471}} \color[HTML]{000000} $21925.2\pm 82.9$ & {\cellcolor[HTML]{FEE28E}} \color[HTML]{000000} $17251.8\pm 122.1$ \\
Linear Ramp$^\star$ & {\cellcolor[HTML]{C64902}} \color[HTML]{F1F1F1} $61009.2\pm 10563.0$ & {\cellcolor[HTML]{FEE28E}} \color[HTML]{000000} $17138.4\pm 42.8$ & {\cellcolor[HTML]{FECD63}} \color[HTML]{000000} $24320.1\pm 45.3$ & {\cellcolor[HTML]{FEE18C}} \color[HTML]{000000} $17464.3\pm 10.6$ \\
Linear Ramp & {\cellcolor[HTML]{DB5D0B}} \color[HTML]{F1F1F1} $54526.2\pm 16538.3$ & {\cellcolor[HTML]{FEE392}} \color[HTML]{000000} $16413.5\pm 0.2$ & {\cellcolor[HTML]{FECE65}} \color[HTML]{000000} $23860.1\pm 2.0$ & {\cellcolor[HTML]{FEE390}} \color[HTML]{000000} $16919.7\pm 7.4$ \\
Linear Ramp$^\dagger$ & {\cellcolor[HTML]{FEEAA1}} \color[HTML]{000000} $12773.3\pm 5743.0$ & {\cellcolor[HTML]{FEE799}} \color[HTML]{000000} $14894.2\pm 0.0$ & {\cellcolor[HTML]{FEDE86}} \color[HTML]{000000} $18578.1\pm 0.0$ & {\cellcolor[HTML]{FEE697}} \color[HTML]{000000} $15067.9\pm 13.3$ \\
Recursive TS$^\star$ & {\cellcolor[HTML]{B44103}} \color[HTML]{F1F1F1} $64715.2\pm 2579.0$ & {\cellcolor[HTML]{FEE392}} \color[HTML]{000000} $16341.1\pm 239.8$ & {\cellcolor[HTML]{FECF67}} \color[HTML]{000000} $23642.7\pm 120.3$ & {\cellcolor[HTML]{FEE28E}} \color[HTML]{000000} $17237.7\pm 19.7$ \\
TQA$^\star$ & {\cellcolor[HTML]{F3801C}} \color[HTML]{F1F1F1} $44900.5\pm 13578.2$ & {\cellcolor[HTML]{FEE28E}} \color[HTML]{000000} $17190.4\pm 15.8$ & {\cellcolor[HTML]{FED069}} \color[HTML]{000000} $23112.6\pm 0.4$ & {\cellcolor[HTML]{FEE18C}} \color[HTML]{000000} $17295.8\pm 27.9$ \\
TQA & {\cellcolor[HTML]{FED471}} \color[HTML]{000000} $21946.0\pm 12600.0$ & {\cellcolor[HTML]{FEE392}} \color[HTML]{000000} $16300.9\pm 0.7$ & {\cellcolor[HTML]{FEDE86}} \color[HTML]{000000} $18395.7\pm 0.0$ & {\cellcolor[HTML]{FEE390}} \color[HTML]{000000} $16773.6\pm 16.5$ \\
TQA$^\dagger$ & {\cellcolor[HTML]{FFFFE5}} \color[HTML]{000000} $-4975.9\pm 5540.3$ & {\cellcolor[HTML]{FEE697}} \color[HTML]{000000} $15206.8\pm 0.0$ & {\cellcolor[HTML]{FEE18C}} \color[HTML]{000000} $17444.2\pm 0.0$ & {\cellcolor[HTML]{FEE390}} \color[HTML]{000000} $16737.9\pm 12.1$ \\
\cline{1-5}
\multicolumn{5}{c}{\rule{0pt}{8pt}PP} \\
\midrule
Fourier$^\star$ & {\cellcolor[HTML]{004F2D}} \color[HTML]{F1F1F1} $78103.2\pm 7976.1$ & {\cellcolor[HTML]{D7EFA2}} \color[HTML]{000000} $17172.0\pm 0.0$ & {\cellcolor[HTML]{B3E091}} \color[HTML]{000000} $25853.4\pm 0.0$ & {\cellcolor[HTML]{D5EEA1}} \color[HTML]{000000} $17937.1\pm 32.6$ \\
Interp.$^\star$ & {\cellcolor[HTML]{2F944E}} \color[HTML]{F1F1F1} $55315.5\pm 32067.7$ & {\cellcolor[HTML]{DAF0A4}} \color[HTML]{000000} $16369.2\pm 0.0$ & {\cellcolor[HTML]{B6E192}} \color[HTML]{000000} $25263.7\pm 0.0$ & {\cellcolor[HTML]{D5EEA1}} \color[HTML]{000000} $17782.2\pm 19.1$ \\
Linear Ramp$^\star$ & {\cellcolor[HTML]{389F55}} \color[HTML]{F1F1F1} $52197.1\pm 5099.0$ & {\cellcolor[HTML]{D7EFA2}} \color[HTML]{000000} $17171.5\pm 0.0$ & {\cellcolor[HTML]{B5E092}} \color[HTML]{000000} $25640.2\pm 0.0$ & {\cellcolor[HTML]{D5EEA1}} \color[HTML]{000000} $17780.5\pm 36.5$ \\
Linear Ramp & {\cellcolor[HTML]{B6E192}} \color[HTML]{000000} $25132.2\pm 3849.8$ & {\cellcolor[HTML]{DAF0A4}} \color[HTML]{000000} $16555.8\pm 0.0$ & {\cellcolor[HTML]{B8E293}} \color[HTML]{000000} $24967.6\pm 0.0$ & {\cellcolor[HTML]{D7EFA2}} \color[HTML]{000000} $17039.4\pm 14.4$ \\
Linear Ramp$^\dagger$ & {\cellcolor[HTML]{FCFED3}} \color[HTML]{000000} $-523.6\pm 6928.0$ & {\cellcolor[HTML]{DFF3A8}} \color[HTML]{000000} $14551.0\pm 0.0$ & {\cellcolor[HTML]{D0EC9F}} \color[HTML]{000000} $18733.9\pm 0.0$ & {\cellcolor[HTML]{DFF3A8}} \color[HTML]{000000} $14501.0\pm 1.5$ \\
Recursive TS$^\star$ & {\cellcolor[HTML]{004529}} \color[HTML]{F1F1F1} $81317.8\pm 13668.3$ & {\cellcolor[HTML]{DAF0A4}} \color[HTML]{000000} $16529.8\pm 0.0$ & {\cellcolor[HTML]{B8E293}} \color[HTML]{000000} $24697.8\pm 0.0$ & {\cellcolor[HTML]{D5EEA1}} \color[HTML]{000000} $17612.9\pm 40.0$ \\
TQA$^\star$ & {\cellcolor[HTML]{CCEA9D}} \color[HTML]{000000} $19827.1\pm 3918.3$ & {\cellcolor[HTML]{D7EFA2}} \color[HTML]{000000} $17075.8\pm 0.0$ & {\cellcolor[HTML]{B6E192}} \color[HTML]{000000} $25357.7\pm 0.0$ & {\cellcolor[HTML]{D5EEA1}} \color[HTML]{000000} $17632.6\pm 96.0$ \\
TQA & {\cellcolor[HTML]{F9FDC4}} \color[HTML]{000000} $3285.8\pm 753.9$ & {\cellcolor[HTML]{DBF1A4}} \color[HTML]{000000} $15990.3\pm 0.0$ & {\cellcolor[HTML]{D0EC9F}} \color[HTML]{000000} $18897.2\pm 0.0$ & {\cellcolor[HTML]{D9F0A3}} \color[HTML]{000000} $16842.6\pm 4.4$ \\
TQA$^\dagger$ & {\cellcolor[HTML]{FAFDC9}} \color[HTML]{000000} $2065.4\pm 863.9$ & {\cellcolor[HTML]{DCF1A5}} \color[HTML]{000000} $15752.8\pm 0.0$ & {\cellcolor[HTML]{D3EDA0}} \color[HTML]{000000} $18189.6\pm 0.0$ & {\cellcolor[HTML]{D9F0A3}} \color[HTML]{000000} $16836.1\pm 4.2$ \\
\cline{1-5}
\multicolumn{5}{c}{\rule{0pt}{8pt}SV} \\
\midrule
Fourier$^\star$ & {\cellcolor[HTML]{9EBAD9}} \color[HTML]{000000} $4716.6\pm 742.4$ & {\cellcolor[HTML]{F5EFF6}} \color[HTML]{000000} $2212.8\pm 0.0$ & {\cellcolor[HTML]{023858}} \color[HTML]{F1F1F1} $9290.9\pm 4162.0$ & {\cellcolor[HTML]{0569A5}} \color[HTML]{F1F1F1} $7664.8\pm 3014.5$ \\
Interp.$^\star$ & {\cellcolor[HTML]{BBC7E0}} \color[HTML]{000000} $4089.5\pm 607.9$ & {\cellcolor[HTML]{FFF7FB}} \color[HTML]{000000} $1711.2\pm 0.0$ & {\cellcolor[HTML]{94B6D7}} \color[HTML]{000000} $4905.3\pm 3834.9$ & {\cellcolor[HTML]{C4CBE3}} \color[HTML]{000000} $3880.6\pm 2462.0$ \\
Linear Ramp$^\star$ & {\cellcolor[HTML]{97B7D7}} \color[HTML]{000000} $4829.3\pm 792.9$ & {\cellcolor[HTML]{F5EEF6}} \color[HTML]{000000} $2231.4\pm 0.0$ & {\cellcolor[HTML]{023858}} \color[HTML]{F1F1F1} $9282.9\pm 3910.1$ & {\cellcolor[HTML]{04649D}} \color[HTML]{F1F1F1} $7918.1\pm 3162.0$ \\
Linear Ramp & {\cellcolor[HTML]{9AB8D8}} \color[HTML]{000000} $4777.5\pm 779.0$ & {\cellcolor[HTML]{F7F0F7}} \color[HTML]{000000} $2107.4\pm 0.0$ & {\cellcolor[HTML]{3991C1}} \color[HTML]{F1F1F1} $6402.1\pm 2074.9$ & {\cellcolor[HTML]{358FC0}} \color[HTML]{F1F1F1} $6448.6\pm 2419.1$ \\
Linear Ramp$^\dagger$ & {\cellcolor[HTML]{D2D3E7}} \color[HTML]{000000} $3542.3\pm 359.6$ & {\cellcolor[HTML]{FDF5FA}} \color[HTML]{000000} $1847.2\pm 0.0$ & {\cellcolor[HTML]{97B7D7}} \color[HTML]{000000} $4839.5\pm 1178.5$ & {\cellcolor[HTML]{A9BFDC}} \color[HTML]{000000} $4470.3\pm 970.5$ \\
Recursive TS$^\star$ & {\cellcolor[HTML]{A5BDDB}} \color[HTML]{000000} $4578.2\pm 720.8$ & {\cellcolor[HTML]{F8F1F8}} \color[HTML]{000000} $2094.1\pm 0.0$ & {\cellcolor[HTML]{023E62}} \color[HTML]{F1F1F1} $9099.4\pm 3870.4$ & {\cellcolor[HTML]{0567A1}} \color[HTML]{F1F1F1} $7803.1\pm 3117.3$ \\
TQA$^\star$ & {\cellcolor[HTML]{B4C4DF}} \color[HTML]{000000} $4239.2\pm 561.7$ & {\cellcolor[HTML]{F5EEF6}} \color[HTML]{000000} $2228.6\pm 0.0$ & {\cellcolor[HTML]{509AC6}} \color[HTML]{F1F1F1} $6051.2\pm 1720.6$ & {\cellcolor[HTML]{2F8BBE}} \color[HTML]{F1F1F1} $6584.1\pm 2329.4$ \\
TQA & {\cellcolor[HTML]{EDE7F2}} \color[HTML]{000000} $2644.2\pm 630.6$ & {\cellcolor[HTML]{F7F0F7}} \color[HTML]{000000} $2126.2\pm 0.0$ & {\cellcolor[HTML]{E8E4F0}} \color[HTML]{000000} $2784.5\pm 831.0$ & {\cellcolor[HTML]{BDC8E1}} \color[HTML]{000000} $4028.6\pm 1065.1$ \\
TQA$^\dagger$ & {\cellcolor[HTML]{FEF6FA}} \color[HTML]{000000} $1798.4\pm 190.6$ & {\cellcolor[HTML]{F7F0F7}} \color[HTML]{000000} $2106.5\pm 0.0$ & {\cellcolor[HTML]{FDF5FA}} \color[HTML]{000000} $1836.2\pm 963.9$ & {\cellcolor[HTML]{FDF5FA}} \color[HTML]{000000} $1834.2\pm 1234.4$ \\
\cline{1-5}
\bottomrule
\end{tabular}

    \caption{Average Energy for Different Methods and Graph Types, solving the MIS problem, with depth-$6$ QAOA.
    Graph types shown are Erd\H{o}s-R\'{e}nyi (ER) with $20~\%$ edge-probability, heavy-hex (HH), Line-to-Full (LB) with $1$ through $10$ swap layers, and Random-Regular (RR) with degree $3$ through $9$, for a total of $37$ graphs per row.
    ER, LB, and RR graphs consist of $20$ ($100$) nodes for SV (MPS and PP) results, while $HH$ graphs consist of $21$ ($144$) nodes.
    The average energies are shown, with their standard deviation, over all graph instances of a given type.
    Purple, orange, green, and blue denote Matrix Product State (MPS) with Qiskit Aer, MPS with Quimb, Pauli-Propagation (PP), and Statevector (SV) energy-evaluation methods, respectively.
    Cells are colored so higher energies and more instances are darker.
    The range of energies defining the darkest and lightest colors are shared between MPS and PP; the range for SV is independent.
    }
    \label{tab:summary_p06_mis}
\end{table*}

\subsection{Interp. versus Linear Ramp in the Best-Methods Analysis\label{app:best_method_distribution}}

Sec.~\ref{sec:results_quality} shows that Interp.$^\star$ and Linear Ramp$^\star$ perform well across all energy evaluation methods and depths, when run on the $37$-graph dataset.
In particular, Interp.$^\star$ appears better with MPS whereas Linear Ramp$^\star$ is better with SV.
The split is approximately ``50/50'' for PP.
Here we analyze the difference between these two methods on the instances where either of them are the best.
We define the relative advantage of Linear Ramp$^\star$ over Interp. using the energies $E$ as
\begin{equation}
    \text{Relative Advantage} = \frac{
    E(\text{Linear Ramp}^\star) - E(\text{Interp.}^\star)
    }{
    E(\text{Linear Ramp}^\star)
    }.
\end{equation}
A positive value $x$ indicates that the energy obtained with Interp.$^\star$ is $x\%$ smaller than the energy obtained with Linear Ramp$^\star$.

\begin{figure}[t]
    \centering
    \includegraphics[]{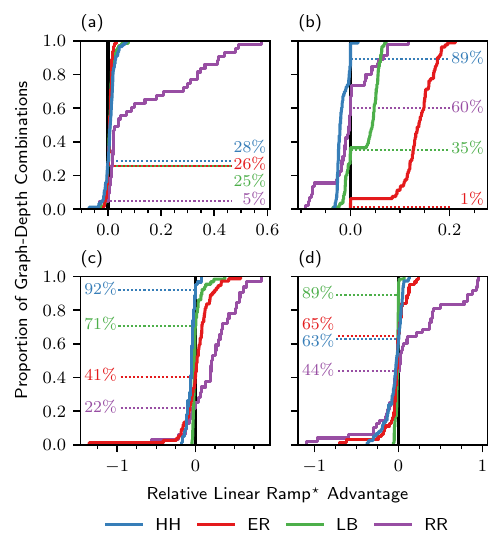}
    \caption{
    Cumulative distribution of the relative advantage of Linear Ramp$^\star$ over Interp.$^\star$ for $37$ graphs and depths $2,3,\ldots,10$.
    Axes are for energy evaluation methods (a) SV, (b) PP, (c) MPS Aer, and (d) MPS Quimb.
    The proportion of graph-depth combinations which are better with Interp.$^\star$ than Linear Ramp$^\star$ are shown with dotted lines.
    }
    \label{fig:best_methods_advantage}
\end{figure}

Though Linear Ramp$^\star$ for SV in Fig.~\ref{fig:best_methods} is better for all graph types, the advantage over Interp.$^\star$ is largest for random-regular graphs, see Fig.~\ref{fig:best_methods_advantage}.
This is across all depths and degrees.
These data do not indicate if this generalizes to all random-regular graphs since the dataset comprises only $7$ such graphs.
The distributions for both MPS energy evaluation methods are quite similar, with similar ranges of relative advantage values and proportion of data where Interp.$^\star$ is better than Linear Ramp$^\star$.
The behavior for both training methods is different for PP. Here, Linear Ramp$^\star$ has the largest advantage over Interp.$^\star$ on Erd\H{o}s-R\'{e}nyi graphs and not random-regular graphs.
This may be a consequence of how PP truncates terms and how the graph connectivity affects this truncation.
Additionally, the maximum advantage with PP is about $25\%$, far higher than with SV when ignoring random-regular graphs.
Fig.~\ref{fig:best_methods_advantage} and Sec.~\ref{sec:results_quality} indicate that Linear Ramp$^\star$ and Interp.$^\star$ are more likely the better methods, their performance is heavily dependent on the graph type and particular instance.

\subsection{LABS\label{app:labs}}

LABS, defined in App.~\ref{app:problem_models}, is a notoriously difficult combinatorial optimization problem even at small sizes. 
The cost landscape is extremely rugged; single spin-flips in a near-optimal sequence typically degrade the energy to values indistinguishable from the ensemble average~\cite{bernasconi1987, Mertens_1996}. 
The classical state of the art exact solver is a branch-and-bound algorithm~\cite{Packebusch_2016}, which exhaustively enumerates all optimal sequences up to $n = 66$ and scales as $O(1.73^n)$. 
For larger instances, heuristic methods dominate.
For example, Boskovic et al. \cite{Bo_kovi__2017} achieve a $O(1.34^n)$ scaling with a memetic Tabu method but do not provide an optimality-guarantee.
Recently, GPU-accelerated memetic Tabu search has pushed best-known Merit Factors to sizes $n = 92$--$120$~\cite{zhang2025}. 
Any claimed quantum advantage on LABS must thus be measured against continually improving classical baselines.

LABS has emerged as a compelling benchmark for quantum algorithms \cite{koch2025quantum, Shaydulin2024, apte2025, sciorilli2026competitivenisqqubitefficientsolver, cadavid2025scalingadvantagequantumenhancedmemetic} since it has a unique instance per size $n$ and is already challenging at the qubit counts available on current hardware.
Noiseless simulations up to $n = 40$ qubits with fixed-angle schedules, transferred across system sizes, provid empirical evidence that QAOA may scale as $O(1.21^n)$ when combined with quantum minimum finding \cite{Shaydulin2024, vanApeldoorn2020quantumsdpsolvers, durr1996quantum}. 
Crucially, this scaling relies on angles set by maximizing the overlap with the known ground state for small $n$ and then transferred to larger problems, rather than optimizing an experimentally accessible objective such as the Energy or mean Merit Factor $\langle \text{MF} \rangle = \mathbb{E}[n^2 / 2 H_C]$. 

The existing LABS QAOA studies do not compare the diverse angle setting strategies, such as, LR, TQA, and RTS. 
Also, the interplay between the optimization objective, e.g., energy minimization versus merit factor maximization, and the resulting solution quality could be further explored since these metrics may not perfectly correlate. 
Therefore, we benchmark QAOA angle setting strategies on LABS by evaluating their performance in terms of energy, Merit Factor, and Time-to-Solution $\mathrm{TTS}=p_\text{opt}^{-1}$.
Here, $p_\text{opt}$ is the probability to sample any optimal bitstring from the QAOA state.
We optimize the QAOA angles by minimizing the LABS energy $\langle H_C \rangle$. 
While alternative objectives exist, such as maximizing $\langle \text{MF} \rangle$ or the ground state overlap~\cite{Shaydulin2024, apte2025}, we minimize $\langle H_C \rangle$ to be consistent with our MaxCut and MIS benchmarks. 
Small scale experiments indicated that this results in competitive solutions compared to optimizing $\langle \text{MF} \rangle$.
We evaluated the Fourier, Interp., RTS, TQA, and LR-QAOA protocols. 
To the best of our knowledge, fixed angles are not known for LABS. 
We include the trivial layer-extension method, introduced in App.~\ref{app:trivial}, which exhibits a high empirical performance. 
Since LABS involves many four-local interactions that generate high circuit entanglement, classical simulation via MPS or PP is computationally prohibitive. 
Furthermore, the high connectivity and gate depth pose significant challenges for current quantum hardware. 
Consequently, our LABS results rely on statevector simulations.

\begin{figure}
    \centering
    \includegraphics[width=\columnwidth]{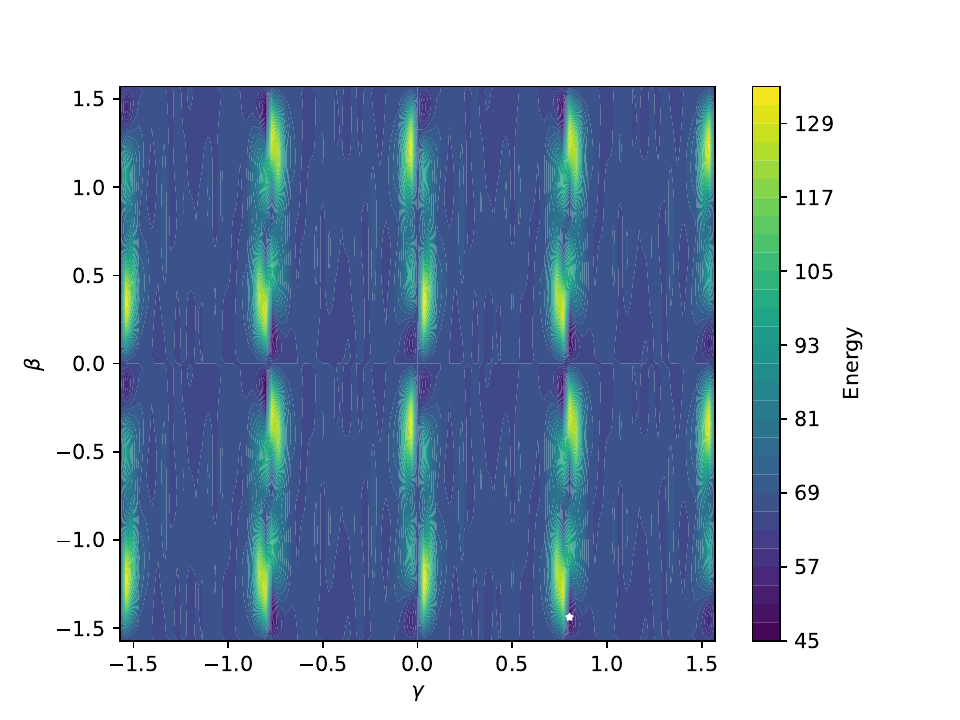}
    \caption{Depth-one energy landscape of LABS with 12 spins. The landscape is symmetric and contains many identical global minima.}
    \label{fig:labs_energy_landscape}
\end{figure}

\begin{table*}[t]
\centering
\caption{
LABS energy expectation values $\langle H_C \rangle$ of the different angle setting methods as a function of the number of variables $n$ for a $p=50$ QAOA. 
RTS is only run up to $p=10$.
Shaded cells indicate the method achieving the lowest energy per problem size.
The last row indicates the energy of the optimal solution.
}
\label{tab:labs_energy}
\footnotesize
\setlength{\tabcolsep}{3.3pt} 
\begin{tabular}{l *{12}{c}}
\hline\hline
& \multicolumn{12}{c}{Node count ($n$)} \\
Method & 10 & 11 & 12 & 13 & 14 & 15 & 16 & 17 & 18 & 19 & 20 & 21 \\ 
\midrule
Fourier     & 13.6 & \cellcolor[HTML]{C0C0C0}5.9 & 12.7 & \cellcolor[HTML]{C0C0C0}15.8 & 20.8 & \cellcolor[HTML]{C0C0C0}24.8 & \cellcolor[HTML]{C0C0C0}31.1 & \cellcolor[HTML]{C0C0C0}38.1 & \cellcolor[HTML]{C0C0C0}38.5 & \cellcolor[HTML]{C0C0C0}43.3 & \cellcolor[HTML]{C0C0C0}49.1 & \cellcolor[HTML]{C0C0C0}54.4 \\
Interp.      & \cellcolor[HTML]{C0C0C0}13.4 & 6.1 & \cellcolor[HTML]{C0C0C0}12.4 & 16.1 & \cellcolor[HTML]{C0C0C0}20.6 & 25.0 & 31.5 & 38.2 & 38.8 & 43.6 & 49.7 & 55.3 \\
Trivial     & 14.2 & 17.6 & 16.8 & 22.8 & 25.3 & 30.4 & 35.5 & 42.3 & 45.1    & 51.1 & 56.2 & 64.8 \\
RTS         & 18.4 & 22.4 & 25.1 & 27.9 & 32.2 & 38.5 & 44.9 & 52.1 & 58.1    & 65.0 & 71.6 & 78.6 \\
TQA         & 42.0 & 51.9 & 65.5 & 73.5 & 89.0 & 103.6 & 116.8 & 135.0 & 152.3 & 170.3 & 189.4 & 209.2 \\
TQA$^\star$ & 31.7 & 39.9 & 55.0 & 65.9 & 83.4 & 98.4 & 112.9 & 132.0 & 149.8 & 168.4 & 187.3 & 207.0 \\
LR          & 15.0 & 16.7 & 18.2 & 21.4 & 25.3 & 29.8 & 35.2 & 41.9 & 45.5 & 51.0 & 56.6 & 62.6 \\
LR$^\star$  & 14.0 & 15.7 & 16.5 & 20.5 & 24.4 & 28.7 & 34.2 & 41.0 & 43.8 & 49.2 & 54.8 & 60.7 \\ 
\hline
Optimal     & 13   & 5    & 10   & 6    & 19   & 15   & 24   & 32   & 25   & 29   & 26   & 26 \\ 
\hline \hline
\end{tabular}
\end{table*}

We evaluate the angle setting methods up to depth $p=50$ on system sizes up to $n=21$, with the exception of RTS, which we limit to $p=10$ due to its long runtime. 
The iterative methods are initialized from a 100-point grid search of the $p=1$ energy landscape. 
Since the LABS phase separation operator is $\pi$-periodic, i.e., $\gamma = \pi + \epsilon$ is physically equivalent to $\gamma=\epsilon$, we restrict this search to $(\beta,\gamma)\in [-\pi/8, \pi/8]^2$, see Fig.~\ref{fig:labs_energy_landscape}.
This is necessary because iterative  methods are agnostic to the Hamiltonian's periodicity. 
If seeded with an angle approaching $\pi$ the extrapolation that initializes the $p+1$ layer may induce a severe phase over-rotation, thus displacing the QAOA state from the optimal basin into a high-energy state. 
For example, the performance of the Fourier initialization at later depths is very sensitive to the choice of parameters at depth $p=1$.
After the grid scan we run the COBYLA optimizer with a maximum of 500 iteration per depth and an initial trust-region radius \texttt{rhobeg} of $0.01$. 
Crucially, this small step size is necessary to navigate the characteristically narrow basins of attraction in the LABS energy landscape, see Fig.~\ref{fig:labs_energy_landscape}.

Our experiments show that increasing depth $p$ at the cost of fewer iterations per depth is a superior tactic. 
For example, running up to depth $p=25$ and with 2000 iterations per depth requires the same compute budget as the current approach with 500 iteration per depth, but resulted in significantly worse final states.
For the TQA and LR protocols, we sweep initial slopes of $\{0.01, 0.05, 0.1, 0.2\}$. 
Smaller ramps consistently yield superior performance across all metrics.

\begin{figure*}[t!]
    \centering
    \includegraphics[width=\textwidth]{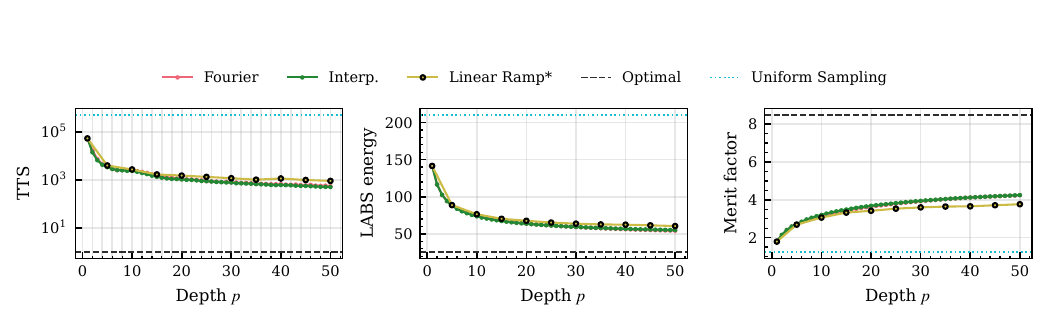}
    \caption{
    Time-to-solution (left), LABS energy (centre), and Merit Factor (right) for the 21-qubit LABS instance as a function of QAOA depth $p$. 
    The performance, for the angle setting methods, consistently improves with depth $p$.
    Three angle-setting strategies are compared: Fourier, Interp. and Linear Ramp ($LR^\star$). 
    Dashed and dotted lines correspond to the optimal and uniform sampling baselines, respectively. 
    The performance of Interp. and Fourier is nearly identical across all depths $p$.}
    \label{fig:labs21_comparison}
\end{figure*}

Tab.~\ref{tab:labs_energy} shows the average energy, i.e., the optimization target, while Tab.~\ref{tab:labs_mftts} of the main text shows the Merit Factor and the TTS of the QAOA angles.  
The Fourier and Interp. methods yield the best angles, with a near-identical performance, see also Fig.~\ref{fig:labs21_comparison}.
Fourier achieves slightly lower energies in most cases, whereas Interp. has marginally higher $\langle \textrm{MF}\rangle$ and lower TTS. 
Crucially, two angle schedules may have a comparable energy, but different distributions near the optimal bitstring.
Indeed, $\langle H_C \rangle$ is an average over all solutions in the QAOA state while the TTS depends only on the probability of sampling the ground state. 
The average Merit Factor is sensitive to both the ground state and the first few excited states.
Therefore, minimizing $\langle H_C \rangle$ or TTS and maximizing $\langle \textrm{MF}\rangle$ are all different objectives. 

Among the non-iterative methods, LR$^\star$ is the most competitive, see Fig.~\ref{fig:labs21_comparison} as well as Tab.~\ref{tab:labs_mftts} (main text) and~\ref{tab:labs_energy}.
It bypasses the iterative depth-by-depth re-optimization resulting in a lower computational cost compared to Fourier and Interp. 
For example, the total compute time of the iterative methods reached up to approximately 24 hours on a single NVIDIA A100 GPU for the largest instances.
By comparioson, LR$^\star$ ran in 1 hour for the same instance.
Despite this, LR$^\star$ achieves very competitive performance across all system sizes considered. 
Since the slope initialization is neither fine-grained nor asymmetric in $\Delta_\beta$ and $\Delta_\gamma$ our results likely underestimate its potential.
Initializing LR$^\star$ from $\Delta_\beta\neq\Delta_\gamma$ points may improve our results.  

\section{Utility-scale hardware runs\label{app:example_hw}}

\subsection{Quantum circuit creation\label{app:circuit_creation}}

Our hardware experiments run on IBM Quantum devices.
They are based on superconducting qubit quantum processing units (QPU) with a heavy-hexagonal qubit coupling topology.
The native gate set is built from controlled-$Z$ rotations and single-qubit gates.

To execute QAOA on quantum hardware we must decompose $e^{-i\gamma H_C}$ into the QPU's native gate set.
The topology of the cost Hamiltonian $H_C$ is given by the edges of the graph $G$ of the underlying MaxCut problem.
Since the Hamiltonians are quadratic, i.e., built from linear combinations of $Z_iZ_j$ Paulis, the unitary $e^{-i\gamma H_C}$ is made of commuting $R_{ZZ}(\theta, i,j)=e^{-i\theta Z_iZ_j/2}$ gates.
SWAP gates are inserted into the QAOA circuit if there is no physical coupling between qubits $i$ and $j$ on the QPU.
We add SWAP gates following a predetermined network of SWAP gates~\cite{Weidenfeller2022}.
In particular, for a graph with $n$ nodes we select a line of $n$ qubits on the hardware and apply $\ell$ layers of a line SWAP strategy.
This SWAP strategy alternates between layers of simultaneously applicable SWAP gates on qubits $(2i, 2i+1)$ and $(2i+1, 2i+2)$.
Full connectivity is reached after $n-2$ layers.
Furthermore, the number of SWAP layers $\ell$ can be minimized by appropriately choosing the mapping between the decision variables and the physical qubits.
Here, the initial mapping problem with a fixed number of SWAP layers $\ell$ is formulated as a boolean satisfiability problem (SAT) and a binary search over $\ell$ finds the smallest number of layers for which a satisfiable initial mapping exists~\cite{Matsuo2023}.
We refer to this method as the SAT mapping.

The QAOA circuits of the four graph families discussed in App.~\ref{app:graphs} are mapped to hardware as follows.
Heavy-hex graphs do not require any SWAPs since they are hardware native.
By design, line-based graphs $G_\ell$ are routed in $\ell$ layers of the line SWAP strategy.
Random regular graphs and Erd\H{o}s-R\'{e}nyi graphs are routed with a line SWAP strategy with the SAT mapping to minimize the number of SWAP layers.

\subsection{Training QAOA angles on hardware\label{app:hardware_train}}

Here, we provide details on the hardware training example of Sec.~\ref{sec:hw_training}.
The graph considered is built from a line of 100 nodes on which two layers of SWAP gates are applied.
Each edge is assigned a weight of $\pm1$ with $50\%$ probability.
We use COBYLA to minimize $\langle H_C\rangle_\alpha$ computed over the $\alpha=5\%$ best samples ranked by the MaxCut objective function from Eq.~(\ref{eqn:max_cut}).
This average over the tail of the sample distribution is known as the conditional value at risk at a level $\alpha$ ($\text{CVaR}_\alpha$)~\cite{Barron2024}.
I.e., only the $5\%$ lowest energy samples out of the total of 20 thousand samples per iteration are considered.
This 5\% is loosely based on the $\smash F_\text{CZ}=99.872\%$ median $CZ$-gate fidelity of \emph{ibm\_boston}.
Indeed, by approximating $\sqrt{\gamma}$ as $\smash{F_\text{CZ}^{N_\text{CZ}/2}=28.1\%}$, where the number of CZ gates is $N_{\text{CZ}=1980}$, our $\alpha$ of 5\% is slightly too conservative.

During the variational optimization we carefully tracked the time of each step with the code below.
\begin{lstlisting}
with Session(backend=backend) as session:
  sampler = Sampler(mode=session)
        
  def objective(x):
    start = time()
    qc = circuit.assign_parameters(
        x, 
        inplace=False,
    )
    job = sampler.run([qc], shots=shots)
    res = job.result()
    counts = res[0].data.c.get_counts()
    sampling_time = time() - start

    return cvar(counts, 0.05)

  result = minimize(objective, ...)
\end{lstlisting}
Here, \texttt{sampling\_time} measures the duration the backend takes to return a new set of samples.
This quantity is referred to in Fig.~\ref{fig:hardware_conv}(b) as the \emph{QPU time}.
Averaged over each iteration, the user's wall clock time, i.e., \texttt{sampling\_time}, to generate $20{\rm k}$ samples was $8.2\pm0.9$~seconds. 
The backend reported an average completion time of $5.8\pm0.5$ seconds to generate the 20 thousand samples.
The difference can be attributed in part to the time it took to send the circuits and samples over the internet.
On top of this, the total duration of the variational loop must account for any classical processing time, for instance in \texttt{cvar(counts, 0.05)}, at each iteration which is here negligible.

\subsection{Angle validation}

\begin{figure*}[t]
    \centering
    \includegraphics[width=0.95\textwidth]{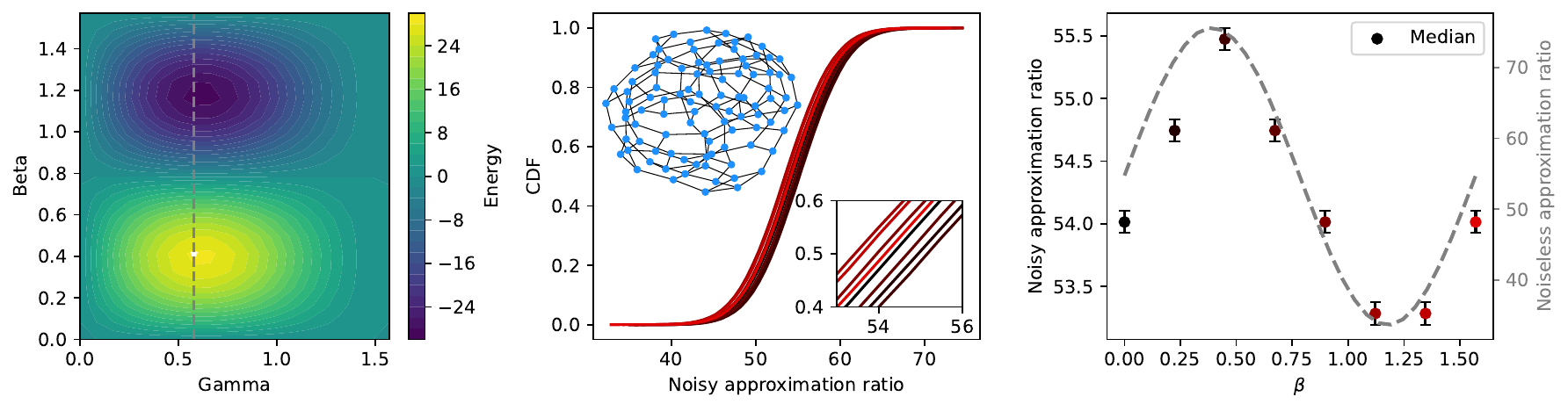}
    \caption{
    Example of a utility-scale hardware test.
    (a) Depth-one QAOA energy landscape efficiently and exactly computed.
    The dashed gray line shows the angles that we test.
    (b) Cumulative distribution function of the samples from \emph{ibm\_fez}.
    The insets show the random-three-regular graph that we evaluated, and the inset shows a zoom of the CDFs around the median.
    (c) Comparison between the median approximation ratio of the noisy samples and the noiseless energy.
    }
    \label{fig:hardware_test}
\end{figure*}

Here, we further exemplify the quantum hardware validation by evaluating the energy on a QPU of a depth-one QAOA for a random three-regular graph with 100 nodes.
An efficient and exact computation of the depth-one energy, shown in Fig.~\ref{fig:hardware_test}(a), serves as a ground truth.
We create a hardware-native QAOA circuit with SWAP networks and a SAT mapping as discussed in App.~\ref{app:circuit_creation}.
The circuits of eight QAOA angle sets are sampled $65\,536$ times from \emph{ibm\_fez} where $\gamma$ is always optimal and $\beta$ ranges from $0$ to $\pi/2$, see the dashed line in Fig.~\ref{fig:hardware_test}(a).
Despite the 4926 two-qubit gates we observe a clear signal in the cumulative distribution function of the samples for the different $\beta$ values, see Fig.~\ref{fig:hardware_test}(b).
The approximation ratio measured by the hardware undergoes the same oscillatory behavior as the noiseless exact energy but with a smaller amplitude, see Fig.~\ref{fig:hardware_test}(c).
We can thus leverage the hardware to identify which QAOA angles perform best.
Crucially, if two sets of QAOA angles are near to a local extrema the hardware may conclude that their approximation ratio is statistically identical.
We observe this for example in the $p=5$ and $10$ heavy-hex graph instances shown in the main text.

\subsection{Statistical analysis\label{app:stat_test}}

\begin{figure*}
    \centering
    \includegraphics[width=0.85\textwidth, clip, trim=0 0 0 30]{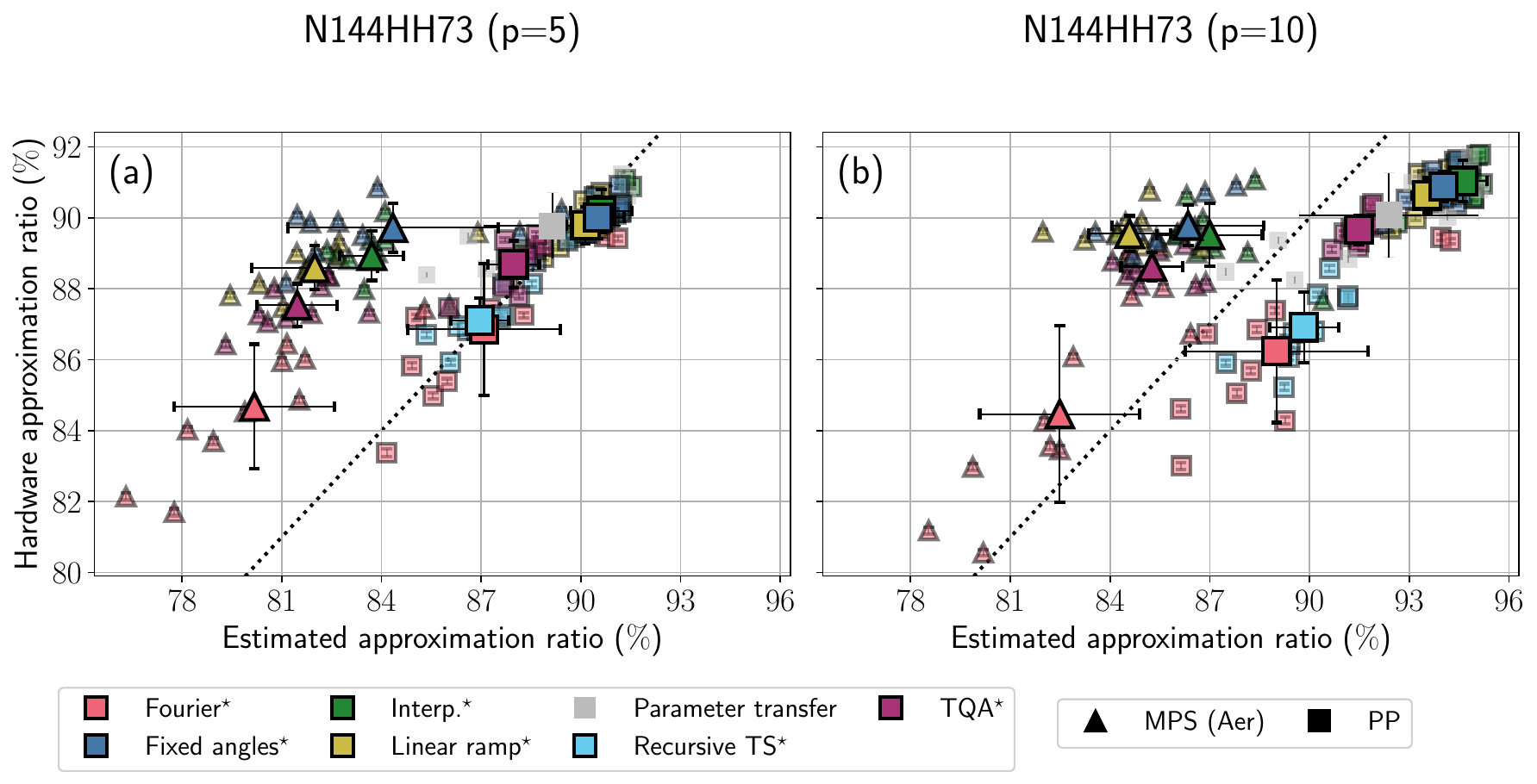}
    \caption{Hardware validation of the top 5\% shots on \emph{ibm\_boston} for ten 144-node heavy-hex graphs using QAOA at depths $p=5$ (a) and $p=10$ (b), evaluated with the Qiskit Aer MPS with a max bond dimension of 40 and Pauli propagation with a max weight of 6 (fixed angles interpret the graphs as having a degree of three). The dashed line indicates where the hardware approximation ratio is equal to the estimated approximation ratio.
    For the MPS, the correlation coefficients between the hardware and the estimated approximation ratio are 0.69 and 0.67 for $p=5$ and $10$, respectively.
    For PP, these numbers are 0.90 and 0.90.
    }
    \label{fig:144hh510_5pc_hw}
\end{figure*}

Here, we describe the analysis used to determine whether the differences between the approximation ratios measured in Sec.~\ref{sec:results_validation} are statistically meaningful.
A significant Kruskal-Wallis result indicates that at least one method is different from the others without revealing which one. 
To resolve this, the Conover-Iman post hoc test provides pairwise comparisons that remain non‑parametric and robust to unequal sample sizes, which can be important when looking at methods with significantly different amounts of training time.
By comparing the rank distributions of each pair of methods, the Conover–Iman test identifies where performance gaps truly exist, such as whether Interp. consistently yields higher estimated ratios than other methods.
Because multiple pairwise tests inflate the risk of false positives, the Holm correction is applied to the resulting $p$‑values.
This step‑down adjustment controls the family‑wise error rate while retaining more statistical power than conservative alternatives like Bonferroni.
In practice, the Holm-corrected Conover-Iman results provide a map of which QAOA methods are genuinely superior under hardware noise and which differences are statistically indistinguishable.
For the hardware results in Sec.~\ref{sec:hardware}, we can investigate them one step further by looking at only the top 5\% of the shots per method, see Fig.~\ref{fig:144hh510_5pc_hw}.

There is an overall increase in the hardware approximation ratio for both $p=5$ and $p=10$, and most of the methods still have a similar qualitative performance relative to each other compared to their counterparts in Fig.~\ref{fig:144hh510_hw}. We then run the Conover-Iman test on these shots, see Tab.~\ref{tab:N144HH73_5_10_5pc_hw}. Compared to Tab.~\ref{tab:N144HH73_5_10_100pc_hw}, there is a roughly 8\% increase in the medians across all methods, and more methods are statistically distinguishable from each other in $p=5$. Furthermore, the rankings of the methods are almost identical, with the exception of Interp. with MPSAer, although it is still statistically indistinguishable from other methods around it, such as linear ramp with Pauli propagation and MPSAer. For $p=10$,
similar to $p=5$, the medians increase by roughly 8\% across all methods, although there is no significant increase in the number of statistical distinguishable pairs. The overall ranking of the methods also remains largely unchanged, with the exception of parameter‑transfer with Pauli propagation and TQA with Pauli propagation; nonetheless, both methods are still statistically indistinguishable from their neighboring methods.

\begin{table*}[ht]
\centering
\resizebox{\textwidth}{!}{%
\begin{tabular}{l|ccccccccccccc|cccccccccccc|}
\cline{2-26}
& \multicolumn{13}{c|}{Estimated} & \multicolumn{12}{c|}{Hardware} \\
\hline

Methods(p=5) & 1. & 2. & 3. & 4. & 5. & 6. & 7. & 8. & 9. & 10. & 11. & 12. & CVaR & 2. & 3. & 4. & 5. & 6. & 7. & 8. & 9. & 10. & 11. & 12. & CVaR \\
\hline

1. (I, PP)$^\star$ & -- & 1 & 1 & 3.4e-1 & {\cellcolor[HTML]{00441B}} \color[HTML]{F1F1F1} 7.4e-4 & {\cellcolor[HTML]{00441B}} \color[HTML]{F1F1F1} 1.7e-6 & {\cellcolor[HTML]{00441B}} \color[HTML]{F1F1F1} 4.1e-5 & {\cellcolor[HTML]{00441B}} \color[HTML]{F1F1F1} 7.4e-13 & {\cellcolor[HTML]{00441B}} \color[HTML]{F1F1F1} 1.9e-11 & {\cellcolor[HTML]{00441B}} \color[HTML]{F1F1F1} 1.2e-17 & {\cellcolor[HTML]{00441B}} \color[HTML]{F1F1F1} 7.5e-19 & {\cellcolor[HTML]{00441B}} \color[HTML]{F1F1F1} 7.6e-21 & 91 & 1 & 1 & 1 & {\cellcolor[HTML]{00441B}} \color[HTML]{F1F1F1} 7.5e-5 & {\cellcolor[HTML]{00441B}} \color[HTML]{F1F1F1} 3.8e-13 & {\cellcolor[HTML]{00441B}} \color[HTML]{F1F1F1} 3.8e-11 & {\cellcolor[HTML]{309950}} \color[HTML]{F1F1F1} 1.4e-3 & 1 & {\cellcolor[HTML]{00441B}} \color[HTML]{F1F1F1} 2.2e-5 & {\cellcolor[HTML]{00441B}} \color[HTML]{F1F1F1} 4.2e-11 & {\cellcolor[HTML]{00441B}} \color[HTML]{F1F1F1} 6.7e-17 & 90.2 \\
2. (FA, PP)$^\star$ & -- & -- & 1 & 6.2e-1 & {\cellcolor[HTML]{309950}} \color[HTML]{F1F1F1} 2.6e-3 & {\cellcolor[HTML]{00441B}} \color[HTML]{F1F1F1} 7.4e-6 & {\cellcolor[HTML]{00441B}} \color[HTML]{F1F1F1} 1.6e-4 & {\cellcolor[HTML]{00441B}} \color[HTML]{F1F1F1} 4.3e-12 & {\cellcolor[HTML]{00441B}} \color[HTML]{F1F1F1} 1.0e-10 & {\cellcolor[HTML]{00441B}} \color[HTML]{F1F1F1} 7.0e-17 & {\cellcolor[HTML]{00441B}} \color[HTML]{F1F1F1} 4.4e-18 & {\cellcolor[HTML]{00441B}} \color[HTML]{F1F1F1} 4.4e-20 & 90.8 & -- & 1 & 1 & {\cellcolor[HTML]{00441B}} \color[HTML]{F1F1F1} 6.0e-4 & {\cellcolor[HTML]{00441B}} \color[HTML]{F1F1F1} 5.7e-12 & {\cellcolor[HTML]{00441B}} \color[HTML]{F1F1F1} 5.2e-10 & {\cellcolor[HTML]{309950}} \color[HTML]{F1F1F1} 7.9e-3 & 1 & {\cellcolor[HTML]{00441B}} \color[HTML]{F1F1F1} 2.0e-4 & {\cellcolor[HTML]{00441B}} \color[HTML]{F1F1F1} 5.7e-10 & {\cellcolor[HTML]{00441B}} \color[HTML]{F1F1F1} 1.0e-15 & 90.1 \\
3. (LR, PP)$^\star$ & -- & -- & -- & 1 & 6.2e-2 & {\cellcolor[HTML]{00441B}} \color[HTML]{F1F1F1} 5.1e-4 & {\cellcolor[HTML]{309950}} \color[HTML]{F1F1F1} 6.9e-3 & {\cellcolor[HTML]{00441B}} \color[HTML]{F1F1F1} 7.6e-10 & {\cellcolor[HTML]{00441B}} \color[HTML]{F1F1F1} 1.6e-8 & {\cellcolor[HTML]{00441B}} \color[HTML]{F1F1F1} 1.5e-14 & {\cellcolor[HTML]{00441B}} \color[HTML]{F1F1F1} 9.2e-16 & {\cellcolor[HTML]{00441B}} \color[HTML]{F1F1F1} 9.0e-18 & 90.3 & -- & -- & 1 & {\cellcolor[HTML]{309950}} \color[HTML]{F1F1F1} 7.2e-3 & {\cellcolor[HTML]{00441B}} \color[HTML]{F1F1F1} 2.1e-10 & {\cellcolor[HTML]{00441B}} \color[HTML]{F1F1F1} 1.7e-8 & 6.8e-2 & 1 & {\cellcolor[HTML]{309950}} \color[HTML]{F1F1F1} 3.0e-3 & {\cellcolor[HTML]{00441B}} \color[HTML]{F1F1F1} 1.9e-8 & {\cellcolor[HTML]{00441B}} \color[HTML]{F1F1F1} 4.4e-14 & 89.8 \\
4. (PT, PP) & -- & -- & -- & -- & 4.9e-1 & {\cellcolor[HTML]{B5E1AE}} \color[HTML]{000000} 1.2e-2 & 1.0e-1 & {\cellcolor[HTML]{00441B}} \color[HTML]{F1F1F1} 6.8e-8 & {\cellcolor[HTML]{00441B}} \color[HTML]{F1F1F1} 1.3e-6 & {\cellcolor[HTML]{00441B}} \color[HTML]{F1F1F1} 1.8e-12 & {\cellcolor[HTML]{00441B}} \color[HTML]{F1F1F1} 1.2e-13 & {\cellcolor[HTML]{00441B}} \color[HTML]{F1F1F1} 1.1e-15 & 89.8 & -- & -- & -- & {\cellcolor[HTML]{B5E1AE}} \color[HTML]{000000} 1.3e-2 & {\cellcolor[HTML]{00441B}} \color[HTML]{F1F1F1} 5.9e-10 & {\cellcolor[HTML]{00441B}} \color[HTML]{F1F1F1} 4.6e-8 & 1.2e-1 & 1 & {\cellcolor[HTML]{309950}} \color[HTML]{F1F1F1} 6.1e-3 & {\cellcolor[HTML]{00441B}} \color[HTML]{F1F1F1} 5.0e-8 & {\cellcolor[HTML]{00441B}} \color[HTML]{F1F1F1} 1.3e-13 & 89.9 \\
5. (TQA, PP)$^\star$ & -- & -- & -- & -- & -- & 1 & 1 & {\cellcolor[HTML]{00441B}} \color[HTML]{F1F1F1} 5.5e-4 & {\cellcolor[HTML]{309950}} \color[HTML]{F1F1F1} 5.3e-3 & {\cellcolor[HTML]{00441B}} \color[HTML]{F1F1F1} 6.1e-8 & {\cellcolor[HTML]{00441B}} \color[HTML]{F1F1F1} 4.7e-9 & {\cellcolor[HTML]{00441B}} \color[HTML]{F1F1F1} 5.6e-11 & 88 & -- & -- & -- & -- & {\cellcolor[HTML]{309950}} \color[HTML]{F1F1F1} 4.0e-3 & 6.2e-2 & 1 & {\cellcolor[HTML]{B5E1AE}} \color[HTML]{000000} 1.3e-2 & 1 & 6.4e-2 & {\cellcolor[HTML]{00441B}} \color[HTML]{F1F1F1} 5.3e-6 & 88.9 \\
6. (RTS, PP)$^\star$ & -- & -- & -- & -- & -- & -- & 1 & 6.4e-2 & 3.1e-1 & {\cellcolor[HTML]{00441B}} \color[HTML]{F1F1F1} 4.2e-5 & {\cellcolor[HTML]{00441B}} \color[HTML]{F1F1F1} 4.3e-6 & {\cellcolor[HTML]{00441B}} \color[HTML]{F1F1F1} 7.6e-8 & 86.9 & -- & -- & -- & -- & -- & 1 & {\cellcolor[HTML]{00441B}} \color[HTML]{F1F1F1} 2.4e-4 & {\cellcolor[HTML]{00441B}} \color[HTML]{F1F1F1} 5.5e-10 & {\cellcolor[HTML]{309950}} \color[HTML]{F1F1F1} 9.1e-3 & 1 & 1 & 87.1 \\
7. (F, PP)$^\star$ & -- & -- & -- & -- & -- & -- & -- & {\cellcolor[HTML]{309950}} \color[HTML]{F1F1F1} 7.3e-3 & 5.1e-2 & {\cellcolor[HTML]{00441B}} \color[HTML]{F1F1F1} 1.7e-6 & {\cellcolor[HTML]{00441B}} \color[HTML]{F1F1F1} 1.5e-7 & {\cellcolor[HTML]{00441B}} \color[HTML]{F1F1F1} 2.2e-9 & 86.6 & -- & -- & -- & -- & -- & -- & {\cellcolor[HTML]{309950}} \color[HTML]{F1F1F1} 6.2e-3 & {\cellcolor[HTML]{00441B}} \color[HTML]{F1F1F1} 4.2e-8 & 1.2e-1 & 1 & 2.1e-1 & 87.2 \\
8. (I, MPSAer)$^\star$ & -- & -- & -- & -- & -- & -- & -- & -- & 1 & 4.9e-1 & 1.5e-1 & {\cellcolor[HTML]{B5E1AE}} \color[HTML]{000000} 1.3e-2 & 83.7 & -- & -- & -- & -- & -- & -- & -- & 1.2e-1 & 1 & {\cellcolor[HTML]{309950}} \color[HTML]{F1F1F1} 6.6e-3 & {\cellcolor[HTML]{00441B}} \color[HTML]{F1F1F1} 1.6e-7 & 89 \\
9. (FA, MPSAer)$^\star$ & -- & -- & -- & -- & -- & -- & -- & -- & -- & 1.2e-1 & {\cellcolor[HTML]{B5E1AE}} \color[HTML]{000000} 3.0e-2 & {\cellcolor[HTML]{309950}} \color[HTML]{F1F1F1} 1.7e-3 & 83.1 & -- & -- & -- & -- & -- & -- & -- & -- & {\cellcolor[HTML]{309950}} \color[HTML]{F1F1F1} 5.8e-3 & {\cellcolor[HTML]{00441B}} \color[HTML]{F1F1F1} 4.6e-8 & {\cellcolor[HTML]{00441B}} \color[HTML]{F1F1F1} 1.2e-13 & 89.9 \\
10. (LR, MPSAer)$^\star$ & -- & -- & -- & -- & -- & -- & -- & -- & -- & -- & 1 & 1 & 81.7 & -- & -- & -- & -- & -- & -- & -- & -- & -- & 1.2e-1 & {\cellcolor[HTML]{00441B}} \color[HTML]{F1F1F1} 1.8e-5 & 88.6 \\
11. (TQA, MPSAer)$^\star$ & -- & -- & -- & -- & -- & -- & -- & -- & -- & -- & -- & 1 & 81.5 & -- & -- & -- & -- & -- & -- & -- & -- & -- & -- & 2.1e-1 & 87.3 \\
12. (F, MPSAer)$^\star$ & -- & -- & -- & -- & -- & -- & -- & -- & -- & -- & -- & -- & 80.4 & -- & -- & -- & -- & -- & -- & -- & -- & -- & -- & -- & 84.7 \\
\hline

Methods(p=10) & 1. & 2. & 3. & 4. & 5. & 6. & 7. & 8. & 9. & 10. & 11. & 12. & CVaR & 2. & 3. & 4. & 5. & 6. & 7. & 8. & 9. & 10. & 11. & 12. & CVaR \\
\hline

1. (I, PP)$^\star$ & -- & 7.6e-1 & 7.9e-2 & {\cellcolor[HTML]{B5E1AE}} \color[HTML]{000000} 1.8e-2 & {\cellcolor[HTML]{00441B}} \color[HTML]{F1F1F1} 1.9e-5 & {\cellcolor[HTML]{00441B}} \color[HTML]{F1F1F1} 3.7e-9 & {\cellcolor[HTML]{00441B}} \color[HTML]{F1F1F1} 1.7e-10 & {\cellcolor[HTML]{00441B}} \color[HTML]{F1F1F1} 2.4e-16 & {\cellcolor[HTML]{00441B}} \color[HTML]{F1F1F1} 1.6e-18 & {\cellcolor[HTML]{00441B}} \color[HTML]{F1F1F1} 5.5e-22 & {\cellcolor[HTML]{00441B}} \color[HTML]{F1F1F1} 7.4e-24 & {\cellcolor[HTML]{00441B}} \color[HTML]{F1F1F1} 1.5e-26 & 94.9 & 1 & 1 & {\cellcolor[HTML]{B5E1AE}} \color[HTML]{000000} 4.8e-2 & {\cellcolor[HTML]{00441B}} \color[HTML]{F1F1F1} 2.7e-4 & {\cellcolor[HTML]{00441B}} \color[HTML]{F1F1F1} 5.3e-16 & {\cellcolor[HTML]{00441B}} \color[HTML]{F1F1F1} 1.0e-15 & {\cellcolor[HTML]{00441B}} \color[HTML]{F1F1F1} 5.9e-5 & {\cellcolor[HTML]{309950}} \color[HTML]{F1F1F1} 2.3e-3 & {\cellcolor[HTML]{00441B}} \color[HTML]{F1F1F1} 2.5e-11 & {\cellcolor[HTML]{00441B}} \color[HTML]{F1F1F1} 6.3e-5 & {\cellcolor[HTML]{00441B}} \color[HTML]{F1F1F1} 1.0e-18 & 91 \\
2. (FA, PP)$^\star$ & -- & -- & 1 & 7.5e-1 & {\cellcolor[HTML]{309950}} \color[HTML]{F1F1F1} 9.9e-3 & {\cellcolor[HTML]{00441B}} \color[HTML]{F1F1F1} 8.7e-6 & {\cellcolor[HTML]{00441B}} \color[HTML]{F1F1F1} 5.5e-7 & {\cellcolor[HTML]{00441B}} \color[HTML]{F1F1F1} 1.5e-12 & {\cellcolor[HTML]{00441B}} \color[HTML]{F1F1F1} 9.8e-15 & {\cellcolor[HTML]{00441B}} \color[HTML]{F1F1F1} 3.1e-18 & {\cellcolor[HTML]{00441B}} \color[HTML]{F1F1F1} 3.7e-20 & {\cellcolor[HTML]{00441B}} \color[HTML]{F1F1F1} 6.3e-23 & 94.1 & -- & 1 & 2.3e-1 & {\cellcolor[HTML]{309950}} \color[HTML]{F1F1F1} 2.4e-3 & {\cellcolor[HTML]{00441B}} \color[HTML]{F1F1F1} 1.1e-14 & {\cellcolor[HTML]{00441B}} \color[HTML]{F1F1F1} 2.2e-14 & {\cellcolor[HTML]{00441B}} \color[HTML]{F1F1F1} 5.9e-4 & {\cellcolor[HTML]{B5E1AE}} \color[HTML]{000000} 1.6e-2 & {\cellcolor[HTML]{00441B}} \color[HTML]{F1F1F1} 4.9e-10 & {\cellcolor[HTML]{00441B}} \color[HTML]{F1F1F1} 6.3e-4 & {\cellcolor[HTML]{00441B}} \color[HTML]{F1F1F1} 2.2e-17 & 90.6 \\
3. (LR, PP)$^\star$ & -- & -- & -- & 1 & 2.4e-1 & {\cellcolor[HTML]{309950}} \color[HTML]{F1F1F1} 1.1e-3 & {\cellcolor[HTML]{00441B}} \color[HTML]{F1F1F1} 1.0e-4 & {\cellcolor[HTML]{00441B}} \color[HTML]{F1F1F1} 6.6e-10 & {\cellcolor[HTML]{00441B}} \color[HTML]{F1F1F1} 5.4e-12 & {\cellcolor[HTML]{00441B}} \color[HTML]{F1F1F1} 1.7e-15 & {\cellcolor[HTML]{00441B}} \color[HTML]{F1F1F1} 2.1e-17 & {\cellcolor[HTML]{00441B}} \color[HTML]{F1F1F1} 3.2e-20 & 93.7 & -- & -- & 9.5e-1 & {\cellcolor[HTML]{B5E1AE}} \color[HTML]{000000} 1.7e-2 & {\cellcolor[HTML]{00441B}} \color[HTML]{F1F1F1} 2.6e-13 & {\cellcolor[HTML]{00441B}} \color[HTML]{F1F1F1} 5.1e-13 & {\cellcolor[HTML]{309950}} \color[HTML]{F1F1F1} 5.2e-3 & 8.3e-2 & {\cellcolor[HTML]{00441B}} \color[HTML]{F1F1F1} 9.8e-9 & {\cellcolor[HTML]{309950}} \color[HTML]{F1F1F1} 5.5e-3 & {\cellcolor[HTML]{00441B}} \color[HTML]{F1F1F1} 5.3e-16 & 90.6 \\
4. (PT, PP) & -- & -- & -- & -- & 6.8e-1 & {\cellcolor[HTML]{309950}} \color[HTML]{F1F1F1} 7.0e-3 & {\cellcolor[HTML]{00441B}} \color[HTML]{F1F1F1} 7.6e-4 & {\cellcolor[HTML]{00441B}} \color[HTML]{F1F1F1} 8.2e-9 & {\cellcolor[HTML]{00441B}} \color[HTML]{F1F1F1} 7.6e-11 & {\cellcolor[HTML]{00441B}} \color[HTML]{F1F1F1} 2.5e-14 & {\cellcolor[HTML]{00441B}} \color[HTML]{F1F1F1} 3.0e-16 & {\cellcolor[HTML]{00441B}} \color[HTML]{F1F1F1} 4.7e-19 & 93.6 & -- & -- & -- & 1 & {\cellcolor[HTML]{00441B}} \color[HTML]{F1F1F1} 5.1e-9 & {\cellcolor[HTML]{00441B}} \color[HTML]{F1F1F1} 9.8e-9 & 9.5e-1 & 1 & {\cellcolor[HTML]{00441B}} \color[HTML]{F1F1F1} 6.1e-5 & 9.5e-1 & {\cellcolor[HTML]{00441B}} \color[HTML]{F1F1F1} 1.3e-11 & 90.4 \\
5. (TQA, PP)$^\star$ & -- & -- & -- & -- & -- & 6.8e-1 & 2.1e-1 & {\cellcolor[HTML]{00441B}} \color[HTML]{F1F1F1} 3.9e-5 & {\cellcolor[HTML]{00441B}} \color[HTML]{F1F1F1} 6.2e-7 & {\cellcolor[HTML]{00441B}} \color[HTML]{F1F1F1} 3.7e-10 & {\cellcolor[HTML]{00441B}} \color[HTML]{F1F1F1} 5.1e-12 & {\cellcolor[HTML]{00441B}} \color[HTML]{F1F1F1} 7.9e-15 & 91.5 & -- & -- & -- & -- & {\cellcolor[HTML]{00441B}} \color[HTML]{F1F1F1} 6.6e-6 & {\cellcolor[HTML]{00441B}} \color[HTML]{F1F1F1} 1.2e-5 & 1 & 1 & {\cellcolor[HTML]{B5E1AE}} \color[HTML]{000000} 1.6e-2 & 1 & {\cellcolor[HTML]{00441B}} \color[HTML]{F1F1F1} 2.7e-8 & 89.6 \\
6. (RTS, PP)$^\star$ & -- & -- & -- & -- & -- & -- & 1 & {\cellcolor[HTML]{B5E1AE}} \color[HTML]{000000} 2.8e-2 & {\cellcolor[HTML]{309950}} \color[HTML]{F1F1F1} 1.3e-3 & {\cellcolor[HTML]{00441B}} \color[HTML]{F1F1F1} 2.4e-6 & {\cellcolor[HTML]{00441B}} \color[HTML]{F1F1F1} 4.8e-8 & {\cellcolor[HTML]{00441B}} \color[HTML]{F1F1F1} 1.1e-10 & 89.8 & -- & -- & -- & -- & -- & 1 & {\cellcolor[HTML]{00441B}} \color[HTML]{F1F1F1} 3.4e-5 & {\cellcolor[HTML]{00441B}} \color[HTML]{F1F1F1} 5.2e-7 & 7.3e-1 & {\cellcolor[HTML]{00441B}} \color[HTML]{F1F1F1} 3.1e-5 & 1 & 86.8 \\
7. (F, PP)$^\star$ & -- & -- & -- & -- & -- & -- & -- & 1.6e-1 & {\cellcolor[HTML]{B5E1AE}} \color[HTML]{000000} 1.1e-2 & {\cellcolor[HTML]{00441B}} \color[HTML]{F1F1F1} 3.5e-5 & {\cellcolor[HTML]{00441B}} \color[HTML]{F1F1F1} 8.6e-7 & {\cellcolor[HTML]{00441B}} \color[HTML]{F1F1F1} 2.4e-9 & 88.3 & -- & -- & -- & -- & -- & -- & {\cellcolor[HTML]{00441B}} \color[HTML]{F1F1F1} 5.8e-5 & {\cellcolor[HTML]{00441B}} \color[HTML]{F1F1F1} 9.7e-7 & 9.5e-1 & {\cellcolor[HTML]{00441B}} \color[HTML]{F1F1F1} 5.3e-5 & 1 & 86.2 \\
8. (I, MPSAer)$^\star$ & -- & -- & -- & -- & -- & -- & -- & -- & 1 & 2.1e-1 & {\cellcolor[HTML]{B5E1AE}} \color[HTML]{000000} 2.0e-2 & {\cellcolor[HTML]{00441B}} \color[HTML]{F1F1F1} 2.8e-4 & 86.7 & -- & -- & -- & -- & -- & -- & -- & 1 & {\cellcolor[HTML]{B5E1AE}} \color[HTML]{000000} 4.8e-2 & 1 & {\cellcolor[HTML]{00441B}} \color[HTML]{F1F1F1} 1.6e-7 & 89.5 \\
9. (FA, MPSAer)$^\star$ & -- & -- & -- & -- & -- & -- & -- & -- & -- & 8.5e-1 & 2.4e-1 & {\cellcolor[HTML]{309950}} \color[HTML]{F1F1F1} 8.8e-3 & 85.8 & -- & -- & -- & -- & -- & -- & -- & -- & {\cellcolor[HTML]{309950}} \color[HTML]{F1F1F1} 2.6e-3 & 1 & {\cellcolor[HTML]{00441B}} \color[HTML]{F1F1F1} 1.8e-9 & 89.7 \\
10. (TQA, MPSAer)$^\star$ & -- & -- & -- & -- & -- & -- & -- & -- & -- & -- & 1 & 4.4e-1 & 84.8 & -- & -- & -- & -- & -- & -- & -- & -- & -- & {\cellcolor[HTML]{B5E1AE}} \color[HTML]{000000} 4.8e-2 & {\cellcolor[HTML]{B5E1AE}} \color[HTML]{000000} 3.3e-2 & 88.6 \\
11. (LR, MPSAer)$^\star$ & -- & -- & -- & -- & -- & -- & -- & -- & -- & -- & -- & 1 & 84.7 & -- & -- & -- & -- & -- & -- & -- & -- & -- & -- & {\cellcolor[HTML]{00441B}} \color[HTML]{F1F1F1} 1.5e-7 & 89.5 \\
12. (F, MPSAer)$^\star$ & -- & -- & -- & -- & -- & -- & -- & -- & -- & -- & -- & -- & 82.3 & -- & -- & -- & -- & -- & -- & -- & -- & -- & -- & -- & 83.9 \\
\hline

\end{tabular}
}
\caption{Conover–Iman post-hoc test with Holm correction on estimated (left) and hardware (right) ratios generated from the top 5\% shots of 10 144-node heavy-hex graphs at $p=5$ (top) and $p=10$ (bottom). \raisebox{0.7ex}{\colorbox[HTML]{B5E1AE}{\hspace{1em}}} indicates p-values less than 0.05, \raisebox{0.7ex}{\colorbox[HTML]{309950}{\hspace{1em}}} indicates p-values less than 0.01, and \raisebox{0.7ex}{\colorbox[HTML]{00441B}{\hspace{1em}}} indicates p-values less than 0.001.}
\label{tab:N144HH73_5_10_5pc_hw}
\end{table*}

\subsection{Line-based graphs\label{app:hw_line_based}}

\begin{figure*}
    \centering
    \includegraphics[width=0.85\textwidth, clip, trim=0 0 0 30]{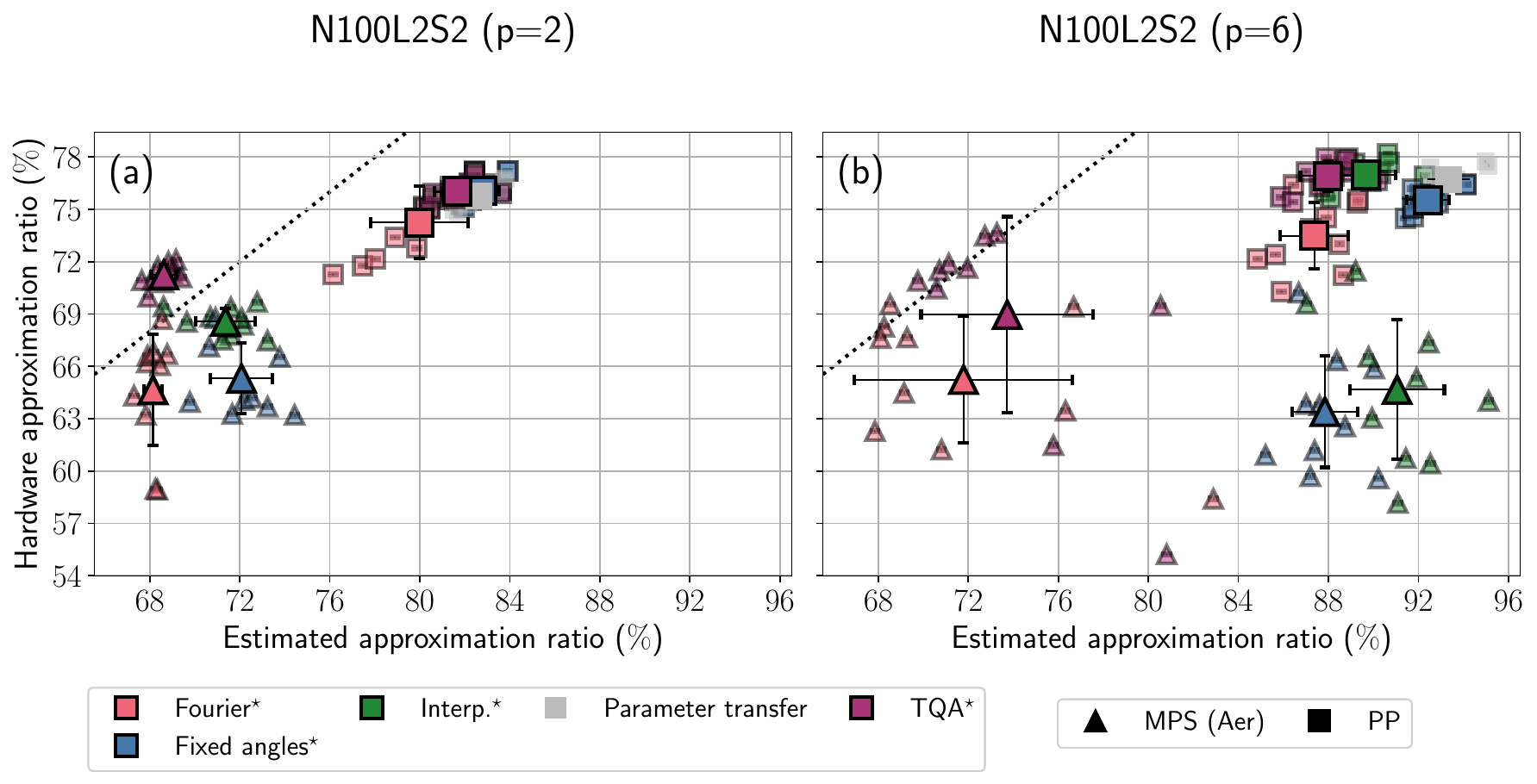}
    \caption{
    Hardware validation on \emph{ibm\_boston} for ten 100-node graphs built from applying two layers of SWAP gates to a line of 100 qubits. 
    Panels (a) and (b) show depth-two and depth-six QAOA respectively, evaluated with the Qiskit Aer MPS with a max bond dimension of 40 and Pauli propagation with a max weight of 6. 
    The dashed line indicates where the hardware approximation ratio is equal to the estimated approximation ratio.
    For the MPS, the correlation coefficients between the hardware and the estimated approximation ratio are -0.15 and -0.43 for $p=2$ and $6$, respectively.
    For PP, these numbers are 0.87 and 0.40.
    }
    \label{fig:100l2s26_hw}
\end{figure*}

Here, we repeat the analysis of Sec.~\ref{sec:hardware}, but on ten 100-node line-based graphs.
The edges are generated by an even and an odd SWAP layer applied to a line of 100 qubits.
Each edge has a $\pm 1$ weight chosen with a $50\%$ probability. 
Following the same procedure as in Sec.~\ref{sec:hardware}, we report the hardware approximation ratio against the training approximation ratio. 
The hardware ratios are lower than their estimated counterparts due to noise, see data for $p = 2$ and $6$ in Fig.~\ref{fig:100l2s26_hw}. 
Compared to Fig~\ref{fig:144hh510_hw}, it is harder to tell which method performs the best. 
Most of the methods exhibit low variance, except reoptimized TQA$^\star$ with MPSAer and Fourier$^\star$ with MPSAer (purple and red triangles, respectively). 
Parameter Transfer, TQA$^\star$, and Interp.$^\star$ with PP all perform very well.
As discussed in the main text, the inaccuracy of the MPS evaluator can be explained by the use of a relatively small bond dimension that -- due to the presence of small cycles in the graph -- yield inaccurate simulations.

\begin{table*}[ht]
\centering
\resizebox{\textwidth}{!}{%
\begin{tabular}{l|cccccccccc|ccccccccc|}
\cline{2-20}
& \multicolumn{10}{c|}{Estimated} & \multicolumn{9}{c|}{Hardware} \\
\hline

Methods(p=2) & 1. & 2. & 3. & 4. & 5. & 6. & 7. & 8. & 9. & Median & 2. & 3. & 4. & 5. & 6. & 7. & 8. & 9. & Median \\
\hline

1. (FA, PP)$^\star$ & -- & 1 & {\cellcolor[HTML]{B5E1AE}} \color[HTML]{000000} 2.1e-2 & {\cellcolor[HTML]{B5E1AE}} \color[HTML]{000000} 2.5e-2 & {\cellcolor[HTML]{00441B}} \color[HTML]{F1F1F1} 1.6e-4 & {\cellcolor[HTML]{00441B}} \color[HTML]{F1F1F1} 1.1e-13 & {\cellcolor[HTML]{00441B}} \color[HTML]{F1F1F1} 5.2e-15 & {\cellcolor[HTML]{00441B}} \color[HTML]{F1F1F1} 4.5e-21 & {\cellcolor[HTML]{00441B}} \color[HTML]{F1F1F1} 4.0e-23 & 82.5 & 1 & 1 & 1 & 2.2e-1 & {\cellcolor[HTML]{00441B}} \color[HTML]{F1F1F1} 2.3e-16 & {\cellcolor[HTML]{00441B}} \color[HTML]{F1F1F1} 9.5e-12 & {\cellcolor[HTML]{00441B}} \color[HTML]{F1F1F1} 2.8e-7 & {\cellcolor[HTML]{00441B}} \color[HTML]{F1F1F1} 1.3e-16 & 75.9 \\
2. (PT, PP) & -- & -- & 2.1e-1 & 2.3e-1 & {\cellcolor[HTML]{309950}} \color[HTML]{F1F1F1} 3.4e-3 & {\cellcolor[HTML]{00441B}} \color[HTML]{F1F1F1} 6.0e-12 & {\cellcolor[HTML]{00441B}} \color[HTML]{F1F1F1} 2.8e-13 & {\cellcolor[HTML]{00441B}} \color[HTML]{F1F1F1} 2.0e-19 & {\cellcolor[HTML]{00441B}} \color[HTML]{F1F1F1} 1.6e-21 & 82.3 & -- & 1 & 1 & 1 & {\cellcolor[HTML]{00441B}} \color[HTML]{F1F1F1} 1.7e-14 & {\cellcolor[HTML]{00441B}} \color[HTML]{F1F1F1} 7.6e-10 & {\cellcolor[HTML]{00441B}} \color[HTML]{F1F1F1} 1.5e-5 & {\cellcolor[HTML]{00441B}} \color[HTML]{F1F1F1} 8.7e-15 & 75.8 \\
3. (I, PP)$^\star$ & -- & -- & -- & 1 & 6.6e-1 & {\cellcolor[HTML]{00441B}} \color[HTML]{F1F1F1} 1.3e-7 & {\cellcolor[HTML]{00441B}} \color[HTML]{F1F1F1} 6.8e-9 & {\cellcolor[HTML]{00441B}} \color[HTML]{F1F1F1} 3.7e-15 & {\cellcolor[HTML]{00441B}} \color[HTML]{F1F1F1} 2.2e-17 & 81.7 & -- & -- & 1 & 2.9e-1 & {\cellcolor[HTML]{00441B}} \color[HTML]{F1F1F1} 5.7e-16 & {\cellcolor[HTML]{00441B}} \color[HTML]{F1F1F1} 2.4e-11 & {\cellcolor[HTML]{00441B}} \color[HTML]{F1F1F1} 6.2e-7 & {\cellcolor[HTML]{00441B}} \color[HTML]{F1F1F1} 2.9e-16 & 75.9 \\
4. (TQA, PP)$^\star$ & -- & -- & -- & -- & 6.6e-1 & {\cellcolor[HTML]{00441B}} \color[HTML]{F1F1F1} 9.4e-8 & {\cellcolor[HTML]{00441B}} \color[HTML]{F1F1F1} 4.9e-9 & {\cellcolor[HTML]{00441B}} \color[HTML]{F1F1F1} 2.6e-15 & {\cellcolor[HTML]{00441B}} \color[HTML]{F1F1F1} 1.6e-17 & 81.7 & -- & -- & -- & 2.0e-1 & {\cellcolor[HTML]{00441B}} \color[HTML]{F1F1F1} 1.7e-16 & {\cellcolor[HTML]{00441B}} \color[HTML]{F1F1F1} 7.2e-12 & {\cellcolor[HTML]{00441B}} \color[HTML]{F1F1F1} 2.1e-7 & {\cellcolor[HTML]{00441B}} \color[HTML]{F1F1F1} 9.2e-17 & 75.9 \\
5. (F, PP)$^\star$ & -- & -- & -- & -- & -- & {\cellcolor[HTML]{00441B}} \color[HTML]{F1F1F1} 6.6e-5 & {\cellcolor[HTML]{00441B}} \color[HTML]{F1F1F1} 4.5e-6 & {\cellcolor[HTML]{00441B}} \color[HTML]{F1F1F1} 3.1e-12 & {\cellcolor[HTML]{00441B}} \color[HTML]{F1F1F1} 1.8e-14 & 80.2 & -- & -- & -- & -- & {\cellcolor[HTML]{00441B}} \color[HTML]{F1F1F1} 9.0e-12 & {\cellcolor[HTML]{00441B}} \color[HTML]{F1F1F1} 3.4e-7 & {\cellcolor[HTML]{309950}} \color[HTML]{F1F1F1} 2.6e-3 & {\cellcolor[HTML]{00441B}} \color[HTML]{F1F1F1} 4.8e-12 & 74.5 \\
6. (FA, MPSAer)$^\star$ & -- & -- & -- & -- & -- & -- & 1 & {\cellcolor[HTML]{309950}} \color[HTML]{F1F1F1} 2.2e-3 & {\cellcolor[HTML]{00441B}} \color[HTML]{F1F1F1} 3.5e-5 & 71.9 & -- & -- & -- & -- & -- & 2.2e-1 & {\cellcolor[HTML]{00441B}} \color[HTML]{F1F1F1} 1.4e-4 & 1 & 64.1 \\
7. (I, MPSAer)$^\star$ & -- & -- & -- & -- & -- & -- & -- & {\cellcolor[HTML]{B5E1AE}} \color[HTML]{000000} 1.8e-2 & {\cellcolor[HTML]{00441B}} \color[HTML]{F1F1F1} 4.2e-4 & 71.6 & -- & -- & -- & -- & -- & -- & 2.2e-1 & 1.7e-1 & 68.6 \\
8. (TQA, MPSAer)$^\star$ & -- & -- & -- & -- & -- & -- & -- & -- & 9.9e-1 & 68.7 & -- & -- & -- & -- & -- & -- & -- & {\cellcolor[HTML]{00441B}} \color[HTML]{F1F1F1} 8.3e-5 & 71.2 \\
9. (F, MPSAer)$^\star$ & -- & -- & -- & -- & -- & -- & -- & -- & -- & 68.2 & -- & -- & -- & -- & -- & -- & -- & -- & 66.1 \\
\hline

Methods(p=6) & 1. & 2. & 3. & 4. & 5. & 6. & 7. & 8. & 9. & Median & 2. & 3. & 4. & 5. & 6. & 7. & 8. & 9. & Median \\
\hline

1. (PT, PP) & -- & 6.7e-1 & {\cellcolor[HTML]{309950}} \color[HTML]{F1F1F1} 2.7e-3 & {\cellcolor[HTML]{00441B}} \color[HTML]{F1F1F1} 1.3e-6 & {\cellcolor[HTML]{00441B}} \color[HTML]{F1F1F1} 3.0e-12 & {\cellcolor[HTML]{00441B}} \color[HTML]{F1F1F1} 9.3e-14 & {\cellcolor[HTML]{00441B}} \color[HTML]{F1F1F1} 1.2e-12 & {\cellcolor[HTML]{00441B}} \color[HTML]{F1F1F1} 4.6e-23 & {\cellcolor[HTML]{00441B}} \color[HTML]{F1F1F1} 1.4e-24 & 93 & {\cellcolor[HTML]{B5E1AE}} \color[HTML]{000000} 2.4e-2 & {\cellcolor[HTML]{00441B}} \color[HTML]{F1F1F1} 9.0e-17 & 1 & 1 & {\cellcolor[HTML]{00441B}} \color[HTML]{F1F1F1} 1.2e-5 & {\cellcolor[HTML]{00441B}} \color[HTML]{F1F1F1} 4.2e-18 & {\cellcolor[HTML]{00441B}} \color[HTML]{F1F1F1} 2.3e-11 & {\cellcolor[HTML]{00441B}} \color[HTML]{F1F1F1} 2.9e-16 & 76.8 \\
2. (FA, PP)$^\star$ & -- & -- & 1.6e-1 & {\cellcolor[HTML]{00441B}} \color[HTML]{F1F1F1} 3.5e-4 & {\cellcolor[HTML]{00441B}} \color[HTML]{F1F1F1} 2.4e-9 & {\cellcolor[HTML]{00441B}} \color[HTML]{F1F1F1} 7.5e-11 & {\cellcolor[HTML]{00441B}} \color[HTML]{F1F1F1} 9.7e-10 & {\cellcolor[HTML]{00441B}} \color[HTML]{F1F1F1} 2.2e-20 & {\cellcolor[HTML]{00441B}} \color[HTML]{F1F1F1} 5.5e-22 & 91.9 & -- & {\cellcolor[HTML]{00441B}} \color[HTML]{F1F1F1} 1.1e-10 & {\cellcolor[HTML]{309950}} \color[HTML]{F1F1F1} 2.9e-3 & {\cellcolor[HTML]{309950}} \color[HTML]{F1F1F1} 4.7e-3 & 2.2e-1 & {\cellcolor[HTML]{00441B}} \color[HTML]{F1F1F1} 4.3e-12 & {\cellcolor[HTML]{00441B}} \color[HTML]{F1F1F1} 1.5e-5 & {\cellcolor[HTML]{00441B}} \color[HTML]{F1F1F1} 3.7e-10 & 75.6 \\
3. (I, MPSAer)$^\star$ & -- & -- & -- & 2.3e-1 & {\cellcolor[HTML]{00441B}} \color[HTML]{F1F1F1} 3.2e-5 & {\cellcolor[HTML]{00441B}} \color[HTML]{F1F1F1} 1.6e-6 & {\cellcolor[HTML]{00441B}} \color[HTML]{F1F1F1} 1.4e-5 & {\cellcolor[HTML]{00441B}} \color[HTML]{F1F1F1} 4.9e-16 & {\cellcolor[HTML]{00441B}} \color[HTML]{F1F1F1} 9.8e-18 & 91.3 & -- & -- & {\cellcolor[HTML]{00441B}} \color[HTML]{F1F1F1} 4.2e-18 & {\cellcolor[HTML]{00441B}} \color[HTML]{F1F1F1} 8.3e-18 & {\cellcolor[HTML]{00441B}} \color[HTML]{F1F1F1} 1.5e-6 & 1 & 5.7e-2 & 1 & 64.7 \\
4. (I, PP)$^\star$ & -- & -- & -- & -- & {\cellcolor[HTML]{B5E1AE}} \color[HTML]{000000} 3.0e-2 & {\cellcolor[HTML]{309950}} \color[HTML]{F1F1F1} 3.1e-3 & {\cellcolor[HTML]{B5E1AE}} \color[HTML]{000000} 1.8e-2 & {\cellcolor[HTML]{00441B}} \color[HTML]{F1F1F1} 5.7e-12 & {\cellcolor[HTML]{00441B}} \color[HTML]{F1F1F1} 1.1e-13 & 89.7 & -- & -- & -- & 1 & {\cellcolor[HTML]{00441B}} \color[HTML]{F1F1F1} 7.3e-7 & {\cellcolor[HTML]{00441B}} \color[HTML]{F1F1F1} 1.9e-19 & {\cellcolor[HTML]{00441B}} \color[HTML]{F1F1F1} 9.3e-13 & {\cellcolor[HTML]{00441B}} \color[HTML]{F1F1F1} 1.3e-17 & 77 \\
5. (TQA, PP)$^\star$ & -- & -- & -- & -- & -- & 1 & 1 & {\cellcolor[HTML]{00441B}} \color[HTML]{F1F1F1} 2.1e-6 & {\cellcolor[HTML]{00441B}} \color[HTML]{F1F1F1} 5.9e-8 & 88.1 & -- & -- & -- & -- & {\cellcolor[HTML]{00441B}} \color[HTML]{F1F1F1} 1.4e-6 & {\cellcolor[HTML]{00441B}} \color[HTML]{F1F1F1} 3.8e-19 & {\cellcolor[HTML]{00441B}} \color[HTML]{F1F1F1} 1.9e-12 & {\cellcolor[HTML]{00441B}} \color[HTML]{F1F1F1} 2.6e-17 & 77 \\
6. (F, PP)$^\star$ & -- & -- & -- & -- & -- & -- & 1 & {\cellcolor[HTML]{00441B}} \color[HTML]{F1F1F1} 4.1e-5 & {\cellcolor[HTML]{00441B}} \color[HTML]{F1F1F1} 1.5e-6 & 87.6 & -- & -- & -- & -- & -- & {\cellcolor[HTML]{00441B}} \color[HTML]{F1F1F1} 7.2e-8 & {\cellcolor[HTML]{B5E1AE}} \color[HTML]{000000} 2.7e-2 & {\cellcolor[HTML]{00441B}} \color[HTML]{F1F1F1} 4.3e-6 & 73.5 \\
7. (FA, MPSAer)$^\star$ & -- & -- & -- & -- & -- & -- & -- & {\cellcolor[HTML]{00441B}} \color[HTML]{F1F1F1} 4.9e-6 & {\cellcolor[HTML]{00441B}} \color[HTML]{F1F1F1} 1.4e-7 & 87.5 & -- & -- & -- & -- & -- & -- & {\cellcolor[HTML]{309950}} \color[HTML]{F1F1F1} 8.1e-3 & 1 & 63.2 \\
8. (TQA, MPSAer)$^\star$ & -- & -- & -- & -- & -- & -- & -- & -- & 1 & 72.3 & -- & -- & -- & -- & -- & -- & -- & 1.1e-1 & 71.2 \\
9. (F, MPSAer)$^\star$ & -- & -- & -- & -- & -- & -- & -- & -- & -- & 69.2 & -- & -- & -- & -- & -- & -- & -- & -- & 66 \\
\hline

\end{tabular}
}
\caption{Conover–Iman post-hoc test with Holm correction on estimated (left) and hardware (right) ratios generated from 10 100-node line-based graphs with 2 swap layers at $p=2$ (top) and $p=6$ (bottom). \raisebox{0.7ex}{\colorbox[HTML]{B5E1AE}{\hspace{1em}}} indicates p-values less than 0.05, \raisebox{0.7ex}{\colorbox[HTML]{309950}{\hspace{1em}}} indicates p-values less than 0.01, and \raisebox{0.7ex}{\colorbox[HTML]{00441B}{\hspace{1em}}} indicates p-values less than 0.001.}
\label{tab:N100L2S2_2_6_100pc_hw}
\end{table*}

To look at the statistical impact of the different training methods on the estimated and the hardware approximation ratios, we again perform a Kruskal–Wallis statistical test followed by Conover-Iman posthoc test with Holm correction.
At $p=2$, the test distinguishes most of the methods in terms of estimated ratios, with the exception of several top performing methods due to low depth, see the top-left subtable of Tab~\ref{tab:N100L2S2_2_6_100pc_hw}. 
At $p=6$, the test distinguishes more methods in terms of the estimated ratios than at $p=2$, see the bottom-left subtable of Tab.~\ref{tab:N100L2S2_2_6_100pc_hw}. 
Once executed on hardware, the top performing methods, Fixed Angles$^\star$, TQA$^\star$, and Interp.$^\star$, become indistinguishable from one another.

\section{Resource cost analysis\label{app:cost_analysis}}

\begin{figure*}[t!]
\centering
\includegraphics[width=\textwidth]{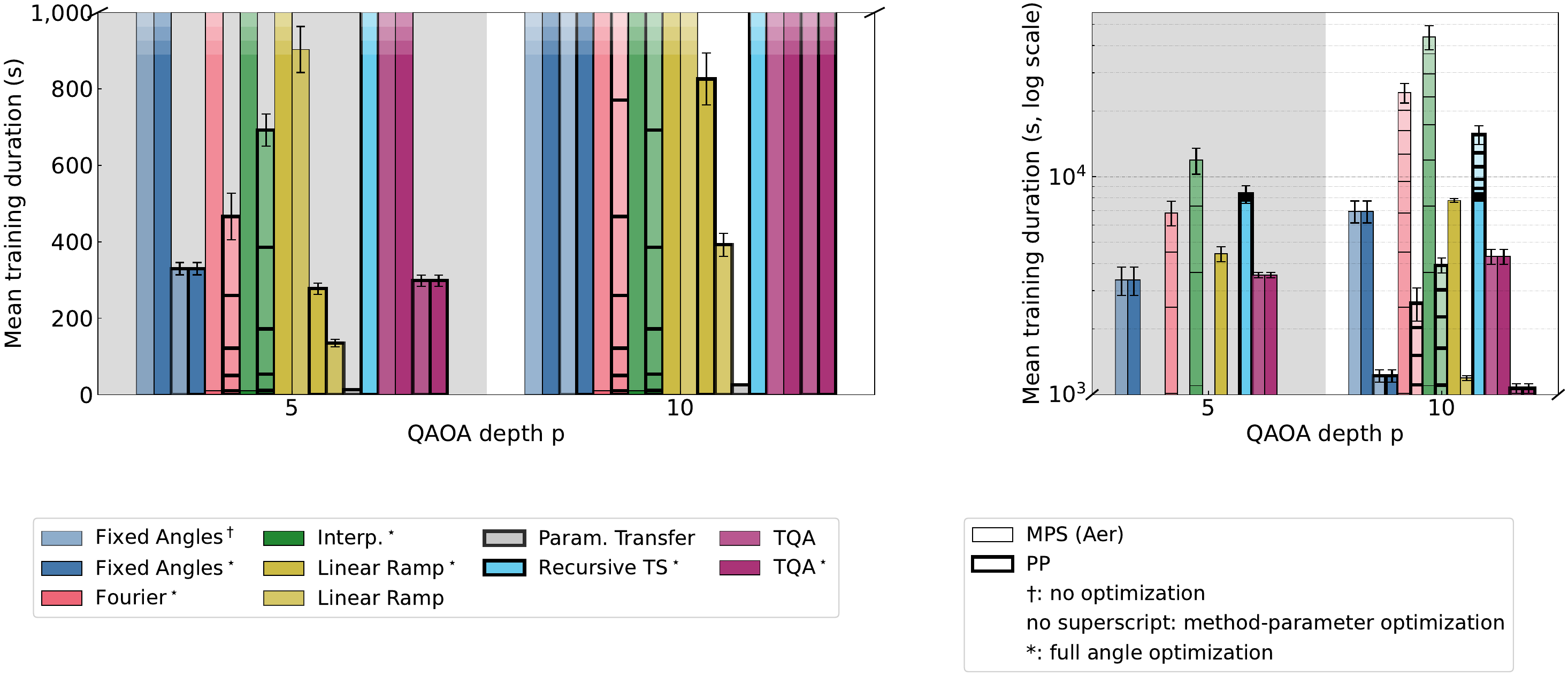}
\caption{
Mean training duration for different angle-setting strategies at QAOA depths $p=5$ and $p=10$. 
Each horizontal line within the bar represents the cumulative classical training time required to determine QAOA angles up to the corresponding circuit depth.
The left panel shows all the methods we tested while the right one focuses on the most time-consuming methods.
\label{fig:training_bricks}}
\end{figure*}

The training duration analysis shown in Fig.~\ref{fig:training_bricks} highlights the cost scaling of different angle setting strategies on the 144-node heavy-hex instances. 
Methods such as parameter transfer, linear ramp, and fixed angles incur consistently low training costs, as they rely on precomputed angles or structured re-optimization.
This advantage becomes particularly pronounced at the utility scale, where these approaches introduce only limited additional overhead.
In contrast, iterative methods exhibit a strong increase in training time with both circuit depth and problem size due to repeated circuit evaluations.
Consequently, increasing the QAOA depth from $p=5$ to $p=10$ on these instances significantly increases the training cost while yielding only marginal improvements in approximation ratios, on quantum hardware.

\begin{figure*}[t!]
\centering
\includegraphics[width=\textwidth]{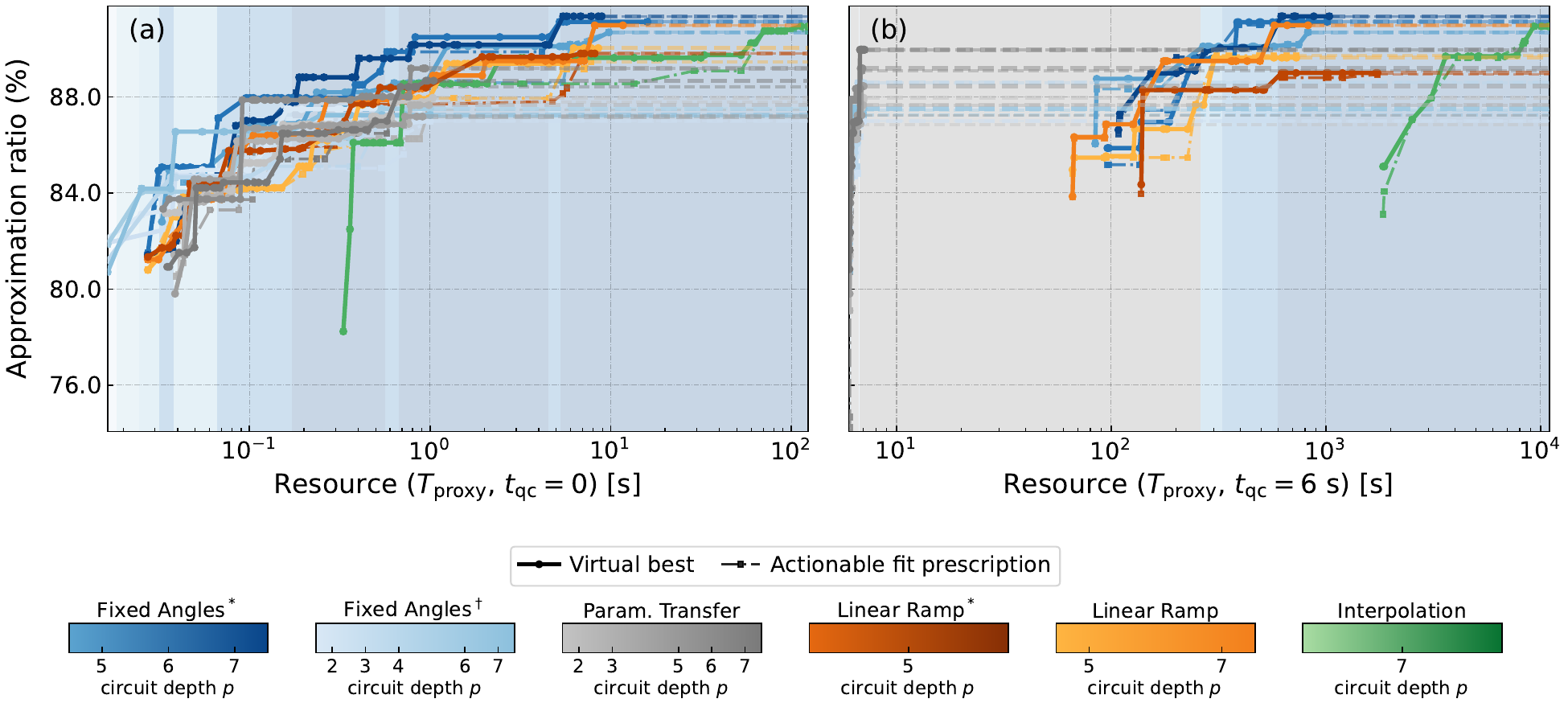}
\caption{
Multi-strategy window sticker on the ten test instances under the two cost models of Fig.~\ref{fig:pareto-overlay}: (a) $t_{qc} = 0$ and (b) $t_{qc} = 6s$.
Solid lines are the virtual best and dashed the actionable prescription. 
The QAOA depths tested for each strategy are represented with a color gradient.
Background shading marks the method configuration holding the frontier at each budget.
\label{fig:window-sticker-cost}}
\end{figure*}

Figure~\ref{fig:pareto-overlay} shows only the Pareto frontier, i.e., which angle setting strategy the actionable prescription recommends at each budget.
Figure~\ref{fig:window-sticker-cost} shows every curve on which the Pareto frontier was constructed: on the ten test instances with (a) $t_\text{qc}=0~\mathrm{s}$ and (b) $t_\text{qc}=6~\mathrm{s}$, as before.
Solid lines are the virtual best, which gives the best approximation ratio reached by a strategy at that budget, and the dashed lines are the actionable prescription, derived from the values of $(N, M, Q)$ chosen from the training instances alone.
The background shading depicts which family holds the actionable frontier, i.e., what Fig.~\ref{fig:pareto-overlay} draws, at each budget.

In Fig.~\ref{fig:window-sticker-cost}(a), several methods closely follow Fixed Angles$^\star$, with differences narrowing at larger budgets.
Including submission latency, in Fig.~\ref{fig:window-sticker-cost}(b), favors training-free methods at smaller budgets and delays improvements from methods requiring parameter optimization. 
Parameter Transfer therefore leads the frontier over an extended budget range before Fixed Angles$^\star$ overtakes it, while Interp. improves only at substantially larger budgets although it never appears on the frontier within the tested range. 

\clearpage
\bibliography{refs}

@article{Abbas2024,
  title        = {Challenges and opportunities in quantum optimization},
  author       = {Abbas, Amira and Ambainis, Andris and Augustino, Brandon and B{\"a}rtschi, Andreas and Buhrman, Harry and Coffrin, Carleton and Cortiana, Giorgio and Dunjko, Vedran and Egger, Daniel J and Elmegreen, Bruce G and others},
  year         = {2024},
  journal      = {Nat. Rev. Phys.},
  publisher    = {Nature Publishing Group UK London},
  volume       = {6},
  pages        = {718–735},
  doi          = {https://doi.org/10.1038/s42254-024-00770-9}
}

@inproceedings{Acampora2023,
  title        = {Fuzzy Clustering for {QAOA} Complexity Reduction},
  author       = {Acampora, Giovanni and Chiatto, Angela and Vitiello, Autilia},
  year         = {2023},
  booktitle    = {2023 {IEEE} International Conference on Fuzzy Systems ({FUZZ})},
  pages        = {1--7},
  doi          = {10.1109/FUZZ52849.2023.10309767}
}

@article{Agirre2025,
  title        = {A Monte Carlo Tree Search approach to {QAOA}: finding a needle in the haystack},
  author       = {Agirre, Andoni and van Nieuwenburg, Evert and Wauters, Matteo M},
  year         = {2025},
  month        = {apr},
  journal      = {New J. Phys.},
  publisher    = {IOP Publishing},
  volume       = {27},
  number       = {4},
  pages        = {043014},
  doi          = {10.1088/1367-2630/adc765}
}

@article{Akshay2021,
  title        = {Parameter concentrations in quantum approximate optimization},
  author       = {Akshay, V. and Rabinovich, D. and Campos, E. and Biamonte, J.},
  year         = {2021},
  month        = {Jul},
  journal      = {Phys. Rev. A},
  publisher    = {American Physical Society},
  volume       = {104},
  pages        = {L010401},
  doi          = {10.1103/PhysRevA.104.L010401},
  url          = {https://link.aps.org/doi/10.1103/PhysRevA.104.L010401},
  issue        = {1},
}

@inproceedings{Alam2020,
  title        = {Accelerating quantum approximate optimization algorithm using machine learning},
  author       = {Alam, Mahabubul and Ash-Saki, Abdullah and Ghosh, Swaroop},
  year         = {2020},
  booktitle    = {2020 Design, Automation \& Test in Europe Conference \& Exhibition ({DATE})},
  location     = {Grenoble, France},
  publisher    = {EDA Consortium},
  address      = {San Jose, CA, USA},
  series       = {DATE '20},
  pages        = {686–689},
  isbn         = {9783981926347},
  url          = {https://ieeexplore.ieee.org/document/9116348},
  doi          = {10.23919/DATE48585.2020.9116348}}

@article{Amosy2024-xm,
  title        = {{Iteration-free quantum approximate optimization algorithm using neural networks}},
  author       = {Amosy, Ohad and Danzig, Tamuz and Lev, Ohad and Porat, Ely and Chechik, Gal and Makmal, Adi},
  year         = {2024},
  journal      = {Quantum Machine Intelligence},
  publisher    = {Springer},
  volume       = {6},
  number       = {2},
  pages        = {38},
  doi          = {https://doi.org/10.1007/s42484-024-00159-y}
}

@misc{apte2025,
  title        = {Iterative Interpolation Schedules for Quantum Approximate Optimization Algorithm},
  author       = {Anuj Apte and Shree Hari Sureshbabu and Ruslan Shaydulin and Sami Boulebnane and Zichang He and Dylan Herman and James Sud and Marco Pistoia},
  year         = {2025},
  doi          = {10.48550/arXiv.2504.01694},
  eprint       = {2504.01694},
  archiveprefix = {arXiv}
}

@article{Ayral2023_DMRG-FiniteFidelity,
  title        = {Density-Matrix Renormalization Group Algorithm for Simulating Quantum Circuits with a Finite Fidelity},
  author       = {Ayral, Thomas and Louvet, Thibaud and Zhou, Yiqing and Lambert, Cyprien and Stoudenmire, E. Miles and Waintal, Xavier},
  year         = {2023},
  journal      = {PRX Quantum},
  volume       = {4},
  pages        = {020304},
  doi          = {10.1103/PRXQuantum.4.020304},
  url          = {https://link.aps.org/doi/10.1103/PRXQuantum.4.020304},
  issue        = {2}
}

@article{bapat2018bang,
  title        = {Bang-bang control as a design principle for classical and quantum optimization algorithms},
  author       = {Bapat, Aniruddha and Jordan, Stephen},
  year         = {2019},
  month        = {may},
  journal      = {Quantum Info. Comput.},
  publisher    = {Rinton Press, Incorporated},
  address      = {Paramus, NJ},
  volume       = {19},
  number       = {5–6},
  pages        = {424–446},
  url          = {https://dl.acm.org/doi/10.5555/3370251.3370255},
  issn         = {1533-7146},
}

@article{Barron2024,
  title        = {Provable bounds for {noise-free} expectation values computed from noisy samples},
  author       = {Barron, Samantha V. and Egger, Daniel J. and Pelofske, Elijah and B\"{a}rtschi, Andreas and Eidenbenz, Stephan and Lehmkuehler, Matthis and Woerner, Stefan},
  year         = {2024},
  month        = {Nov},
  journal      = {Nat. Comput. Sci.},
  publisher    = {Springer Science and Business Media LLC},
  volume       = {4},
  number       = {11},
  pages        = {865–875},
  doi          = {10.1038/s43588-024-00709-1},
  issn         = {2662-8457}
}

@article{Begusic2025SparsePauli,
  title        = {Real-Time Operator Evolution in Two and Three Dimensions via Sparse Pauli Dynamics},
  author       = {Tomislav Begu{\v{s}}i{\'c} and Garnet Kin-Lic Chan},
  year         = {2025},
  journal      = {PRX Quantum},
  publisher    = {American Physical Society},
  volume       = {6},
  number       = {2},
  pages        = {020302},
  doi          = {10.1103/PRXQuantum.6.020302},
  url          = {https://doi.org/10.1103/PRXQuantum.6.020302}
}

@article{bernasconi1987,
  title        = {Low autocorrelation binary sequences: statistical mechanics and configuration space analysis},
  author       = {Bernasconi, J.},
  year         = {1987},
  journal      = {J. Phys. France},
  volume       = {48},
  number       = {4},
  pages        = {559--567},
  doi          = {10.1051/jphys:01987004804055900}
}

@article{Bittel_2021,
  title        = {Training Variational Quantum Algorithms Is {NP}-Hard},
  author       = {Lennart Bittel and Martin Kliesch},
  year         = {2021},
  month        = {sep},
  journal      = {Phys. Rev. Lett.},
  publisher    = {American Physical Society ({APS})},
  volume       = {127},
  pages        = {120502},
  doi          = {10.1103/PhysRevLett.127.120502},
  url          = {https://link.aps.org/doi/10.1103/PhysRevLett.127.120502},
  issue        = {12},
}

@inproceedings{bittelQCMA,
  title        = {The Optimal Depth of Variational Quantum Algorithms Is {QCMA}-Hard to Approximate},
  author       = {Lennart Bittel and Sevag Gharibian and Martin Kliesch},
  year         = {2023},
  booktitle    = {38th Computational Complexity Conference ({CCC} 2023)},
  publisher    = {Schloss Dagstuhl -- Leibniz-Zentrum f{\"u}r Informatik},
  series       = {Leibniz International Proceedings in Informatics (LIPIcs)},
  volume       = {264},
  pages        = {34:1--34:24},
  doi          = {10.4230/LIPICS.CCC.2023.34}
}

@article{tate2023warm,
  title={Warm-started {QAOA} with custom mixers provably converges and computationally beats {G}oemans-{W}illiamson's max-cut at low circuit depths},
  author={Tate, Reuben and Moondra, Jai and Gard, Bryan and Mohler, Greg and Gupta, Swati},
  journal={Quantum},
  volume={7},
  pages={1121},
  year={2023},
  publisher={Verein zur F{\"o}rderung des Open Access Publizierens in den Quantenwissenschaften},
  doi={	https://doi.org/10.22331/q-2023-09-26-1121}
}

@misc{sakos2026global,
  title        = {Global Optimization for Parametrized Quantum Circuits},
  author       = {Sakos, Iosif and Varvitsiotis, Antonios and Korpas, Georgios and Lin, Wayne},
  year         = {2026},
  eprint       = {2603.21757},
  archivePrefix= {arXiv},
  url          = {https://arxiv.org/abs/2603.21757}, 
}

@article{truger2024warm,
  title={Warm-starting and quantum computing: A systematic mapping study},
  author={Truger, Felix and Barzen, Johanna and Bechtold, Marvin and Beisel, Martin and Leymann, Frank and Mandl, Alexander and Yussupov, Vladimir},
  journal={ACM Comput. Surv.},
  volume={56},
  number={9},
  pages={1--31},
  year={2024},
  publisher={ACM New York, NY},
  doi={https://doi.org/10.1145/3652510}
}

@inproceedings{tate2025warm,
  title={Warm-Started {QAOA} with Aligned Mixers Converges Slowly Near the Poles of the {Bloch} Sphere},
  author={Tate, Reuben and Eidenbenz, Stephan},
  booktitle={International Conference on Current Trends in Theory and Practice of Computer Science},
  pages={311--323},
  year={2025},
  organization={Springer},
  doi={https://doi.org/10.1007/978-3-031-82697-9_23}
}

@article{Hatami2014,
	author = {Hatami, Hamed and Lov{\'a}sz, L{\'a}szl{\'o} and Szegedy, Bal{\'a}zs},
	date = {2014/02/01},
	doi = {10.1007/s00039-014-0258-7},
	isbn = {1420-8970},
	journal = {Geom. Funct. Anal.},
	number = {1},
	pages = {269--296},
	title = {Limits of locally--globally convergent graph sequences},
	volume = {24},
	year = {2014}
}

@article{Bo_kovi__2017,
  title        = {Low-autocorrelation binary sequences: On improved merit factors and runtime predictions to achieve them},
  author       = {Bošković, Borko and Brglez, Franc and Brest, Janez},
  year         = {2017},
  month        = jul,
  journal      = {Appl. Soft Comput.},
  publisher    = {Elsevier BV},
  volume       = {56},
  pages        = {262–285},
  doi          = {10.1016/j.asoc.2017.02.024},
  issn         = {1568-4946},
  url          = {http://dx.doi.org/10.1016/j.asoc.2017.02.024}
}

@misc{boulebnane2025,
  title        = {{Quantum Approximate Optimization Algorithm} in Finite Size and Large Depth and Equivalence to Quantum Annealing},
  author       = {Sami Boulebnane and James Sud and Ruslan Shaydulin and Marco Pistoia},
  year         = {2025},
  url          = {https://arxiv.org/abs/2503.09563},
  eprint       = {2503.09563},
  archiveprefix = {arXiv}
}

@article{Brady2021BangAnnealBang,
  title        = {{Optimal Protocols in Quantum Annealing and Quantum Approximate Optimization Algorithm Problems}},
  author       = {Brady, Lucas T. and Baldwin, Christopher L. and Bapat, Aniruddha and Kharkov, Yaroslav and Gorshkov, Alexey V.},
  year         = {2021},
  journal      = {Phys. Rev. Lett.},
  volume       = {126},
  pages        = {070505},
  doi          = {10.1103/PhysRevLett.126.070505}
}

@misc{brandao2018,
  title        = {For Fixed Control Parameters the Quantum Approximate Optimization Algorithm's Objective Function Value Concentrates for Typical Instances},
  author       = {Fernando G. S. L. Brandao and Michael Broughton and Edward Farhi and Sam Gutmann and Hartmut Neven},
  year         = {2018},
  url          = {https://arxiv.org/abs/1812.04170},
  eprint       = {1812.04170},
  archiveprefix = {arXiv}
}

@article{Bravyi2020,
  title        = {Obstacles to Variational Quantum Optimization from Symmetry Protection},
  author       = {Bravyi, Sergey and Kliesch, Alexander and Koenig, Robert and Tang, Eugene},
  year         = {2020},
  month        = {Dec},
  journal      = {Phys. Rev. Lett.},
  publisher    = {American Physical Society},
  volume       = {125},
  pages        = {260505},
  doi          = {10.1103/PhysRevLett.125.260505},
  url          = {https://link.aps.org/doi/10.1103/PhysRevLett.125.260505},
  issue        = {26},
}

@article{11389907,
  author={Mastropietro, Daniel and Korpas, Georgios and Kungurtsev, Vyacheslav and Marecek, Jakub},
  journal={IEEE Transactions on Quantum Engineering}, 
  title={Parallel Variational Quantum Algorithms With Gradient-Informed Restart to Speed Up Optimization in the Presence of Barren Plateaus}, 
  year={2026},
  volume={7},
  pages={1-17},
  doi={10.1109/TQE.2026.3663507},
}

@inproceedings{khot2010unique,
  title={On the unique games conjecture (invited survey)},
  author={Khot, Subhash},
  booktitle={2010 {IEEE} 25th annual conference on computational complexity},
  pages={99--121},
  year={2010},
  organization={IEEE Computer Society},
  doi={10.1109/CCC.2010.19}
}

@article{goemans1995improved,
  title={Improved approximation algorithms for maximum cut and satisfiability problems using semidefinite programming},
  author={Goemans, Michel X and Williamson, David P},
  journal={Journal of the ACM (JACM)},
  volume={42},
  number={6},
  pages={1115--1145},
  year={1995},
  publisher={ACM New York, NY, USA},
  url={https://doi.org/10.1145/227683.227684},
  doi={10.1145/227683.227684},
}

@article{BravyiKlieschKoenigTang2022,
  title        = {Hybrid quantum-classical algorithms for approximate graph coloring},
  author       = {Bravyi, Sergey and Kliesch, Alexander and Koenig, Robert and Tang, Eugene},
  year         = {2022},
  journal      = {Quantum},
  volume       = {6},
  pages        = {678},
  doi          = {10.22331/q-2022-03-30-678}
}

@inproceedings{BarakMarwaha2022,
  title        = {Classical algorithms and quantum limitations for maximum cut on high-girth graphs},
  author       = {Barak, Boaz and Marwaha, Kunal},
  year         = {2022},
  booktitle    = {13th Innovations in Theoretical Computer Science Conference ({ITCS} 2022)},
  series       = {LIPIcs},
  volume       = {215},
  pages        = {14:1--14:21},
  publisher    = {Schloss Dagstuhl -- Leibniz-Zentrum f{\"u}r Informatik},
  doi          = {10.4230/LIPIcs.ITCS.2022.14}
}

@InProceedings{ChouLoveSandhuShi2021,
  author =	{Chou, Chi-Ning and Love, Peter J. and Sandhu, Juspreet Singh and Shi, Jonathan},
  title =	{{Limitations of Local Quantum Algorithms on Random MAX-k-XOR and Beyond}},
  booktitle =	{49th International Colloquium on Automata, Languages, and Programming ({ICALP} 2022)},
  pages =	{41:1--41:20},
  series =	{Leibniz International Proceedings in Informatics (LIPIcs)},
  ISBN =	{978-3-95977-235-8},
  ISSN =	{1868-8969},
  year =	{2022},
  volume =	{229},
  editor =	{Boja\'{n}czyk, Miko{\l}aj and Merelli, Emanuela and Woodruff, David P.},
  publisher =	{Schloss Dagstuhl -- Leibniz-Zentrum f{\"u}r Informatik},
  address =	{Dagstuhl, Germany},
  URL =		{https://drops.dagstuhl.de/entities/document/10.4230/LIPIcs.ICALP.2022.41},
  doi =		{10.4230/LIPIcs.ICALP.2022.41}
}

@misc{ChenHuangMarwaha2023,
  title        = {Local algorithms and the failure of log-depth quantum advantage on sparse random {CSPs}},
  author       = {Chen, Antares and Huang, Neng and Marwaha, Kunal},
  year         = {2023},
  url          = {https://arxiv.org/abs/2310.01563},
  eprint       = {2310.01563},
  archiveprefix = {arXiv}
}

@misc{cadavid2025scalingadvantagequantumenhancedmemetic,
  title        = {{Scaling advantage with quantum-enhanced memetic tabu search for LABS}},
  author       = {Alejandro Gomez Cadavid and Pranav Chandarana and Sebastián V. Romero and Jan Trautmann and Enrique Solano and Taylor Lee Patti and Narendra N. Hegade},
  year         = {2025},
  url          = {https://arxiv.org/abs/2511.04553},
  eprint       = {2511.04553},
  archiveprefix = {arXiv}
}

@article{Campos2021,
  title        = {Training saturation in layerwise quantum approximate optimization},
  author       = {Campos, E. and Rabinovich, D. and Akshay, V. and Biamonte, J.},
  year         = {2021},
  month        = {Sep},
  journal      = {Phys. Rev. A},
  publisher    = {American Physical Society},
  volume       = {104},
  pages        = {L030401},
  doi          = {10.1103/PhysRevA.104.L030401},
  url          = {https://link.aps.org/doi/10.1103/PhysRevA.104.L030401},
  issue        = {3},
}

@misc{cepaite2025quantum,
  title        = {Quantum-Enhanced Optimization by Warm Starts},
  author       = {Ieva \v{C}epait\.{e} and Niam Vaishnav and Leo Zhou and Ashley Montanaro},
  year         = {2025},
  url          = {https://arxiv.org/abs/2508.16309},
  eprint       = {2508.16309},
  archiveprefix = {arXiv}
}

@misc{Chen2025,
  title        = {Learning to Learn with Quantum Optimization via Quantum Neural Networks},
  author       = {Kuan-Cheng Chen and Hiromichi Matsuyama and Wei-Hao Huang},
  year         = {2025},
  url          = {https://arxiv.org/abs/2505.00561},
  eprint       = {2505.00561},
  archiveprefix = {arXiv}
}

@article{Cheng2024,
  title        = {Quantum approximate optimization via learning-based adaptive optimization},
  author       = {Cheng, Lixue and Chen, Yu-Qin and Zhang, Shi-Xin and Zhang, Shengyu},
  year         = {2024},
  month        = {Mar},
  day          = {06},
  journal      = {Commun. Phys.},
  volume       = {7},
  number       = {1},
  pages        = {83},
  doi          = {10.1038/s42005-024-01577-x},
  issn         = {2399-3650},
  url          = {https://doi.org/10.1038/s42005-024-01577-x}
}

@article{Das2005,
  title        = {Quantum annealing in a kinetically constrained system},
  author       = {Das, Arnab and Chakrabarti, Bikas K. and Stinchcombe, Robin B.},
  year         = {2005},
  month        = {Aug},
  journal      = {Phys. Rev. E},
  publisher    = {American Physical Society},
  volume       = {72},
  pages        = {026701},
  doi          = {10.1103/PhysRevE.72.026701},
  url          = {https://link.aps.org/doi/10.1103/PhysRevE.72.026701},
  issue        = {2},
}

@misc{Dowling2024Magic,
  title        = {{Magic Resources of the Heisenberg Picture}},
  author       = {Neil Dowling and Pavel Kos and Xhek Turkeshi},
  year         = {2025},
  doi          = {10.48550/arXiv.2408.16047},
  url          = {https://doi.org/10.48550/arXiv.2408.16047},
  eprint       = {2408.16047},
  archiveprefix = {arXiv}
}

@article{Dragoi2025,
  title        = {Approximate quadratization of high-order Hamiltonians for combinatorial quantum optimization},
  author       = {Dr\u{a}goi, Sabina and Baiardi, Alberto and Egger, Daniel J.},
  journal      = {Phys. Rev. Res.},
  volume       = {8},
  issue        = {2},
  pages        = {023159},
  year         = {2026},
  month        = {May},
  publisher    = {American Physical Society},
  doi          = {10.1103/9rc4-vj11},
  url          = {https://link.aps.org/doi/10.1103/9rc4-vj11}
}

@misc{durr1996quantum,
  title        = {{A Quantum Algorithm for Finding the Minimum}},
  author       = {Dürr, Christoph and Høyer, Peter},
  year         = {1996},
  url          = {https://arxiv.org/abs/quant-ph/9607014},
  eprint       = {quant-ph/9607014},
  archiveprefix = {arXiv}
}

@article{Egger2021warmstartingquantum,
  title        = {Warm-starting quantum optimization},
  author       = {Egger, Daniel J. and Mare{\v{c}}ek, Jakub and Woerner, Stefan},
  year         = {2021},
  month        = jun,
  journal      = {{Quantum}},
  publisher    = {{Verein zur F{\"{o}}rderung des Open Access Publizierens in den Quantenwissenschaften}},
  volume       = {5},
  pages        = {479},
  doi          = {10.22331/q-2021-06-17-479},
  issn         = {2521-327X},
  url          = {https://doi.org/10.22331/q-2021-06-17-479}
}

@misc{eichenseher2025pattern,
  title        = {Going off Pattern? {QAOA} Parameter Heuristics and Potentials of Parsimony},
  author       = {Vincent Eichenseher and Maja Franz and Christian Wolff and Wolfgang Mauerer},
  year         = {2026},
  url          = {https://arxiv.org/abs/2510.08153},
  eprint       = {2510.08153},
  archiveprefix = {arXiv}
}

@inproceedings{ester1996density,
  title        = {A density-based algorithm for discovering clusters in large spatial databases with noise},
  author       = {Ester, Martin and Kriegel, Hans-Peter and Sander, J{\"o}rg and Xu, Xiaowei},
  year         = {1996},
  booktitle    = {{KDD'96}},
  pages        = {226--231},
  url          = {https://dl.acm.org/doi/10.5555/3001460.3001507}
}

@misc{Farhi2014,
  title        = {A Quantum Approximate Optimization Algorithm},
  author       = {Farhi, Edward and Goldstone, Jeffrey and Gutmann, Sam},
  year         = {2014},
  month        = {11},
  journal      = {arXiv preprint arXiv:1411.4028},
  url          = {https://arxiv.org/abs/1411.4028},
  eprint       = {1411.4028},
  archiveprefix = {arXiv}
}

@inproceedings{Galda2021,
  title        = {Transferability of optimal {QAOA} parameters between random graphs},
  author       = {Galda, Alexey and Liu, Xiaoyuan and Lykov, Danylo and Alexeev, Yuri and Safro, Ilya},
  year         = {2021},
  booktitle    = {2021 {IEEE} International Conference on Quantum Computing and Engineering ({QCE})},
  pages        = {171--180},
  doi          = {10.1109/QCE52317.2021.00034}
}

@article{galda2023similarity,
  title        = {Similarity-based parameter transferability in the quantum approximate optimization algorithm},
  author       = {Galda, Alexey and Gupta, Eesh and Falla, Jose and Liu, Xiaoyuan and Lykov, Danylo and Alexeev, Yuri and Safro, Ilya},
  year         = {2023},
  journal      = {Front. Quantum Sci. Technol.},
  publisher    = {Frontiers Media SA},
  volume       = {2},
  pages        = {1200975},
  url          = {https://www.frontiersin.org/journals/quantum-science-and-technology/articles/10.3389/frqst.2023.1200975/full},
  doi          = {10.3389/frqst.2023.1200975},
}

@inproceedings{Golden_2023,
  title        = {{JuliQAOA}: Fast, Flexible {QAOA} Simulation},
  author       = {Golden, John and Baertschi, Andreas and O'Malley, Dan and Pelofske, Elijah and Eidenbenz, Stephan},
  year         = {2023},
  month        = nov,
  booktitle    = {Proceedings of the {SC’23} Workshops of the International Conference on High Performance Computing, Network, Storage, and Analysis},
  publisher    = {ACM},
  series       = {SC-W 2023},
  pages        = {1454--1459},
  doi          = {10.1145/3624062.3624220},
  url          = {http://dx.doi.org/10.1145/3624062.3624220},
}

@inproceedings{Golden2023b,
  title        = {Numerical Evidence for Exponential Speed-Up of {QAOA} over Unstructured Search for Approximate Constrained Optimization},
  author       = {Golden, John and B{\"a}rtschi, Andreas and O'Malley, Daniel and Eidenbenz, Stephan},
  year         = {2023},
  month        = sep,
  booktitle    = {2023 {IEEE} International Conference on Quantum Computing and Engineering ({QCE})},
  organization = {IEEE},
  pages        = {496--505},
  publisher    = {IEEE},
  volume       = {1},
  doi          = {10.1109/qce57702.2023.00063},
}

@article{Gray2018,
  title        = {quimb: A python package for quantum information and many-body calculations},
  author       = {Gray, Johnnie},
  year         = {2018},
  journal      = {J. Open Source Softw.},
  publisher    = {The Open Journal},
  volume       = {3},
  number       = {29},
  pages        = {819},
  doi          = {10.21105/joss.00819},
  url          = {https://doi.org/10.21105/joss.00819}
}

@inproceedings{HPC:ASU23,
  title        = {{The Sol Supercomputer at Arizona State University}},
  author       = {Jennewein, Douglas M. and Lee, Johnathan and Kurtz, Chris and Dizon, Will and Shaeffer, Ian and Chapman, Alan and Chiquete, Alejandro and Burks, Josh and Carlson, Amber and Mason, Natalie and others},
  year         = {2023},
  month        = {Jul},
  booktitle    = {Practice and Experience in Advanced Research Computing},
  location     = {Portland, OR, USA},
  publisher    = {Association for Computing Machinery},
  address      = {New York, NY, USA},
  series       = {PEARC '23},
  pages        = {296--301},
  doi          = {10.1145/3569951.3597573},
  isbn         = {9781450399852},
}

@article{JuwelsClusterBooster,
  title        = {{JUWELS Cluster and Booster: Exascale Pathfinder with Modular Supercomputing Architecture at Juelich Supercomputing Centre}},
  author       = {Alvarez, Damian},
  year         = {2021},
  journal      = {J. Large-Scale Res. Facil.},
  volume       = {7},
  pages        = {A183},
  doi          = {10.17815/jlsrf-7-183},
  issn         = {2364-091X},
  url          = {http://dx.doi.org/10.17815/jlsrf-7-183},
}

@article{Katial2025,
  title        = {On the Instance Dependence of Parameter Initialization for the Quantum Approximate Optimization Algorithm: Insights via Instance Space Analysis},
  author       = {Katial, Vivek and Smith-Miles, Kate and Hill, Charles and Hollenberg, Lloyd},
  year         = {2025},
  journal      = {INFORMS J. Comput.},
  volume       = {37},
  number       = {1},
  pages        = {146--171},
  doi          = {10.1287/ijoc.2024.0564}
}

@article{Khairy2020,
  title        = {Learning to Optimize Variational Quantum Circuits to Solve Combinatorial Problems},
  author       = {Khairy, Sami and Shaydulin, Ruslan and Cincio, Lukasz and Alexeev, Yuri and Balaprakash, Prasanna},
  year         = {2020},
  month        = {Apr.},
  journal      = {Proceedings of the AAAI Conference on Artificial Intelligence},
  volume       = {34},
  number       = {03},
  pages        = {2367--2375},
  doi          = {10.1609/aaai.v34i03.5616},
  url          = {https://ojs.aaai.org/index.php/AAAI/article/view/5616}
}

@article{Kim2023,
  title        = {Evidence for the utility of quantum computing before fault tolerance},
  author       = {Kim, Youngseok and Eddins, Andrew and Anand, Sajant and Wei, Ken Xuan and van den Berg, Ewout and Rosenblatt, Sami and Nayfeh, Hasan and Wu, Yantao and Zaletel, Michael and Temme, Kristan and Kandala, Abhinav},
  year         = {2023},
  month        = {Jun},
  day          = {01},
  journal      = {Nature},
  volume       = {618},
  number       = {7965},
  pages        = {500--505},
  doi          = {10.1038/s41586-023-06096-3},
  issn         = {1476-4687}
}

@misc{koch2025quantum,
  title        = {Quantum Optimization Benchmark Library--The Intractable Decathlon},
  author       = {Koch, Thorsten and Neira, David E Bernal and Chen, Ying and Cortiana, Giorgio and Egger, Daniel J and Heese, Raoul and Hegade, Narendra N and Cadavid, Alejandro Gomez and Huang, Rhea and Itoko, Toshinari and others},
  year         = {2025},
  doi          = {10.48550/arXiv.2504.03832},
  eprint       = {2504.03832},
  archiveprefix = {arXiv}
}

@misc{korpas2025undecidable,
  title        = {Undecidable problems associated with variational quantum algorithms},
  author       = {Korpas, Georgios and Kungurtsev, Vyacheslav and Mare{\v{c}}ek, Jakub},
  year         = {2025},
  url          = {https://arxiv.org/abs/2503.23723},
  eprint       = {2503.23723},
  archiveprefix = {arXiv}
}

@article{Kotil_2025,
  title        = {{Quantum approximate multi-objective optimization}},
  author       = {Kotil, Ayse and Pelofske, Elijah and Riedmüller, Stephanie and Egger, Daniel J. and Eidenbenz, Stephan and Koch, Thorsten and Woerner, Stefan},
  year         = {2025},
  month        = oct,
  journal      = {Nat. Comput. Sci.},
  publisher    = {Springer Science and Business Media LLC},
  volume       = {5},
  number       = {12},
  pages        = {1168–1177},
  doi          = {10.1038/s43588-025-00873-y},
  issn         = {2662-8457},
  url          = {http://dx.doi.org/10.1038/s43588-025-00873-y}
}

@article{Krause2019,
  title        = {{JUWELS: Modular Tier-0/1 Supercomputer at the J{\"{u}}lich Supercomputing Centre}},
  author       = {Krause, Dorian},
  year         = {2019},
  journal      = {J. Large-Scale Res. Facil.},
  volume       = {5},
  pages        = {A135},
  doi          = {10.17815/jlsrf-5-171},
}

@article{Mezard2005,
  title = {Clustering of Solutions in the Random Satisfiability Problem},
  author = {M\'ezard, M. and Mora, T. and Zecchina, R.},
  journal = {Phys. Rev. Lett.},
  volume = {94},
  issue = {19},
  pages = {197205},
  year = {2005},
  month = {May},
  publisher = {American Physical Society},
  doi = {10.1103/PhysRevLett.94.197205},
  url = {https://link.aps.org/doi/10.1103/PhysRevLett.94.197205}
}

@article{Achlioptas2011,
author = {Achlioptas, Dimitris and Coja-Oghlan, Amin and Ricci-Tersenghi, Federico},
title = {On the solution-space geometry of random constraint satisfaction problems},
journal = {Random Structures \& Algorithms},
volume = {38},
number = {3},
pages = {251-268},
doi = {https://doi.org/10.1002/rsa.20323},
year = {2011}
}

@inproceedings{Gamarnik2014,
author = {Gamarnik, David and Sudan, Madhu},
title = {Limits of local algorithms over sparse random graphs},
year = {2014},
isbn = {9781450326988},
publisher = {Association for Computing Machinery},
address = {New York, NY, USA},
doi = {10.1145/2554797.2554831},
booktitle = {Proceedings of the 5th Conference on Innovations in Theoretical Computer Science},
pages = {369–376},
location = {Princeton, New Jersey, USA},
series = {ITCS '14}
}

@article{Gamarnik2021,
author = {David Gamarnik },
title = {The overlap gap property: A topological barrier to optimizing over random structures},
journal = {Proc. Natl. Acad. Sci.},
volume = {118},
number = {41},
pages = {e2108492118},
year = {2021},
doi = {10.1073/pnas.2108492118},
}

@article{Gamarnik2019,
author = {Wei-Kuo Chen and David Gamarnik and Dmitry Panchenko and Mustazee Rahman},
title = {{Suboptimality of local algorithms for a class of max-cut problems}},
volume = {47},
journal = {Ann. Probab.},
number = {3},
publisher = {Institute of Mathematical Statistics},
pages = {1587 -- 1618},
year = {2019},
doi = {10.1214/18-AOP1291},
URL = {https://doi.org/10.1214/18-AOP1291}
}

@article{kungurtsev2024iteration,
  doi = {10.22331/q-2024-10-10-1495},
  url = {https://doi.org/10.22331/q-2024-10-10-1495},
  title = {Iteration {C}omplexity of {V}ariational {Q}uantum {A}lgorithms},
  author = {Kungurtsev, Vyacheslav and Korpas, Georgios and Marecek, Jakub and Zhu, Elton Yechao},
  journal = {{Quantum}},
  issn = {2521-327X},
  publisher = {{Verein zur F{\"{o}}rderung des Open Access Publizierens in den Quantenwissenschaften}},
  volume = {8},
  pages = {1495},
  month = oct,
  year = {2024}
}

@article{Li2024FALQONSim,
  title        = {Simulation of a feedback-based algorithm for quantum optimization for a realistic neutral-atom system with an optimized small-angle controlled-phase gate},
  author       = {Li, S. X. and Mu, W. L. and You, J. B. and Shao, X. Q.},
  year         = {2024},
  journal      = {Phys. Rev. A},
  volume       = {109},
  pages        = {062603},
  doi          = {10.1103/PhysRevA.109.062603},
  url          = {https://link.aps.org/doi/10.1103/PhysRevA.109.062603}
}

@inproceedings{Liang2024,
  title        = {Invited: Graph Learning for Parameter Prediction of Quantum Approximate Optimization Algorithm},
  author       = {Liang, Zhiding and Liu, Gang and Liu, Zheyuan and Cheng, Jinglei and Hao, Tianyi and Liu, Kecheng and Ren, Hang and Song, Zhixin and Liu, Ji and Ye, Fanny and Shi, Yiyu},
  year         = {2024},
  booktitle    = {Proceedings of the 61st {ACM}/{IEEE} Design Automation Conference},
  location     = {San Francisco, CA, USA},
  publisher    = {Association for Computing Machinery},
  address      = {New York, NY, USA},
  series       = {DAC '24},
  doi          = {10.1145/3649329.3663523},
  isbn         = {9798400706011}
}

@misc{Luchnikov2024_BP-Annealing,
  title        = {{Large-scale quantum annealing simulation with tensor networks and belief propagation}},
  author       = {Ilia A. Luchnikov and Egor S. Tiunov and Tobias Haug and Leandro Aolita},
  year         = {2024},
  url          = {https://arxiv.org/abs/2409.12240},
  eprint       = {2409.12240},
  archiveprefix = {arXiv}
}

@article{Lyngfelt2025,
  title        = {{Symmetry-informed transferability of optimal parameters in the Quantum Approximate Optimization Algorithm}},
  author       = {Lyngfelt, Isak and Garc\'{\i}a-\'Alvarez, Laura},
  year         = {2025},
  month        = {Feb},
  journal      = {Phys. Rev. A},
  publisher    = {American Physical Society},
  volume       = {111},
  pages        = {022418},
  doi          = {10.1103/PhysRevA.111.022418},
  url          = {https://link.aps.org/doi/10.1103/PhysRevA.111.022418},
  issue        = {2},
}

@article{Magann2022FALQON,
  title        = {Feedback-Based Quantum Optimization},
  author       = {Magann, Alicia B. and Rudinger, Kenneth M. and Grace, Matthew D. and Sarovar, Mohan},
  year         = {2022},
  journal      = {Phys. Rev. Lett.},
  volume       = {129},
  pages        = {250502},
  doi          = {10.1103/PhysRevLett.129.250502},
  url          = {https://link.aps.org/doi/10.1103/PhysRevLett.129.250502}
}

@article{Matsuo2023,
  title        = {A {SAT} Approach to the Initial Mapping Problem in {SWAP} Gate Insertion for Commuting Gates},
  author       = {Atsushi Matsuo and Shigeru Yamashita and Daniel J. Egger},
  year         = {2023},
  month        = {November},
  journal      = {IEICE TRANSACTIONS on Fundamentals},
  volume       = {E106-A},
  number       = {11},
  pages        = {1424--1431},
  doi          = {10.1587/transfun.2022EAP1159},
  issn         = {1745-1337},
}

@misc{Mcdowall2026,
  title        = {Spectral Gap Informed Parameter Schedule for {QAOA}},
  author       = {Kieran McDowall and Konstantinos Georgopoulos and Petros Wallden},
  year         = {2026},
  url          = {https://arxiv.org/abs/2604.24580},
  eprint       = {2604.24580},
  archiveprefix = {arXiv}
}

@misc{Medina2024,
  title        = {{A Recursive Lower Bound on the Energy Improvement of the Quantum Approximate Optimization Algorithm}},
  author       = {Raimel A. Medina and Maksym Serbyn},
  year         = {2024},
  url          = {https://arxiv.org/abs/2405.10125},
  eprint       = {2405.10125},
  archiveprefix = {arXiv}
}

@misc{meng2024parameter,
  title        = {Parameter Generation of Quantum Approximate Optimization Algorithm with Diffusion Model},
  author       = {Meng, Fanxu and Zhou, Xiangzhen},
  year         = {2024},
  url          = {https://arxiv.org/abs/2407.12242v3},
  eprint       = {2407.12242v3},
  archiveprefix = {arXiv}
}

@article{Mertens_1996,
  title        = {Exhaustive search for low-autocorrelation binary sequences},
  author       = {Mertens, S},
  year         = {1996},
  month        = sep,
  journal      = {J. Phys. A: Math. Gen.},
  publisher    = {IOP Publishing},
  volume       = {29},
  number       = {18},
  pages        = {L473–L481},
  doi          = {10.1088/0305-4470/29/18/005},
  issn         = {1361-6447},
  url          = {http://dx.doi.org/10.1088/0305-4470/29/18/005}
}

@article{Montanezbarrera2024,
  title        = {Toward a linear-ramp {QAOA} protocol: evidence of a scaling advantage in solving some combinatorial optimization problems},
  author       = {Monta\~{n}ez-Barrera, J. A. and Michielsen, Kristel},
  year         = {2025},
  month        = aug,
  journal      = {npj Quantum Inf.},
  publisher    = {Springer Science and Business Media LLC},
  volume       = {11},
  number       = {1},
  pages        = {131},
  doi          = {10.1038/s41534-025-01082-1},
  issn         = {2056-6387},
  url          = {http://dx.doi.org/10.1038/s41534-025-01082-1}
}

@article{Montanezbarrera2024Transfer,
  title        = {Transfer learning of optimal {QAOA} parameters in combinatorial optimization},
  author       = {Monta{\~{n}}ez-Barrera, J. A. and Willsch, Dennis and Michielsen, Kristel},
  year         = {2025},
  month        = {May},
  day          = {09},
  journal      = {Quantum Inf. Process.},
  volume       = {24},
  number       = {5},
  pages        = {129},
  doi          = {10.1007/s11128-025-04743-4},
  issn         = {1573-1332}
}

@article{Moussa2022,
  title        = {Unsupervised strategies for identifying optimal parameters in Quantum Approximate Optimization Algorithm},
  author       = {Moussa, Charles and Wang, Hao and B{\"a}ck, Thomas and Dunjko, Vedran},
  year         = {2022},
  month        = {May},
  day          = {06},
  journal      = {EPJ Quantum Technol.},
  volume       = {9},
  number       = {1},
  pages        = {11},
  doi          = {10.1140/epjqt/s40507-022-00131-4},
  issn         = {2196-0763}
}

@article{Mukherjee2015,
  title        = {Multivariable optimization: Quantum annealing and computation},
  author       = {Mukherjee, S. and Chakrabarti, B. K.},
  year         = {2015},
  month        = {Feb},
  day          = {01},
  journal      = {Eur. Phys. J. Spec. Top.},
  volume       = {224},
  number       = {1},
  pages        = {17--24},
  doi          = {10.1140/epjst/e2015-02339-y},
  issn         = {1951-6401}
}

@misc{Neira2024,
  title        = {Benchmarking the Operation of Quantum Heuristics and Ising Machines: Scoring Parameter Setting Strategies on Optimization Applications},
  author       = {David E. Bernal Neira and Robin Brown and Pratik Sathe and Filip Wudarski and Marco Pavone and Eleanor G. Rieffel and Davide Venturelli},
  year         = {2024},
  url          = {https://arxiv.org/abs/2402.10255},
  eprint       = {2402.10255},
  archiveprefix = {arXiv}
}

@article{Oleary2025,
  title        = {Efficient online quantum circuit learning with no upfront training},
  author       = {O'Leary, Tom and Czarnik, Piotr and Pelofske, Elijah and Sornborger, Andrew T. and McKerns, Michael and Cincio, Lukasz},
  year         = {2025},
  month        = {Nov},
  day          = {20},
  journal      = {Commun. Phys.},
  volume       = {8},
  number       = {1},
  pages        = {514},
  doi          = {10.1038/s42005-025-02423-4},
  issn         = {2399-3650},
  url          = {https://doi.org/10.1038/s42005-025-02423-4}
}

@article{Ozaeta_2022,
  title        = {Expectation values from the single-layer quantum approximate optimization algorithm on Ising problems},
  author       = {Ozaeta, Asier and van Dam, Wim and McMahon, Peter L},
  year         = {2022},
  month        = sep,
  journal      = {Quantum Sci. Technol.},
  publisher    = {IOP Publishing},
  volume       = {7},
  number       = {4},
  pages        = {045036},
  doi          = {10.1088/2058-9565/ac9013},
  issn         = {2058-9565},
  url          = {http://dx.doi.org/10.1088/2058-9565/ac9013}
}

@article{Packebusch_2016,
  title        = {Low autocorrelation binary sequences},
  author       = {Packebusch, Tom and Mertens, Stephan},
  year         = {2016},
  month        = mar,
  journal      = {J. Phys. A: Math. Gen.},
  publisher    = {IOP Publishing},
  volume       = {49},
  number       = {16},
  pages        = {165001},
  doi          = {10.1088/1751-8113/49/16/165001},
  issn         = {1751-8121},
  url          = {http://dx.doi.org/10.1088/1751-8113/49/16/165001}
}

@misc{Parry2025,
  title        = {{QAOA-PCA}: Enhancing Efficiency in the Quantum Approximate Optimization Algorithm via Principal Component Analysis},
  author       = {Owain Parry and Phil McMinn},
  year         = {2025},
  url          = {https://arxiv.org/abs/2504.16755},
  eprint       = {2504.16755},
  archiveprefix = {arXiv}
}

@article{Patel2024,
  title        = {Reinforcement learning assisted recursive {QAOA}},
  author       = {Patel, Yash J and Jerbi, Sofiene and B{\"a}ck, Thomas and Dunjko, Vedran},
  year         = {2024},
  journal      = {EPJ Quantum Technol.},
  volume       = {11},
  number       = {1},
  pages        = {6},
  doi          = {https://doi.org/10.1140/epjqt/s40507-023-00214-w}
}

@article{pedregosa2011scikit,
  title        = {Scikit-learn: Machine Learning in {Python}},
  author       = {Fabian Pedregosa and Ga{\"e}l Varoquaux and Alexandre Gramfort and Vincent Michel and Bertrand Thirion and Olivier Grisel and Mathieu Blondel and Peter Prettenhofer and Ron Weiss and Vincent Dubourg and Jake VanderPlas},
  year         = {2011},
  journal      = {J. Mach. Learn. Res.},
  volume       = {12},
  pages        = {2825--2830},
  url          = {https://www.jmlr.org/papers/v12/pedregosa11a.html}
}

@article{Pelofske_2024_QAOA_scaling,
  title        = {{Scaling whole-chip QAOA for higher-order ising spin glass models on heavy-hex graphs}},
  author       = {Pelofske, Elijah and B\"{a}rtschi, Andreas and Cincio, Lukasz and Golden, John and Eidenbenz, Stephan},
  year         = {2024},
  month        = nov,
  journal      = {npj Quantum Inf.},
  publisher    = {Springer Science and Business Media LLC},
  volume       = {10},
  number       = {1},
  pages        = {109},
  doi          = {10.1038/s41534-024-00906-w},
  issn         = {2056-6387},
  url          = {http://dx.doi.org/10.1038/s41534-024-00906-w}
}

@article{pelofske2026evaluatinglimitsqaoaparameter,
  title        = {{Evaluating the limits of Quantum Approximate Optimization Algorithm parameter transfer at high rounds on sparse Ising models with geometrically local cubic terms}},
  author       = {Pelofske, Elijah and Rams, Marek M. and B\"{a}rtschi, Andreas and Czarnik, Piotr and Braccia, Paolo and Cincio, Lukasz and Eidenbenz, Stephan},
  year         = {2026},
  month        = {Mar},
  journal      = {Phys. Rev. Res.},
  publisher    = {American Physical Society},
  volume       = {8},
  number       = {2},
  pages        = {023023},
  doi          = {10.1103/p2lg-z4kn},
  url          = {https://journals.aps.org/prresearch/abstract/10.1103/p2lg-z4kn},
}

@misc{Qiskit,
  title        = {Quantum computing with {Q}iskit},
  author       = {Javadi-Abhari, Ali and Treinish, Matthew and Krsulich, Kevin and Wood, Christopher J. and Lishman, Jake and Gacon, Julien and Martiel, Simon and Nation, Paul D. and Bishop, Lev S. and Cross, Andrew W. and Johnson, Blake R. and Gambetta, Jay M.},
  year         = {2024},
  doi          = {10.48550/arXiv.2405.08810},
  eprint       = {2405.08810},
  archiveprefix = {arXiv}
}

@inproceedings{Rahman2024FALQONQCBO,
  title        = {Feedback-based quantum algorithm for constrained optimization problems},
  author       = {Abdul Rahman, Salahuddin and Karabacak, {\"O}zkan and Wisniewski, Rafal},
  year         = {2024},
  booktitle    = {International Conference on Parallel Processing and Applied Mathematics},
  pages        = {277--289},
  doi          = {https://doi.org/10.1007/978-3-031-85700-3_20},
  organization = {Springer}
}

@article{rall2019simulation,
  title        = {Simulation of qubit quantum circuits via Pauli propagation},
  author       = {Rall, Patrick and Liang, Daniel and Cook, Jeremy and Kretschmer, William},
  year         = {2019},
  month        = {Jun},
  journal      = {Phys. Rev. A},
  publisher    = {American Physical Society},
  volume       = {99},
  pages        = {062337},
  doi          = {10.1103/PhysRevA.99.062337},
  url          = {https://link.aps.org/doi/10.1103/PhysRevA.99.062337},
  issue        = {6},
}

@misc{Rudolph2025_BP-QuantumCircuit,
  title        = {{Simulating and Sampling from Quantum Circuits with 2D Tensor Networks}},
  author       = {Manuel S. Rudolph and Joseph Tindall},
  year         = {2025},
  journal      = {arXiv},
  url          = {https://arxiv.org/abs/2507.11424},
  eprint       = {2507.11424},
  archiveprefix = {arXiv}
}

@misc{rudolph2025pauli,
  title        = {Pauli Propagation: A Computational Framework for Simulating Quantum Systems},
  author       = {Manuel S. Rudolph and Tyson Jones and Yanting Teng and Armando Angrisani and Zoë Holmes},
  year         = {2025},
  url          = {https://arxiv.org/abs/2505.21606},
  eprint       = {2505.21606},
  archiveprefix = {arXiv}
}

@article{sack2021quantum,
  title        = {Quantum annealing initialization of the quantum approximate optimization algorithm},
  author       = {Sack, Stefan H and Serbyn, Maksym},
  year         = {2021},
  journal      = {Quantum},
  publisher    = {Verein zur F{\"o}rderung des Open Access Publizierens in den Quantenwissenschaften},
  volume       = {5},
  pages        = {491},
  doi          = {10.22331/q-2021-07-01-491}
}

@article{Sack2023,
  title        = {Recursive greedy initialization of the quantum approximate optimization algorithm with guaranteed improvement},
  author       = {Sack, Stefan H. and Medina, Raimel A. and Kueng, Richard and Serbyn, Maksym},
  year         = {2023},
  month        = {Jun},
  journal      = {Phys. Rev. A},
  publisher    = {American Physical Society},
  volume       = {107},
  pages        = {062404},
  doi          = {10.1103/PhysRevA.107.062404},
  url          = {https://link.aps.org/doi/10.1103/PhysRevA.107.062404},
  issue        = {6},
}

@article{Sack2024,
  title        = {Large-scale quantum approximate optimization on nonplanar graphs with machine learning noise mitigation},
  author       = {Sack, Stefan H. and Egger, Daniel J.},
  year         = {2024},
  month        = {Mar},
  journal      = {Phys. Rev. Res.},
  publisher    = {American Physical Society},
  volume       = {6},
  pages        = {013223},
  doi          = {10.1103/PhysRevResearch.6.013223},
  issue        = {1},
}

@misc{Sakai2025,
  title        = {{Transferring linearly fixed QAOA angles: performance and real device results}},
  author       = {Ryo Sakai and Hiromichi Matsuyama and Wai-Hong Tam and Yu Yamashiro},
  year         = {2025},
  url          = {https://arxiv.org/abs/2504.12632},
  eprint       = {2504.12632},
  archiveprefix = {arXiv}
}

@article{Schollwoeck2011_DMRG,
  title        = {The density-matrix renormalization group in the age of matrix product states},
  author       = {Schollw\"{o}ck, Ulrich},
  year         = {2011},
  journal      = {Ann. Phys.},
  volume       = {326},
  number       = {1},
  pages        = {96–192},
  doi          = {10.1016/j.aop.2010.09.012},
  url          = {http://dx.doi.org/10.1016/j.aop.2010.09.012}
}

@misc{sciorilli2026competitivenisqqubitefficientsolver,
  title        = {{A competitive NISQ and qubit-efficient solver for the LABS problem}},
  author       = {Marco Sciorilli and Giancarlo Camilo and Thiago O. Maciel and Askery Canabarro and Lucas Borges and Leandro Aolita},
  year         = {2026},
  url          = {https://arxiv.org/abs/2506.17391},
  eprint       = {2506.17391},
  archiveprefix = {arXiv}
}

@inproceedings{Shaydulin2019,
  title        = {{Multistart Methods for Quantum Approximate optimization}},
  author       = {Shaydulin, Ruslan and Safro, Ilya and Larson, Jeffrey},
  year         = {2019},
  booktitle    = {2019 {IEEE} High Performance Extreme Computing Conference ({HPEC})},
  pages        = {1--8},
  doi          = {10.1109/HPEC.2019.8916288}
}

@article{Shaydulin2023,
  title        = {Parameter Transfer for Quantum Approximate Optimization of Weighted {MaxCut}},
  author       = {Ruslan Shaydulin and Phillip C. Lotshaw and Jeffrey Larson and James Ostrowski and Travis S. Humble},
  year         = {2023},
  month        = apr,
  journal      = {ACM Trans. Quantum Comput.},
  publisher    = {Association for Computing Machinery},
  address      = {New York, NY, USA},
  volume       = {4},
  number       = {3},
  pages        = {1--15},
  doi          = {10.1145/3584706}
}

@article{Shaydulin2024,
  title        = {Evidence of scaling advantage for the quantum approximate optimization algorithm on a classically intractable problem},
  author       = {Ruslan Shaydulin  and Changhao Li  and Shouvanik Chakrabarti  and Matthew DeCross  and Dylan Herman  and Niraj Kumar  and Jeffrey Larson  and Danylo Lykov  and Pierre Minssen  and Yue Sun  and Yuri Alexeev  and Joan M. Dreiling  and John P. Gaebler  and Thomas M. Gatterman  and Justin A. Gerber  and Kevin Gilmore  and Dan Gresh  and Nathan Hewitt  and Chandler V. Horst  and Shaohan Hu  and Jacob Johansen  and Mitchell Matheny  and Tanner Mengle  and Michael Mills  and Steven A. Moses  and Brian Neyenhuis  and Peter Siegfried  and Romina Yalovetzky  and Marco Pistoia },
  year         = {2024},
  journal      = {Sci. Adv.},
  volume       = {10},
  number       = {22},
  pages        = {eadm6761},
  doi          = {10.1126/sciadv.adm6761}
}

@article{Stoudenmire2020_Limits-QC,
  title        = {What Limits the Simulation of Quantum Computers?},
  author       = {Zhou, Yiqing and Stoudenmire, E. Miles and Waintal, Xavier},
  year         = {2020},
  journal      = {Phys. Rev. X},
  volume       = {10},
  pages        = {041038},
  doi          = {10.1103/PhysRevX.10.041038},
  url          = {https://link.aps.org/doi/10.1103/PhysRevX.10.041038},
  issue        = {4}
}

@article{Sud2024,
  title        = {Parameter-setting heuristic for the quantum alternating operator ansatz},
  author       = {Sud, James and Hadfield, Stuart and Rieffel, Eleanor and Tubman, Norm and Hogg, Tad},
  year         = {2024},
  month        = {May},
  journal      = {Phys. Rev. Res.},
  publisher    = {American Physical Society},
  volume       = {6},
  pages        = {023171},
  doi          = {10.1103/PhysRevResearch.6.023171},
  url          = {https://link.aps.org/doi/10.1103/PhysRevResearch.6.023171},
  issue        = {2},
}

@article{Sureshbabu2024,
  title        = {Parameter Setting in Quantum Approximate Optimization of Weighted Problems},
  author       = {Shree Hari Sureshbabu and Dylan Herman and Ruslan Shaydulin and Joao Basso and Shouvanik Chakrabarti and Yue Sun and Marco Pistoia},
  year         = {2024},
  month        = jan,
  journal      = {Quantum},
  publisher    = {{Verein zur F{\"o}rderung des Open Access Publizierens in den Quantenwissenschaften}},
  volume       = {8},
  pages        = {1231},
  doi          = {10.22331/q-2024-01-18-1231},
  issn         = {2521-327X}
}

@article{Temme2017,
  title        = {Error Mitigation for Short-Depth Quantum Circuits},
  author       = {Temme, Kristan and Bravyi, Sergey and Gambetta, Jay M.},
  year         = {2017},
  month        = {Nov},
  journal      = {Phys. Rev. Lett.},
  publisher    = {American Physical Society},
  volume       = {119},
  pages        = {180509},
  doi          = {10.1103/PhysRevLett.119.180509},
  issue        = {18},
}

@article{Tindall2023_BP-TN,
  title        = {{Gauging tensor networks with belief propagation}},
  author       = {Joseph Tindall and Matt Fishman},
  year         = {2023},
  journal      = {SciPost Phys.},
  volume       = {15},
  pages        = {222},
  doi          = {10.21468/SciPostPhys.15.6.222},
  url          = {https://scipost.org/10.21468/SciPostPhys.15.6.222}
}

@misc{training_pipeline,
  title        = {{QAOA Training Pipeline}},
  year         = {2026},
  url          = {https://github.com/qiskit-community/qaoa_training_pipeline}
}

@misc{Tyagin2025,
  title        = {{QAOA-GPT}: Efficient Generation of Adaptive and Regular Quantum Approximate Optimization Algorithm Circuits},
  author       = {Ilya Tyagin and Marwa H. Farag and Kyle Sherbert and Karunya Shirali and Yuri Alexeev and Ilya Safro},
  year         = {2025},
  url          = {https://arxiv.org/abs/2504.16350},
  eprint       = {2504.16350},
  archiveprefix = {arXiv}
}

@article{vanApeldoorn2020quantumsdpsolvers,
  title        = {Quantum {SDP}-{S}olvers: {B}etter upper and lower bounds},
  author       = {van Apeldoorn, Joran and Gily{\'{e}}n, Andr{\'{a}}s and Gribling, Sander and de Wolf, Ronald},
  year         = {2020},
  month        = feb,
  journal      = {{Quantum}},
  publisher    = {{Verein zur F{\"{o}}rderung des Open Access Publizierens in den Quantenwissenschaften}},
  volume       = {4},
  pages        = {230},
  doi          = {10.22331/q-2020-02-14-230},
  issn         = {2521-327X},
  url          = {https://doi.org/10.22331/q-2020-02-14-230}
}

@misc{venturelli2024transfer,
  title        = {{Investigating layer-selective transfer learning of QAOA parameters for Max-Cut problem}},
  author       = {Venturelli, Francesco Aldo and Das, Sreetama and Caruso, Filippo},
  year         = {2024},
  url          = {https://arxiv.org/abs/2412.21071},
  eprint       = {2412.21071},
  archiveprefix = {arXiv}
}

@article{Venuti2021PontryaginOpen,
  title        = {Optimal Control for Quantum Optimization of Closed and Open Systems},
  author       = {Venuti, Lorenzo Campos and D'Alessandro, Domenico and Lidar, Daniel A.},
  year         = {2021},
  journal      = {Phys. Rev. Applied},
  volume       = {16},
  pages        = {054023},
  doi          = {10.1103/PhysRevApplied.16.054023}
}

@article{Vidal2012_PerfectSampling,
  title        = {{Perfect sampling with unitary tensor networks}},
  author       = {Ferris, Andrew J. and Vidal, Guifre},
  year         = {2012},
  journal      = {Phys. Rev. B},
  volume       = {85},
  pages        = {165146},
  doi          = {10.1103/PhysRevB.85.165146},
  url          = {https://link.aps.org/doi/10.1103/PhysRevB.85.165146}
}

@misc{Vijendran2025,
  title        = {Near-Optimal Parameter Tuning of Level-1 {QAOA} for Ising Models},
  author       = {V Vijendran and Dax Enshan Koh and Eunok Bae and Hyukjoon Kwon and Ping Koy Lam and Syed M Assad},
  year         = {2025},
  url          = {https://arxiv.org/abs/2501.16419},
  eprint       = {2501.16419},
  archiveprefix = {arXiv}
}

@misc{Watanabe2026,
  title        = {Tensor network surrogate models for variational quantum computation},
  author       = {Ryo Watanabe and Dries Sels and Joseph Tindall},
  year         = {2026},
  url          = {https://arxiv.org/abs/2604.20180},
  eprint       = {2604.20180},
  archiveprefix = {arXiv}
}

@article{Weidenfeller2022,
  title        = {Scaling of the quantum approximate optimization algorithm on superconducting qubit based hardware},
  author       = {Weidenfeller, Johannes and Valor, Lucia C. and Gacon, Julien and Tornow, Caroline and Bello, Luciano and Woerner, Stefan and Egger, Daniel J.},
  year         = {2022},
  month        = dec,
  journal      = {{Quantum}},
  publisher    = {{Verein zur F{\"{o}}rderung des Open Access Publizierens in den Quantenwissenschaften}},
  volume       = {6},
  pages        = {870},
  doi          = {10.22331/q-2022-12-07-870},
  issn         = {2521-327X},
  url          = {https://doi.org/10.22331/q-2022-12-07-870}
}

@article{Willsch2022,
  title        = {{GPU-accelerated simulations of quantum annealing and the quantum approximate optimization algorithm}},
  author       = {Willsch, Dennis and Willsch, Madita and Jin, Fengping and Michielsen, Kristel and De Raedt, Hans},
  year         = {2022},
  month        = sep,
  journal      = {Comput. Phys. Commun.},
  publisher    = {Elsevier BV},
  volume       = {278},
  pages        = {108411},
  doi          = {10.1016/j.cpc.2022.108411},
  issn         = {0010-4655},
  url          = {http://dx.doi.org/10.1016/j.cpc.2022.108411}
}

@article{Wurtz2021,
  title        = {{Fixed-angle conjectures for the Quantum Approximate Optimization Algorithm on regular MaxCut graphs}},
  author       = {Wurtz, Jonathan and Lykov, Danylo},
  year         = {2021},
  month        = {Nov},
  journal      = {Phys. Rev. A},
  publisher    = {American Physical Society},
  volume       = {104},
  pages        = {052419},
  doi          = {10.1103/PhysRevA.104.052419},
  issue        = {5},
}

@article{Wurtz2021a,
  title        = {{MaxCut} quantum approximate optimization algorithm performance guarantees for $p>1$},
  author       = {Wurtz, Jonathan and Love, Peter},
  year         = {2021},
  month        = {Apr},
  journal      = {Phys. Rev. A},
  publisher    = {American Physical Society},
  volume       = {103},
  pages        = {042612},
  doi          = {10.1103/PhysRevA.103.042612},
  url          = {https://link.aps.org/doi/10.1103/PhysRevA.103.042612},
  issue        = {4},
}

@misc{WurtzGithub,
  title        = {Fixed angles {QAOA}},
  author       = {Wurtz, Jonathan and Lykov, Danylo},
  year         = {2021},
  note         = {[Online; accessed 27 October 2025]},
  howpublished = {\url{https://github.com/danlkv/fixed-angle-QAOA}}
}

@article{WurtzLove2022CDQAOA,
  title        = {Counterdiabaticity and the quantum approximate optimization algorithm},
  author       = {Wurtz, J. and Love, P. J.},
  year         = {2022},
  journal      = {Quantum},
  volume       = {6},
  pages        = {635},
  doi          = {10.22331/q-2022-01-27-635},
  url          = {https://quantum-journal.org/papers/q-2022-01-27-635/}
}

@article{Wybo2025,
  title        = {Missing {P}uzzle {P}ieces in the {P}erformance {L}andscape of the {Q}uantum {A}pproximate {O}ptimization {A}lgorithm},
  author       = {Wybo, Elisabeth and Leib, Martin},
  year         = {2025},
  month        = oct,
  journal      = {{Quantum}},
  publisher    = {{Verein zur F{\"{o}}rderung des Open Access Publizierens in den Quantenwissenschaften}},
  volume       = {9},
  pages        = {1892},
  doi          = {10.22331/q-2025-10-22-1892},
  issn         = {2521-327X},
  url          = {https://doi.org/10.22331/q-2025-10-22-1892}
}

@article{Xie2023,
  title        = {Quantum approximate optimization algorithm parameter prediction using a convolutional neural network},
  author       = {Xie, Ningyi and Lee, Xinwei and Cai, Dongsheng and Saito, Yoshiyuki and Asai, Nobuyoshi},
  year         = {2023},
  month        = {sep},
  journal      = {J. Phys. Conf. Ser.},
  publisher    = {IOP Publishing},
  volume       = {2595},
  number       = {1},
  pages        = {012001},
  doi          = {10.1088/1742-6596/2595/1/012001},
  url          = {https://dx.doi.org/10.1088/1742-6596/2595/1/012001}
}

@misc{xu2025qaoa,
  title        = {{QAOA} Parameter Transferability for Maximum Independent Set using Graph Attention Networks},
  author       = {Hanjing Xu and Xiaoyuan Liu and Alex Pothen and Ilya Safro},
  year         = {2025},
  url          = {https://arxiv.org/abs/2504.21135},
  eprint       = {2504.21135},
  archiveprefix = {arXiv}
}

@misc{zhang2025,
  title        = {{New Improvements in Solving Large LABS Instances Using Massively Parallelizable Memetic Tabu Search}},
  author       = {Zhiwei Zhang and Jiayu Shen and Niraj Kumar and Marco Pistoia},
  year         = {2025},
  url          = {https://arxiv.org/abs/2504.00987},
  eprint       = {2504.00987},
  archiveprefix = {arXiv}
}

@article{Zhou2020,
  title        = {Quantum Approximate Optimization Algorithm: Performance, Mechanism, and Implementation on Near-Term Devices},
  author       = {Leo Zhou and Sheng-Tao Wang and Soonwon Choi and Hannes Pichler and Mikhail D. Lukin},
  year         = {2020},
  month        = jun,
  journal      = {Phys. Rev. X},
  publisher    = {American Physical Society},
  volume       = {10},
  number       = {2},
  pages        = {021067},
  doi          = {10.1103/PhysRevX.10.021067},
  issn         = {2160-3308}
}

@article{zhu2022adaptive,
  title        = {Adaptive quantum approximate optimization algorithm for solving combinatorial problems on a quantum computer},
  author       = {Zhu, Linghua and Tang, Ho Lun and Barron, George S. and Calderon-Vargas, F. A. and Mayhall, Nicholas J. and Barnes, Edwin and Economou, Sophia E.},
  year         = {2022},
  month        = {Jul},
  journal      = {Phys. Rev. Res.},
  publisher    = {American Physical Society},
  volume       = {4},
  pages        = {033029},
  doi          = {10.1103/PhysRevResearch.4.033029},
  url          = {https://link.aps.org/doi/10.1103/PhysRevResearch.4.033029},
  issue        = {3},
}
\FloatBarrier

\end{document}